\documentclass[english]{article}
\usepackage[T1]{fontenc}
\usepackage[latin9]{inputenc}
\usepackage{geometry}
\usepackage{color}
\usepackage{graphicx}
\usepackage{verbatim}
\usepackage{amssymb}
\usepackage{epstopdf}
\usepackage{amsmath}
\usepackage{mathtools}
\usepackage{caption}
\newcommand{\comments}[1]{}   

\makeatletter

\usepackage{indentfirst}
\usepackage{enumerate}

\makeatother

\begin{document}

\title{Quenched dynamics of classical isolated systems:\\
the spherical spin model with two-body random interactions or \\
the Neumann integrable model}

\author{Leticia F. Cugliandolo$^1$, Gustavo S. Lozano$^2$, Nicol\'as Nessi$^2$, Marco Picco$^1$ and Alessandro Tartaglia$^1$\\
$^1$  Laboratoire  de  Physique  Th\'eorique  et  Hautes  Energies, UMR 7589, \\
Sorbonne Universit\'es et CNRS, 4 place Jussieu, 75252 Paris Cedex 05, France\\
$^2$ Departamento de F\'{\i}sica, \\
FCEYN Universidad de Buenos Aires
\& IFIBA CONICET,\\
Pabell\'on 1 Ciudad Universitaria, 1428 Buenos Aires, Argentina
}

\maketitle

\abstract{
We study the Hamiltonian dynamics of the spherical spin model with fully-connected two-body interactions drawn
from a zero-mean Gaussian probability distribution. In the statistical physics framework, the potential
energy is of the so-called $p=2$ spherical disordered kind, closely linked to the $O(N)$ scalar field theory.
Most importantly for our setting, the energy conserving dynamics are equivalent to the ones of the Neumann integrable system.
We take initial conditions from the Boltzmann equilibrium measure
at a temperature that can be above or below the static phase transition, typical of a disordered (paramagnetic)  or
of an ordered (disguised ferromagnetic) equilibrium phase.  We subsequently evolve the configurations with Newton dynamics
dictated by a different Hamiltonian, obtained from
an instantaneous global rescaling of the elements in the interaction random matrix. In the limit of infinitely many
degrees of freedom, $N\to\infty$, we identify three dynamical phases depending on the parameters that characterise the initial
state and the final Hamiltonian. We obtain the {\it global} dynamical observables (energy density, self correlation function, linear
response function, static susceptibility, etc.) with numerical and analytic methods and we show that, in most cases, they
are  out of thermal equilibrium. We note, however, that for shallow quenches from the condensed phase the dynamics  are close to (though not at)
thermal equilibrium {\it \`a la} Gibbs-Boltzmann.  Surprisingly enough, in the $N\to\infty$ limit and for a particular relation between parameters  the
global observables comply Gibbs-Boltzmann equilibrium. We next set the analysis of  the system with finite number of degrees of freedom
in terms of $N$ non-linearly coupled modes. These are the projections of the vector spin (or particle's position on the sphere) on
the eigenvectors of the interaction
matrix, the most relevant being those linked to the eigenvalues at the edge of the
spectrum. We argue that in a system with infinite size the modes decouple at long times.
We evaluate the mode temperatures and we relate them to the frequency-dependent effective
temperature measured with the fluctuation-dissipation relation in the frequency domain, similarly to what was recently proposed for quantum integrable cases.
Finally, we analyse the $N-1$ integrals of motion, notably, their scaling with $N$, and we use them to
show that the system is out of equilibrium in all phases, even for parameters that show an apparent Gibbs-Boltzmann behaviour of global observables. We elaborate on the
role played by these constants of motion in the post-quench dynamics and we
briefly discuss the possible description of the asymptotic dynamics in terms of a Generalised Gibbs Ensemble.
}

\newpage
\tableofcontents

\clearpage

\section{Introduction}
\label{sec:introduction}

In the past decade, atomic physics experiments have been able to test the global coherent quantum dynamics of interacting
systems. This achievement has boosted research on the dynamics and possible equilibration of many-body isolated systems~\cite{Bloch08}.
Some of the quantum instances realised in the laboratory are low dimensional and considered to be integrable. Therefore they are not
able to act as a bath on themselves and questions on how to describe their asymptotic dynamics pose naturally. With the
aim of describing their asymptotic states, the Generalised Gibbs Ensemble (GGE), an extension of the canonical Gibbs-Boltzmann
density operator that aims to include the effect of all relevant conserved charges, was proposed~\cite{Rigol07,Rigol08} (see, {\it e.g.},  the review
articles~\cite{Polkovnikov10,Pasquale-ed,Gogolin}).

Similar equilibration problems can arise in classical isolated systems. A first study of the dynamics of isolated interacting mean field disordered models
appeared in~\cite{CuLoNe17}. We continue developing this project and we analyse, in this paper, the quench dynamics of a classical integrable
system with (weak) frozen randomness. Both models belong to the class of $p$ spin spherical disordered models with,
however, properties that render their constant energy dynamics very different, as we will show here.

The spherical fully-connected $p$-spin disordered models are paradigms in the mean-field description of glassy
physics. They are solvable models (in the thermodynamic limit) that successfully
mimic the physics of (fragile) glasses for
$p\geq 3$ and domain growth for $p=2$.  The connection between the $p=2$ model, in its classical and quantum formulations,
with coarsening phenomena is made {\it via} its relation to the celebrated $O(N)$ $\lambda \phi^4$ model in the infinite $N$ limit.
Furthermore, the model has recently appeared in a semiclassical study of the Sachdev-Ye-Kitaev model~\cite{Altman}.
The literature on the static, metastable and
stochastic dynamics of the $p$ spin spherical systems is vast. Numerous  aspects of their behaviour are
very well characterised, even analytically (see, {\it e.g.}, the review articles~\cite{LesHouches,Ca09,BeBi11}).

In Ref.~\cite{CuLoNe17} we studied the Hamiltonian dynamics of the $p=3$ spherical
disordered spin model. By adding a kinetic term to the standard potential energy
we induced energy conserving dynamics to the real valued spins. In this setting, the
dynamics correspond to the motion of a particle on an $N-1$ dimensional
sphere under the effect of a complex quenched random potential~\cite{En93,FrMe94,CuLe96}. Here we will
focus on the Newtonian dynamics of the particle under conservative forces arising from
a quadratic potential, the $p=2$ case.

The Hamiltonian $p=2$ disordered model turns out to be equivalent to the Neumann integrable system of classical mechanics~\cite{Neumann},
the constrained motion of a classical particle on $S_{N-1}$ under a harmonic potential, for a special choice of the spring constants.
The only difference is that in the $p=2$ model one imposes the spherical constraint on average while in Neumann's model one does strictly, on each trajectory. This
difference, however, should not be important in the $N\to\infty$ limit. We will exploit results from the
Integrable Systems literature, notably the exact expressions of the $N-1$ conserved charges in involution~\cite{Uhlenbeck,AvTa90,BaTa92}. With these
at hand, we will be able to study the statistical properties in depth and construct a candidate GGE to describe the long-time dynamics.


We perform instantaneous quenches towards a post-quench disordered potential that keeps memory of
the pre-quench one, mimicking in this way the ``quantum quench'' procedure in a classical setting.
The change in the potential energy landscape induces finite injection or extraction of energy density in the
sample. The subsequent dynamics conserve the total energy. We sample the initial conditions from canonical equilibrium at
a tuned temperature, choosing in this way initial configurations typical of a paramagnetic equilibrium state at high temperature
or a condensed, ferromagnetic-like, state at low temperature. The control parameters in the dynamic phase diagram that we will establish are
the amount of energy injected or extracted and the initial temperature of the system, both measured with respect to the same
energy scale.

The dynamic evolution in the different phases of the phase diagram will be pretty different, with cases in which the infinite size
system remains confined (condensed, in the statistical physics language) and cases in which it does not. In none of them
a Gibbs-Boltzmann equilibrium measure is reached, contrary to what happens in the strongly interacting
$p=3$ case. The role played by the $N-1$ integrals of motion on the lack of equilibration of the infinite size
system will be discussed.

The reader just interested in a summary of our results and not so much in the way in which we obtained them can go directly to Sec.~\ref{sec:conclusions}
where we sum up our findings and we present a thorough comparison between the dynamics of the isolated $p=2$ and $p=3$ cases.

 The paper is organised as follows. In Sec.~\ref{sec:background} we recall the main features of the $p=2$ spherical disordered model
 (static, metastable and relaxational  dynamic properties) studied from the statistical mechanics point of view.
 In the following Section we explain the relation between the disordered
spin system and the integrable Neumann model of classical mechanics. We also explain in this Section the statistical  description of the long-term dynamics of
 integrable systems provided by the Generalised Gibbs Ensemble proposal. In Sec.~\ref{sec:quenches}  we present our analytic results
 for the dynamics of the model in the $N\to\infty$ limit and in Sec.~\ref{sec:analytic-finiteN} we go further and we set the analysis of the evolution of the
 finite $N$ case. Section~\ref{sec:numerical} is devoted to the numerical study of the
 $N\to\infty$ and finite $N$ dynamic equations. In Sec.~\ref{sec:integrals-of-motion} we investigate the behaviour of the $N-1$ integrals of motion
 in the various sectors of the phase diagram and we discuss their influence against the equilibration of the system. Finally, Sec.~\ref{sec:conclusions} presents our conclusions.
 Several Appendices complement the presentation in the main part of the paper.

\section{Background}
\label{sec:background}

This Section presents a short account of the equilibrium properties and relaxation dynamics of the spherical $p=2$ disordered model, first
introduced and studied by Kosterlitz, Thouless and Jones~\cite{KoThJo76} as the simplest
possible magnetic model with quenched random interactions. This model, as we will explain below,
shares many points in common with the $O(N)$ model of ferromagnetism when treated in the
infinite $N$ limit. Its static properties have been derived with a direct calculation and using the replica
trick. Its relaxational dynamics are also analytically solvable. The reader familiar with this
model can jump over this Section and go directly to the next one where the relation with Neumann's model
and integrability are discussed.

\subsection{The Hamiltonian spherical $p=2$ spin model}

The $p=2$ spin model is a system with two-spin interactions mediated by
quenched random couplings $J_{ij}$.
The potential energy is
\begin{eqnarray}
H_{\rm pot}[\{s_i\}] &=& - \frac{1}{2} \sum^N_{i\neq j} J_{ij} s_{i} s_{j}
\; .
\label{eq:pspin-pot}
\end{eqnarray}
The coupling exchanges are independent identically distributed random variables taken from a Gaussian distribution with average and variance
\begin{equation}
[ J_{ij } ] =0
\; , \qquad\qquad
[ J^2_{ij } ] = \frac{J^2}{N}
\; .
\label{eq:disorder-statistics}
\end{equation}
The parameter $J$ characterises the width of the Gaussian distribution.
In its standard spin-glass setting the spins are Ising variables and the model is the Sherrington-Kirkpatrick spin-glass.
We will, instead, use continuous
variables, $-\sqrt{N} \leq s_i \leq \sqrt{N}$ with $i=1, \dots, N$, globally forced to satisfy (on average) a spherical
constraint, $\sum_{i=1}^N s_i^2 = N$, with $N$ the total number of spins~\cite{KoThJo76}.
Such spherical constraint is imposed by adding a term
\begin{equation}
H_{\rm constr} = \frac{z}{2} \ \left(\sum_{i=1}^N s_i^2 - N \right)
\end{equation}
to the Hamiltonian, with $z$ a Lagrange multiplier. The spins thus defined do not
have intrinsic  dynamics. In statistical physics applications their temporal evolution
is given by the coupling to a thermal bath, {\it via} a Monte Carlo rule or a
Langevin equation~\cite{CuKu93}.

The quadratic model is a particular case of the family of $p$-spin models, the celebrated
mean-field model for glasses, with potential
energy $H_{\rm pot} = - \sum_{i_1\dots i_p} J_{i_1 \dots i_p} s_{i_1} \dots s_{i_p}$ and $p$ integer. Even more
generally, the form (\ref{eq:pspin-pot}) is one instance of a generic random potential $V(\{s_i\})$
with zero mean and correlations~\cite{En93,FrMe94,CuLe96}
\begin{equation}
[ V(\{s_i\})  V(\{s'_i\}) ] = -N {\cal V}(|\vec s-\vec s'|/N)
\end{equation}
with $ {\cal V}(|\vec s-\vec s'|/N) =-  \frac{J^2}{2} (\vec s \cdot \vec s'/N)^2$.

Similarly to what was done in~\cite{CuLoNe17} in the study of the $p\geq 3$ Hamiltonian dynamics,
the model can be endowed with conservative dynamics by changing the
``spin'' interpretation into a ``particle'' one. In this way, a kinetic energy~\cite{CuLo98,CuLo99}
\begin{eqnarray}
H_{\rm kin}[\{\dot s_i\}] = \frac{m}{2} \sum_{i=1}^N (\dot s_i)^2
\; ,
\label{eq:pspin-kin}
\end{eqnarray}
can be added to the potential energy.
The total energy of the {\it Hamiltonian spherical $p$-spin model}  is then
\begin{equation}
H_{\rm syst} = H_{\rm kin} + H_{\rm pot} + H_{\rm constr}
\; .
\label{eq:pspin-total-energy}
\end{equation}
This model represents a particle constrained to move on an $N$-dimensional hyper-sphere
with radius $\sqrt{N}$. The position of the particle is given by an N-dimensional vector $\vec s=(s_1, \dots, s_N)$ and its velocity
by another $N$-dimensional vector
$\dot{\vec s}=(\dot s_1, \dots, \dot s_N)$.
The $N$ coordinates $s_i$  are globally constrained
to lie, as a vector, on the hypersphere with radius $\sqrt{N}$. The velocity vector $\dot {\vec s}$ is, on average, perpendicular to
$\vec s$, due to the spherical constraint. The parameter $m$ is the mass of the particle.

The generic set of $N$ equations of motion for the isolated system is
\begin{equation}
m \ddot s_i(t_1) + z(t_1) s_i(t_1) =  \sum_{j (\neq i)} J_{ij} s_{j}(t_1)
\; ,
\label{eq:dynamic-p2}
\end{equation}
$i=1, \dots, n$,
where the Lagrange multiplier needs to be time-dependent to enforce the spherical constraint in the course of time.

The initial condition will be taken to be $\{s_i^0, {\dot s}_i^0\} \equiv \{ s_i(0), {\dot s}_i(0)\}$ and chosen in ways that we specify below.
We will be interested in using equilibrium initial states drawn from a Gibbs-Boltzmann measure at different temperatures
$T'$.

From Eq.~(\ref{eq:dynamic-p2}) one derives an identity between the  energy density and the Lagrange multiplier. By multiplying the equation by
$s_i(t_2)$ and taking $t_2 \to t_1$
\begin{equation}
\lim_{t_2\to t_1^-} m \partial_{t_1^2} C(t_1,t_2) + z(t_1) = - 2 e_{\rm pot}(t_1)
\; .
\end{equation}
The first term can be rewritten as $m\lim_{t_2\to t_1^-} \partial_{t_1^2} C(t_1,t_2) = - m \lim_{t_2\to t_1^-}  \sum_{i=1}^N \dot s_i(t_1) \dot s_i(t_2) = - m  \sum_{i=1}^N (\dot s_i(t_1) )^2$.
Therefore
\begin{equation}
 z(t_1) = - 2 e_{\rm pot}(t_1) + 2 e_{\rm kin}(t_1)
 \; .
 \label{eq:relation-z-energies}
\end{equation}
The Lagrange multiplier takes the form of an action density, as a difference between kinetic and potential
energy densities.
Using now the conservation of the total energy, $e_f=e_{\rm pot}(t_1) + e_{\rm kin}(t_1)$,
\begin{equation}
 z(t_1) =2 e_f - 4 e_{\rm pot}(t_1) = - 2 e_f +  4 e_{\rm kin}(t_1)
 \; .
 \label{eq:relation-z-energies2}
\end{equation}

The $p=2$ model belongs to a different {\it universality class} from the one of the $p\geq 3$ cases, in the
sense that its free-energy landscape and relaxation dynamics are much simpler. It is, indeed, a model that resembles
strongly the large $N$, $ O(N)$ model for ferromagnetism.  A hint on the simpler properties of its potential energy landscape is given by the
fact that the equations derived for general $p$  simplify considerably for $p=2$. For example, the static and dynamic transitions occur at
the same temperature $T_c=T_d$, and the number of metastable states is drastically reduced. We recall these properties in the rest of this
Section.

\subsection{The statics}
\label{subsec:statics}

The static properties of the $p=2$ spherical model were elucidated in~\cite{KoThJo76}. The trick is to project the
spin vector $\vec s$ onto the basis of eigenvectors of the interaction matrix. One calls
$\lambda_\mu$ and
$\vec v_\mu$ the $\mu$-th eigenvalue and eigenvector of the matrix $J_{ij}$, and
$s_\mu = \vec s \cdot \vec v_\mu$ the projection of $\vec s$ on the eigenvector $\vec v_\mu$. In
terms of the latter the Hamiltonian is not only quadratic but also diagonal. The extrema of the potential
energy landscape and the partition function
can then be easily computed. In the thermodynamic limit, $N\to\infty$, the eigenvalues are
distributed according to the Wigner-Dyson semi-circle form~\cite{Mehta}
\begin{equation}
\rho(\lambda_\mu) = \frac{1}{2\pi J^2} \sqrt{(2J)^2 - \lambda_\mu^2} \; \theta(2J -|\lambda_\mu|)
\; .
\end{equation}
For finite $N$ the distance between the largest and next to largest eigenvalues is order
$N^{-1/6}/N^{1/2}=N^{-2/3}$.

\subsubsection{The potential energy landscape}
\label{sect:p2-statics}

Let us label the eigenvalues of $J_{ij}$ in such a way that they are ordered:
$\lambda_1 \leq \lambda_2 \leq
\dots \leq \lambda_N$. We call their associated eigenvectors $\vec v_\mu$
with $\mu=1,\dots,N$ and we take them to be orthonormal,  such that $v_\mu^2 = 1$.
We consider the potential energy landscape $H_J(\{s_i\},z)$ with  $z$ taken as a variable.

In the absence of a magnetic field, all eigenstates of the interaction
matrix are stationary points of the potential energy hyper-surface,
\begin{displaymath}
\left. \frac{\partial H_J}{\partial s_i}\right|_{\vec s^*} =
- \sum_{j (\neq i)}^N  J_{ij} s_j + z s_i {\large |}_{\vec s^*,z^*} =
0 \;\;\;\;
\forall i \; ,
\;\;\;\;
\Rightarrow
\;\;\;\;
\vec s^*=\pm \sqrt{N} \vec v_\mu \; , \; z^*=\lambda_\mu \;, \forall \mu
\; .
\end{displaymath}
These stationary points are metastable states at zero temperature, apart from two of them that are the
marginally stable ground states, and their number is linear in $N$, the number of spins. (The role of marginal stability
in the physical behaviour of condensed matter systems was recently summarised in~\cite{MullerWyart15}.)
These statements are shown in the way described in the next paragraph.

The Hessian of the potential energy surface on each stationary point is
\begin{equation}
\left. \frac{\partial H_J}{\partial s_i \partial s_j}\right|_{\vec s^*, z^*}
=
\left. - J_{ij} + z \delta_{ij}\right|_{\vec s^*, z^*}  =
- J_{ij} + \lambda_\mu  \delta_{ij}
\; .
\end{equation}
This matrix can be easily diagonalised and  one finds
$D_{\nu\eta}=(-\lambda_{\nu}+\lambda_\eta)\delta_{\nu\eta}$.  Thus, on
the stationary point, $\vec s^*=\pm\sqrt{N} \vec v_\mu$, the Hessian has one
vanishing eigenvalue (for $\nu=\mu$), $\mu-1$ positive eigenvalues
(for $\nu <\mu$), and $N-\mu$ negative eigenvalues (for $\nu >
\mu$). Positive (negative) eigenvalues of the Hessian indicate stable
(unstable) directions. This implies that each saddle point labeled by $\mu$ has one
marginally stable direction, $\mu-1$ stable directions and $N-\mu$
unstable directions.  (In other words, the number of stable directions plus the
marginally stable one is given by the index $\mu$ labelling the
eigenvalue associated to the stationary state.)  In conclusion, there are
two maxima, $\vec s^* = \pm \sqrt{N} \vec v_1$, in general two saddles $\vec s^* = \pm \sqrt{N} \vec v_I$
with $I=\mu-1$ stable directions and $N-I$ unstable ones, with $I$ running with $\mu$ as $I=\mu-1$ and $\mu=3, \dots, N$,
and finally two (marginally stable) minima, $\vec s^* = \pm \sqrt{N} \vec v_N$.
In the large $N$ limit the density of eigenvalues of the Hessian at each metastable
state $\mu$ is a translated semi circle law~\cite{LaKu}.

The zero temperature energy of a generic configuration under no applied field is
\begin{equation}
H_J = - \frac{1}{2} \sum_{ij} J_{ij} s_i s_j + \frac{z}{2} \left( \sum_i s_i^2 - N\right) =
-\frac{1}{2} \sum_\mu (\lambda_\mu -z) s_\mu^2   - \frac{z}{2} N
\; .
\end{equation}
At each stationary point ${\vec s}^* = \pm \sqrt{N}\vec v_\mu $  and $z^*=\lambda_\mu$ this energy is
\begin{equation}
H_\mu^* \equiv H_J(\vec s^* = \vec v_\mu)
=
-\frac{1}{2}
\sum_{i} v_i^\mu \sum_j J_{ij} v_j^\mu + \frac{z^*}{2} \left( \sum_i (v_i^\mu)^2 - N \right)
=
-\frac{1}{2} \lambda_\mu N
\; .
\end{equation}
Here we used the notation $v_i^\mu$ to indicate the $i$th component of the $\mu$th eigenvector $\vec v_\mu$.
The energy difference between the minima and the lowest saddles
depends on the distribution of eigenvalues, a semi-circle law for the
Gaussian distributed interaction matrices that we consider here.

A magnetic field reduces the number of stationary points from a
macroscopic number to just two. Indeed, the stationary state equation
now reads
\begin{displaymath}
\left. \frac{\partial H_J}{\partial s_i}\right|_{\vec s^*} =
- \sum_{j(\neq i)}^N  J_{ij} s_j + z s_i - h_i {\large |}_{\vec s^*,z^*} =
0 \; , \;
\forall i \; ,
\;\;\;\;
\Rightarrow
\;\;\;\;
s_i^*= (z^*-J)^{-1}_{ij} h_j
\end{displaymath}
and $z^*$ is fixed by imposing the spherical constraint on $\vec
s^*$. One then finds two solutions for the Lagrange multiplier that
lie outside the interval of variation of the eigenvalues of the
$J_{ij}$ matrix: $|z^*|>\lambda_N$. The stability analysis shows that
the stationary points are just one fully stable minimum and a fully
unstable maximum. The elimination of saddle-points by an external field
has important consequences on the dynamics of the system~\cite{CuDe95b}.
In this paper we do not apply any external field.

The analysis of large dimensional random potential energy landscapes~\cite{Adler81,BrDe07,Fyodorov15,BenArous15}
is a research topic in itself with implications in condensed matter physics, notably in glass
theory~\cite{Wales04,MullerWyart15}, but also claimed to play a role in string theory~\cite{Susskind03,Douglas04}, evolution~\cite{FyKo16} or
other fields.
The $p=2$ spherical model provides a particularly simple case in which the potential energy landscape can be completely
elucidated.

\subsubsection{The free-energy density}
\label{subsubsec:free-energy}

This special (almost) quadratic model allows for the complete evaluation of its
free-energy density for a typical realisation of the disordered exchanges.
The traditional derivation of the disorder averaged free-energy density
can also be done using the replica method and a simple replica symmetric {\it Ansatz}
solves this problem completely. We recall how the two methods~\cite{KoThJo76,FiHe} work in this
Section.

The partition function reads
\begin{displaymath}
Z_J = \prod_{i=1}^N \int_{-\infty}^\infty ds_i \;
e^{\frac{\beta}{2} \sum_{i\neq j} J_{ij} s_i s_j}
\; \frac{1}{2\pi i} \int_{c-i\infty}^{c+i\infty} dz
\; e^{-\frac{\beta z}{2} \left(\sum_{i=1}^N s_i^2 -N \right)}
\label{eq:part-funct-p2}
\end{displaymath}
where $c$ is a real constant to be fixed below.

It is convenient to diagonalise the matrix
$J_{ij}$ with an orthogonal transformation and
write the exponent
in terms of the projection of the spin vector $\vec s$ on
the eigenvectors of $J_{ij}$,
$s_\mu \equiv \vec s \cdot \vec v_\mu$.
This operation can be done for
any particular realisation of the interaction matrix.
The new variables $s_\mu$ are also continuous and
unbounded and the partition function can be recast as
\begin{equation}
Z_J =  \prod_{\mu=1}^N \int_{-\infty}^\infty ds_\mu \;
\frac{1}{2\pi i} \int_{c-i\infty}^{c+i\infty} dz
\; e^{\sum_{\mu=1}^N \beta (\lambda_\mu-z) s_\mu^2/2 +
 \beta z N/2 }
 \; .
\label{eq:part-funct-p2-2}
\end{equation}
 Assuming that one can
exchange the quadratic integration over $s_\mu$
with the one over the Lagrange multiplier,
and that $c$ is such that
the influence of eigenvalues $\lambda_\mu>c$ is
negligible,
one obtains
\begin{eqnarray}
Z_J &=&
\frac{1}{2\pi i} \int_{c-i\infty}^{c+i\infty} dz \;
e^{-N \left[ -\beta z/2 +(2N)^{-1}
\sum_\mu \ln [\beta (z-\lambda_\mu)/2]
\right]}
\; .
\label{eq:part-funct-p2-3}
\end{eqnarray}
In the saddle-point approximation valid for $N\to\infty$  the Lagrange multiplier is
given by
\begin{equation}
1 = \langle\langle \;
k_BT \left(z_{\rm sp} - \lambda_\mu \right)^{-1}
\; \rangle\rangle
\label{eq:z-sp-statics}
\end{equation}
and this equation determines the different phases in the
model. We indicate with double brackets the
sum over the eigenvalues of the matrix $J_{ij}$
that in the limit $N\to\infty$ can be traded for an integration over its density:
\begin{equation}\label{eq:sum_int}
\frac{1}{N} \sum_{\mu=1}^N g(\lambda_\mu)
=
\int d\lambda_\mu \; \rho(\lambda_\mu)\,  g(\lambda_\mu)
\equiv \langle \langle \, g(\lambda_\mu) \, \rangle \rangle
\; .
\end{equation}

Let us discuss the problem in the absence of a magnetic field.
The high temperature solution to Eq.~(\ref{eq:z-sp-statics})
\begin{equation}\label{eq:z_eq_pm_mf}
z_{\rm sp} = z_{\rm eq} = T + \frac{J^2}{T}
\end{equation}
can be smoothly continued
to lower temperatures until the critical point
\begin{equation}
(k_BT_c)^{-1} = \langle\langle \; (z_{\rm sp}-\lambda_\mu)^{-1}
\; \rangle\rangle
\label{eq:p2-Ts}
\end{equation}
is reached where $z_{\rm sp}$ touches the
maximum eigenvalue of the $J_{ij}$ matrix,
and it sticks to it for all $T<T_c=J$:
\begin{equation}
z_{\rm sp}= z_{\rm eq} = \lambda_{\rm max} = 2J \;\;\;\;\;\;\;\;
T\leq T_c
\; .
\end{equation}
$T_c$ is the static critical temperature.
(A magnetic field with a component on the largest eigenvalue,
$\vec h \cdot \vec v_{\rm max} \neq 0$, acts as
an ordering field and erases the phase transition.)

If one now checks whether the spherical constraint is satisfied
by  these saddle-point Lagrange multiplier
values, one verifies that it is in the high temperature phase,
but it is not in the low temperature phase, where
\begin{equation}
\sum_{\mu=1}^N \langle s_\mu^2 \rangle = \frac{T}{T_c} N
\; .
\end{equation}
The way out is to give a macroscopic weight to the projection of the spin
in the direction of the eigenvector that corresponds to the largest eigenvalue:
\begin{equation}
s_N = m_0 \sqrt{N} + \delta s_N = \sqrt{\left( 1-\frac{T}{T_c} \right) N} + \delta s_N
\end{equation}
with $\langle \delta s_N\rangle =0$ so that
\begin{equation}
\langle s_N^2 \rangle + \sum^{N-1}_{\nu=1} \langle s_\nu^2 \rangle
= \left( 1-\frac{T}{T_c} \right) \, N + \frac{T}{T_c} N = N
\; .
\end{equation}

The thermal average of the projection of the spin vector
on each eigenvalue vanishes in the
high temperature phase and reads
\begin{eqnarray}
\langle s_{\mu} \rangle &=&
\left\{
\begin{array}{ll}
[N ( 1- T/T_c)]^{\frac12} &
 \;\;\;\;\;\;
\lambda_\mu = \lambda_{\rm max}
\; ,
\\
0 & \;\;\;\;\;\;
\lambda_\mu < \lambda_{\rm max}
\; ,
\end{array}
\right.
\end{eqnarray}
below the phase transition (once we have chosen one of the ergodic components with the spontaneous symmetry breaking
of the $\vec s \to - \vec s$ invariance). The configuration {\it condenses}
onto the eigenvector associated to the largest eigenvalue
of the exchange matrix that carries a weight proportional
to $\sqrt{N}$. Going back to the original spin basis, the mean magnetisation per site is zero
at all temperatures but the thermal average of the square of the
local magnetisation, that defines the Edwards-Anderson parameter,
is not when $T<T_c$:
\begin{eqnarray}
\langle m^2_i \rangle &=&
1-T/T_c
\;\;\;\;
\Rightarrow
\;\;\;\;
q_{\rm EA} \equiv [\langle m^2_i \rangle]_J = 1-T/T_c
\;\;\;\;
\mbox{with}
\;\;\;\;
T_c=J
\; .
\label{eq:p2-qea}
\end{eqnarray}
The order parameter $q_{\rm EA}$
vanishes at $T_c$ and the static transition is of second
order.

The condensation phenomenon occurs for
any distribution of exchanges with a finite support. If the
distribution has long tails, as when the model is defined on a sparse random graph~\cite{EdJo76,SeCu02,Slanina11},
the energy density diverges and
the behaviour is more subtle~\cite{SeCu03,Kuhn15}.


The disorder averaged free-energy density can also be
computed using the replica trick~\cite{MePaVi} and a replica symmetric {\it Ansatz}.
This  {\it Ansatz} corresponds to an overlap matrix between replicas
$Q_{ab} = \delta_{ab} + q_{\rm EA} \epsilon_{ab}$ with $\epsilon_{ab} =1 $
for $a\neq b$ and $\epsilon_{ab} =0$ for $a=b$.
When $N\to\infty$ the saddle point equations fixing the parameter
$q_{\rm EA}$ yield $0$ above $T_c$ and a marginally stable solution with
$q_{\rm EA}=1-T/T_c$ and identical physical properties to the ones discussed above
below $T_c$.

The  equilibrium energy is given by
\begin{eqnarray}
e^{\rm pot}_{\rm eq}
&=&
\left\{
\begin{array}{ll}
 - \frac{J^2}{2T}
\left[
1-
\left( 1-\frac{T}{J}\right)^2
\right]
=
\frac{1}{2} (k_BT - \lambda_{\rm max})
& \qquad\qquad T<T_c
\; ,
\\
-\frac{J^2}{2T}
& \qquad\qquad T>T_c
\; .
\end{array}
\right.
\label{eq:epot-equil}
\end{eqnarray}
We added a superscript ${\rm pot}$ since in the modified model that we will study in this paper the total energy will also
have a kinetic energy contribution. The entropy diverges at low temperatures as $\ln T$,
just as for the classical ideal gas, as usual in classical continuous
spin models.

\subsection{Relaxation dynamics}
\label{subsubsec:relaxation-dynamics}

The over-damped relaxation dynamics of the spherical $p=2$ spin model (coupled to a Markovian bath) were
studied in~\cite{ShSi81,CidePa88,CuDe95a,CuDe95b,FyPeSc15}.
One of the settings considered in these papers evolve a completely random initial condition, $\{s_i^0\}$,
that corresponds  (formally)  to an infinite temperature initial
state. The system is then subject to an instantaneous temperature quench by changing the temperature of  the bath to a final value $T< + \infty$.
Initial conditions drawn from equilibrium at temperature $T'<T_c$, and evolving in contact with a bath at the same temperature, $T=T'$,
were considered in~\cite{CuDe95a}
and it was shown in this paper that equilibrium at the same temperature is maintained ever after. A quench of the
dissipative system from equilibrium at $T'<T_c$ to another subcritical temperature $T<T_c$ was also studied in~\cite{CuDe95a}
and it was there shown that equilibrium at the target temperature is achieved.

The coupling to the bath is modeled with a stochastic equation of Langevin kind.
This equation can be exactly solved in the basis of eigenvectors of the interaction
matrix
\begin{equation}
s_\mu(t) = s_\mu(0) e^{-\lambda_\mu t - \int_0^t dt' \, z(t')}
+
\int_0^t dt' \; e^{-\lambda_\mu (t-t') - \int_{t'}^t dt' \, z(t'')} \; [ \xi_\mu(t') + h_\mu(t')]
\end{equation}
and the Lagrange multiplier $z(t)$ can be fixed by imposing the spherical constraint
\begin{equation}
C(t,t) = \frac{1}{N} \sum_\mu s_\mu^2(t) =1
\end{equation}
at all times.
This yields a self-consistent equation for $z(t)$. Notably, the asymptotic solution depends on the
choice of initial state, as we expose below.
The applied field $h_\mu$ is used to compute the linear response function
\begin{equation}
R(t_1,t_2) = \frac{1}{N} \left. \sum_{i=1}^N \frac{\delta \langle s_i(t_1)\rangle_h}{\delta h_i(t_2)}\right|_{h=0} =
\frac{1}{N} \left. \sum_{\mu=1}^N \frac{\delta \langle s_\mu(t_1)\rangle_h}{\delta h_\mu(t_2)}\right|_{h=0}
\; .
\end{equation}

For quenches of  initial conditions drawn from equilibrium at $T'\to\infty$, and evolution in contact with a bath at temperature $T>T_c$,
the dynamics quickly approach equilibrium at the new temperature. The Lagrange multiplier quickly converges to
$z_{\rm eq} = T+J^2/T$. The correlation and linear response are invariant under translations of
time and they are related by the fluctuation dissipation theorem~\cite{CuDe95a,DeGuMa07}, see Eq.~(\ref{eq:FDT-equil}) below.

For quenches of  initial conditions drawn from equilibrium at $T'\to\infty$, and evolution in contact with a bath at temperature
$T<T_c$, the correlation and linear response functions behave as in {\it coarsening systems}~\cite{Bray94,Onuki02,Puri09,CorberiPoliti},
decaying in two time regimes, one stationary for short
time differences, $(t_1-t_2)/t_2 \ll 1$, and one non-stationary for long time differences$(t_1-t_2)/t_2 \gg 1$. The detailed time-dependence in
the two regimes can be extracted using the procedure sketched above. It yields the behaviour of the self correlation and
linear response that scale in the same way as these do in the $O(N)$ model of ferromagnetism studied in the large $N$ limit, see Sec.~\ref{subsec:ON}.
The progressive condensation of the spin ``vector'' in the direction of the eigenvector corresponding to the largest eigenvalue
of the interaction matrix is the equivalent of the ordering process in the $O(N)$ model, that is to say, the condensation on
the zero wave-vector mode. Complete alignment with an overlap of order $\sqrt{N}$ is not reached in finite times with
respect to $N$.

For low temperature quenches from random initial conditions, the Lagrange multiplier approaches $z_f = 2J$ as a power law,
$z(t) - z_f \simeq -3/(4t)$.  The slow approach to the asymptotic value is determinant to allow for the non-stationary slow relaxation.
The global correlation and linear response are computed from the spin solution $s_\mu(t)$ and they
can be cast as~\cite{CuDe95a}
\begin{eqnarray}
C(t_1,t_2) &=& C_{\rm st} (t_1 - t_2) + C_{\rm ag}(t_1,t_2)
\; ,
\label{eq:C-sep}
\\
R(t_1,t_2) &=& R_{\rm st} (t_1 - t_2) + R_{\rm ag}(t_1,t_2)
\; ,
\label{eq:R-sep}
\end{eqnarray}
with the stationary and a non-stationary terms linked by the FDT at the temperature of the
bath \begin{eqnarray}
R_{\rm st}(t_1-t_2) &=& - \frac{1}{T} \frac{d C_{\rm st}(t_1-t_2)}{d(t_1-t_2)}
\label{eq:FDT-equil}
\; ,
\end{eqnarray}
and a modified FDT at an effective temperature $T_{\rm eff}$~\cite{CuKuPe97,Cu11} selected by the dynamics,
\begin{eqnarray}
R_{\rm ag}(t_1-t_2) &=& \frac{1}{T_{\rm eff}(t_1,t_2)} \frac{\partial C_{\rm ag}(t_1,t_2)}{\partial t_2}
\; ,
\end{eqnarray}
always with $t_1\geq t_2$. In the asymptotic limit, the two terms added to form $C$ and $R$ evolve in different regimes in the
sense that when one changes the other one is constant and vice versa.
The limiting values of the two contributions to the correlation function
are
\begin{eqnarray}
C_{\rm st}(0) = 1-q \; , \qquad\qquad &
\lim_{t_1-t_2 \to \infty} C_{\rm st}(t_1-t_2) = 0
\; ,
\label{eq:lim-stat}
\\
\lim_{t_2\to t_1^-} C_{\rm ag}(t_1,t_2) = q \; , \qquad\qquad &
\lim_{t_1\gg t_2} C_{\rm ag}(t_1,t_2) = 0
\; ,
\label{eq:lim-aging}
\end{eqnarray}
with the parameter $q$  being equal to
\begin{eqnarray}
q  &=& 1-\frac{T}{J}
\; .
\label{eq:qth}
\end{eqnarray}
This is the correct expression of the Edwards-Anderson parameter for the equilibrium low temperature solution,
see Eq.~(\ref{eq:p2-qea}), and once again one finds $T_c=J$ from $q=0$.
The complete solution of the Langevin equations allows one to deduce the exact scaling forms of the
stationary and ageing contributions to the correlation and linear response. These are
\begin{eqnarray}
C_{\rm ag}(t_1,t_2) = f_C\left(\frac{t_2}{t_1}\right)
\qquad
\qquad
R_{\rm ag}(t_1,t_2) = t_2^{-3/2} \; f_R\left(\frac{t_2}{t_1}\right)
\end{eqnarray}
with $f_C(x)$ and $f_R(x)$ known analytically.
The behaviour of the effective temperature is special in the $p=2$ model in the sense that contrary to what
happens in the $p\geq 3$ cases~\cite{CuKu93} it is not constant but grows with time. More precisely, it
scales as $T_{\rm eff}(t_1,t_2) \simeq t^{1/2} f_T(t_1/t_2)$ and it diverges asymptotically as $t_1^{1/2}$. This implies, in particular, that the ageing
regime does not contribute to the asymptotic potential energy that, after a quench to $T<T_c$, reads
\begin{equation}
e^{\rm pot}_{\rm asymp}
= - \frac{J^2}{2} \left[ \frac{1}{T} (1-q^2)  \right]
= e^{\rm pot}_{\rm eq}
\label{eq:e-th2}
\end{equation}
and is identical to the equilibrium one, see Eq.~(\ref{eq:epot-equil}), once $q=1-T/J$ is used.

For quenches within the ordered phase, $T'<T_c \to T<T_c$,
for example taking initial conditions in equilibrium at zero temperature, $s_\mu(0)=\sqrt{N} \delta_{\mu, N}$ and
evolving them at $T<T_c$,  the Lagrange multiplier approaches $z_f=2J$ faster than any power law and the system
rapidly equilibrates to the after quench conditions~\cite{CuDe95a}.

The same technique, based on the projection of the spin vector on the eigenvectors of $J_{ij}$,
can also be implemented in the case in which there is inertia and the
differential equation has a second order time derivative.
The dynamics are  recast  into the ones of harmonic oscillators coupled by a self-consistent
time-dependent Lagrange multiplier. We will use this formulation in Sec.~\ref{sec:analytic-finiteN}.
Although a full analytical solution is not possible with the second time derivative, a performant
numerical algorithm will allow us to monitor the evolution of the different modes.

\subsection{Relation with the $O(N)$ $\lambda \phi^4$ model in the large $N$ limit}
\label{subsec:ON}

The $\lambda \phi^4$ scalar field theory in $d$ dimensions is defined by the Hamiltonian
\begin{equation}
H = \int d^d x \left[ \frac{1}{2} (\nabla \phi)^2 + \frac{r}{2} \phi^2 + \frac{\lambda}{4} \phi^4 \right]
\; ,
\end{equation}
where $r$ and $\lambda$ are two parameters. This model is the Ginzburg-Landau free energy for the local
order parameter of the paramagnetic-ferromagnetic transition controlled by the parameter $r$ going
from $r>0$ to $r<0$. When the field is upgraded to a vector with $N$ components and the limit $N$ $\to\infty$
is taken the quartic term, first conveniently normalised by $N$, can be approximated by
$\frac{\lambda}{4{\small{N}}} \phi^4 \mapsto \frac{\lambda}{4{\small{N}}} \langle \phi^2\rangle \phi^2$. The
quantity $z(t) = \langle \phi^2\rangle $ is not expected to fluctuate and plays the role of the Lagrange multiplier
in the spherical disordered model. Once this approximation made,
the model becomes quadratic in the field and its statics and relaxation dynamics can be easily
studied. The only difficulty lies in imposing the self-consistent constraint that determines $z(t)$. The condensation
phenomenon that we discussed in the disordered model is also present in the field theory and it
corresponds to a condensation on the zero wave vector mode. In the dynamic problem this corresponds to the
progressive approach to the homogeneous field configuration~\cite{LesHouches}.

The conserved energy dynamics of the $\lambda \phi^4$ model, especially after sudden quenches,
has been studied by a number of groups. Details on the behaviour of the scalar problem, as well as
a review of general equilibration and pre-equilibration issues can be found
in J. Berges' Les Houches Lecture Notes~\cite{Berges15}, see also~\cite{BoDeDe04}.
The dynamics of the large $N$ limit of the $O(N)$ model was analysed
in~\cite{BoDeHoSa99,ScBi10,ScBi11,ScBi13,ChNaGuSo13,MaCHMiGa15}.
More recent works use renormalisation group techniques to study the short time dynamics~\cite{ChTaGaMi16,ChGaDiMa17}
at the dynamic phase transition.


\section{Neumann's model, integrability and equilibration}
\label{sec:Neumann-sec}

In this Section we explain the relation between the Hamiltonian $p=2$ disordered model and
the integrable model of Neumann~\cite{Neumann}. We start by recalling some basic properties of classical
integrable systems in the sense of Liouville~\cite{BaBeTa09,Dunajski}. We then recall the definition of Neumann's model and we compare it to
the $p=2$ one.  Finally, we explain the  ideas behind the Generalised Gibbs
Ensemble. Later, in Sec.~\ref{sec:integrals-of-motion}, we will use this formalism to analyse certain aspects of the
 long time dynamics of the system, and we set the stage for a future study of the eventual approach to a
 GGE ensemble.

\subsection{Integrable systems}

In classical mechanics, systems are said to be Liouville integrable if there exist
sufficiently many well-behaved first integrals or constant of motions in involution such that the
problem can then be solved by quadratures~\cite{BaBeTa09,Dunajski}, in other words,
the solution can be reduced to a finite number of algebraic operations and integrations.
In more concrete terms, an integrable dynamical system consists of a $2N$-dimensional phase space $\Gamma$ together
with $N$ independent functions\footnote{In the sense that the gradients $\vec \nabla O_i$ are linearly independent vectors
on a tangent space to any point in $\Gamma$} $O_1, \dots, O_N$: $\Gamma \to \mathbb{R}$, such that the mutual Poisson brackets vanish,
\begin{equation}
\{ O_j, O_l \} = 0 \qquad \mbox{for all} \;\; j, l
\; .
\end{equation}
We will assume henceforth that the $O_i$ do not depend explicitly on time and
that $dO_i/dt = 0$ is equivalent to $\{ H, O_i\}=0$. Conventionally, the first function $O_1$ is the Hamiltonian itself and
the first constant of motion is the energy. All other $O_i$ with $i \neq 1$ are also constants of motion since
their Poisson bracket with $H$ vanishes. The dynamics of the system can then be seen as the motion in a manifold of
dimension $2N-N=N$ in which all configurations share the initial values of all the conserved quantities $O_i(t)=O_i(0)$.
Under these conditions Hamilton's equations of motion
are solvable by performing a canonical transformation into action-angle
variables $(I_i,\phi_i)$ with $i=1, \dots, N$ such that the Hamiltonian is rewritten as $\tilde H(I)$ and
\begin{equation}
I_i(t) = I_i(0) \; , \qquad\qquad \phi_i(t) = \phi_i(0) + t \; \frac{\partial \tilde H(I)}{\partial I_i}  =
\phi_k(0) + t \; \omega_i(I) \; .
\label{eq:solution-integrable}
\end{equation}
The action functions $I_i$ are conserved quantities and we collected them in $I$ in the dependence
of the frequencies $\omega_i$ and the Hamiltonian $\tilde H$. The remaining evolution is given by
$N$ circular motions with constant angular velocities.
Both deciding whether a system is integrable and finding the canonical transformation that
leads to the pairs $(I_i, \phi_i)$
are in practice very difficult questions. Whenever the system is integrable, and one knows the action-angle
pairs, the statement in Eq.~(\ref{eq:solution-integrable}) is part of the Liouville-Arnold theorem~\cite{Arnold78}.

\subsection{Neumann's model and its integrals of motion}
\label{sec:Neumann}

The model proposed by Neumann in 1850 describes the dynamics of a particle constrained to move on the $N-1$ dimensional sphere
under the effect of harmonic forces~\cite{Neumann}. The Hamiltonian is
\begin{equation}
H = \frac{1}{4N} \sum_{k\neq l} L_{kl}^2 + \frac{1}{2}\sum_k a_k x^2_k
\label{eq:Neumann}
\end{equation}
where the $L_{kl}$ are the elements of an angular momentum anti-symmetric matrix
\begin{equation}
\sqrt{m} \, L_{kl} = x_k p_l - p_k x_l
\; ,
\end{equation}
and $p_k$ and  $x_k$ are phase space variables with canonical Poisson brackets $\{ x_k, p_l \} = \delta_{kl}$.
The global spherical constraint
\begin{equation}
\sum_{k=1}^N x_k^2 =N
\end{equation}
ensures that the motion takes place on $S_{N-1}$.
Using the fact that $L_{kk}=0$ to rewrite the double sum in the first term in $H$
as an unconstrained sum, and replacing $L_{kl}$ by its explicit form in terms of $x_k$ and $p_k$,
one derives
$
m\sum_{k\neq l} L_{kl}^2 =
m\sum_{k, l}  L^2_{kl} = 2 \sum_{k} x_k^2 \sum_{l} p_l^2 - 2 \sum_{k} x_k p_k \sum_{l} x_l p_l
$.
Imposing next the spherical constraint, that also implies $\sum_k x_k p_k =0$, the sum simply becomes
\begin{equation}
\sum_{k\neq l} L_{kl}^2 =
\frac{2N}{m} \sum_{k} p_k^2
\; .
\end{equation}
We note that we added a factor $1/N$ in ifront of the kinetic energy in Eq.~(\ref{eq:Neumann}) in order to ensure that the two
terms in $N$ be extensive and the thermodynamic limit non-trivial.

It is quite clear that Neumann's model is therefore identical to the Hamiltonian $p=2$ model once the latter is written in the basis of
eigenvectors of the interaction matrix $J_{ij}$.

The $N-1$ integrals of motion of this problem were constructed by K. Uhlenbeck~\cite{Uhlenbeck} and more recently rederived
by Babelon \& Talon~\cite{BaTa92} with a separation of variables method. In a notation that is convenient for our application they read
\begin{equation}
I_k =x_k^2 +\frac{1}{N} \sum_{l(\neq k)} \frac{L_{kl}^2}{a_k-a_l} =  x_k^2 +\frac{1}{mN} \sum_{l(\neq k)} \frac{x_k^2 p_l^2 + x_l^2 p_k^2 - 2 x_k p_l x_l p_k}{a_k-a_l}
\end{equation}
and satisfy $\sum_k I_k=N$ and $\frac{1}{2} \sum_k a_k I_k = H$.
In the definition of our Hamiltonian and equations of motion we used a convention such that $a_k \mapsto -\lambda_\mu$ (note the minus
sign). After a trivial translation to the variables of the $p=2$ spherical model we then have
\begin{equation}
I_\mu  = s_\mu^2  +\frac{1}{mN} \sum_{\nu(\neq \mu)}  \frac{s_\mu^2 p_\nu^2  +  s_\nu^2 p_\mu^2
- 2 s_\mu p_\nu s_\nu p_\mu}{\lambda_\nu-\lambda_\mu}
\; ,
\label{eq:ImuUhlenbeck}
\end{equation}
$\sum_\mu I_\mu =1$ and $\sum_\mu \lambda_\mu I_\mu =- 2H$.

\subsection{Statistical measures for integrable systems}

Let $\vec X=(x_1,p_1,\dots, x_N,p_N)$ be a generic point in phase space.
The fact that the microcanonical measure
\begin{equation}
\rho_{\rm GME}(\vec X) =  c^{-1} \prod_{j=1}^N \delta(I_j(\vec X)-i_j)
\; ,
\end{equation}
with
$c = \int d\vec X \; \prod_{j=1}^N \delta(I_j(\vec X)-i_j)$
be sampled asymptotically is ensured by the Liouville-Arnold theorem~\cite{Arnold78},
if the frequencies of the periodic motion on the
torus are independent, that is, $\vec k \cdot \vec \omega = 0$ for $\vec k =
(k_1, \dots, k_N)$ with integer $k_j$ has the unique solution $\vec k=0$.
One can call this ensemble the Generalized Microcanonical Ensemble.

In principle, the Generalized Canonical Ensemble, commonly
called Generalized Gibbs Ensemble (GGE), can now be constructed from the Generalized Microcanonical Ensemble
following the usual steps.
The idea is to look for the joint probability distribution of $N$ extensive (as for the
Hamiltonian in the usual case) constants of motion of a subsystem
$P(i_1, \dots, i_N) di_1 \dots di_N$.
As in cases with just one conserved quantity, it is convenient to interpret $P$ as a probability over position and momenta
variables, and write
\begin{equation}
P_{\rm GGE}(\vec X) =  Z_{\rm GGE}^{-1}(\lambda_1, \dots, \lambda_N) \;  e^{-\sum_j \zeta_j {I_j}(\vec X)}
\; .
\label{eq:probGGE}
\end{equation}
This form can be derived under the same kind of assumptions used in the derivation of the canonical measure
from the microcanonical one, that is
(i) independence of the chosen subsystem with respect to the rest, in other words, the factorisation of the density
of states $g(\{i_1,\dots, i_N\}) = g(\{i^{(1)}_1,\dots, i^{(1)}_N\}) g(\{i^{(2)}_1,\dots, i^{(2)}_N\})$,
(ii) additivity of the conserved quantities
$i_j^{(2)} = i_j - i_j^{(1)}$,
(iii) small system 1 $(i_j^{(1)} \ll   i_j^{(2)})$,
(iv) constant inverse `temperatures'
$\zeta_j\equiv \partial_{i_j} S_2(\{ i_1, \dots, i_N\}) = k_B \partial_{i_j} \ln g_2(\{ i_1, \dots, i_N\})$.
An inspiring discussion along these lines appeared in~\cite{Yuzbashyan}. The conditions just listed imply a locality requirement
on the $i_j$s, otherwise (ii) and (iii) would be violated. This is similar to the requirement of having short-range interactions
to derive the equivalence between the canonical and microcanonical ensembles in standard statistical mechanics.

In quenching procedures,
the parameters $\zeta_j$ should be determined by requiring that the expectation value of each
conserved quantity $I_j$ calculated on $\rho_{\rm GGE}$
matches the (conserved) initial value $I_j(0^+)$ (right after the quench):
\begin{equation}
I_j(0^+) = \int d\vec X \; I_j(\vec X) \, P_{\rm GGE}(\vec X)
\; .
\end{equation}
The $\zeta_j$  are then the Lagrange multipliers that enforce this
set of $N$ constraints.

In the $p=2$ or Neumann model a set of conserved quantities in involution are the $I_\mu$ defined in Eq.~(\ref{eq:ImuUhlenbeck}).
We will study them in Sec.~\ref{sec:integrals-of-motion}.

We note that if the Lagrange multipliers
became, under some special conditions $-\lambda_\mu \beta_f/2$, with $\lambda_\mu$ the
eigenvalues of the random interaction matrix, the GGE measure would be
\begin{equation}
P_{\rm GGE}(\vec X) = Z_{\rm GGE}^{-1} \ e^{-\beta_f \frac{(-1)}{2} \sum_\mu \lambda_\mu I_\mu} = Z_{\rm GGE}^{-1} \ e^{-\beta_f H} = P_{\rm GB}(\vec X)
\; ,
\end{equation}
the Gibbs-Boltzmann one.

\subsection{Averages in the long time limit}

Take now a generic function of the phase space variables $A(\vec X)$ that does not depend
explicitly on time and is not conserved. Birkhoff's theorem~\cite{Khinchin} states that  its  long-time average exists and reaches a
constant,
\begin{equation}
\overline A \equiv \lim_{\tau \to\infty} \frac{1}{\tau} \int_{t_{\rm st}}^{t_{\rm st}+\tau} dt' \, A(\vec X(t')) = \mbox{cst}
\label{eq:time-average}
\end{equation}
for $\tau$ sufficiently long and $t_{\rm st}$ a reference transient time.
We will use this fact at various points in our study.

The claim of equilibration to a Generalised Gibbs Ensemble is that the long time averages should also be given by the
averages over the statistical measure $P_{\rm GGE}$:
\begin{equation}
\overline A = \int {\cal D}\vec X \; A(\vec X) \ P_{\rm GGE}(\vec X)
\; .
\end{equation}
Which are the observables for which this result should hold is an interesting question that needs to be
answered with care.

The GGE proposal~\cite{Rigol07,Rigol08} and most, if not all, of its discussion appeared in the treatment and study of quantum
isolated systems and, especially, the dynamics following an instantaneous quench performed as
a sudden change in a parameter of the system's Hamiltonian. A series of review articles
are~\cite{Polkovnikov10,Pasquale-ed,Gogolin}. The main motivation for our research project is to ask similar
questions in the classical context, with the aim of  disentangling the quantum aspects from the bare consequences of
isolation and integrability.

\subsection{The GGE temperatures and the fluctuation dissipation theorem}

The fluctuation-dissipation theorem (FDT)~\cite{Kubo} is a model independent equilibrium relation between the time-delayed linear response
of a chosen observable and its companion correlation function.
In Gibbs-Boltzmann equilibrium this relation is independent of the specific system and observable
and it only involves the inverse temperature of the system. For classical systems it admits simple
expressions in the time and frequency domains\footnote{When dealing with the numerical data we used a Fourier transform convention
such that $R(\omega, t_2) = \sum_{ k } R(t_k + t_2, t_2) e^{i \omega t_k}$
and $C(\omega, t_2) = \sum_{ k } ( C(t_k + t_2, t_2) - q ) \cos(\omega t_k)$
with $t_k$ the discrete times on which we collect the data points.
The Fourier transform of the correlation is computed for $t_k > 0$ only,
taking advantage of the long-time stationarity property $C_{\rm st}(-t) =C_{\rm st}(t)$. For this reason, there is
no factor $2$ in the left-hand-side of the FDT in the frequency domain.}
\begin{equation}
R_{AB}(t_1-t_2) = \beta \partial_{t_2} C_{AB}(t_1-t_2) \theta(t_1-t_2)
\qquad
\mbox{or}
\qquad
\mbox{Im} \hat R_{AB}(\omega) = - \beta \omega \hat C_{AB}(\omega)
\; .
\end{equation}

Out of canonical equilibrium, the fluctuation-dissipation relations (FDR) between the linear
response and the correlation function have been
used to quantify the departure from equilibrium~\cite{CuKu93}. Indeed, the possibly time and observable
dependent parameter that replaces $\beta$ in far form equilibrium systems yields
an effective temperature that in certain cases with slow dynamics admits the interpretation
of a proper temperature~\cite{CuKuPe97,Cu11}. Specially useful for our purposes
is the fluctuation dissipation relation (FDR) in the frequency domain
\begin{equation}
\frac{\mbox{Im} \hat R_{AB}(\omega)}{\omega \hat C_{AB}(\omega)} = - \beta^{AB}_{\rm eff}(\omega)
\end{equation}
that concretely defines the frequency dependent, and also possibly observable dependent,  inverse effective temperature $\beta^{AB}_{\rm eff}$.

It was shown in~\cite{FoGaKoCu16,deNardis17}  that  the Lagrange multipliers $\zeta_j$
of  the  GGE, seen as inverse temperatures $\beta_j$,  of  a  number  of  isolated   integrable quantum systems which
reach a stationary state can be read from the FDR's of properly chosen observables
\begin{equation}
\beta_j = \beta_{\rm eff}(\omega_j) \qquad\mbox{for all} \; j
\; .
\end{equation}
In this paper we will show that this statement also applies to the classical
integrable system that we analyse.


\section{Analytic results for the dynamics of the infinite size system}
\label{sec:quenches}

We now enter the heart of our study and we consider the dynamics of the isolated system after different kinds of quenches.
In this Section we use an analytic treatment of the global dynamics in the thermodynamic limit. Long time
regimes will be considered only after the diverging number of degrees of freedom:
\begin{equation}
\lim_{t\to\infty} \lim_{N\to\infty}
\end{equation}

\subsection{Dynamical equations}
\label{subsec:dynamic-eq}

We start by giving a short description of steps that lead to the dynamic equations that couple linear response and correlation function and fully
characterise the evolution of the model in the $N\to\infty$ limit.

\subsubsection{The Schwinger-Dyson equations}

In the $N\to\infty$ limit, the only relevant correlation and linear response functions are
\begin{eqnarray}
\qquad\qquad
C_{ab}(t_1,t_2) &=&  N^{-1} \sum_{i=1}^N \; [\langle s^a_i(t_1) s^b_i(t_2)\rangle] \; ,
\\
\qquad\qquad
C_{ab}(t_1,0) &=&  N^{-1} \sum_{i=1}^N \; [\langle s^a_i(t_1) s^b_i(0)\rangle] \; ,
\\
\qquad\qquad
R_{ab}(t_1,t_2) &=&  \left. N^{-1} \frac{\delta \;\;\;}{\delta h_b(t_2)} \sum_{i=1}^N \; [\langle {s^a}^{(h)}_{\!\!\! i}(t_1)\rangle]
\right|_{h=0} \; ,
\end{eqnarray}
for $t_1, \ t_2 > 0$,
where the infinitesimal perturbation $h$ is linearly coupled to the spin $H\mapsto H - h \sum_{i=1}^N s_i$ at time $t_2$
and the upperscript ${(h)}$ indicates that the configuration is measured after having applied the field $h$.
Since causality is respected, the linear response is non-zero only for $t_1>t_2$.
The square brackets denote here and everywhere in the paper the average over quenched disorder. The angular brackets indicate the average over
thermal noise if the system is coupled to an environment, and over the initial conditions of the dynamics sampled, say, with a
probability distribution. When the coupling to the bath is set to zero, as we do in this paper, the last average is the only one remaining in the
angular brackets operation. The meaning of the indices $a, b$ is given in the next paragraph.

The dynamical equations starting from a random state  are well-known and can be found in
Refs.~\cite{Ba97,CuKu93,CuKu95,CaCa05}. They are usually derived from the
dynamical Martin-Siggia-Rose-Janssen-deDominicis generating function. The method has  been
modified to take into account the effect of equilibrium initial conditions in~\cite{HoJaYo83} and it was
applied to the relaxational $p$ spin model in~\cite{FrPa95,BaBuMe96}.
The average over disorder now becomes non-trivial and needs the use of the
replica trick. The scripts $a, b$ indicate then the replica indices $a, b = 1, \dots, N$.
For initial conditions in equilibrium the replica structure is replica symmetric (see Sec.~\ref{subsec:statics} and
\cite{KoThJo76}), with
\begin{equation}
Q_{aa}=1 \qquad\mbox{and} \qquad Q_{a\neq b}=q_{\rm in}
\end{equation}
and $q_{\rm in}$ in the paramagnetic state while $q_{\rm in}\neq 0$ in the condensed phase.
This structure has an effect on the equation for the time-dependent correlation function that will  keep the
initial replica structure. There will be two kinds of correlations with the initial condition
\begin{equation}
C_1(t,0) \qquad\mbox{and}\qquad C_{a\neq 1}(t,0)
\; ,
\end{equation}
where we singled out the replica $a=1$. Since there is no reason to think that the replicas that are not $a=1$
behave differently, we follow the dynamics of
\begin{equation}
C_2(t_1,t_2)
\end{equation}
as a representative of this group. The interpretation of the correlations $C_1(t_1,t_2)$ and $C_2(t_1,0)$ can be given in
terms of real replicas. The former is the self-correlation between the  configuration of one replica of the system
$\{s_i\}(t_2)$ and the same replica evolved until a later time $t_1$, $\{s_i\}(t_1)$. For this reason, we will eliminate
the subscript $_1$ and call $C_1(t_1,t_2) \mapsto C(t_1,t_2)$ in most places hereafter. The latter is the correlation between one
replica of the system $\{\sigma_i\}(0)$ at the initial time $0$ and another replica of the system evolved until
time $t_1$ and represented by $\{s_i\}(t_1)$.
Although we could also write an evolution equation for the two-time $C_2(t_1,t_2)$ we do not need it here since
only $C_2(t_1,0)$ intervenes in the other equations.

In the $N\to\infty$ limit one derives the dynamical equations that read
\begin{eqnarray}
\left(m\partial_{t_{1}}^{2}+z(t_{1})\right)R(t_{1},t_{2}) & = &
J^2 \int_{t_{2}}^{t_{1}}dt'R(t_{1},t') R(t',t_{2})
+\delta(t_1-t_2)
\; ,
\label{eq:dyn-eqs-R}\\
\left(m\partial_{t_{1}}^{2}+z(t_{1})\right)C(t_{1},t_{2}) & = &
J^2\int_{0}^{t_{1}}dt'R(t_{1},t') C(t',t_{2})
\nonumber\\
 &  &
+ J^2\int_{0}^{t_{2}}dt'R(t_{2},t')C(t_{1},t')
+\frac{J J_0}{T'} \sum_a C_a(t_{1},0) C_a(t_2,0)
\; ,
\label{eq:dyn-eqs-C-0}
\\
\left(m\partial_{t_{1}}^{2}+z(t_{1})\right)C_a(t_{1},0) & = &
J^2\int_{0}^{t_{1}}dt' \; R(t_{1},t') C_a(t',0)
+\frac{J J_0}{T'} \sum_b C_b(t_{1},0) Q_{ab}
\; ,
\label{eq:dyn-eqs-C1-0}
\\
z(t_{1}) & = &
-m\partial_{t_{1}}^{2}C(t_{1},t_2)\vert_{t_{2}\rightarrow t_{1}^{-}}
 +
2J^2\int_{0}^{t_{1}}dt'R(t_{1},t')C(t_{1},t')
\nonumber\\
 &  &
+
\frac{J J_0}{T'} \sum_a (C_a(t_{1},0))^2
\label{eq:dyn-eqs-z-0}
\; .
\end{eqnarray}
The border conditions are
\begin{eqnarray}
\begin{array}{ll}
&C_1(t_1,0)= C(t_1,0)
\qquad
\mbox{implying} \qquad
C_1(0,0)= C(0,0) =1
\; ,
\\
&
C_2(0,0) = Q_{1,2} \qquad\qquad \!
\mbox{implying} \qquad C_{2}(0,0) = q_{\rm in}
\; .
\end{array}
\end{eqnarray}
Note that the initial condition is not the same for all $C_b(t_1,0)$. It is equal to 1 for $b=1$ and equal  to
$q_{\rm in}$ for $b\neq 1$.
One can check that these equations coincide with the ones in \cite{FrPa95,BaBuMe96} when inertia is neglected, $p=2$ and $J=J_0$,
and a coupling to a bath is introduced.
With respect to the equations studied in~\cite{CuLoNe17}, they correspond to $p=2$ and they have the extra ingredient
of the influence of equilibrium initial conditions with a non-trivial  replica structure, allowing for condensed initial states in proper
thermal equilibrium.

The sums over the replica indices appearing in Eqs.~(\ref{eq:dyn-eqs-C-0}), (\ref{eq:dyn-eqs-C1-0}) and (\ref{eq:dyn-eqs-z-0}) can be readily computed
in the $n\to 0$ limit; they read
\begin{eqnarray}
\frac{J J_0}{T'} \sum_a C_a(t_{1},0) C_a(t_2,0) &=&
\frac{J J_0}{T'} \left[  C_1(t_1,0) C_1(t_2,0) + (n-1) C_2(t_1,0) C_2(t_2,0)  \right]
\nonumber\\
&\underset{n\to 0}{\to} &
\frac{J J_0}{T'} \left[  C_1(t_1,0) C_1(t_2,0) - C_2(t_1,0) C_2(t_2,0) \right]
\; ,
\label{sum1}
\\
\frac{J J_0}{T'} \sum_b C_b(t_{1},0) Q_{1b} &=&
\frac{J J_0}{T'} \left[  C_1(t_1,0) \, 1 + (n-1) C_{b(\neq 1)}(t_1,0) q_{\rm in} \right] \nonumber\\
&\underset{n\to 0}{\to} &
\frac{J J_0}{T'} \left[  C_a(t_1,0) - C_{b(\neq 1)}(t_1,0) q_{\rm in} \right]
\; ,
\label{sum2}
\\
\frac{J J_0}{T'} \sum_b C_b(t_{1},0) Q_{(a\neq 1)b} &=&
\frac{J J_0}{T'} \left[  C_{1}(t_1,0) q_{\rm in} + C_2(t_1,0) \, 1 + (n-2) C_{b(\neq (1,2))}(t_1,0) q_{\rm in} \right] \nonumber\\
&\underset{n\to 0}{\to} &
\frac{J J_0}{T'} \left[  q_{\rm in} C_1(t_1,0)  + (1-2q_{\rm in})  C_2(t_1,0)   \right]
\; .
\label{sum3}
\end{eqnarray}
Consequently,  the terms induced in the equation for $C(t_1,t_2)$ and $C_2(t_1,0)$ are different.

With inertia and no coupled bath, the equal-time conditions are
\begin{eqnarray}
C(t_1,t_1) & = & 1 \; ,
\nonumber\\
R(t_1,t_1) & = & 0 \; ,
\nonumber\\
\partial_{t_1}C(t_1,t_2)\vert_{t_2\rightarrow t_1^{-}}=\partial_{t_1}C(t_1,t_2)\vert_{t_2\rightarrow t_1^{+}} & = & 0 \; ,
\nonumber\\
\partial_{t_1}C_2(t_1,t_2)\vert_{t_2\rightarrow t_1^{-}}=\partial_{t_1}C_2(t_1,t_2)\vert_{t_2\rightarrow t_1^{+}} & = & 0 \; ,
\\
\partial_{t_1}C_2(t_1,0)\vert_{t_1\rightarrow 0^+} & = & 0 \; ,
\nonumber\\
\partial_{t_1}R(t_1,t_2)\vert_{t_2\rightarrow t_1^{-}} & = & \frac{1}{m} \; ,
\nonumber\\
R(t_1,t_2)\vert_{t_2\rightarrow t_1^{+}} & = & 0 \; ,
\nonumber
\end{eqnarray}
for all times $t_1, t_2$ larger than or equal to  $0^+$, when the dynamics start.

We found convenient to numerically integrate the integro-differential equations using
an expression of the Lagrange multiplier that trades the second-time
derivative of the correlation function into the total conserved energy after the quench.
Following the same steps explained in~\cite{CuLoNe17} we deduce
\begin{equation}
z(t_{1})=2e(t_1)+  4J^2 \int_{0}^{t_{1}}dt' \; R(t_{1},t')C(t_{1},t')+ \frac{2JJ_0}{T'} \ \left[ C^2(t_{1},0) -C_2^2(t_{1},0)\right]
\; ,
\label{eq:z-Nicolas}
\end{equation}
where we used Eq.~(\ref{sum1}) evaluated at $t_1=t_2$.
It seems that we have simply traded $z(t_1)$ by $e(t_1)$.
Indeed, taking advantage of the fact that for an isolated system $e(t_1)=e_f$, a constant,
the numerical solution of the evolution equations becomes now easier since it does not involve the second time derivative of the correlation function.
In practice, in the
numerical algorithm we fix the total energy $e_f$ to its post-quench
value derived in Sec.~\ref{sec:energy-before-after},
and we then integrate the set of coupled integro-differential
equations with a standard Runge-Kutta method.

In short, the set of equations that fully determine the evolution of the system from an initial condition in
canonical Boltzmann equilibrium at any temperature $T'$ are
\begin{eqnarray}
\left(m\partial_{t_{1}}^{2}+z(t_{1})\right)R(t_{1},t_{2}) & = &
\delta(t_1-t_2)
+
J^2 \int_{t_{2}}^{t_{1}}dt'R(t_{1},t') R(t',t_{2})
\; ,
\label{eq:dyn-eqs-R}\\
\left(m\partial_{t_{1}}^{2}+z(t_{1})\right)C(t_{1},t_{2}) & = &
J^2\int_{0}^{t_{1}}dt'R(t_{1},t') C(t',t_{2})
+ J^2\int_{0}^{t_{2}}dt'R(t_{2},t')C(t_{1},t')
\;
\nonumber\\
&&
+\frac{J J_0}{T'} \; [C(t_{1},0) C(t_2,0) - C_2(t_1,0) C_2(t_2,0)]
\; ,
\label{eq:dyn-eqs-C}
\\
\left(m\partial_{t_{1}}^{2}+z(t_{1})\right)C_2(t_{1},0) & = &
J^2\int_{0}^{t_{1}}dt' \; R(t_{1},t') C_2(t',0)
+ \frac{J J_0}{T'} \left[  q_{\rm in} C(t_1,0)  + (1-2q_{\rm in})  C_2(t_1,0)  \right]
\; ,
\label{eq:dyn-eqs-C1}
\\
z(t_{1}) & = &
-m\partial_{t_{1}}^{2}C(t_{1},t_2)\vert_{t_{2}\rightarrow t_{1}^{-}}
+
2J^2\int_{0}^{t_{1}}dt'R(t_{1},t')C(t_{1},t')
\nonumber\\
 &  &
 +
\frac{J J_0}{T'} \; [C^2(t_{1},0) - C^2_2(t_1,0)]
\label{eq:dyn-eqs-z}
\; .
\end{eqnarray}
High and low temperature initial states are distinguished by $q_{\rm in}=0$ for $T'>J_0$, and $q_{\rm in} =1-T'/J_0$
for $T'<J_0$, respectively.
The equation for $C(t_1,0)$ is just the one for $C(t_1,t_2)$ evaluated at $t_2=0$ so we do not write it explicitly.

\subsection{Constant energy dynamics}
\label{sec:constant energy}

In order to ensure constant energy dynamics we set $J=J_0$ and $m=m_0$ in this Subsection. We verify that the equations
consistently conserve the equilibrium conditions. Moreover, we derive a number of properties of the linear response function
that will be useful in the analysis of the instantaneous quenches as well.

\subsubsection{Consistency with equilibrium parameters}

The equation for $C_2(t,0)$ admits the solution $C_2(t_1,0)=q_{\rm in} = \mbox{cst}$ in the case in which no
quench is performed. Indeed, setting $C_2(t_1,0)=q_{\rm in}$ in Eq.~(\ref{eq:dyn-eqs-C1}) one has
\begin{eqnarray}
z(t_1) q_{\rm in} = J^2 q_{\rm in} \int_0^{t_1} dt' \, R(t_1,t') + \frac{J^2}{T'} [ q_{\rm in} C(t_1,0) + (1-2 q_{\rm in}) q_{\rm in}]
\; .
\end{eqnarray}
This equation has the solution $q_{\rm in}=0$, the one of the paramagnetic phase, and a non-vanishing $q_{\rm in}\neq 0$ solution
relevant in the ordered phase. Let us now focus on the case $q_{\rm in}\neq 0$.
Using FDT, a property of equilibrium, the integral can be performed, the contribution from $t'=0$ cancels the first term in the square brackets, and the
one from $t'=t_1$ combines with the second term in the square bracket;
the ensuing equation simplifies to read
\begin{equation}
z(t_1) = \frac{2J^2}{T'}  (1 - q_{\rm in}) = z_f
\; .
\end{equation}
Therefore, $z(t_1)$ is also a constant. The equation for $z(t_1)$, using FDT, becomes
\begin{eqnarray}
z(t_1) \!\!\! &=& \!\!\! T' + \frac{J^2}{T'} [C^2(t_1,t_1) - C^2(t_1,0)] + \frac{J^2}{T'} [C^2(t_1,0) - q_{\rm in}^2]
\nonumber\\
\!\!\! &=& \!\!\! T' + \frac{J^2}{T'} (1 - q_{\rm in}^2) \ = \ z_f
\; .
\end{eqnarray}
 The two equations yield  the low-temperature values
$z_f=2J$ and $q_{\rm in}=1-T'/J$.

In equilibrium $C_2(t,0)$ and $z(t_1)$ remain constant and equal to
their initial values, $q_{\rm in}$ and $z_f$.

\subsubsection{The linear response in the frequency domain}

Knowing that $z(t_1)$ remains constant in equilibrium, one can easily analyse the response equation in
Fourier space. The equation that determines its dynamical evolution is transformed into
\begin{equation}
\hat R(\omega)
=
\displaystyle{
\frac{1}{-m \omega^2 + z_f -  \Sigma(\omega)}
}
\; .
\label{Rstat}
\end{equation}
For this model $\Sigma(\omega)=J^2 \, \hat R(\omega)$ and then
\begin{equation}
\hat R(\omega)
=
\frac{1}{2 J^2} \left[ (-m\omega^2 + z_f) \pm \sqrt{(-m\omega^2 + z_f)^2 - 4 J^2} \right].
\label{eq:response_fourier_transform-text}
\end{equation}
$\hat R(\omega)$, and also $R(t)$,  are independent of temperature for $T'<T_c=J$ while they depend on temperature
through $z_f$ for $T'>T_c=J$.

The terms under the square root can be more conveniently written as functions of the special values
\begin{equation}
m\omega_{\pm}^2 = z_f\pm 2J
\label{eq:omega_pm}
\end{equation}
and the linear response is recast as
\begin{equation}
\hat R(\omega)
=
\frac{1}{2 J^2} \left[ -m\omega^2 + z_f \pm m \sqrt{(\omega_-^2-\omega^2) (\omega_+^2-\omega^2)} \right].
\label{eq:response_fourier_transform-text1}
\end{equation}
The imaginary and real parts of $\hat R(\omega)$ are then
\begin{eqnarray}
\mbox{Im} \hat R(\omega) &=&
\frac{1}{2J^2}
\left\{
\begin{array}{ll}
\ -m \sqrt{(\omega^2-\omega^2_-)(\omega_+^2-\omega^2) }
\qquad & \qquad\qquad\;\; \mbox{for} \;\; \qquad \omega_-\leq \omega \leq \omega_+
\\
0 \qquad\qquad & \qquad\qquad\;\; \mbox{otherwise}
\end{array}
\right.
\\
\mbox{Re} \hat R(\omega) &=&
\frac{1}{2J^2}
\left\{
\begin{array}{ll}
z_f - m\omega^2 -  m \sqrt{(\omega^2_- -\omega^2)(\omega_+^2-\omega^2)}
& \qquad \mbox{for} \;\; \omega \leq \omega_-
\\
z_f - m\omega^2 & \qquad \mbox{for} \;\; \omega_-\leq \omega \leq \omega_+
\\
z_f - m\omega^2+ m \sqrt{(\omega^2 - \omega^2_- )(\omega^2 - \omega_+^2)}
& \qquad \mbox{for}  \;\; \omega_+ \leq \omega
\end{array}
\right.
\end{eqnarray}
(note the unusual choice of sign for the imaginary part that we adopted.)
In terms of the physical parameters,
$\hat R(\omega)$ is real for $|-m\omega^2+z_f|> 2J$.
In the low temperature phase, since $z_f=2J$, this implies $\omega_-=0$ and the imaginary part
of the linear response is gapless.
\comments{
This condition implies
\begin{eqnarray}
\mbox{Im} \hat R(\omega) = 0 \qquad\mbox{for}\qquad
m\omega^2 \geq 4J \qquad \mbox{if} \qquad T'\leq T_c=J
\end{eqnarray}
and
\begin{eqnarray}
\mbox{Im} \hat R(\omega) = 0 \qquad\mbox{for}\qquad
\left\{
\begin{array}{l}
m\omega^2 \leq \displaystyle{\left(\sqrt{T'} - \frac{J}{\sqrt{T'}}\right)^2} \equiv m{\omega_c^+}^2
\\
m\omega^2 \geq \displaystyle{\left(\sqrt{T'} + \frac{J}{\sqrt{T'}}\right)^2} \equiv m{\omega_c^-}^2
\end{array}
\right.
\qquad \mbox{if} \qquad T' \geq T_c=J
\; .
\end{eqnarray}
}
One can easily check that $|\hat R(\omega)|^2 = 1/J^2$ in the interval $\omega_-\leq \omega \leq \omega_+$.
Away from this interval the modulus of the linear response is a complicated function of the frequency.

The zero frequency linear response
\begin{equation}
\hat R(\omega=0) = \int_0^\infty d\tau \, R(\tau)
=
\displaystyle{\frac{1}{2 J^2}}
\left[ z_f \pm \sqrt{z_f^2 - 4 J^2} \right]
\; ,
\label{eq:Romega0}
\end{equation}
with $\tau$ a time delay, must be a real quantity
and this form is a manifestation of the condition $z_f \geq 2J$.
The static susceptibility is then given by
\begin{eqnarray}
\chi_{\rm st} =
\left\{
\begin{array}{ll}
1/J & \qquad \mbox{for} \quad T' < T_c=J
\\
1/T' & \qquad \mbox{for} \quad T' > T_c=J
\end{array}
\right.
\end{eqnarray}
with the result in the second line being ensured by the choice of minus sign in front of the square root.
The frequency-dependent linear response can then be transformed back to real-time and thus get its
full time-evolution.

In the lower limit of the spectrum the imaginary part of the linear response goes as

\begin{align}
\mbox{Im} \hat R(\omega_{-} + \epsilon) \sim
\left\{
\begin{array}{ll}
\displaystyle{ - \frac{1}{J} \sqrt{\frac{m}{J}}} \; \ \epsilon \qquad & \quad \mbox{at} \quad T<T_c
\\
\displaystyle{ - \frac{( 2 m J \omega_{-})^{1/2}}{J^2} }
\sqrt{\epsilon} & \quad \mbox{at} \quad T>T_c
\end{array}
\right. \quad & \qquad \mbox{as} \qquad \epsilon \rightarrow 0^{+}
\end{align}
while in the upper limit it vanishes as
\begin{eqnarray}
\mbox{Im} \hat R(\omega_{+} - \epsilon) \sim
\displaystyle{
- \frac{( 2 m J \omega_{+})^{1/2}}{J^2} } \sqrt{\epsilon}
\qquad & \mbox{as} \qquad \epsilon \rightarrow 0^{+}
\end{eqnarray}
with the corresponding $\omega_{\pm}$ at $T>T_c$ or $T<T_c$.

\subsubsection{The correlation functions}

The FDT in the frequency domain $\mbox{Im} \hat R(\omega) = - \beta \omega \hat C(\omega)$
implies that $\hat C(\omega)$ should vanish in the same frequency intervals in which the linear response is real.
In the $\omega=0$ case the linear response is real and the FDT as written above only
imposes that $\hat C(\omega=0)=\int_{-\infty}^\infty dt \, C(t)$ must be finite. One has to bear in mind that in the
cases in which the correlation function approaches a non-vanishing constant $q$ asymptotically
the Fourier transform to be computed is the one of $C_{\rm st}(t_1-t_2) = C(t_1-t_2)-q$ with respect to $t_1-t_2$ and the integral
should then yield $C_{\rm st}(\omega=0)=\int_{-\infty}^\infty dt \, C_{\rm st}(t)= 1-q$; more details
are given in App.~\ref{app:asymptotic}.

Consistently with the FDT constraints discussed in the previous paragraph,
the time-dependent equation for $C_{\rm st}(t_1-t_2) = C(t_1-t_2)-q_{\rm in}$
can be treated following the steps explained in~\cite{LesHouches} and App.~\ref{app:asymptotic-corr} (that do not assume FDT)
and one finds
\begin{eqnarray}
C_{\rm st}(\omega)=J^2 C_{\rm st}(\omega) |\hat R(\omega)|^2 \; .
\end{eqnarray}
This equation has two solutions, either $C_{\rm st}(\omega)=0$ or  $|\hat R(\omega)|^2 = 1/J^2$. The latter
holds for frequencies in the interval $\omega \in [\omega_-, \omega_+]$. Outside of this interval
$C_{\rm st}(\omega)$ must vanish.

By taking the derivative of Eq.~(\ref{eq:dyn-eqs-C}) with respect to $t_2$ one readily checks that it equals
Eq.~(\ref{eq:dyn-eqs-R}) if the FDT between $R$ and $C$ is satisfied for all times and $\partial_{t_2}C_2(t_2,0)=0$ implying
\begin{equation}
C_2(t_2,0)=q_{\rm in}
\; .
\end{equation}
This last condition is a property of equilibrium as we have already discussed.

Having established that $C_2$ is a constant, Eq.~(\ref{eq:dyn-eqs-C1}) enforces $q_{\rm in}=0$,
the high temperature initial value, or
\begin{equation}
z_f = \frac{J^2}{T'} [1-C(t_1,0)] + \frac{J^2}{T'} [C(t_1,0) + (1-2q_{\rm in})] =  2J \; ,
\end{equation}
the low temperature Lagrange multiplier.

Concerning the correlation function $C(t_1,0)$,  we write it as $C(t_1,0)=C_{\rm st}(t_1,0)+q_0$ allowing for a
non-vanishing asymptotic value, $q_0$, and taking $C_{\rm st}(t_1,0)$ such that it vanishes in the long $t_1$ limit.
The equation~(\ref{eq:dyn-eqs-C}) evaluated at $t_2=0$ is then rewritten as
\begin{eqnarray}
\left(m\partial_{t_{1}}^{2}+z_f\right)[C_{\rm st}(t_{1},0) + q_0] =
J^2\int_{0}^{t_{1}}dt' \; R(t_{1},t') [C_{\rm st}(t',0)+q_0]
+\frac{J^2}{T'} \; [C_{\rm st}(t_{1},0) +q_0 - q_{\rm in}^2]
\; .
\label{eq:dyn-eqs-Cbis}
\end{eqnarray}
This equation has three terms that do not depend on $C_{\rm st}(t_1,0)$ explicitly
\begin{equation}
z_f q_0 - J^2 q_0 \lim_{t_1\to\infty} \int_0^{t_1} dt' \; R(t_1,t') - \frac{ J^2}{T'} (q_0 - q_{\rm in}^2 )
\end{equation}
and their sum should vanish  in the long $t_1$ limit. It trivially does for high temperature initial
states since $q_0=q_{\rm in}=0$ and, in equilibrium at low temperatures, we can
assume $q_0=q_{\rm in}$, use FDT,  and find
\begin{equation}
z_f =\frac{2J^2}{T'} (1-q_{\rm in}) =2J
\end{equation}
confirming the assumption $q_0=q_{\rm in}$. The remaining equation fixes the time-dependence of
$C(t_1,0)$.

\subsection{The energy before and after a quench}
\label{sec:energy-before-after}

For the sake of completeness, we compute the energy variation due to a simultaneous change of the mass
$m_0\to m$ and the variance of the interaction strengths $J_0\to J$. In the applications and numerical tests we will
focus on the latter changes only.

The kinetic energy density  before the quench is
\begin{equation}
e_{\rm kin}(0^-) = \frac{1}{N} \sum_{i=1}^N \frac{m_0}{2} (\dot s_i(0^-))^2 = \frac{T'}{2}
\; ,
\end{equation}
the last equality being due to the fact that we take equilibrium initial conditions at temperature $T'$.
The potential energy density before the quench depends on the system being paramagnetic or condensed initially:
\begin{eqnarray}
\begin{array}{rcll}
e_{\rm pot}(0^-) &=&
\displaystyle{ - \frac{J_0^2}{2T'} \left[ 1 - q_{\rm in}^2(T'/J_0)\right] =
 - \frac{J_0^2}{2T'} \left[ 1 - \left(1-\frac{T'}{J_0}\right)^2\right] }
\qquad\quad
&\mbox{condensed}
\qquad
\\
e_{\rm pot}(0^-) &=&
\displaystyle{ - \frac{J_0^2}{2T'} }
\qquad\qquad\qquad\qquad\qquad\qquad\qquad\qquad\qquad\qquad\qquad\quad
& \mbox{paramagnetic}
\qquad
\end{array}
\end{eqnarray}

The kinetic energy density right after the quench is
\begin{equation}
e_{\rm kin}(0^+) = \frac{1}{N} \sum_{i=1}^N \frac{m}{2} (\dot s_i(0^+))^2
\end{equation}
and, since the velocities do not change in the infinitesimal interval taking from $0^-$ to $0^+$,
\begin{equation}
e_{\rm kin}(0^+) = \frac{1}{N} \sum_{i=1}^N \frac{m}{2} (\dot s_i(0^+))^2 =  \frac{1}{N} \sum_{i=1}^N \frac{m}{2} (\dot s_i(0^-))^2
= \frac{m}{m_0} \frac{T'}{2} \; .
\end{equation}
The post-quench potential energy density can be estimated from the relation between the Lagrange multiplier and the
energy
\begin{equation}
 e_{\rm pot}(0^+) = - \frac{1}{2} \; z(0^+) + e_{\rm kin}(0^+)
\end{equation}
and the equation for $z(t_1)$ (see Sec.~\ref{subsec:dynamic-eq})
\begin{equation}
z(t_1) = - m \lim_{t_2\to t_1^-} \partial^2_{t_1} C(t_1,t_2) + \frac{J J_0}{T'} [C^2(t_1,0) - C_2^2(t_1,0)]
\; .
\end{equation}
Thus
\begin{equation}
e_{\rm pot}(0^+) = - \frac{J J_0}{2T'} \lim_{t_1\to 0^+}[C^2(t_1,0) - C_2^2(t_1,0)]
\; .
\end{equation}
Assuming that the spin configuration did not change between the infinitesimal time
step going from 0 to $0^+$,
\begin{eqnarray}
\begin{array}{rcl}
e_{\rm pot}(0^+) &=&
\displaystyle{- \frac{JJ_0}{2T'} \left[ 1 - \left(1-\frac{T'}{J_0}\right)^2\right]}
\qquad\qquad
\mbox{condensed}
\\
e_{\rm pot}(0^+) &=&
\displaystyle{ - \frac{JJ_0}{2T'} }
\qquad\qquad\qquad\qquad\qquad\qquad
\mbox{paramagnetic}
\end{array}
\end{eqnarray}
All the values just derived imply the changes in the total energy
\begin{eqnarray}
\begin{array}{rcl}
\Delta e_{\rm tot} &=&
\displaystyle{
\Delta e_{\rm kin} + \Delta e_{\rm pot} =
\left( \frac{m}{m_0}-1\right) \frac{T'}{2}
-
\frac{J_0 (J-J_0)}{2T'}
\left[ 1
-
\left( 1-\frac{T'}{J_0} \right)^2
\right]
}
\; ,
\qquad\qquad
\\
\Delta e_{\rm tot} &=&
\displaystyle{
\Delta e_{\rm kin} + \Delta e_{\rm pot} =
\left( \frac{m}{m_0}-1\right) \frac{T'}{2}
-
\frac{J_0 (J-J_0)}{2T'}
}
\; ,
\qquad\qquad\qquad\qquad\qquad\qquad\;
\end{array}
\end{eqnarray}
respectively.

We will concentrate on potential energy quenches only, and we will trace the phase diagram using the parameters
\begin{equation}
y = \frac{T'}{J_0} \qquad\qquad\qquad x=\frac{J}{J_0} \; .
\end{equation}
$x>1$ corresponds to energy extraction and $x<1$ to energy injection. We will show that the parameter space is split in several sectors
displaying fundamentally different dynamics.

\subsection{Asymptotic analysis of the quench dynamics}
\label{sec:asymptotic}

We now study the full set of equations (\ref{eq:dyn-eqs-R})-(\ref{eq:dyn-eqs-z}), derived in the $N\to\infty$ limit,
that couple the correlation $C$ and linear response $R$ functions.
Using a number of hypotheses that we carefully list below, and that are not always satisfied by the actual evolution
 found with the numerical integration, we deduce some properties of the
Lagrange multiplier, linear response and correlation function, in the long time limit. In this Section we
state the assumptions, we summarise the results, and we leave most details of how these are derived to
App.~\ref{app:asymptotic}.

Consider the system in equilibrium at $T'$ with parameters $J_0,\, m_0$ and let it
evolve in isolation with parameters $J, \, m$.
We will assume that the dynamics approach a long times limit in which
one-time quantities, such as the Lagrange multiplier, reach a constant.
Later we will further suppose that  (in most cases)  the correlation function becomes, itself, invariant under time-translations.
Finally, we will explore in which circumstances a fluctuation-dissipation theorem (FDT) can establish
with respect to a temperature $T_f$ for all time-delays, in stationary cases, or for correlation values that are in the
stationary regime, when we look for ageing solutions.

These assumptions are not obvious and, as we will show analytically in some cases and numerically in the next Section, do
not apply to all quenches.
Still, we find useful to explore their consequences and derive from them a set of relations between the control
parameters for which special behaviour arises,  that we will
later put to the numerical test.

\subsubsection{Asymptotic values in a steady state}

Let us assume that the limiting value of the Lagrange multiplier is a constant
\begin{equation}
\lim_{t_1\to\infty} z(t_1) = z_f
\; .
\end{equation}

The limit of the correlation function can be zero if the system decorrelates completely at sufficiently long
time-delays or non-zero if it remains within a confined state. We therefore call
\begin{equation}
q = \lim_{t_1-t_2\to\infty} \lim_{t_2\to \infty} C(t_1,t_2)
\end{equation}
the asymptotic value of the full two time correlation, or its stationary part in possible ageing cases,
after the quench. Similarly,
\begin{eqnarray}
q_0 &=&  \lim_{t_1\to \infty} C(t_1,0)
\; ,
\\
q_2 &=& \lim_{t_1\to\infty}C_2(t_1,0)
\; .
\end{eqnarray}

\subsubsection{The linear response function}
\label{subsubsec:thelinearresponse}

\vspace{0.25cm}

The equation for the linear response function does not depend explicitly on the pre-quench
parameters, it does only on the post-quench mass $m$ and interaction strength $J$. (An implicit dependence
on the initial state is not excluded, since the value taken by the Lagrange multiplier may depend on it.)
The analysis
that we developed for the constant energy dynamics applies to the sudden quench case too.
The solution of the response equation in
Fourier space yields
\begin{equation}
\hat R(\omega)
=
\frac{1}{2 J^2} \left[ (-m\omega^2 + z_f) \pm \sqrt{(-m\omega^2 + z_f)^2 - 4 J^2} \right].
\label{eq:response_fourier_transform}
\end{equation}
with the zero frequency value
\begin{equation}
\hat R(\omega=0)
=
\displaystyle{\frac{1}{2 J^2}}
\left( z_f - \sqrt{z_f^2 - 4 J^2} \right)
\; .
\label{eq:Romega00}
\end{equation}

From the numerical solution of the full equations that we will present in Sec.~\ref{sec:numerical}
we infer that for $m=m_0$ and $J\neq J_0$
\begin{eqnarray}
z_f
=
\left\{
\begin{array}{ll}
2J \qquad  & \qquad x>y \; ,
\\
T' + J^2/T' \qquad  & \qquad x<y \; ,
\end{array}
\right.
\label{eq:prediction-zf}
\end{eqnarray}
 with, we recall, $x=J/J_0$ and $y=T'/J_0$. Replacing in Eq.~(\ref{eq:Romega00}) one notices that the minus sign has to be
selected for $x<y$ at low frequency and
\begin{eqnarray}
\chi_{\rm st} \equiv \hat R(\omega=0)
=
\left\{
\begin{array}{ll}
1/T'  & \qquad x<y \; ,
\\
1/J & \qquad x>y \; ,
\end{array}
\right.
\label{eq:prediction-Romega}
\end{eqnarray}
where we called $\chi_{\rm st}$, as a static susceptibility, the zero frequency response.
After the quench Im$\hat R(\omega)$ is non-zero in a finite interval of frequencies $[\omega_-,\omega_+]$ with
$m\omega^2_{\pm} = (z_f\pm 2J)$ as in Eq.~(\ref{eq:omega_pm}) and $z_f$ taking the values in Eq.~(\ref{eq:prediction-zf}).
Therefore, Im$\hat R(\omega)$ is gapless for $x>y$ and it is gapped for $x<y$.

These results are exact and do not assume anything apart from a long-time limit in which $z_f$ is time-independent
and given by Eq.~(\ref{eq:prediction-zf}).
We have verified them with the complete numerical solution of the $N\to\infty$ dynamic equations, see Sec.~\ref{sec:numerical},
on all sectors of the phase diagram. We can now Fourier back to real time to get the full time dependence of the linear response
function. In the numerical Section we will compare this functional form, named $R_{\rm st}$, to the outcome of the
full integration of the dynamic equations.

\subsubsection{The asymptotic kinetic and potential energies}
\label{app:energies}

From the relation between $z$ and the energies, the
conservation of energy, and Birkhoff's theorem,
\begin{eqnarray}
\overline z = 2 \overline e_{\rm kin} - 2 \overline e_{\rm pot}
\; ,
\qquad\qquad
e_{\rm tot}(0^+) = \overline e_{\rm kin} + \overline e_{\rm pot}
\; ,
\end{eqnarray}
where the overlines represent a long-time average defined in Eq.~(\ref{eq:time-average}).
Note that the Lagrange multiplier takes the form of an action density, as a difference between
kinetic and potential energy densities.

Using these relations one derives the parameter dependence of the
kinetic and potential energies in the four relevant regions of the phase diagram
parametrised by $x=J/J_0$ and $y=T'/J_0$.

 \vspace{0.25cm}

(I) $y>1$ and $x<y$ ($q_{\rm in}=0$, $z_f=T'+J^2/T'$, $\chi_{\rm st}=1/T'$)
\begin{eqnarray}
\begin{array}{rcl}
4{\overline e_{\rm kin}} \!\! & \! = \! & \!\!
\displaystyle{
 - \frac{JJ_0}{T'} + 2T' + \frac{J^2}{T'} \ = \ J_0 \left( - \frac{x}{y} + 2y + \frac{x^2}{y}\right)
 }
\\
4{\overline e_{\rm pot}} \!\! & \! = \! & \!\!
\displaystyle{
- \frac{JJ_0}{T'} -  \frac{J^2}{T'} \ = \ J_0 \left( - \frac{x}{y} - \frac{x^2}{y}\right)
}
\label{eq:I}
\end{array}
\end{eqnarray}

(II) $y>1$ and $x>y$ ($q_{\rm in}=0$, $z_f=2J$, $\chi_{\rm st}=1/J$)
\begin{eqnarray}
\begin{array}{rcl}
4{\overline e_{\rm kin}} \! & \! = \! & \!
\displaystyle{
 - \frac{JJ_0}{T'} + T' + 2J \ = \ J_0 \left( -\frac{x}{y} + y + 2 x\right)
 }
\\
4{\overline e_{\rm pot}} \! & \! = \! & \!
\displaystyle{
- \frac{JJ_0}{T'} + T' - 2J \ = \  J_0 \left(- \frac{x}{y} + y - 2 x\right)
}
\end{array}
\end{eqnarray}

(III) $y<1$ and $x>y$ ($q_{\rm in}\neq 0$, $z_f=2J$, $\chi_{\rm st}=1/J$)
\begin{eqnarray}
\begin{array}{rcl}
4{\overline e_{\rm kin}}\!\! & \! = \! & \! \!
\displaystyle{
T' + \frac{T'J}{J_0} \ = \  J_0 \left( y + yx \right)
}
\\
4{\overline e_{\rm pot}}\!\! & \! = \! & \!\!
\displaystyle{
 T' + \frac{T'J}{J_0} -4J \ = \ J_0 \left( y + yx - 4x\right)
 }
 \end{array}
 \label{eq:asymptotic-kin-pot-energy-III}
\end{eqnarray}

(IV) $y<1$ and $x<y$ ($q_{\rm in} \neq 0$, $z_f = T'+J^2/T'$, $\chi_{\rm st}=1/T'$)
\begin{eqnarray}
\begin{array}{rcl}
4{\overline e_{\rm kin}} \! & \! = \! & \!
\displaystyle{
\frac{JT'}{J_0} - 2J  + 2 T' + \frac{J^2}{T'} \ = \  J_0 \left( xy - 2 x + 2 y + \frac{x^2}{y} \right)
}
\\
4{\overline e_{\rm pot}} \! & \! = \! & \!
\displaystyle{
\frac{JT'}{J_0} - 2J  - \frac{J^2}{T'} \ = \  J_0 \left( xy - 2 x - \frac{x^2}{y} \right)
}
\label{eq:IV}
\end{array}
\end{eqnarray}
The minimum potential energy density, $e_{\rm pot}=-J$ is realised for $T'=0$ in Sector III.

We stress that we have found these results without using FDT and they can therefore hold out of thermal equilibrium.
We will investigate later which other conditions impose the use of FDT, a strong Gibbs-Boltzmann equilibrium condition.

\subsubsection{Kinetic temperature from the kinetic energy density}

We can identify a kinetic temperature from the kinetic energy densities derived in the previous Subsection by simply imposing
$T_{\rm kin} = 2\overline{e}_{\rm kin}$. This operation leads to
\begin{eqnarray}
\begin{array}{rclrcl}
\displaystyle{T^{(\mbox{\scriptsize I})}_{\rm kin}} &=&
 \displaystyle{- \frac{JJ_0}{2T'} + T' + \frac{J^2}{2T'}}
 \; ,
 \qquad\qquad\qquad\;
  T^{(\mbox{\scriptsize II})}_{\rm kin} &=& \displaystyle{- \frac{JJ_0}{2T'} + \frac{T'}{2} + J }
  \; ,
 \vspace{0.15cm}
 \\
\displaystyle{T^{(\mbox{\scriptsize III})}_{\rm kin}} &=& \displaystyle{\frac{T'}{2} \left(1+\frac{J}{J_0}\right) }
\; ,
\qquad\qquad\qquad\qquad
\displaystyle{T^{(\mbox{\scriptsize IV})}_{\rm kin}} &=& \displaystyle{\frac{JT'}{2J_0} - J  + T' + \frac{J^2}{2T'} }
\; ,
\end{array}
\label{eq:kin-temps}
\end{eqnarray}
in the four Sectors of the phase diagram distinguished in the previous Subsection.

\subsubsection{Final temperature under thermal equilibrium assumption}

In App.~\ref{app:final-temp-anal} we explain how we can exploit the conservation of the total energy, under the assumption that
the asymptotic kinetic and potential energies take the form of Gibbs-Boltzmann equilibrium paramagnetic and condensed equilibrium phases
at a single temperature $T_f$ to fix its value. This means that we require  $2\overline e_{\rm kin} = T_f$ {\it and}
\begin{eqnarray}
\overline{e}_{\rm pot} &=&
\left\{
\begin{array}{lll}
\displaystyle{-\frac{J^2}{2T_f}} \qquad\qquad &  & \mbox{paramagnetic or}
\vspace{0.2cm}
\\
\displaystyle{-\frac{J^2}{2T_f} (1-q^2)} \qquad\qquad &  & \mbox{condensed or two-step} \; ,
\end{array}
\right.
\end{eqnarray}
with $q=1-T_f/J$.
For $x>y$ we find
\begin{eqnarray}
\frac{T_f}{J_0} &=&
\left\{
\begin{array}{ll}
\displaystyle{\frac{J^2}{J_0} \; \left[
J -\frac{T'}{2} - \frac{JJ_0}{2T'} \right]^{-1}
}
\qquad\qquad & \mbox{for} \;\; 1\leq y  \;\; \mbox{(IIa)} \; ,
\vspace{0.2cm}
 \\
\displaystyle{
\frac{J}{J_0} +\frac{T'}{2J_0} - \frac{J}{2T'}
=x +\frac{y}{2} - \frac{x}{2y} }
\qquad\qquad & \mbox{for} \;\; 1\leq  y \;\; \mbox{(IIb)} \; ,
\vspace{0.2cm}
 \\
 \vspace{0.2cm}
\displaystyle{
\frac{T'}{2J_0} \left( 1+\frac{J}{J_0}\right)
=
\frac{y}{2} \left( 1+x \right)
}
\qquad\qquad & \mbox{for} \;\; y\leq 1 \;\; \mbox{(III)} \; .
\end{array}
\right.
\label{eq:Tf-asymptotic-values}
\end{eqnarray}
The roman numbers between parenthesis refer to the
cases listed in Eqs.~(\ref{eq:I})-(\ref{eq:IV}) and to the four sectors in the phase diagram in Fig.~\ref{fig:phase_diagram_new}.
The difference between (IIa) and (IIb) is that in the first case we used a paramagnetic potential energy and in the
second case a condensed one at $T_f$. The values for $x<y$ are given in App.~\ref{app:final-temp-anal}.
One can check that in the cases (IIb) and (III) $T_f=T_{\rm kin}$, the kinetic temperatures given in Eq.~(\ref{eq:kin-temps}). Instead,
final and kinetic temperatures do
not coincide in (IIa) nor in (I) and (IV). This fact excludes the possibility of reaching Gibbs-Boltzmann equilibrium in
(I) and (IV), that is, for $x<y$ and it leaves the possibility open in (II) and (III) at the price of considering a
potential energy density with a non-vanishing $q=1-T_f/J$.

\vspace{0.25cm}

\noindent
{\it Limits of validity}

\vspace{0.25cm}

The two bounds
\begin{equation}
0 \leq T_f \leq  J
\end{equation}
serve to find special curves on the phase diagram.
The first bound is natural since we do not want to have a negative kinetic energy.
The second one ensures that $q\geq 0$. The implications of these bounds, that are
examined in App.~\ref{app:critical-lines}, are
\begin{eqnarray}
y  &\leq&
\left\{
\begin{array}{ll}
\sqrt{x} & \qquad \qquad \mbox{for} \;\; y\geq 1
\; ,
\vspace{0.2cm}
\\
\displaystyle{\frac{2x}{x+1}}  & \qquad \qquad \mbox{for} \;\; y\leq 1
\; .
\end{array}
\right.
\end{eqnarray}
They mean that an asymptotic state with a single temperature $T_f$, or a double
regime with temperature $T_f$ for $C:1 \to q$ and $T_{\rm eff}\to\infty$ for $C:q\to 0$,
could only exist below the piecewise curve $y(x)$. One can simply check that the piece for $y\leq 1$ lies in Sector IV,
since $y<x$, and it is therefore irrelevant given  that we have already shown that there cannot be a single
temperature scenario in this Sector. The limit then moves to $x=y$ for $y<1$.
Parameters on the special curve $y=\sqrt{x}$ will play a special role, as we will
show below.

\vspace{0.25cm}

\noindent
{\it Particular values}

\vspace{0.25cm}

For the moment, a single temperature scenario for the global observables in the $N\to\infty$ limit
seems possible for $y<1$ and $x>y$ (III), and below the curve $y=\sqrt{x}$ for $y>1$ in II.
It is instructive to work out the limiting values of
$T_f/J_0$ and $q=1-T_f/J$ on the borders of the region $y<1$ and $y<x$ (III) of the phase diagram,
and the no-quench case $x=1$:
\begin{eqnarray}
\frac{T_f}{J_0} \ = \
\left\{
\begin{array}{ll}
0 & \qquad y=0
\\
\displaystyle{\frac{x(x+1)}{2} }& \qquad x =y
\vspace{0.1cm}
\\
\displaystyle{ \frac{x+y}{2} } & \qquad y=1
\\
y & \qquad x=1
\end{array}
\right.
\qquad
&
\qquad
q \ = \
\left\{
\begin{array}{ll}
1 & \qquad y=0
\\
\displaystyle{\frac{1-x}{2} }& \qquad x =y
\vspace{0.1cm}
\\
\displaystyle{ \frac{x-1}{2x} } & \qquad y=1
\\
1-y & \qquad x=1
\end{array}
\right.
\qquad\quad
\mbox{(III)}
\end{eqnarray}
These values match at $x=y=1$. $q$ is larger than zero on the lines $x=y$ and $y=1$ that mark the end of what we call Sector III. Moreover, the
approximate asymptotic analysis of the mode dynamics of the finite $N$ system will lead to $T_\mu=T_f$ in Sector III,
see Eq.~(\ref{eq:mode-temp-prediction}).

On the limiting curve in Sector II, $y=\sqrt{x}$ for $y>1$, $T_f=J$ and $q=0$.

As one could have intuitively
expected, $T_f>T'$ for energy injection quenches ($x<1$) and $T_f<T'$ for energy extraction quenches ($x>1$).

\subsubsection{Results under FDT at a single temperature}

We now add one  assumption to the analysis: that
the FDT, at the single temperature $T_f$,
relates the linear response to the correlation
\begin{equation}
R(t_1-t_2) = - \frac{1}{T_f} \; \frac{dC(t_1-t_2)}{d(t_1-t_2)} \ \theta(t_1-t_2)
\label{eq:fdt}
\; .
\end{equation}

\vspace{0.25cm}

\noindent
{\it Static susceptibility}

\vspace{0.25cm}

The values of the zero frequency linear response
\begin{eqnarray}
\hat R(\omega = 0 ) = \int_0^\infty d\tau \; R(\tau) = \frac{1}{T_f} (1-q)
\; ,
\end{eqnarray}
where one must recall that $\tau$ is a time-difference, force
  \begin{eqnarray}
\hat R(\omega = 0 ) =
\left\{
\begin{array}{lllll}
1/T' \quad & \mbox{for} \qquad x<y \qquad & \Rightarrow \qquad T_f=T' \quad & \mbox{if} \quad q=0 \quad \mbox{(I)}  \; \& \; \mbox{(IV)}
\\
1/J \quad & \mbox{for}\qquad x>y \qquad & \Rightarrow \qquad T_f=J \quad & \mbox{if} \quad q=0 \quad \mbox{(IIa)}
\\
1/J \quad & \mbox{for}\qquad x>y \qquad & \Rightarrow \qquad T_f=J/(1-q)\quad & \mbox{if} \quad q\neq 0 \quad \mbox{(IIb)} \; \&  \; \mbox{(III)}
\end{array}
\right.
\end{eqnarray}
The result in the last line, valid for (IIb) and (III), does not put any additional constraint on $T_f$. Instead, the other conditions
in the first two lines are incompatible with the expressions imposed
by the energetic considerations.  They corroborate the impossibility of having a single
$T_f$ in (I) and (IV) or with $q=0$ in (II).

\vspace{0.25cm}

\noindent
{\it Limits of validity of the single temperature scenario}

\vspace{0.25cm}

As explained above and in App.~\ref{app:special-cases}, the consistency between the static susceptibility values
and the $T_f$ derived from conservation of  energy impose that FDT with a single temperature
may only hold for $x>y$ and $y<1$ (sector III) or below the special curve $y=\sqrt{x}$, for $y>1$ (lying inside sector II).
Whether this is realised or not needs to be investigated numerically.

\subsubsection{Two step (possibly ageing) {\it Ansatz}}

One can also look for a two step solution with similar characteristics to the ageing one found for dissipative dynamics~\cite{CuDe95a}
and summarised in Sec.~\ref{subsubsec:relaxation-dynamics}.
Asking for the relation between correlation and linear response in Eq.~(\ref{eq:fdt}) to hold in a stationary regime of relaxation
in which $C$ decays from $1$ to $q$, and that the effective temperature $T_{\rm eff}$ characterising the second regime of decay from $q$ to $0$
diverges, one recovers
 \begin{eqnarray}
\begin{array}{lll}
z_f = 2 J \; , \qquad \chi_{\rm st} = \displaystyle{\frac{1}{J}} \; , \qquad & \displaystyle{q = 1- \frac{T_f}{J}} \quad\qquad\;\; \mbox{and} \qquad & q_0 = q_2=0
\end{array}
\end{eqnarray}
with the same $T_f$ and $q\neq 0$ as in Eq.~(\ref{eq:Tf-asymptotic-values}).

\subsubsection{The correlation function}

From the analysis of the equation ruling the two-time correlation function, assuming stationarity and hence a dependence on
time-difference only, one deduces  (the details of the derivation are
given in App.~\ref{app:asymptotic-corr}, see also~\cite{LesHouches} for a general treatment)
\begin{equation}
\hat C_{\rm st}(\omega) = J^2 \hat C_{\rm st}(\omega) |\hat R(\omega)|^2
\label{eq:SD-corr}
\end{equation}
where $\hat C_{\rm st}(\omega)$ is the Fourier transform of the time-varying part (subtracting the
possible non-vanishing asymptotic value $q$). Note that the relation between the correlation and the linear
response is the same as the one that we derived in the constant energy no-quench problem.
It implies
\begin{eqnarray}
\hat C_{\rm st}(\omega) \neq 0 \;\; \mbox{and} \;\; |\hat R(\omega)|^2 = 1/J^2 \qquad \mbox{or} \qquad
\hat C_{\rm st}(\omega) = 0
\; ,
\end{eqnarray}
independently of the control parameters. Below we check numerically that these relations
are satisfied in various quenches.
In particular, from the analytic form of $\hat R(\omega)$ one can easily see that
$|\hat R(\omega)|^2 = 1/J^2 $ in the frequency interval in which the linear response is complex.

Importantly enough,   we cannot yield an explicit analytic form of $\hat C(\omega)$ since it is factorised on both sides of the
identity (\ref{eq:SD-corr}). We are forced to go back to the full set of dynamic equations and solve them numerically
to get insight on the behaviour of $C(t_1,t_2)$.

\subsubsection{Numerical results preview}

We will see that a state with a single $T_f$ characterising the fluctuation-dissipation relation
is reached numerically in the following two cases only:

(1) the dynamics are run at constant parameters (no quench), $x=1$;

(2) the special relation $y=\sqrt{x}$ between parameters holds and $y>1$ (within sector II).

In all other cases no equilibrium results {\it \`a la} Gibbs-Boltzmann are found for the global observables (correlation functions,
linear response functions, kinetic and potential energies) but a different statistical description,
of a generalised kind, should be adopted. In particular, in Sector III where the conditions derived from the
energy conservation and static susceptibility allowed for a single temperature scenario, the
full solution of the complete set of questions will prove that this is not realised.
A detailed explanation is given in the numerical section of the paper and in the
analysis of the $N-1$ integrals of motion presented in Sec.~7.

We recall that the $p\geq 3$ strongly interacting case behaves very differently~\cite{CuLoNe17}. On the one hand,
equilibrium towards a proper paramagnetic state, and within confining metastable states, were
reached in two sectors of its dynamic phase diagram. On the other hand, an  ageing asymptotic state
in a tuned regime of parameters was also found for more than two spin interactions in the
potential energy. In the $p=2$ integrable model we do not find an ageing asymptotic state. Moreover,
Gibbs-Boltzmann equilibrium is achieved in the two very particular cases listed above only.


\section{Analytic results for the dynamics of the finite size system}
\label{sec:analytic-finiteN}

In this Section we describe how the finite system size dynamics can be solved by
using a convenient basis in which the evolution becomes the one of harmonic
oscillators coupled only through the Lagrange multiplier. We show that these oscillators decouple under the assumption
$z(t) \to z_f$ allowing for a simple approximate solution of the problem that can, however, be relevant for $N\to\infty$ only. We
then explain a way to numerically solve the dynamics for finite $N$.

\subsection{Newton equations in a rotated  basis}

Take a system with finite $N$. The post-quench matrix $J_{ij}$ has $\mu = 1, . . . , N$ eigenmodes with
eigenvalues $\lambda_\mu$. If we denote
\begin{equation}
s_{\mu}(t)=\vec{s}(t) \cdot \vec{v}_{\mu}
\end{equation}
the projection of the spin vector in the direction of the $\mu$-th eigenvector of $J_{ij}$, the $N$ rotated equations of motion read
\begin{equation}\label{eq:newton_eq0}
m \ddot{s}_{\mu}(t)+(z(t)-\lambda_{\mu})s_{\mu}(t)=0
\; .
\end{equation}
This set of equations has to be complemented with the initial conditions $s_\mu(0)$ and $\dot s_\mu(0)$. They are
very similar to the equations for a parametric oscillator, the difference being that, in our case,
the time-dependent frequency depends on the variables {\it via} the Lagrange multiplier. Furthermore, they are identical to the
equations of the Neumann's integrable classical system~\cite{Neumann}, see Sec.~\ref{sec:Neumann}.

Once the equations of motion for the $s_{\mu}$ are solved,
we can recover the trajectories for $\vec{s}$ using $\vec{s}(t)=\sum_{\mu}s_{\mu}(t) \vec{v}_{\mu}$. In particular, the correlation function is given by
\begin{equation}
C_{J}(t_1,t_2)=\frac{1}{N}\sum_{\mu} \langle s_{\mu}(t_1)s_{\mu}(t_2) \rangle \; ,
\end{equation}
where the subscript $J$ means that the result depends, in principle, on the interaction matrix chosen, and
the angular brackets represent an average over initial conditions. One could then perform the disorder average
or analyse the self-averageness properties of the correlation in different time regimes. The $J$ dependence should disappear
in the $N\to\infty$ limit.

Since we are interested in an uniform interaction quench, it is easy to see that
\begin{equation}
\lambda^{(0)}_{\mu}=\frac{J_0}{J}\lambda_{\mu}
\; .
\end{equation}

\subsection{Behaviour under stationary conditions}
\label{sec:st_condtions_finite_n}

Let us assume that the system reaches stationarity and that  the Lagrange multiplier approaches a
constant
\begin{equation}
\lim_{t\to\infty} z(t) = z_f
\; .
\end{equation}
In order to simplify the notation, in the rest of this section we will measure time with respect to a
reference time $t_{\mathrm{st}}$ for which the stationary regime for $z(t)$ has already been established.
(We insist upon the fact that this assumption can only hold for $N\to\infty$.)

The equation of motion of each mode becomes
\begin{equation}
m\ddot s_\mu(t) + (z_f-\lambda_\mu) s_\mu(t) =0
\end{equation}
and can be thought of as Newton's equation for the mode Hamiltonian
\begin{equation}
H_\mu = \frac{1}{2} m {\dot s}_\mu^2 + \frac{1}{2} m \omega_\mu^2 s_\mu^2
\end{equation}
with $\omega_{\mu}^2 \equiv (z_f-\lambda_\mu)/m$.
This equation has three types of solutions depending on the sign of $\omega_\mu^2$:
\begin{eqnarray}
\begin{array}{lll}
\omega_{\mu}^2 > 0
&
\qquad\qquad
s_\mu(t) = s_\mu(t_{\rm st}) \cos \left[\omega_{\mu} (t-t_{\mathrm{st}})\right] + \frac{\dot s_\mu(t_{\rm st})}{\omega{\mu}} \sin\left[\omega_{\mu} (t-t_{\mathrm{st}})\right]
\; ,
\vspace{0.1cm}
\\
\omega_{\mu}^2 =0
&
\qquad\qquad s_\mu(t) =s_\mu(t_{\rm st}) + \dot s_\mu(t_{\rm st}) (t-t_{\mathrm{st}})
\; ,
\vspace{0.1cm}
\\
\omega_{\mu}^2 < 0
&
\qquad\qquad
s_\mu(t) =
\frac{1}{2}
\left( s_\mu(t_{\rm st}) - \frac{\dot s_\mu(t_{\rm st})}{|\omega_{\mu}|} \right) e^{-|\omega_{\mu}| (t-t_{\mathrm{st}})} +
\frac{1}{2}
\left( s_\mu(t_{\rm st}) + \frac{\dot s_\mu(t_{\rm st})}{|\omega_{\mu}|} \right)  e^{|\omega_{\mu}| (t-t_{\mathrm{st}})}
\; ,
\end{array}
\label{eq:harm-osc-const-freq}
\end{eqnarray}
that is to say, oscillatory solutions with constant amplitude in the first case, diffusive behaviour in the
intermediate case and exponentially diverging solutions in the last case. We insist upon the
fact that the initial time here is taken to be $t_{\rm st}$ the time needed to reach the stationary state.

If the Lagrange multiplier approaches, then, a value that is larger than $\lambda_N$, all modes oscillate indefinitely. In
Gibbs-Boltzmann equilibrium in the PM phase, $z_{\rm eq}>\lambda_N$ and such a fully oscillating behaviour is
expected. In equilibrium in the low temperature condensed phase $z_{\rm eq}=\lambda_{\rm max}=\lambda_N$ for $N\to\infty$
and the $\mu=N $ mode  should grow linearly in time while all other modes should oscillate with frequency
$\omega_{\mu} = \sqrt{(z_f-\lambda_\mu)/m}$.
The amplitude of each mode is determined by the initial conditions, that are actually matching conditions at time $t_{\rm st}$,
when stationarity is reached in this case. (Recall that $\lambda_{N-1}$ is at distance
$N^{-2/3}$ from $\lambda_N$~\cite{Mehta}. This means that, under the assumption $z_f\to\lambda_N$,
$z_f-\lambda_{N-1} = \lambda_N - \lambda_{N-1} \simeq N^{-2/3}$ and there will be
almost diffusive modes close to the largest one in the large $N$ limit.) However, the simulations at finite $N$ show that
for finite $N$, $z_f$ is always greater than $\lambda_N$ and all modes are oscillatory. For ``condensed-type" dynamics $z_f$ will still be greater than
$\lambda_{N}$, although very close to it. The diffusive behaviour of the $N$th mode (in the $N\rightarrow\infty$ limit) would be
obtained as the limit of zero frequency of a (finite $N$) oscillating $N$th mode.

\subsection{Mode observables}

At variance with the $N\to\infty$ approach, the finite size study allows to access the details of the dynamics
of each mode. In this section we define some mode-observables that will provide valuable information.
Of particular interest are the mode energies, which can be defined as
\begin{eqnarray}
\begin{array}{rcl}
\epsilon^{\mathrm{kin}}_{\mu}(t)&=&
\displaystyle{ \frac{m}{2}\langle\dot s^2_{\mu}(t)\rangle} \; ,
\vspace{0.15cm}
\\
\epsilon^{\mathrm{pot}}_{\mu}(t)&=&
\displaystyle{ \frac{1}{2}(z(t)-\lambda_{\mu})\langle s^2_{\mu}(t)\rangle} \; ,
\vspace{0.15cm}
\\
\epsilon^{\mathrm{tot}}_{\mu}(t)&=&
\displaystyle{ e^{\mathrm{kin}}_{\mu}(t)+e^{\mathrm{pot}}_{\mu}(t) } \; .
\end{array}
\end{eqnarray}
Note that in the analysis of the $N\to\infty$ model
the potential energy density is $e_{\rm pot} = -1/(2N) \sum_\mu \lambda_\mu \langle \, s_\mu^2 \, \rangle$ without the term proportional to $z(t)$. For this
reason we use here the different symbol $\epsilon_\mu^{\rm pot}$ for the mode potential energies that include the term proportional to $z(t)$.
The values of these energies at $t=0^-$ are given by the fact that all modes are in equilibrium at the same temperature:
\begin{equation}
\epsilon^{\mathrm{tot}}_{\mu}(0^-)=2\epsilon^{\mathrm{kin}}_{\mu}(0^-)=2\epsilon^{\mathrm{pot}}_{\mu}(0^-)=T^{\prime} \; .
\end{equation}
Immediately after the quench, they are
\begin{eqnarray}
\epsilon^{\mathrm{kin}}_{\mu}(0^+)=\frac{m}{2}\langle\dot s^2_{\mu}(0^+)\rangle \; ,\qquad\qquad
\epsilon^{\mathrm{pot}}_{\mu}(0^+)=\frac{T^{\prime}}{2}\frac{z(0^+)-\lambda_{\mu}}{z(0^-)-\lambda^{(0)}_{\mu}} \; .
\end{eqnarray}

In order to study the eventual thermalisation of the system, we can define an effective time dependent mode temperature
through the total mode energy
\begin{equation}
T_{\mu}(t)\equiv \epsilon^{\mathrm{tot}}_{\mu}(t)
\end{equation}
based on the fact that the modes are (quasi) decoupled.
Whenever the system enters a stationary regime in which $z(t)$ is constant, see Section~\ref{sec:st_condtions_finite_n},
the mode temperatures $T_{\mu}$ are independent of time, since the system
behaves as a collection of non-coupled harmonic oscillators. We can also define mode temperatures using
the kinetic and potential mode energies that oscillate around their mean if we take their time-average~\cite{Khinchin}
\begin{equation}
{\overline e}^{\rm kin,pot}_\mu = \lim_{1\ll \tau} \frac{1}{\tau} \int_{t_{\rm st}}^{t_{\rm st}+\tau} dt' \, e^{\rm kin,pot}_\mu(t')
\end{equation}
and propose
\begin{equation}\label{eq:mode_temps2}
T^{\mathrm{kin},\mathrm{pot}}_{\mu}\equiv2\overline{\epsilon}^{\mathrm{kin},\mathrm{pot}}_{\mu}.
\end{equation}
In the stationary regime, as shown in Section~\ref{sec:st_condtions_finite_n}, the different mode temperatures should verify
\begin{equation}
T^{\mathrm{kin}}_{\mu}=T^{\mathrm{pot}}_{\mu}.
\end{equation}

Other useful observables are the time-delayed mode correlation functions
\begin{equation}
C_{\mu}(t_1,t_2)=\langle s_{\mu}(t_1)s_{\mu}(t_2) \rangle
\end{equation}
and the mode linear response functions
\begin{equation}
R_{\mu}(t_1,t_2)= \left. \frac{\delta{\langle s_{\mu}(t_1)\rangle_h}}{\delta h_{\mu}(t_2)} \right|_{h=0}
\end{equation}
that is defined and measured as follows.
If we add an external field $h_{\mu}$ linearly coupled to each mode $s_{\mu}$, the equations of motion are modified into
\begin{equation}\label{eq:newton_inhom}
m \ddot{s}_{\mu}+(z(t)-\lambda_{\mu})s_{\mu}-h_{\mu}=0
\end{equation}
and its solution reads
\begin{equation}
s^{(\mathrm{inhom})}_{\mu}(t)=s_{\mu}^{(\mathrm{hom})}(t)+s_{\mu}^P(t)
\;  ,
\end{equation}
where $s_{\mu}^{(\mathrm{hom})}(t)$ is the solution to the Newton equation without the external field and $s_{\mu}^P(t)$ is
a particular solution of the inhomogeneous problem with initial condition $s_{\mu}^P(0)=0$ and $\dot s_{\mu}^P(0)=0$.
The linear response function $R_{\mu}(t_1,t_2)\equiv \delta s_{\mu}(t_1)/\delta h_{\mu}(t_2)|_{h=0}$ of the mode $\mu$ can be defined equivalently through
\begin{equation}\label{eq:linear_resp}
s_{\mu}^P(t_1)=\int_{0}^{t_1}dt_2\;R_{\mu}(t_1,t_2)h_{\mu}(t_2)
\; .
\end{equation}

In practice, to measure the linear response function numerically we apply a small external field localised in time
$h_{\mu}(t)=h_0\delta(t-t_2)$ and we solve the inhomogeneous problem to obtain
$s^{(\mathrm{inhom})}_{\mu}(t)$. Using Eq.~(\ref{eq:linear_resp}) we obtain the linear response as
\begin{equation}
R_{\mu}(t_1,t_2)=\frac{s^{(\mathrm{inhom})}_{\mu}(t_1)-s_{\mu}^{(\mathrm{hom})}(t_1)}{h_0}
\; ,
\end{equation}
where $s_{\mu}^{(\mathrm{hom})}(t_1)$ must have been calculated independently.
In thermal equilibrium the linear response and correlation function are related by the fluctuation dissipation relation,
\begin{equation}
R_\mu(t_1,t_2) = \frac{1}{T_\mu} \partial_{t_2} C_\mu(t_1,t_2) \theta(t_1-t_2)
\; .
\end{equation}
Whether the time-evolving correlation and linear response satisfy this relation, whether the mode
temperatures are the same as the ones obtained from the energy characteristics of the modes
and, finally, whether they all take the same value, are issues that we will explore.

\subsection{Kinetic and potential mode energies in the stationary state}

As already mentioned,
in a steady state, $z(t) \to z_f$, the modes kinetic and potential energies are
\begin{eqnarray}
\epsilon_\mu^{\rm kin}(t) = \frac{1}{2} m {\dot s}^2_\mu(t)
\; ,
\qquad\qquad
\epsilon_\mu^{\rm pot}(t) = \frac{1}{2} m \omega_{\mu}^2 s^2_\mu(t)
\; .
\label{eq:epsilon-kin-pot}
\end{eqnarray}

Clearly, neither the kinetic nor the potential mode energies are constant,
but, in the steady state limit the  sum of the two, that is to say the total mode energy, is
\begin{equation}
 \epsilon_\mu^{\rm kin}(t) + \epsilon_\mu^{\rm pot}(t)  =
\epsilon_\mu^{\rm kin}(t_{\rm st}) + \epsilon_\mu^{\rm pot}(t_{\rm st}) \equiv \epsilon_\mu^{\rm tot}
\; ,
\end{equation}
with $t_{\rm st}$ the time at which the steady state is established and $t>t_{\rm st}$.

As expected from Birkhoff's theorem~\cite{Khinchin}, the  long-time averages, say taken after $t_{\rm st}$,
should be constants and one can expect them to be equal to half the total energy
\begin{eqnarray}
\begin{array}{lll}
&&
\displaystyle{
\overline{\epsilon_\mu^{\rm kin}} = \frac{1}{2} \epsilon_\mu^{\rm tot}
=
\frac{1}{2} m \left[ \frac{1}{2} \omega_{\mu}^2 s^2_\mu(t_{\rm st}) + \frac{1}{2} {\dot s}^2_\mu(t_{\rm st}) \right]
\; ,
}
\vspace{0.2cm}
\\
&&
\displaystyle{
\overline{\epsilon_\mu^{\rm pot}} = \frac{1}{2} \epsilon_\mu^{\rm tot}
= \frac{1}{2} m \omega_{\mu}^2 \left[ \frac{1}{2}  s^2_\mu(t_{\rm st})
+ \frac{1}{2} \frac{{\dot s}^2_\mu(t_{\rm st})}{\omega_{\mu}^2}
\right]
\; .
}
\end{array}
\end{eqnarray}
(In practice, the average over a few periods is enough to obtain the constant value.)
If one now associates a temperature to these values, arguing equipartition of quadratic degrees of freedom,
one has
\begin{equation}
T_\mu = 2 \overline{\epsilon^{\rm kin}_\mu} = 2 \overline{\epsilon^{\rm pot}_\mu} = \epsilon_\mu^{\rm tot} \; .
\end{equation}
The mode temperatures depend on the averages at the end of the transient, and the mode frequency $\Omega_{f_\mu}^2$ that itself depends
on the asymptotic limit of the Lagrange multiplier $z_f$ and the eigenvalue $\lambda_\mu$.

In the argument above we implicitly assumed that $\omega_{\mu}^2$ does not vanish.
The case $\mu=N$ is tricky.
If one naively sets $\omega^2_{\mu}$ to zero from the outset
$2\epsilon_N^{\rm pot}=\omega^2_{\mu}\langle s^2_{N} \rangle$
apparently vanishes. The correct way of treating the largest mode is to remember that the projection on the largest mode
condenses and that  $\langle s^2_{N} \rangle$ is proportional to $N$. This will ensure that
 $\lim_{N\gg1 }\langle s^2_{N} \rangle \propto N$, in such a way that
 $\lim_{N\rightarrow\infty}\omega^{2}_{\mu}\langle s^2_{N} \rangle=2\overline{\epsilon_\mu^{\rm kin}}$, similarly to what happens in equilibrium, where
$\langle s^2_{N} \rangle=q_{\mathrm{in}}N$ and the Lagrange multiplier is such that $(z_{f}-\lambda_{N})q_{\mathrm{in}}N=T^{\prime}$.

We will see in the next Sections that, in some cases,
the scenario described in this Section is actually realised by the dynamics. Which are the quenches in which such a behaviour is
observed will be determined by the complete solution of Newton's equations with the methods that we will
now describe.

\subsection{Initial conditions: equilibrium averages with finite $N$}
\label{sec:averages_finite_n}

In this section we address the calculation of equilibrium averages at finite $N$ in order to provide suitable initial conditions
for the numerical integration of the mode dynamics explained in Sec.~\ref{subsec:parametric-osc}.

If we were to naively integrate the mode equations,
we would need to draw initial vectors, $\vec s(0) = (s_1, \dots, s_N)$ and $\dot {\vec s}(0) = (\dot s_1, \dots, \dot s_N)$,
mimicking an initial thermal state at finite temperature, be it  $T'>T_c=J_0$ or $T'<T_c=J_0$,
for a given realisation of the $N\times N$ interaction matrix. Averages over these initial states
of the interesting observables should then be computed. This method is computationally expensive as
a large number of initial state should be considered to get smooth and reliable results. Instead,
the numerical method that we will explain in Sec.~\ref{subsec:parametric-osc} is such that only the
averages $\langle \, s_\mu^2 \,\rangle_{\rm eq} $ and $\langle \, {\dot s}_\mu^2 \,\rangle_{\rm eq} $ are needed
as input for the initial conditions. We then focus on determining these averages in a finite size system in
equilibrium.

The canonical equilibrium probability density of the configuration
$\{p_\mu = m \dot s_\mu, s_\mu\}$  at temperature $T^{\prime}$, for a given realization of disorder, is
\begin{equation}
P_{\rm GB}(\{p_\mu, s_\mu\})
=\frac{1}{Z}
\exp\left[-\frac{1}{T^{\prime}}\sum_{\mu}\frac{p_{\mu}^2}{2m} - \frac{1}{T'} \sum_{\mu} (z^{(N)}_{\mathrm{eq}}-\lambda^{(0)}_{\mu})s^2_{\mu}\right],
\end{equation}
with $Z$ the partition function. The statistical averages are computed as integrals over this measure. The integrals over
$p_{\mu}$ range from $-\infty$ to $\infty$.
The quadratic averages of the velocities are thus simply given by
\begin{equation}
\langle \, \dot{s}^2_{\mu} \, \rangle_{\mathrm{eq}}=\frac{T'}{m}\;\qquad \forall\mu, \, N \; ,
\label{eq:equil-initial-kinetic}
\end{equation}
just as for the infinite $N$ case, and the initial conditions will be
$\langle \, {\dot s}^2_{\mu}(0^+) \, \rangle =\langle \, {\dot s}^2_{\mu} \, \rangle_{\mathrm{eq}}$.

\begin{figure}[h!]
\begin{center}
\includegraphics[scale=0.6]{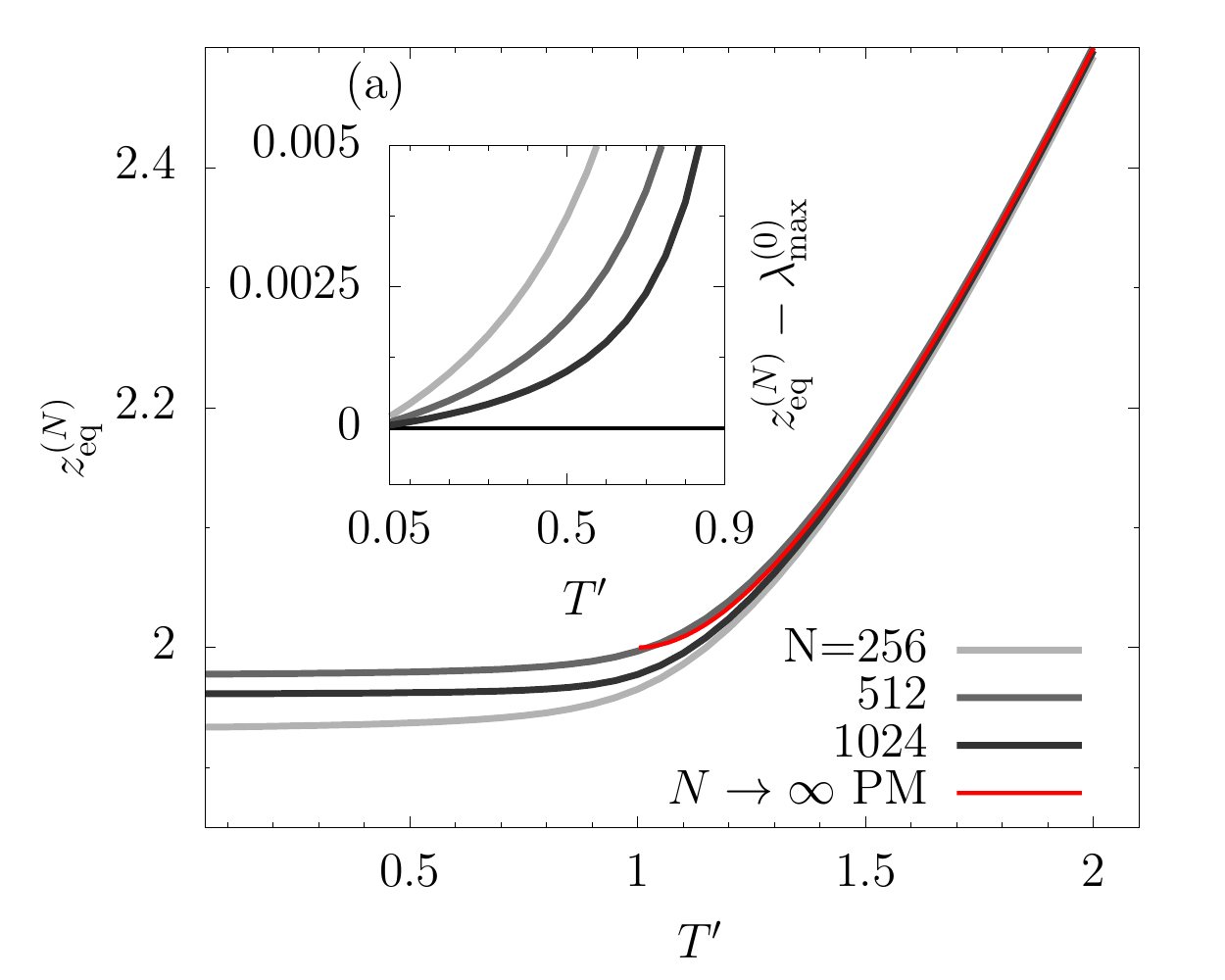}
\includegraphics[scale=0.6]{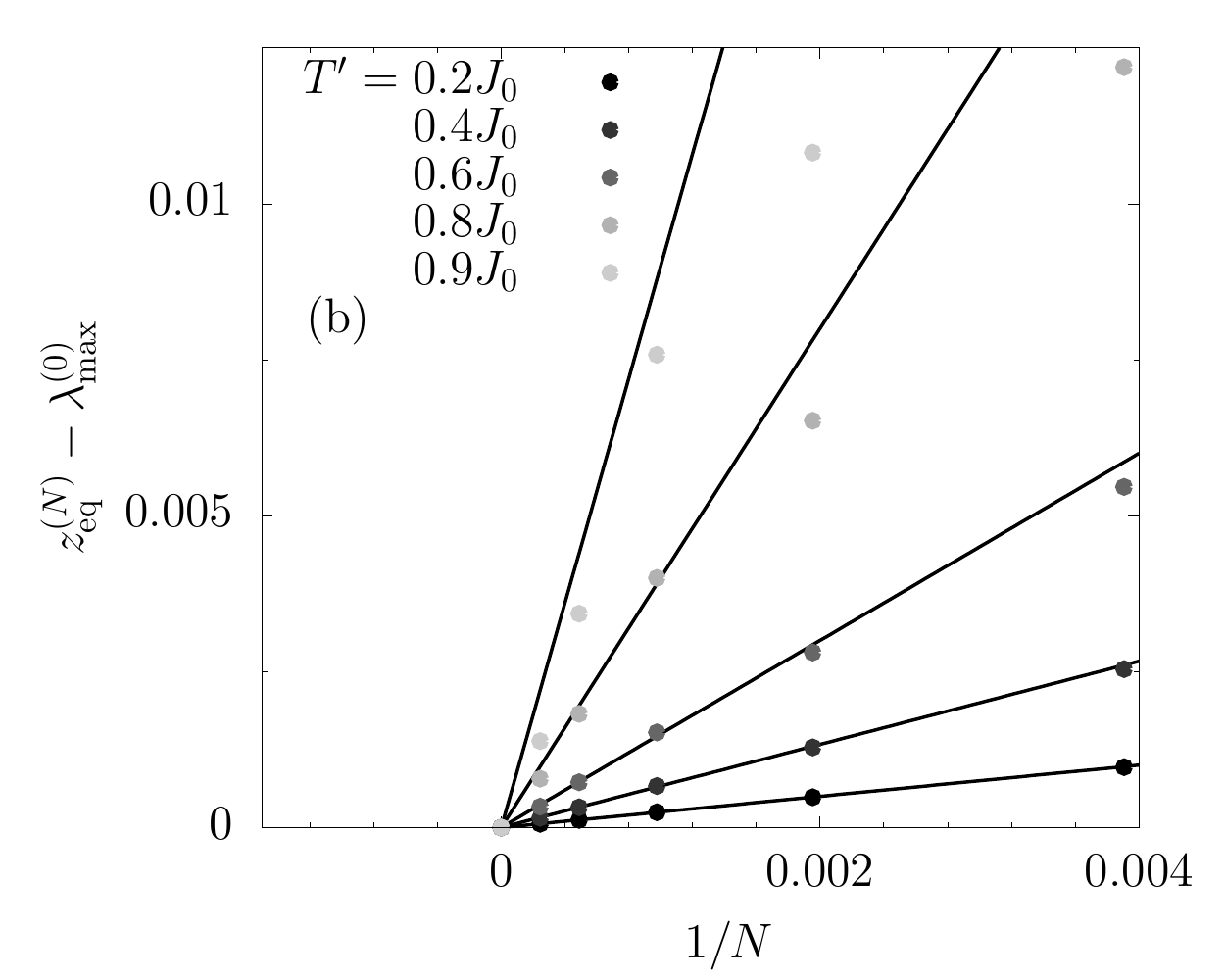}
\end{center}
\caption{\small  {\bf Equilibrium dynamics of the finite $N$ system.} (a) Equilibrium Lagrange multiplier at finite $N$.
Solution to Eq.~(\ref{eq:finite_n_sph_constraint}) as a function of temperature using one particular
realisation of disorder for each size. The non-monotonic $N$ dependence of the plateau
is of the order of magnitude of the variation with $N$ of the largest eigenvalue.
Inset: difference between the Lagrange multiplier and the maximum eigenvalue of the interaction matrix in the condensed region
as a function of temperature. The trend is now monotonic in $N$.
(b) System size scaling of the Lagrange multiplier in the condensed phase.
Difference between the Lagrange multiplier, as obtained from the solution to Eq.~(\ref{eq:finite_n_sph_constraint}),
and the maximum eigenvalue of the interaction matrix as a function of $1/N$ for different system sizes, using one particular realization of disorder for each size.
The dashed lines are $T^{\prime}/(Nq_{\mathrm{in}})$, with $q_{\mathrm{in}}=1-T'/J_0$  the value of the self-overlap in the $N\to\infty$ limit.
}
\label{fig:zeq}
\end{figure}

As long as  the equilibrium value of the Lagrange multiplier be strictly larger than the maximum eigenvalue,
\begin{equation}
z^{(N)}_{\mathrm{eq}}>\lambda^{(0)}_{\mathrm{max}}
\; ,
\end{equation}
the weight of the coordinates $s_\mu$ are well-defined independent Gaussian factors. We will see that the self-consistent
solution complies with this bound.
Relying on the spherical constraint being imposed by the Lagrange multiplier,  we extend the $s_{\mu}$ integrals to $\pm \infty$
and
\begin{equation}
\langle \, {s}^2_{\mu} \, \rangle_{\mathrm{eq}}=\frac{T'}{z^{(N)}_{\mathrm{eq}}-\lambda^{(0)}_{\mu}}\; \qquad\forall\mu, N \; .
\label{eq:equil-initial-pot}
\end{equation}
The difference between the two equilibrium phases will be codified in the value of $z^{(N)}_{\mathrm{eq}}$, which can be obtained as the solution of the spherical
constraint equation
\begin{equation}\label{eq:finite_n_sph_constraint}
\sum_{\mu=1}^{N} \langle s_\mu^2\rangle_{\rm eq} =
\sum_{\mu=1}^{N}\frac{T'}{z^{(N)}_{\mathrm{eq}}-\lambda^{(0)}_{\mu}}=N
\; .
\end{equation}
We solved this equation numerically to determine $z^{(N)}_{\mathrm{eq}}$
and we found that the solution turns out to be always greater than $\lambda^{(0)}_{\mathrm{max}}$, for any value of the temperature
and finite $N$.
In Fig.~\ref{fig:zeq} (a) we show $z^{(N)}_{\mathrm{eq}}$ as a function of temperature for three values of $N$ and a single realisation of
the random matrix in each case. At high temperatures all the curves collapse (on the scale of the figure)
on the paramagnetic curve $z_{\mathrm{eq}}=T^{\prime}+J_0^2/T^{\prime}$,
irrespective of the system size. At low temperatures (inset), $z^{(N)}_{\mathrm{eq}}$ is always larger than $\lambda^{(0)}_{\mathrm{max}}$ and,
as expected, the difference between them decreases with system size.

Once the finite size Lagrange multiplier is obtained, we replace it in Eq.~(\ref{eq:equil-initial-pot}) to obtain the initial
conditions $\langle \, {s}^2_{\mu}(0^+) \, \rangle =\langle \, {s}^2_{\mu} \, \rangle_{\mathrm{eq}}$ for the
mode dynamics.


\vspace{0.5cm}

To gain insight into the scaling with the system size, in Fig.~\ref{fig:zeq} (b)
we plot the difference between $z^{(N)}_{\mathrm{eq}}$ and $\lambda^{(0)}_{\mathrm{max}}$ for temperatures in the condensed
phase as a function of $1/N$. The straight dashed lines have slope
$T^{\prime}/q_{\rm in}$, where $q_{\mathrm{in}}=
1-T^{\prime}/J_0$ is the $N\to\infty$ value of the self-overlap. For temperatures sufficiently below the transition,
the finite size data, obtained for one particular
realisation of the random matrix $J_{ij}$, follow the infinite size results for all system sizes analysed. For temperatures close to the transition, there appear
deviations for the smallest system sizes (largest $1/N$). In conclusion, we find that for large system sizes or temperatures
not too close to the transition, the solution to Eq.~(\ref{eq:finite_n_sph_constraint}) behaves as
\begin{equation}
z^{(N)}_{\mathrm{eq}}\simeq\lambda^{(0)}_{\mathrm{max}}+\frac{T^{\prime}}{Nq_{\mathrm{in}}}
\qquad\qquad T' < J_0 \; .
\end{equation}

\begin{figure}[h!]
\begin{center}
\includegraphics[scale=0.6]{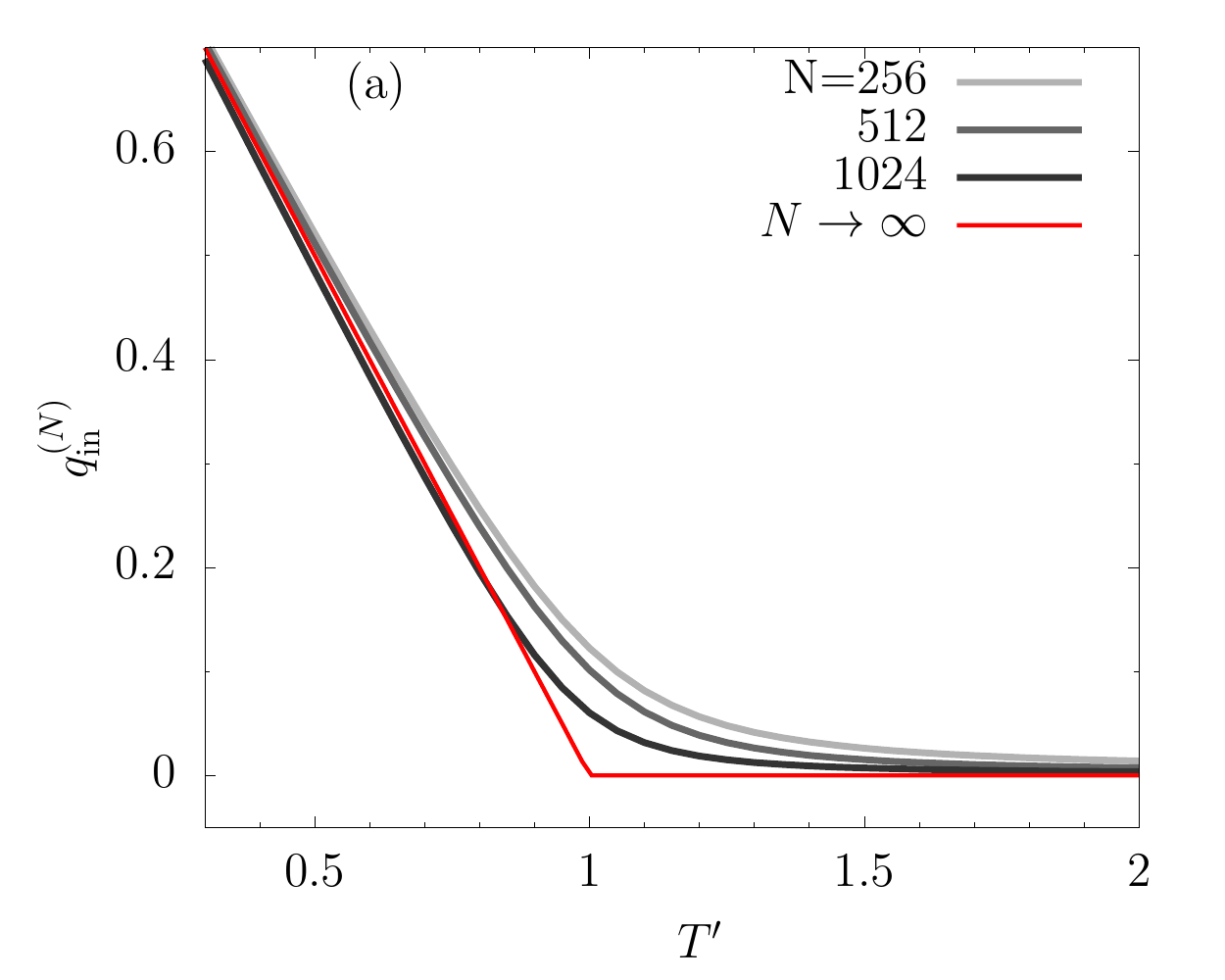}
\includegraphics[scale=0.6]{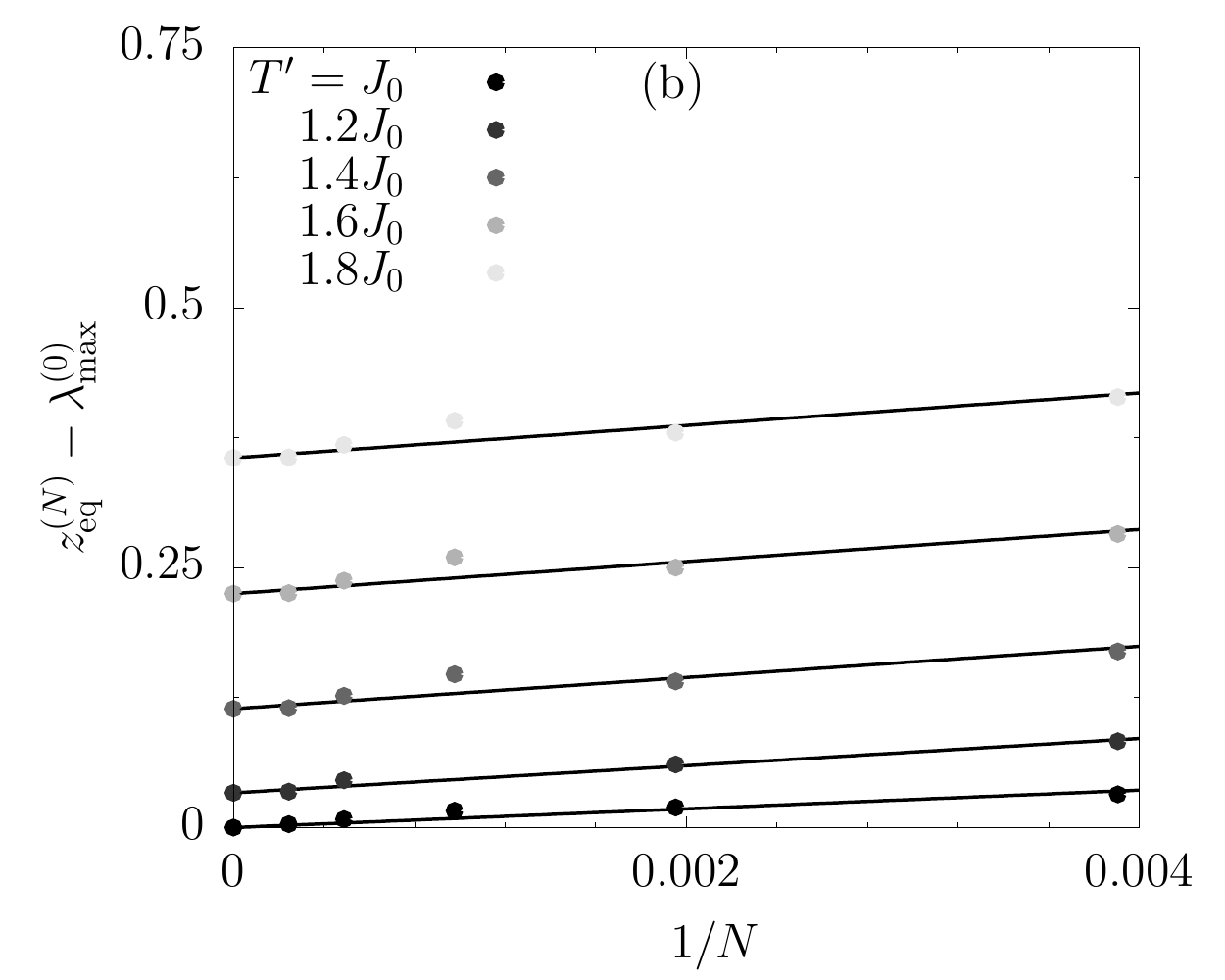}
\end{center}
\caption{\small
 {\bf Equilibrium dynamics of the finite $N$ system.}
(a) Equilibrium $q_{\mathrm{in}}$ at finite $N$.
Self overlap as a function of temperature for different system sizes, using one particular realization of disorder for each size.
In this plot we check the leading finite order of $q_{\rm in}$ and its dependence on $T'$ as $1-T'/J_0$ far from the transition. Close
to the transition there are finite $N$ corrections.
(b) System size scaling of the Lagrange multiplier in the paramagnetic phase.
Difference between the Lagrange multiplier, as obtained from the solution of Eq.~(\ref{eq:finite_n_sph_constraint}),
and the largest eigenvalue of the interaction matrix as a function of $1/N$ for different system sizes, using one particular
realisation of disorder for each size.
The non-vanishing value at $1/N \ll 1$ corresponds to $z_{\rm eq}^{(N\to\infty)}-2J_0$.
}
\label{fig:zeq_N_pm}
\end{figure}

Based on this, we define a finite size version of the equilibrium self-overlap
\begin{equation}\label{eq:q_finite_n}
q^{(N)}_{\mathrm{in}}\equiv\frac{T'}{N(z^{(N)}_{\mathrm{eq}}-\lambda^{(0)}_{N})}
\; ,
\end{equation}
which is finite if the highest mode is macroscopically populated.
For $N\rightarrow\infty$, $q_{\rm in}=1-T^{\prime}/J_0$ for $T^{\prime}<J_0$ and zero for $T^{\prime}>J_0$.
In Fig.~\ref{fig:zeq_N_pm} (a) we show $q^{(N)}_{\mathrm{in}}$ as a function of temperature.
We can observe the convergence of the finite size results towards the $N\to\infty$
predictions as the system size is increased.


\vspace{0.5cm}

Finally, we investigate the finite $N$ corrections to $z^{(N)}_{\mathrm{eq}}$ in the
paramagnetic phase, $T^{\prime}>J_0$.
We find that a linear scaling in $1/N$ also applies here, but  the value of $z_{\rm eq}^{(N)}-\lambda_{\rm max}^{(0)}$ at $N\rightarrow\infty$
does not vanish and it is given by $T^{\prime}+J_0^2/T^{\prime} - 2J_0$. Then, in the paramagnetic phase we find
\begin{equation}\label{eq:pm_finite_size_z}
z^{(N)}_{\mathrm{eq}}-\lambda^{(0)}_{\mathrm{max}} \simeq z_{\rm eq}^{(N\to\infty)}- 2J_0+\frac{s(T^{\prime})}{N}
\qquad\qquad T' > J_0
\end{equation}
where $s(T^{\prime})=s$ is the slope of the dashed lines that we obtained from a fit and turns out to be independent of temperature
(all the dashed curves are parallel straight lines).

Using the definition in Eq.~(\ref{eq:q_finite_n}), we can express the Lagrange multiplier as
$
z^{(N)}_{\mathrm{eq}}=\lambda^{(0)}_{\mathrm{max}}+T^{\prime}/(N q^{(N)}_{\mathrm{in}})
$,
and we can verify that the potential energy of the highest mode (note that we included the term proportional
to the Lagrange multiplier and we therefore compute $\epsilon_N^{\mathrm{pot}}$ instead of $e_N^{\mathrm{pot}}$)
\begin{equation}
\epsilon^{\mathrm{pot}}_{N}=\frac{1}{2}(z^{(N)}_{\mathrm{eq}}-\lambda^{(0)}_{N})\langle \, {s}^2_{N} \, \rangle_{\mathrm{eq}}=\frac{T^{\prime}}{2}
\; ,
\end{equation}
assumes the correct value in equilibrium, i.e., the one consistent with the equipartition theorem, if
\begin{equation}
\langle \, s_N^2 \, \rangle_{\rm eq} = q_{\rm in}^{(N)} N \; .
\end{equation}

\subsection{Energy variation at the quench}

We here compute the  finite-$N$ equilibrium values of the Lagrange multiplier, kinetic and potential mode energies
using the finite size averages for $\langle \, s^{2}_{\mu} \, \rangle_{\mathrm{eq}}$ and $\langle \, \dot s^{2}_{\mu} \, \rangle_{\mathrm{eq}}$
 proposed in the previous Section, and we compare them with the equilibrium $N\to\infty$ results obtained in
 Sec.~\ref{sect:p2-statics}. Analogously to the $N\to\infty$ results in Sec.~\ref{sec:energy-before-after},
 these equilibrium values define the initial condition for the interaction quench and, therefore,
 set the values of the observables at $t=0^+$.

\subsubsection{Pre-quench energies}

We begin with the kinetic energy right before the quench,
\begin{equation}
e^{(N)}_{\mathrm{kin}}(0^-)=\frac{m}{2N}\sum_{\mu=1}^{N}\langle \, \dot s^{2}_{\mu} \, \rangle_{\mathrm{eq}}=\frac{T'}{2}
\; .
\end{equation}
It coincides with the infinite-$N$ mean-field result.

Next we analyze the pre-quench potential energy,
\begin{eqnarray}
\label{eq:e_pot_ic_finite_n}
e^{(N)}_{\rm pot}(0^-)
= - \frac{1}{2N} \sum^{N}_{\mu=1} \lambda^{(0)}_\mu \langle \, s_\mu^2 \, \rangle_{\rm eq}
= -\frac{T'}{2N}\left(\sum^{N}_{\mu=1}\frac{\lambda^{(0)}_{\mu}-z^{(N)}_{\mathrm{eq}}}{z^{(N)}_{\mathrm{eq}}-\lambda^{(0)}_{\mu}} + z^{(N)}_{\mathrm{eq}}\sum^{N}_{\mu=1}\frac{1}{z^{(N)}_{\mathrm{eq}}-\lambda^{(0)}_{\mu}} \right)
=
\frac{T^{\prime}}{2}-\frac{z^{(N)}_{\mathrm{eq}}}{2}
\; ,
\end{eqnarray}
where we used Eq.~(\ref{eq:finite_n_sph_constraint}).
In the equilibrium condensed phase we can rely on $z^{(N)}_{\mathrm{eq}}=\lambda^{(0)}_{\mathrm{max}}+T^{\prime}/(N q^{(N)}_{\mathrm{in}})$
to obtain
\begin{equation}\label{eq:e_pot_finite_n}
e^{(N)}_{\mathrm{pot}}(0^-)=-\frac{\lambda^{(0)}_{\mathrm{max}}}{2}+\frac{T^{\prime}}{2}-\frac{T^{\prime}}{2Nq^{(N)}_{\mathrm{in}}}
\; .
\end{equation}
The $N\to\infty$ result is
$
e_{\mathrm{pot}}(0^-)=-J_0+T'/2
$,
consistent with Eq.~(\ref{eq:e_pot_finite_n}), since $\lim_{N\rightarrow\infty}\lambda^{(N)}_{\mathrm{max}}=2J_0$.

In the paramagnetic phase $q^{(N)}_{\mathrm{in}}\ll 1$ and the third term in Eq.~(\ref{eq:e_pot_finite_n})
induces important corrections. In this case, using Eq.~(\ref{eq:pm_finite_size_z}) we can write
\begin{equation}
e^{(N)}_{\mathrm{pot}}(0^-)\simeq -\frac{J^2_0}{2T^{\prime}}-\left(\frac{\lambda^{(0)}_{\mathrm{max}}}{2}-J_0\right)-\frac{s(T^{\prime})}{2N}
\end{equation}
and one readily recovers the $N\to\infty$ limit
$
e_{\mathrm{pot}}(0^-)=-J_0^2/(2T^{\prime})
$.

\subsubsection{Post-quench energies}

Now we will compute the values of the kinetic and potential energy, and the Lagrange multiplier after an interaction quench
\begin{equation}
\lambda^{(0)}_{\mu}\rightarrow\lambda_{\mu}=\frac{J}{J_0}\lambda^{(0)}_{\mu}
\; .
\end{equation}
The kinetic energy is not affected by the quench in the interaction and, just as in the $N\to\infty$ limit
(see Sec.~\ref{sec:energy-before-after}), we have that
\begin{equation}
e^{(N)}_{\mathrm{kin}}(0^+)=e^{(N)}_{\mathrm{kin}}(0^-)=\frac{T'}{2}
\; .
\end{equation}
For the potential energy it is enough to note that
\begin{eqnarray}
e^{(N)}_{\mathrm{pot}}(0^+)=-\frac{1}{2N}\sum^{N}_{\mu=1}\lambda_{\mu}\langle \, s^2_{\mu} \, \rangle_{\mathrm{eq}}
=
-\frac{1}{2N}\frac{J}{J_0}\sum^{N}_{\mu=1}\lambda^{(0)}_{\mu}\langle \, s^2_{\mu} \, \rangle_{\mathrm{eq}}
=
\frac{J}{J_0}e^{(N)}_{\mathrm{pot}}(0^-)
\; .
\end{eqnarray}

Using that $z^{(N)}(0^+)=2(e^{(N)}_{\mathrm{kin}}(0^+)-e^{(N)}_{\mathrm{pot}}(0^+)) =
T'-2J/J_0 \, e_{\rm pot}^{(N)}(0^-) $ it
is now easy to find the initial value of the Lagrange multiplier. When the initial conditions are taken from the condensed phase,
$q^{(N)}_{\mathrm{in}}=\mathcal{O}(1)$, and we can write
\begin{equation}
z^{(N)}(0^+)=\lambda_{\mathrm{max}}+T^{\prime}\left( 1-\frac{J}{J_0} \right) + \frac{JT^{\prime}}{NJ_0q^{(N)}_{\mathrm{in}}}
\; .
\end{equation}
For initial states in the paramagnetic phase
\begin{eqnarray}
z^{(N)}(0^+)\simeq  \frac{JJ_0}{T^{\prime}} + T^{\prime}+\left(\lambda_{\mathrm{max}}-2J\right)+\frac{J}{J_0}\frac{s(T^{\prime})}{N}
\; .
\end{eqnarray}

\subsection{Independent harmonic oscillators in the asymptotic limit}
\label{subsec:independent-harmonic-oscillators}

We now use the results in App.~\ref{app:harmonic-oscillator}
concerning the quench dynamics of a harmonic oscillator in the context of our non trivial problem.
In equilibrium at time $t=0^-$ the initial frequencies of the modes are
\begin{equation}
{\omega_0^2}_\mu = [z_{\rm eq}(T', J_0) - \lambda_\mu^{(0)}]/m
\; .
\end{equation}
In the asymptotic limit after the quench we identify the frequencies with
\begin{equation}
\omega^2_\mu = [z_f- \lambda_\mu]/m
\end{equation}
where we assumed that the Lagrange multiplier reached the constant $z_f=\lim_{t\to\infty} z(t)$.

The analysis of the harmonic oscillator does not need any long-time  assumption
to set its spring constant, or frequency, to a constant value. In our problem, the dynamics
may approach the ones of independent harmonic oscillators with constant spring constants
only asymptotically. During the
transient evolution the mode energies vary. In reality, we do not know the
values they take at the end of the transient regime. We can make a rough approximation
in which we assume that $z_f$ is reached instantaneously after the quench, $z(0^+)=z_f$, so that we
can use
\begin{equation}\label{eq:finite_N_app}
\langle p_\mu^2(0^+)\rangle/m \approx m\omega^2_{\mu} \langle s_\mu^2(0^+)\rangle \approx T'
\end{equation}
instead of the unknown values at the end of the stationary regime.

Under these assumptions the final mode temperatures are
\begin{equation}
T_\mu^f = \frac{T'}{2} \left(\frac{\omega_\mu^2}{{\omega_0^2}_\mu}+1  \right) = \frac{T'}{2} \left( \frac{z_f- \lambda_\mu}{z_{\rm eq}(T', J_0) - \lambda_\mu^{(0)}} + 1 \right)
\; ,
\end{equation}
see App.~\ref{app:harmonic-oscillator} for the details of the derivation.
It is convenient to replace the post-quench eigenvalues $\lambda_\mu$ by their expression in terms
of the pre-quench ones and the quench parameter $x=J/J_0$,
$\lambda_\mu = x \lambda_\mu^{(0)}$.
We can then distinguish the four cases (I)-(IV) depending on the values of
\begin{equation}
y=\frac{2T'}{\lambda_N^{(0)}} \qquad\mbox{and}\qquad x=\frac{\lambda_N}{\lambda_N^{(0)}}=\frac{J}{J_0}
\; .
\end{equation}
They are
\begin{eqnarray}
\frac{2T_\mu^f}{T'} \ = \
\left\{
\begin{array}{ll}
\displaystyle{
\frac{
T'+ \frac{\lambda_N^2}{4T'}  -\lambda_\mu
}
{
T'+ \frac{(\lambda_N^{(0)})^2}{4T'}  -\lambda^{(0)}_\mu
}
+1
}
&
\qquad
\mbox{for} \;\;\; y>x \;\;\;\mbox{and} \;\;\; y>1 \;\; \mbox{(I)}
\vspace{0.25cm}
\\
\displaystyle{
\frac{
\lambda_N- \lambda_\mu
}
{
T'+ \frac{(\lambda_N^{(0)})^2}{4T'}  -\lambda^{(0)}_\mu
}
+1
}
&
\qquad
\mbox{for} \;\;\; x>y \;\;\;\mbox{and} \;\;\; y>1 \;\; \mbox{(II)}
\vspace{0.25cm}
\\
\vspace{0.25cm}
\displaystyle{
\frac{
\lambda_N- \lambda_\mu
}
{
\lambda^{(0)}_N- \lambda_\mu^{(0)}
}
+1
}
&
\qquad
\mbox{for} \;\;\; x>y \;\;\;\mbox{and} \;\;\; y<1 \;\; \mbox{(III)}
\\
\vspace{0.25cm}
\displaystyle{
\frac{
T'+ \frac{\lambda_N^2}{4T'}  -\lambda_\mu
}
{
\lambda^{(0)}_N- \lambda_\mu^{(0)}
}
+1
}
&
\qquad
\mbox{for} \;\;\; y>x \;\;\;\mbox{and} \;\;\;y<1 \;\; \mbox{(IV)}
\end{array}
\right.
\label{eq:mode-temp-prediction}
\end{eqnarray}
Several comments are in order.
The expression for $x>y$ and $y<1$ (sector III) is the same as the one that we derived from the
analysis of the $N\to\infty$ Schwinger-Dyson equations, see Eq.~(\ref{eq:Tf-asymptotic-values}), simply
$T_\mu^f=T'(x+1)/2=T_f=T_{\rm kin}^{{\scriptsize \rm (III)}}$. The no-quench case $x=1$ in realised in Sectors I and III and one rapidly checks
that $T^f_\mu=T'$ in both cases. On the curve $y=\sqrt{x}$ the mode temperature do not take the same value. We will argue later that
the approximation used in the Section yields a qualitatively erroneous result in this case.
Continuity between sectors I and IV on the one side, and II and III on the
other, are ensured setting $y=1$ that is to say $T'=\lambda_N^{(0)}$.
Finally, continuity across the dynamic transition at $y=x$ or
\begin{equation}
T_{\rm dyn}'=\frac{\lambda_N}{2}
\end{equation}
is also verified.

We would like to know which is the condition satisfied by $z_f$ under this approximation. In order to obtain such equation, we first note that the time-dependent spherical constraint imposes that
\begin{equation}
\sum_{\mu=1}^{N}\langle s^2_{\mu}(t)\rangle=N
\; .
\end{equation}
In particular, this implies that
\begin{equation}
\sum_{\mu=1}^{N}\overline{\langle s^2_{\mu}(t)\rangle}=N
\; .
\end{equation}
Inserting the approximation in Eq.~(\ref{eq:finite_N_app}) in the time-averaged spherical constraint, we find an equation for $z^{(N)}_f$
\begin{eqnarray}
\frac{T^{\prime}}{2m}\sum_{\mu=1}^{N}(\omega_{\mu})^{-2}\left[ \left(\frac{\omega_{\mu}}{\omega^{0}_{\mu}}\right)^2+1 \right]=\frac{T^{\prime}}{2}\sum_{\mu=1}^{N}\left[ \frac{1}{z^{(N)}_{\mathrm{eq}}-\lambda^{(0)}_{\mu}}+\frac{1}{z^{(N)}_{f}-\lambda_{\mu}} \right]=N
\;  .
\end{eqnarray}
Since $z^{(N)}_{\mathrm{eq}}$ is chosen in such a way that
\begin{equation}
\sum_{\mu=1}^{N}\frac{T^{\prime}}{z^{(N)}_{\mathrm{eq}}-\lambda^{(0)}_{\mu}}=N
\; ,
\end{equation}
we find that the equation for $z^{(N)}_f$ simplifies to
\begin{equation}\label{eq:zf_finite_n}
\sum_{\mu=1}^{N}\frac{T^{\prime}}{z^{(N)}_{f}-\lambda_{\mu}}=N
\; .
\end{equation}
In other words, under this approximation, $z^{(N)}_{f}$ is the equilibrium Lagrange multiplier for a system in equilibrium at temperature $T^{\prime}$ with variance of the disorder distribution equal to $J$. In the $N\rightarrow\infty$ limit $z_f=2J$ if $T^{\prime}<J$ and $z_f=T^{\prime}+J^{2}/T^{\prime}$ if $T^{\prime}>J$.

We will put these predictions to the test in Sec.~\ref{sec:numerical} using the
numerical solution to the finite $N$ evolution equations with the numerical method that we
describe in the next Subsection. In various regions of the phase diagram these {\it a priori} approximate forms are in strikingly good
agreement with the numerical data. In others they are not and we discuss why this is so.

\subsection{Exact solution of the mode dynamics}
\label{subsec:parametric-osc}

One possible approach to solve the dynamics of each mode starting from canonical equilibrium initial conditions is to take a large ensemble of initial configurations
drawn from the
Gibbs-Boltzmann distribution, numerically integrate the Newton equations Eq.~(\ref{eq:newton_eq0}) for each initial condition,
and then calculate the observables averaging over the trajectories corresponding to the different initial states.
Such approach is feasible but computationally very demanding. In this Section we describe a more convenient method to solve the dynamics for each mode
that  uses heavily the tools developed to treat a paradigmatic problem in classical mechanics, the one of
{\it parametric oscillators}~\cite{Ermakov,Milne,Pinney}.

In order to solve Eq.~(\ref{eq:newton_eq0}) we propose an amplitude-phase {\it Ansatz}~\cite{Ermakov,Milne,Pinney,SoCa10}
\begin{equation}\label{eq:gen_ansatz0}
s_{\mu}(t)=\frac{A}{\sqrt{\Omega_{\mu}(t)}}\exp\left[-i\int_{0}^{t} dt'\;\Omega_{\mu}(t')\right].
\end{equation}
Inserting this {\it Ansatz} in the $\mu$th mode Newton equation, we obtain an equation for the mode and time dependent
auxiliary function $\Omega_{\mu}(t)$,
\begin{equation}\label{eq:omega0}
\frac{1}{2}\frac{\ddot\Omega_{\mu}(t)}{\Omega_{\mu}(t)}-\frac{3}{4}\left( \frac{\dot\Omega_{\mu}(t)}{\Omega_{\mu}(t)} \right)^2+\Omega^2_{\mu}(t)
= \omega^2_\mu(t)
\; ,
\end{equation}
where $\omega_{\mu}^2(t)\equiv (z(t)-\lambda_{\mu})/m$.
The last equation has to be complemented by the initial values $\omega_\mu(0)$ and $\dot \Omega_\mu(0)$.
If we choose
\begin{equation}
\dot\Omega_{\mu}(0)=0
\; ,
\end{equation}
we find that the projection of the spin configuration is
\begin{equation}
s_\mu(t) = s_\mu(0) \sqrt{\frac{\Omega_{\mu}(0)}{\Omega_{\mu}(t)}}  \; \cos\left( \int_0^t dt' \; \Omega_{\mu}(t')\right)
+  \frac{\dot s_\mu(0)}{\sqrt{\Omega_{\mu}(t)\Omega_{\mu}(0)}} \; \sin\left( \int_0^t dt' \; \Omega_{\mu}(t') \right),
\end{equation}
which is reminiscent of the general solution of the harmonic oscillator problem, see Eq.~(\ref{eq:harm-osc-const-freq}), here with a
time-dependent ``frequency'' $\Omega_\mu(t)$.

We still have to specify the initial condition for $\Omega_{\mu}(0)$. A possible choice is
\begin{equation}
\Omega_{\mu}^2(0)=z(0)-\lambda_{\mu}
\end{equation}
 that enforces $\ddot \Omega_{\mu}(0)=0$~\cite{SoCa10}. However, this choice is consistent with real $\Omega_{\mu}(t)$
 only if $z(0)-\lambda_{\mu}\ge 0$, which is verified uniquely for $J\le J_0$, i.e., uniquely for energy injection.
 An initial condition ensuring real and positive $\Omega_{\mu}(t)$ for all $\mu$ for any quench is
\begin{equation}
\Omega_{\mu}^2(0) = \lambda_N-\lambda_{\mu}
\; .
\end{equation}
We choose this initial condition for the numerical calculations.

In order to solve for $\Omega_{\mu}(t)$ we consider the equal-times mode correlation function
\begin{eqnarray}\label{eq:eqt_mode_corr}
C_{\mu}(t,t)=\langle s^2_{\mu}(t)\rangle
\!\! & \! \! = \!\! & \!\!
\langle s^2_{\mu}(0)\rangle \, \frac{\Omega_{\mu}(0)}{\Omega_{\mu}(t)} \;
\cos^2\left(\int^t_{0}dt'\,\Omega_{\mu}(t')\right)
+
\frac{\langle \dot s^2_{\mu}(0)\rangle}{\Omega_{\mu}(0)\Omega_{\mu}(t)}\
\sin^2\left(\int^t_{0}dt'\,\Omega_{\mu}(t')\right)
\nonumber\\
&&
+
\frac{\langle s_\mu(0) \dot s_{\mu}(0)\rangle}{\Omega_{\mu}(t)}\
\sin\left(\int^t_{0}dt'\,\Omega_{\mu}(t')\right)
\cos\left(\int^t_{0}dt'\,\Omega_{\mu}(t')\right)
\; ,
\end{eqnarray}
in terms of which we write the potential energy as
\begin{equation}\label{eq:epot_corr}
e_{\mathrm{pot}}(t)=-\frac{1}{2N}\sum_{\mu}\lambda_{\mu}C_{\mu}(t,t)
\; .
\end{equation}
Replacing this equation in $z(t) = 2e_f - 4 e_{\rm pot}(t)$,
we find an expression of the Lagrange multiplier as a function of the mode correlations at equal times
\begin{equation}\label{eq:z_corr}
z(t)=2e_f+\frac{2}{N}\sum_{\mu}\lambda_{\mu}C_{\mu}(t,t)
\; .
\end{equation}

Finally, we note that the system conformed by Eqs.~(\ref{eq:omega0}),~(\ref{eq:eqt_mode_corr}) and~(\ref{eq:z_corr}) is
closed and allows to find the time evolution of the Lagrange multiplier and
the auxiliary functions $\Omega_{\mu}(t)$. This set of equations is amenable to numerical integration. Once we obtain $\Omega_{\mu}(t)$,
the most interesting observables can be calculated using the general solution in the form given in Eq.~(\ref{eq:gen_ansatz0}).
The advantage of this method is that we do not need to draw initial states $\{s_\mu(0), \dot s_\mu(0)\}$ but we
only have to specify the initial averages
$\langle s^2_\mu(0)\rangle$ and $\langle {\dot s}^2_\mu(0)\rangle$ that we will take to be the ones
enforced by equilibrium at $T'$, that is to say, the forms given in Eqs.~(\ref{eq:equil-initial-kinetic}) and (\ref{eq:equil-initial-pot}).


\section{Numerical results}
\label{sec:numerical}

This Section summarises what we found numerically by solving the $N\to\infty$ Schwinger-Dyson equations that couple the global correlation and linear response
$C$ and $R$ (see Sec.~\ref{sec:quenches}),
and the finite $N$ ones acting on the mode projections $s_\mu$ (see Sec.~\ref{sec:analytic-finiteN}). Some general considerations about the numerical algorithm
used to integrate the $N\to\infty$ equations are given in App.~\ref{app:discrete-time}.

The finite $N$ results are consistent with the infinite $N$ ones and help us understanding the
mechanism whereby the dynamics take place. We chose to start this Section with the summary of the dynamical phase diagram and the
behaviour of the quantities that determine it.
Later, we give further  details on the dynamics at constant energy (no quench) and in each of the dynamic phases identified after sudden quenches.

We signal here that we will make a special effort to show, in each case considered, that an asymptotic state characterised by the single temperature
$T_f$ that the naive asymptotic analysis of the dynamic equations predicts {\it is not} attained. The investigations that lead to this conclusion
are very
instructive not only because they prove the lack of Gibbs-Boltzmann equilibrium but also because they lead to the evaluation of the mode temperatures that
will play a role in the statistical description of the steady states.

\subsection{The phase diagram}
\label{sec:phase-diagram}

\begin{figure}[h!]
\begin{center}
\includegraphics[scale=0.75]{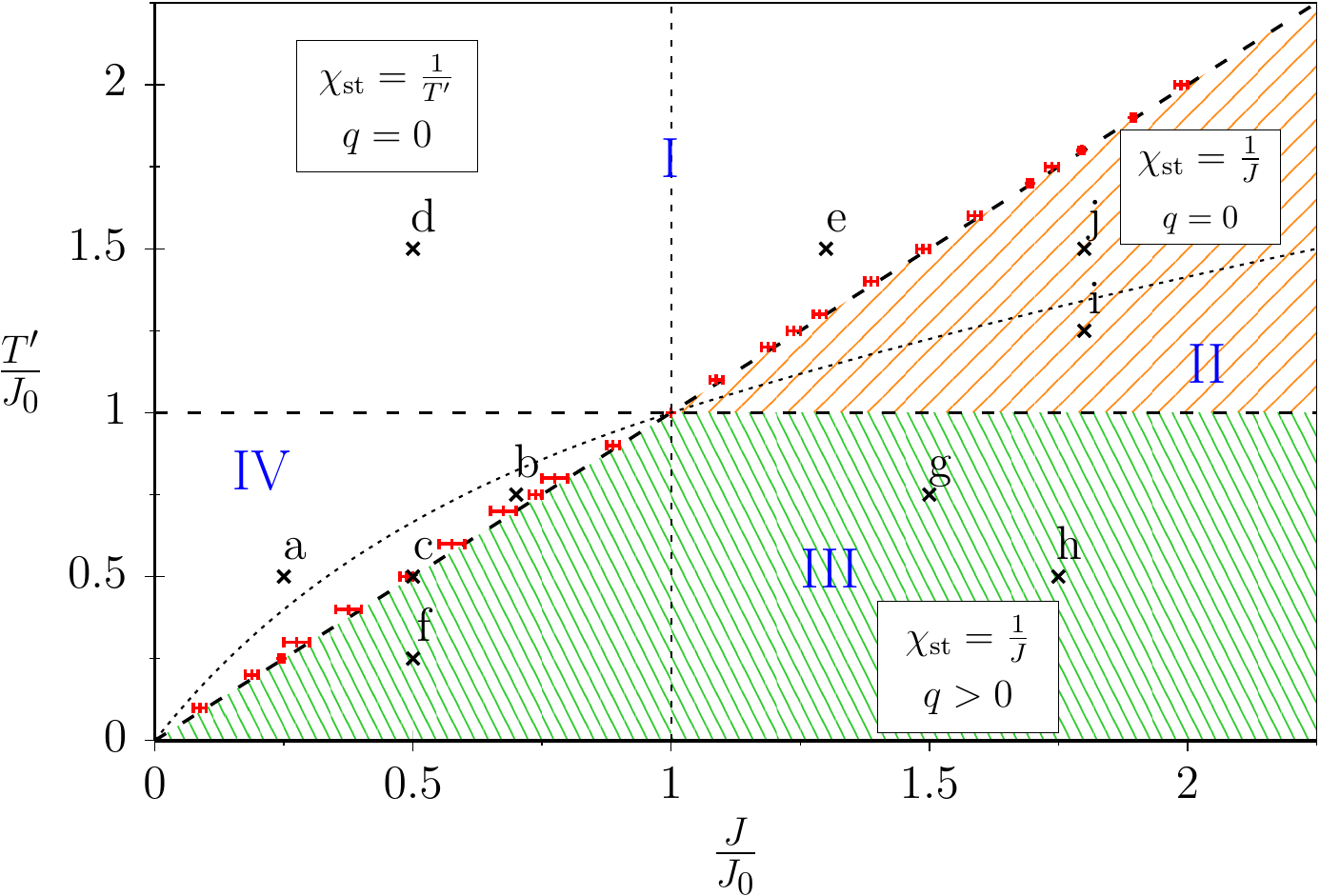}
\end{center}
\caption{\small
{\bf The dynamic phase diagram.} The parameter $x=J/J_0$ controls the energy injection/extraction, with $x < 1$ corresponding to energy injection ($\Delta e >0$),
while $x > 1$ to energy extraction ($\Delta e <0$). The parameter $y = T^{\prime}/J_0$ represents the pre-quench equilibrium temperature.
The dotted lines are the functions $f(x)=\sqrt{x}$ for $x>1$ and $g(x)=2 x/(1+x)$ for $x \le 1$, respectively, discussed in the text.
The three phases that are characterised by different behaviour of $z_f$, $\chi_{\mathrm{st}}$ and the long times limit of
$C(t_1,t_2)$ are highlighted with different colors.
The red data points equipped with error bars indicate the numerical estimate of $x_c$ for several values of $y=T^{\prime}/J_0$
(see the main text for more details). The crosses indicate the cases shown in Fig.~\ref{fig:corr_cmp} with the corresponding labels.}
\label{fig:phase_diagram_new}
\end{figure}

The phase diagram is determined by the asymptotic behaviour of the zero frequency linear response or susceptibility,
$\chi_{\rm st} = \hat R(\omega=0)$, and the asymptotic value of the Lagrange multiplier. We determine their values
through the variation of the parameter $J/J_0$ for fixed $T^{\prime}/J_0$. In the phase diagram presented in
Fig.~\ref{fig:phase_diagram_new} and the ensuing discussion we call $y=T^{\prime}/J_0$ the vertical axis and $x=J/J_0$ the horizontal one.
The former determines the initial state and the latter the kind of quench performed with injection of energy for $x<1$ and extraction of energy
for $x>1$.

We study separately parameters in four Sectors of the phase diagram, although the final results will allow us to distinguish three different phases.
The Sectors are indicated with Roman numbers and the phases with different colours or shades
in Fig.~\ref{fig:phase_diagram_new}. We also mark the line $x=1$ (equilibrium dynamics) and the curve $y=\sqrt{x}$
with $y>1$ where special dynamics are found.

We recall that dynamic phase transitions have been found in the quench dynamics of quantum isolated
systems, see, {\it e.g.}~\cite{EcKoWe09,ScFa10,ScBi10,ScBi11,ScBi13,TsEcWe13,TsWe13,ChTaGaMi16}. Here, and
in~\cite{CuLoNe17}, we
see dynamic phase transitions arise in the Newtonian dynamics of isolated  classical interacting systems.

\begin{figure}[h!]
\begin{center}
\includegraphics[scale=0.65]{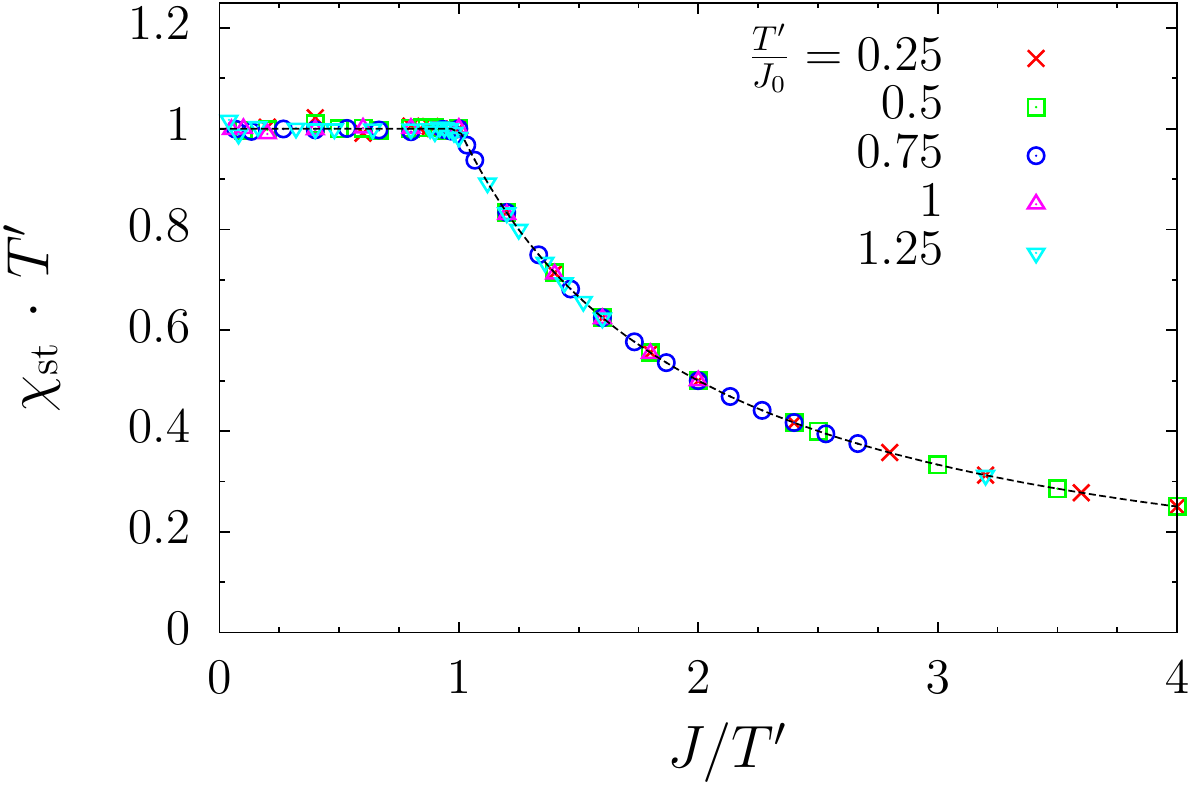}
\end{center}
\caption{\small {\bf The zero frequency linear response}, computed from the Schwinger-Dyson $N\to\infty$ equations,
for several choices of initial conditions given in the key, with both $y<1$ (condensed)
and $y>1$ (paramagnetic) cases, together with the analytic prediction in Eq.~(\ref{eq:prediction-Romega}) plotted with a dashed line.
}
\label{fig:response}
\end{figure}

In Fig.~\ref{fig:response} (a) we check the prediction~(\ref{eq:prediction-Romega}) for the zero frequency linear
response function. We plot $T'\hat R(\omega=0)$ against $J/T'$ and we see the change in behaviour from
$\hat R(\omega=0)=1/T'$ to $\hat R(\omega=0)=1/J$ at $x_c(y)=y$, that is $T'=J$.


\vspace{0.5cm}

\begin{figure}[h!]
\centerline{
\includegraphics[scale=0.56]{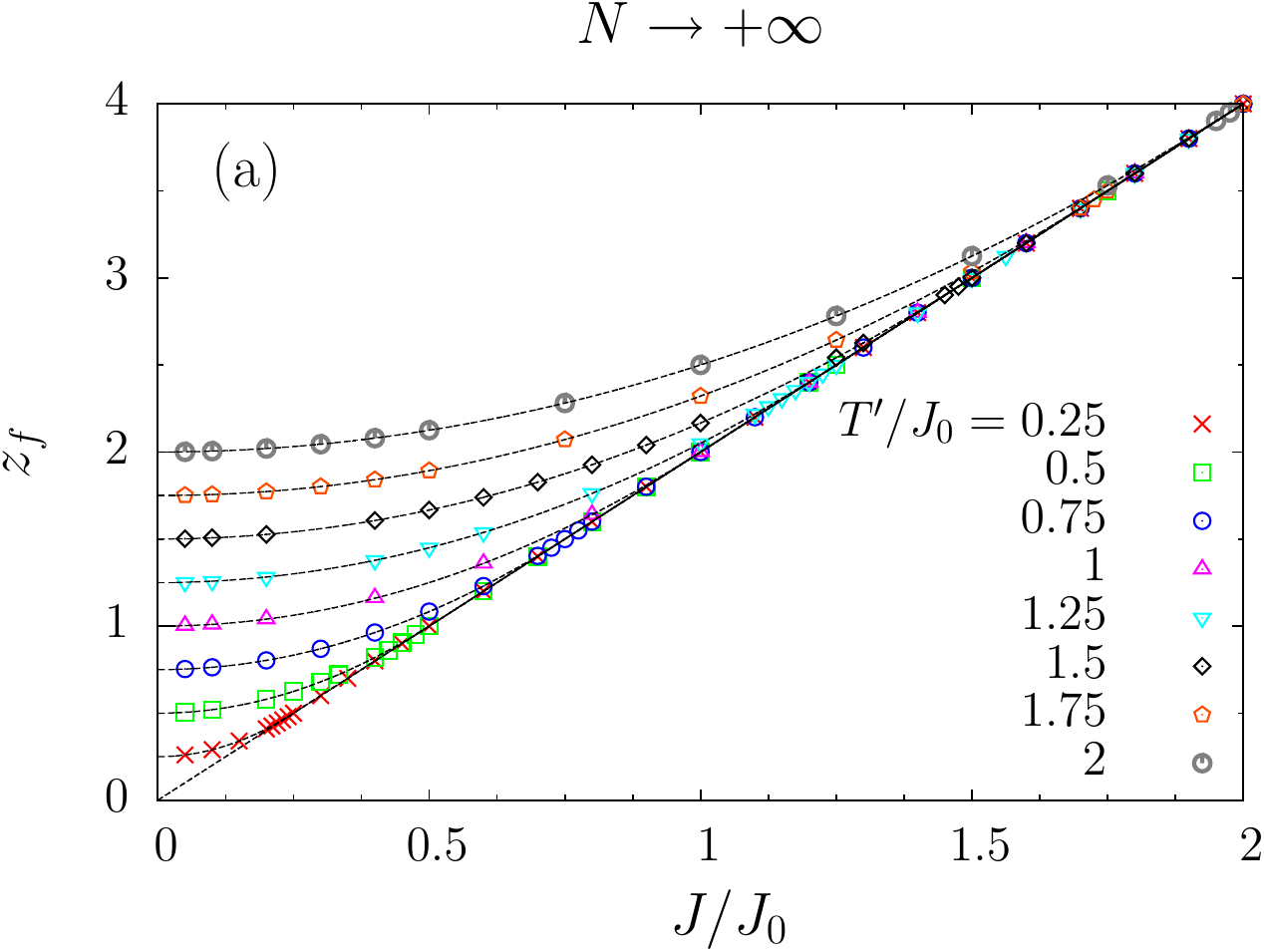}\quad
\includegraphics[scale=0.56]{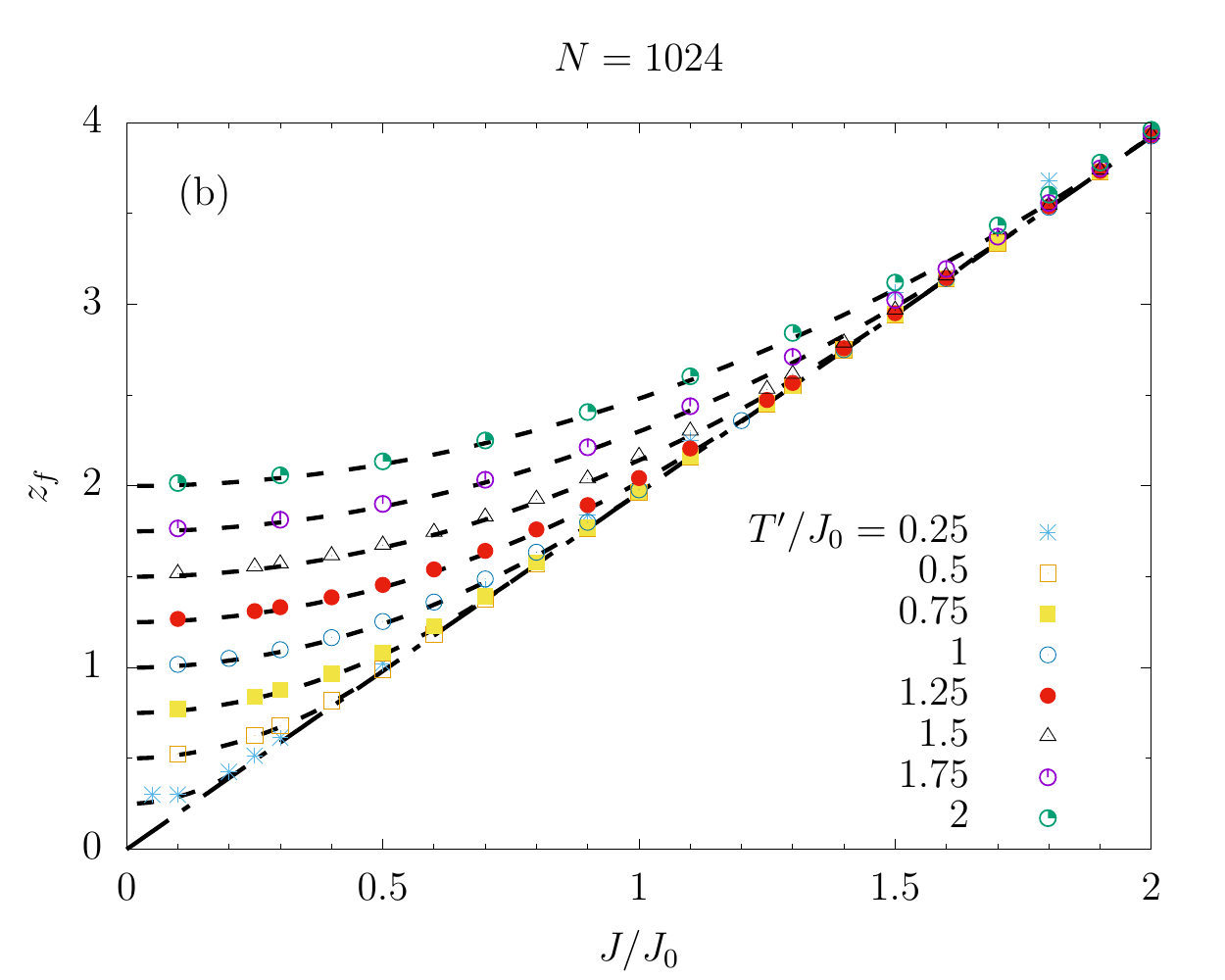}
}
\caption{\small {\bf Estimated asymptotic value of $z(t)$} as function of $J/J_0$, for different values of  $T^{\prime}/J_0$,
as indicated in the keys. (a) $N\to\infty$ results. The dashed curved lines are functions of the form $f(x,y) = y + x^2/y$ for $x<y$
where $x=J/J_0$ and $y=T^{\prime}/J_0$.
We also show the diagonal $2 \,x$ to let the reader see the crossover between the two regimes.
(b) Finite size system with $N=1024$.
The straight dashed line is $J \lambda_{\mathrm{max}}$, the other curves are $T^{\prime}+\lambda_{\max}^2/(4T^{\prime})$, the finite size
version of the infinite $N$ fits.
}
\label{fig:asympt_values}
\end{figure}

The change in $\hat R(\omega=0)$ is accompanied by
a change in the asymptotic value of $z$ as a function of the quench parameter $x = J/J_0$. This fact is confirmed numerically in
Fig.~\ref{fig:asympt_values} where data for $N\to\infty$ and $N$ finite are shown in panels (a) and (b), respectively.

For $x<y$, the numerically estimated $z_f(x)$ for fixed $y$ in the case $N\to\infty$ (a)
were fitted with the polynomial function $f(x) = a\, x^2 + b\, x + c$. We obtained very good results with
$a \simeq 1/y$, $b\simeq 0$ and $c\simeq y$ (the precision of the fit is really very high in terms of reduced $\chi^2$).
These results strongly suggest the following functional dependence of $z_f$ on the parameters $x$ and $y$,
\begin{eqnarray}
z_f(x,y) &=&
J_0
\left\{
\begin{array}{ll}
\displaystyle{\frac{x^2}{y} + y} \qquad & \mbox{for} \;\; x \leq y
\; ,
\vspace{0.1cm}
 \\
2\, x \qquad & \mbox{for} \;\; x\geq y
\; .
\end{array}
\right.
\label{eq:asympt_z_numerical}
\end{eqnarray}
To get a visual confirmation of this argument, in Fig.~\ref{fig:asympt_values} (a) we plotted the functions
$f(x) = x^2/y + y$ for $x< y$, one for each one of the values of $y$ that the numerical data refer to.
The agreement between the data and the prediction is very good.


\vspace{0.5cm}

The analysis of the finite $N$ data was done along the same lines, see Fig.~\ref{fig:asympt_values} (b), with the difference that the
data for $x<y$ were fitted by $T^{\prime}+\lambda_{\max}^2/(4T^{\prime})$ and the ones for
$x>y$ with $J \lambda_{\mathrm{max}}$ finding again very good agreement. (We found an appreciable deviation in the fit for $x>y$ had we
used $2J$ instead of $J \lambda_{\mathrm{max}}$. Regarding the results for $x<y$ we could have used
$T^{\prime}+J^2/T^{\prime}$ with a similar quality for the fit.)
Remarkably, the functional dependence proposed in Eq.~(\ref{eq:asympt_z_numerical}) is the one predicted by the independent harmonic oscillators approximation in Eq.~(\ref{eq:zf_finite_n}).

By using the change in the dependence of $\hat R(\omega=0)$ and $z_f$ on the
quench parameter $x=J/J_0$ as a criterium to track the dynamical phase transition, we obtained the numerical
estimates of the critical curve $x_c(y)$.
In Fig.~\ref{fig:phase_diagram_new} we show the data for $x_c(y)$ (with error bars) for some values of the control parameter $y$.
The data strongly suggest that there is a linear relation between the critical value $x_c$ and the parameter $y$
for any $y$; in short, we confirm $x_c(y)=y$.


\vspace{0.5cm}

Concerning the long-time behaviour of $C(t_1,t_2)$,
it is useful to distinguish the cases $y<1$ and $y>1$, that is, quenches that start from equilibrium in the condensed phase from
quenches that start from equilibrium in the paramagnetic phase.

We observe the following trends:
\begin{itemize}
 \item For $x<x_c(y)$,
       $C(t_1,t_2)$ tends to be stationary, though within the time scales of the numerics it has not reached this limit  yet when $y$ is too
       small. In most instances, $C(t_1,t_2)$ oscillates around $0$, exceptions being the critical quench and the case with
       both $y>1$ and $x>1$ where zero is approached asymptotically from below.
       The time average of $C$ computed on intervals far from the initial transitory regime vanishes in all cases
       suggesting an effective $q=0$.
 \item For $x\ge x_c(y)$,
       $C(t_1,t_2)$ is rapidly stationary and one very clearly identifies the asymptotic constant
       $q=\lim\limits_{\substack{t_2 \gg t_0 \\ t_1 - t_2 \gg t_0}} C(t_1,t_2)$. For $y<1$ it is different from zero while for
       $y>1$ it equals zero. The asymptotic $q_0=\lim_{t_2 \gg t_0} C(t_1,0)$ is different from $q$ in the
       cases in which both are non-vanishing.
\end{itemize}

\begin{figure}[h!]
 $\qquad y=0.50 \;\; x=0.25 \qquad y=0.75 \;\; x=0.70 \qquad y=0.50 \;\; x=0.50 \qquad y=1.50 \;\; x=0.50 \qquad y=1.50 \;\; x=1.30$
\\
$\;$
\\
\centerline{
\includegraphics[scale=0.4]{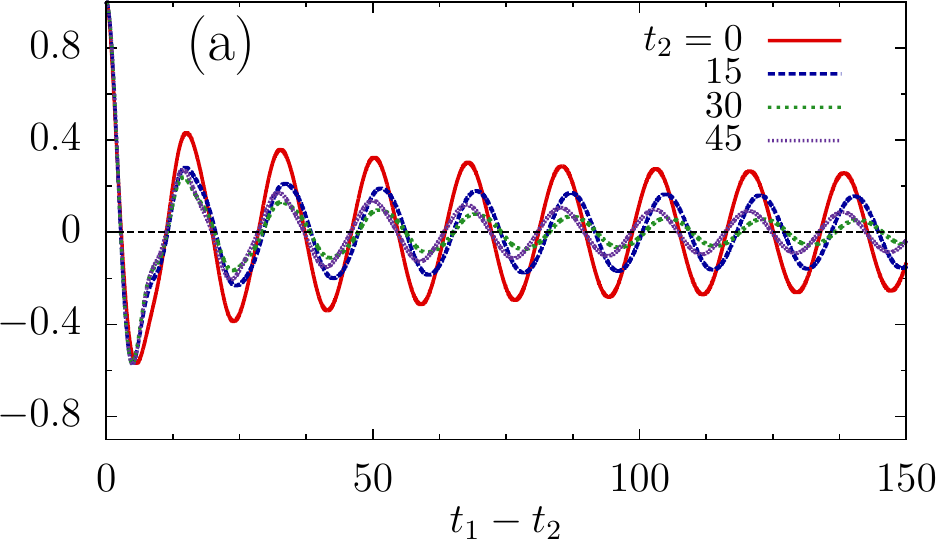} 
\includegraphics[scale=0.4]{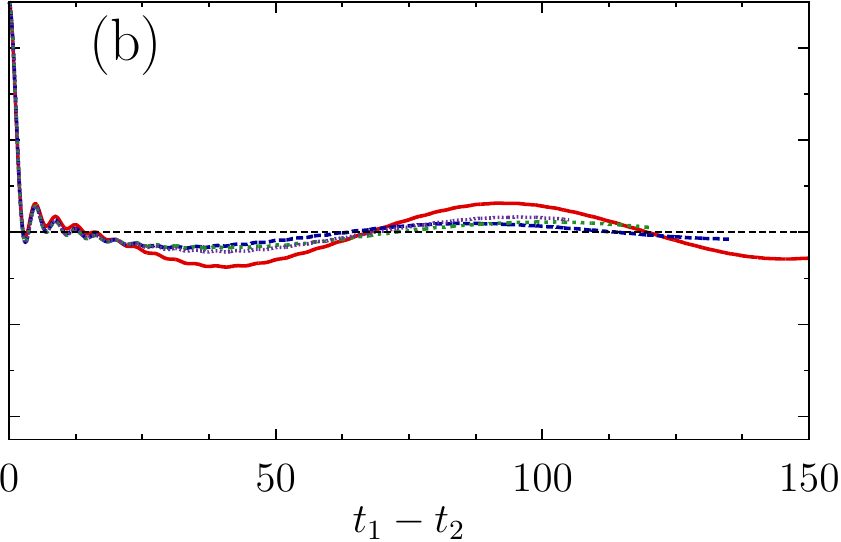} 
\includegraphics[scale=0.4]{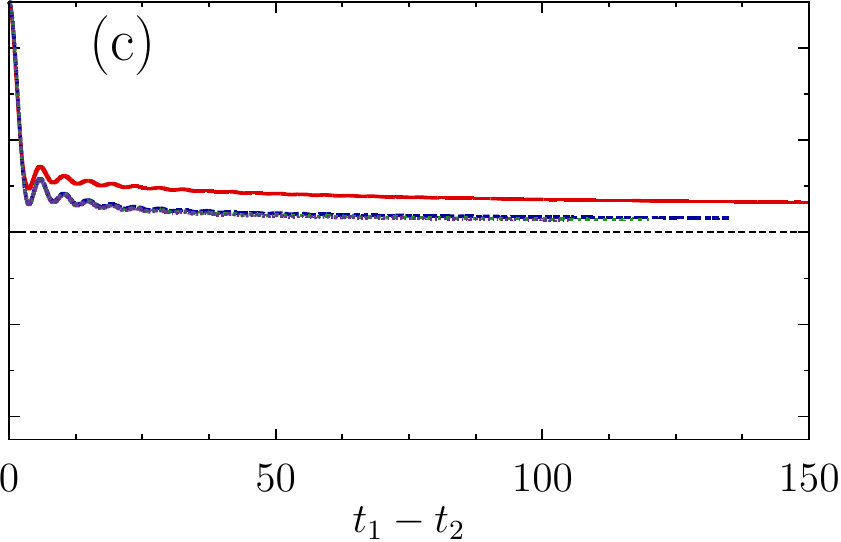} 
\includegraphics[scale=0.4]{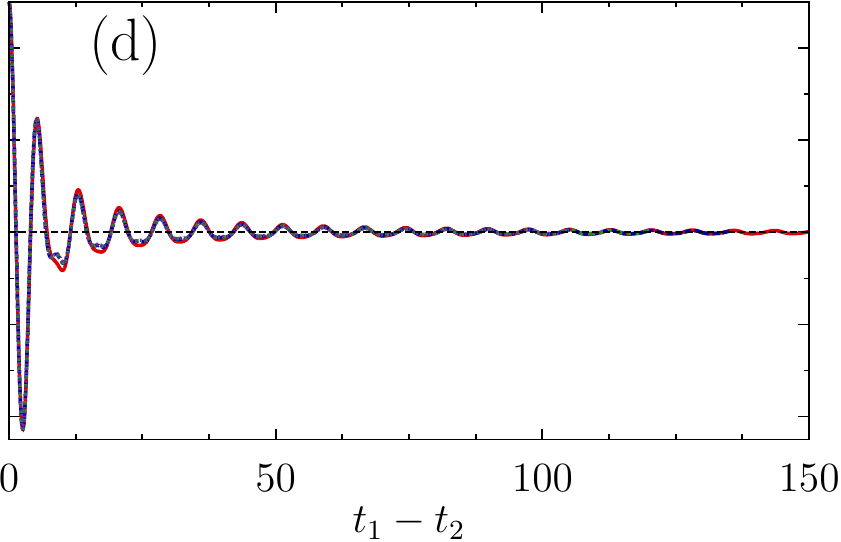} 
\includegraphics[scale=0.4]{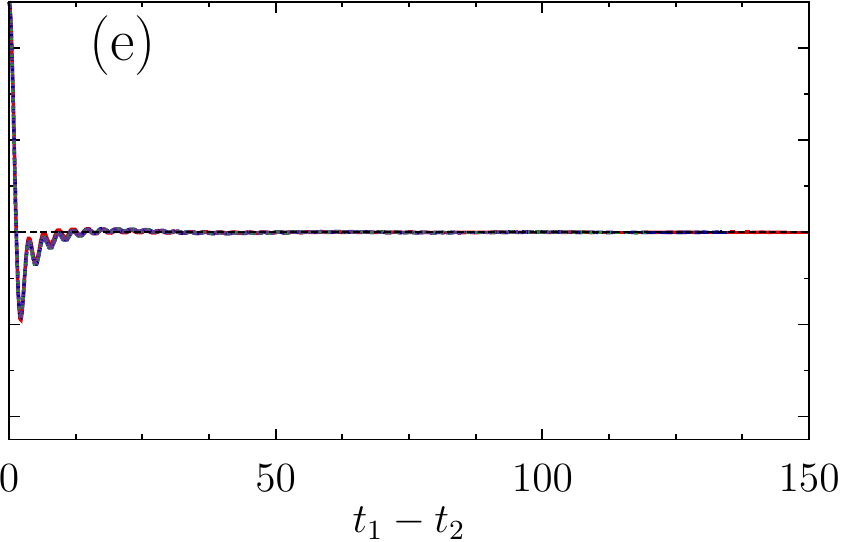} 
}
\\
$\;$
\\
\centerline{
$\;\;\;\;\; y=0.25 \;\; x=0.50 \qquad y=0.75 \;\; x=1.50 \qquad y=0.50 \;\; x=1.75 \qquad y=1.25 \;\; x=1.80 \qquad y=1.50 \;\; x=1.80$
}
\\
$\;$
\\
\centerline{
\includegraphics[scale=0.4]{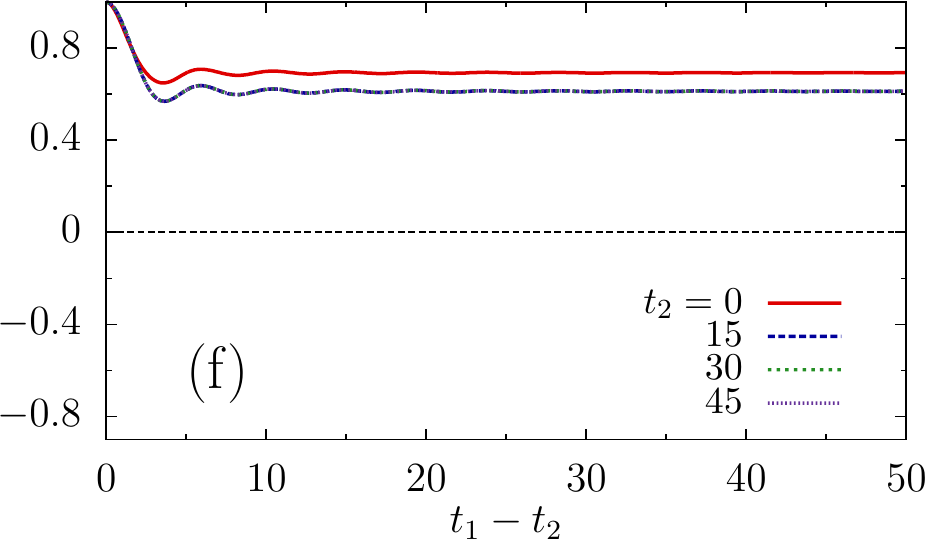}  
\includegraphics[scale=0.4]{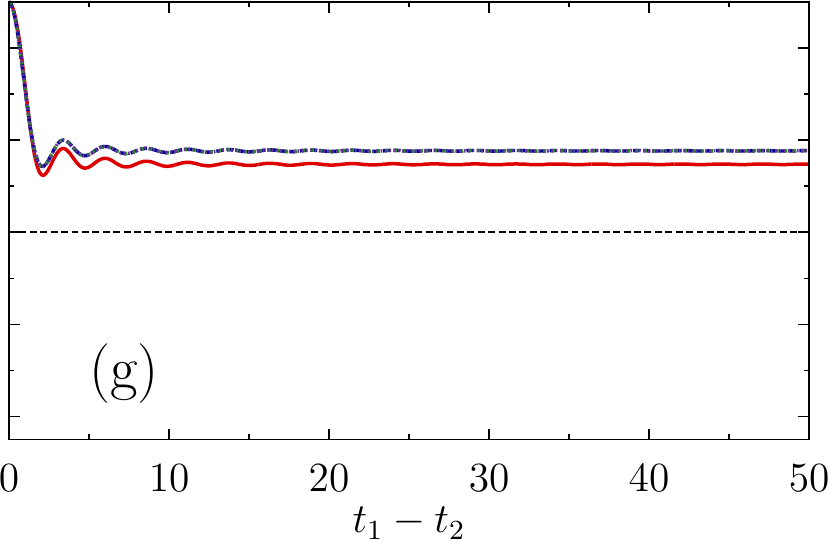} 
\includegraphics[scale=0.4]{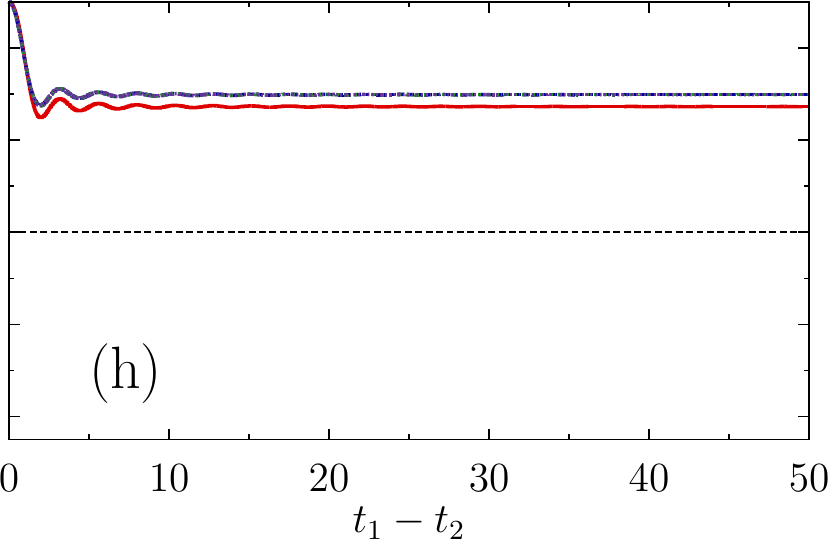} 
\includegraphics[scale=0.4]{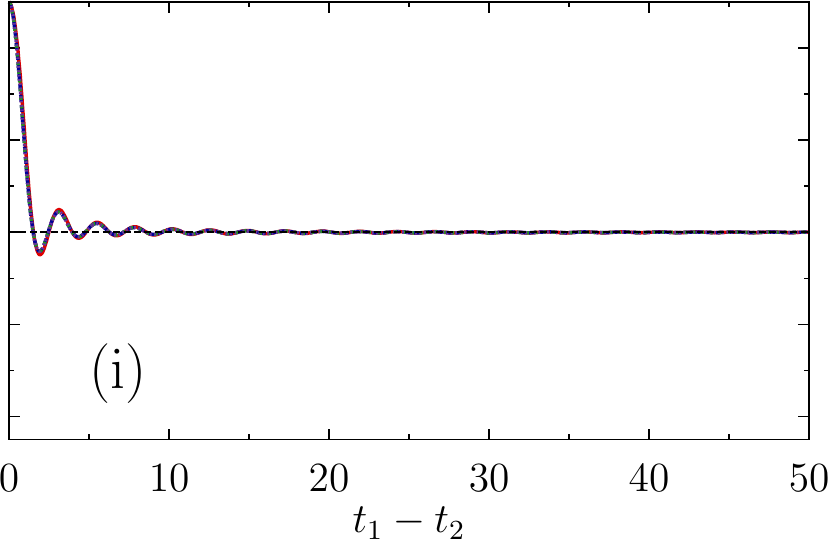} 
\includegraphics[scale=0.4]{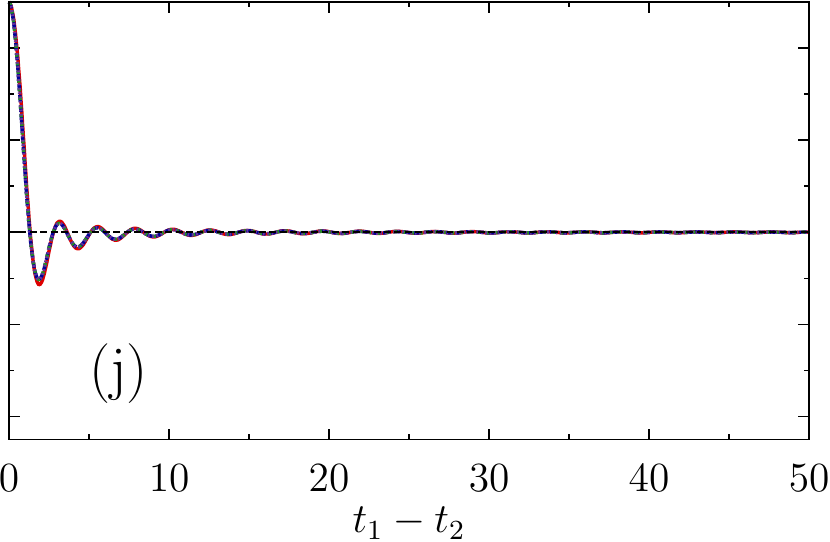} 
}
\vspace{0.15cm}
\caption{\small {\bf The time-delayed correlation function $C(t_1,t_2)$} from the numerical integration of the Schwinger-Dyson equations, for different
values of the parameters $y = T^{\prime}/J_0$ and $x = J/J_0$, as indicated above the plots. Crosses accompanied by the labels (a)-(j) are marked at the corresponding
location in the phase diagram in Fig.~\ref{fig:phase_diagram_new}. In the first row $x\leq y$ while in the second
$x > y$. The first three panels in both rows correspond to $y<1$ and the last two to $y>1$. The third panel in the first row
is on the critical line $x=y$.
}
\label{fig:corr_cmp}
\end{figure}

These facts can be appreciated in Fig.~\ref{fig:corr_cmp} where we display
the decay of the correlation function for several choices of the parameters in different regions of the phase
diagram, marked with crosses in Fig.~\ref{fig:phase_diagram_new}.


\vspace{0.5cm}

Having announced the main features of the dynamics after different types of quenches, in the rest of this Section
we will support these claims with the detailed study of all relevant observables.

\subsection{Constant energy dynamics}
\label{sec:numerical-const-energy}

We first checked that for $J=J_0$ and $m=m_0$, that is to say $\Delta e=0$, and for all initial conditions,
the system has stationary evolution and the total energy as well as other conserved quantities are indeed conserved.
As the dynamics of generic Hamiltonian systems is
hard to control numerically, we  included this analysis to validate our algorithms. Moreover, this study
allowed us to know which is the order of the numerical error incurred into.
A time discretisation step $\delta=0.001$ in the integration of the $N\to\infty$ Schwinger-Dyson equations
was sufficient to assure numerical convergence of our results. In the integration of the finite $N$ problem we
found a weak  dependence on $\delta$ but the value $\delta=0.001$ gave acceptable results.

We studied the equilibrium dynamics for initial paramagnetic configurations ($T'>J_0$) and condensed ones ($T'<J_0$).
We present and discuss the results in two Subsections. As already said, they serve as benchmarks for the more interesting quenching cases
that we put forward later.

\subsubsection{Dynamics in the paramagnetic phase}

\begin{figure}[h!]
\vspace{0.5cm}
\begin{center}
 \includegraphics[scale=0.65]{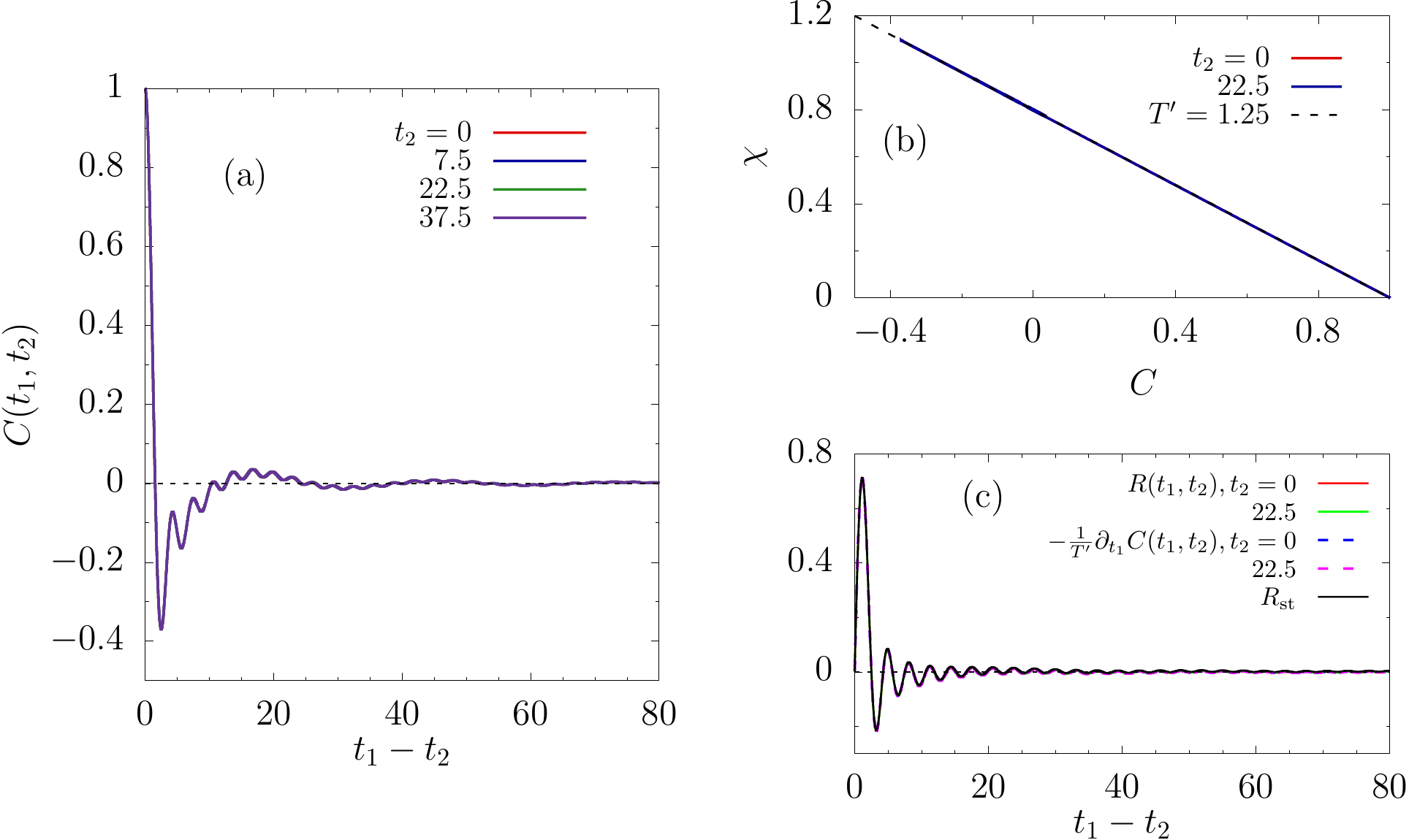}%

 \includegraphics[scale=0.65]{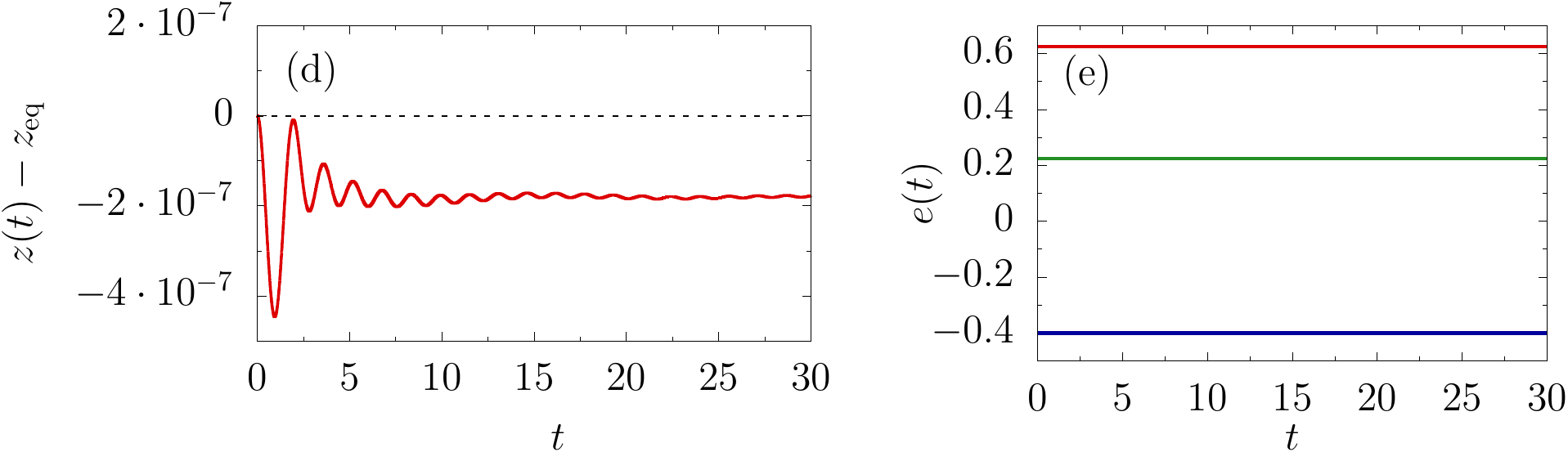}%

 \includegraphics[scale=0.65]{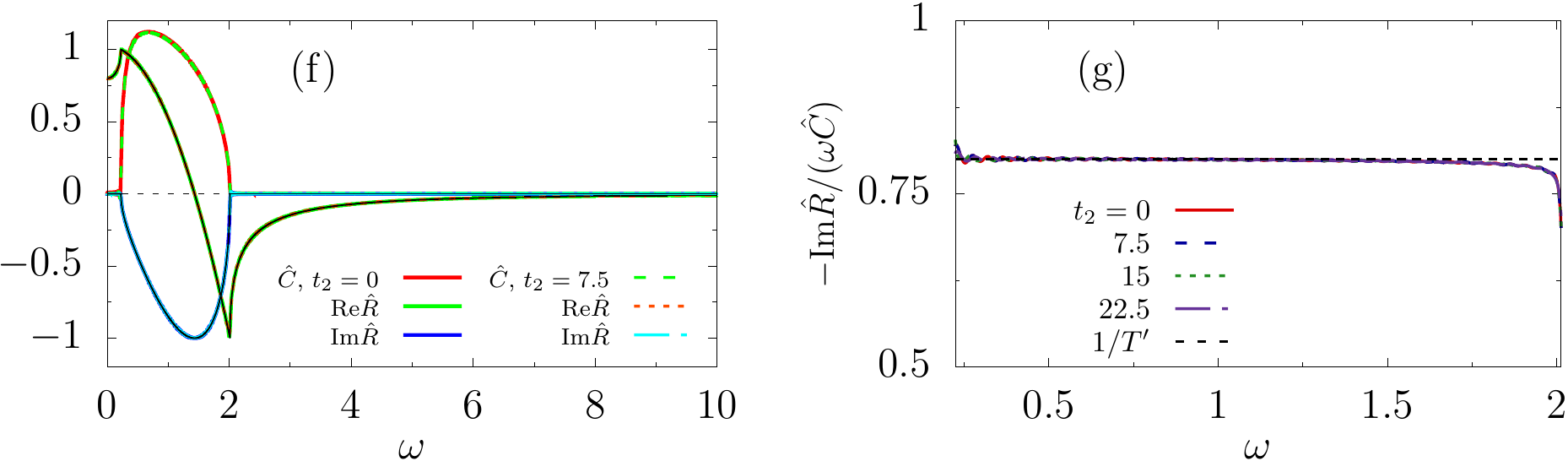}%
\end{center}
\caption{\small {\bf Constant energy dynamics of the $N\to\infty$ system in the PM phase}. $\Delta e=0$ is ensured by $J=J_0$ and $m=m_0$.
$T'=1.25>T^0_c=1$ and the initial condition is in the paramagnetic phase.
(a)~Dynamics of the correlation function for various choices of the waiting time given in the key.
(b)~Linear-response vs. correlation parametric plot for two values of the waiting time $t_2$.
The dashed line shows the FDT with the initial temperature.
(c)~The response function, $R(t_1,t_2)$,   $- (1/T^{\prime}) \, \partial_{t_1}C(t_1,t_2)$,
and $R_{\mathrm{st}}$, the (numerical)  inverse Fourier  transform of the theoretical
prediction given by Eq.~(\ref{eq:response_fourier_transform-text}) with parameters $m=1$, $J=1$ and
$z_{f}=z_{\mathrm{eq}}=T^{\prime}+J^2/T^{\prime}=2.05$,
against $t_1-t_2$, for two values of $t_2$.
 (d) The difference between the numerical Lagrange multiplier, $z(t)$, and the expected value at equilibrium,
$z_{\mathrm{eq}}$.  (e) Time evolution of the kinetic energy density, $e_{\rm kin}$ (red line),
the potential energy density, $e_{\rm pot}$ (blue line) and the total one, $e_{f}$ (green line).
(f) Fourier transforms of the correlation (real part) and the response, for two values of $t_2$.
The black solid lines represent the theoretical predictions for $\hat{R}(\omega)$, given by Eq.~(\ref{eq:response_fourier_transform-text}),
the inverse Fourier  transform of which
is plotted in (c).  We recall that we chose to use a convention such that the imaginary part of $\hat R$ is negative.
(g) The ratio $-\mathrm{Im}{\hat{R}(\omega)}/(\omega \hat{C}(\omega))$ together with
$1/T^{\prime}$ indicated by a dashed horizontal line.
}
\label{fig:T1p25_eq}
\end{figure}

\begin{figure}[h!]
\begin{center}
\includegraphics[scale=0.42]{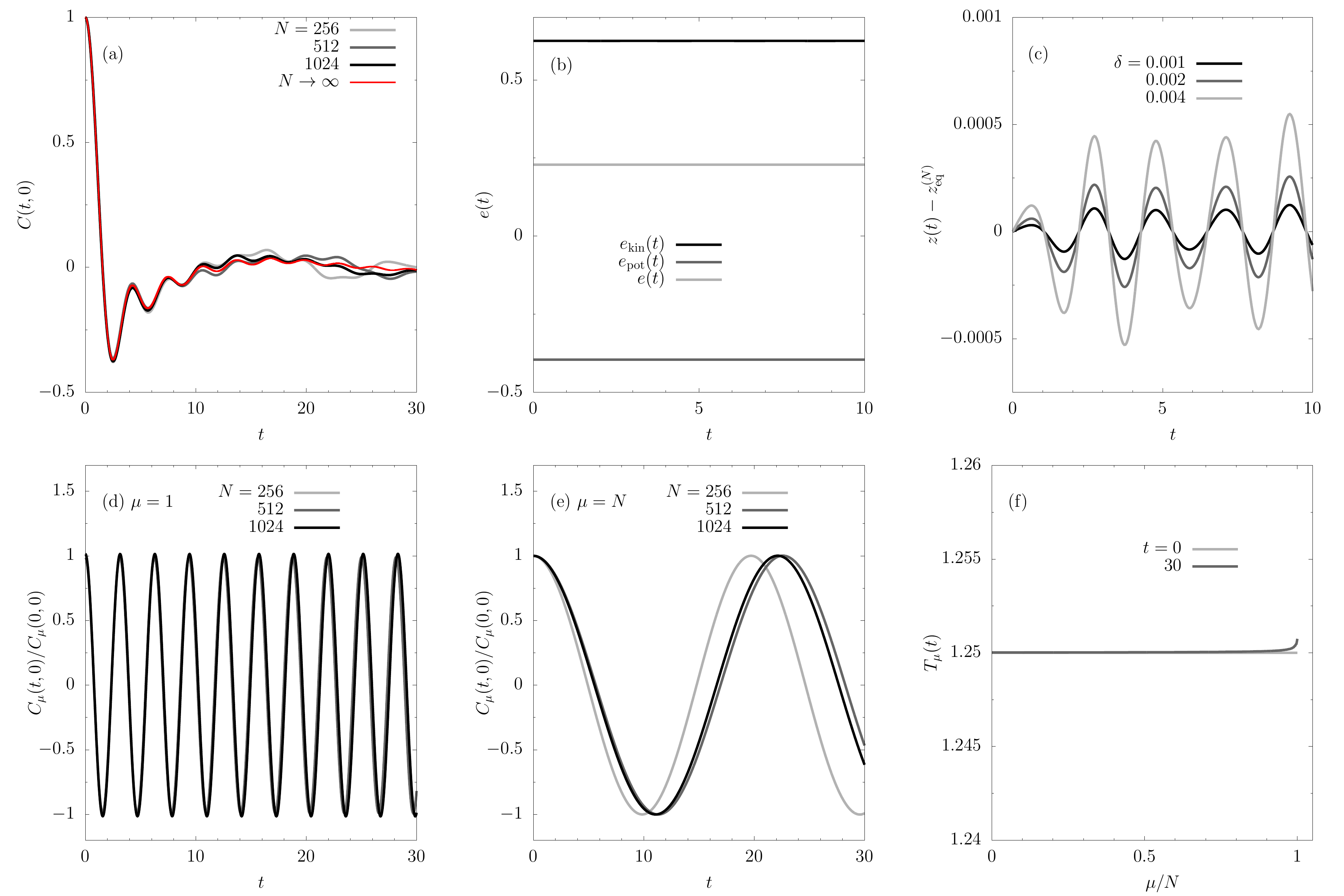}
\end{center}
\caption{\small
{\bf Constant energy dynamics of a finite $N=1024$ system in the PM phase}. $T'=1.25$ and all
other parameters are set to $1$.
(a) The correlation function between a configuration at time $t$ and the initial condition
for different system sizes (grey curves) and the one for $N\to\infty$ (red curve).
(b) The kinetic, potential and total energies as a function of time.
(c) Dynamics of the Lagrange multiplier referred to the finite $N$ equilibrium value
$z^{(N)}_{\mathrm{eq}}$, for different discretisation steps used in the numerical code.
(d) and (e) Mode correlation for different system sizes for $\mu=1$ and $\mu=N$, respectively.
(f) Mode temperatures at different times. In (a), (b), (c)-(e) $\delta=0.001$, the discretisation step
that we adopt in all further studies.
}
\label{fig:cardy_T125_J1}
\end{figure}

In this Section we use parameters in the paramagnetic (PM) equilibrium phase.
We fix $J_0=m_0=m=1$, we equilibrate the initial conditions at $T^{\prime}=1.25$ and we evolve with parameters
$J=J_0=1$ and $m=m_0=1$ (no quench).
In Figs.~\ref{fig:T1p25_eq} and \ref{fig:cardy_T125_J1}  we show the numerical results for finite $N$ and infinite $N$,
respectively. The system should remain in Gibbs-Boltzmann equilibrium in the PM phase in both cases.

In Fig.~\ref{fig:T1p25_eq} we analyse the results of the numerical integration of the $N\to\infty$  Schwinger-Dyson equations
for $T'=1.25$ and all other parameters set to $1$. The figure shows the time evolution of the correlation function~(a),
the fluctuation dissipation parametric plot (b), the response function, $R(t_1,t_2)$ together with $-(1/T^{\prime}) \, \partial_{t_1}C(t_1,t_2)$~(c),
the deviation of the Lagrange multiplier from its equilibrium value~(d), the potential and kinetic contributions to the
energy density~(e), and two studies of the fluctuation-dissipation theorem in the frequency domain~(f) and~(g).

The correlation with the initial condition, $C(t_1,0)$, and the ones between two different times, $C(t_1,t_2)$, are identical,
meaning that time translation invariance is satisfied. All the correlations relax to zero $q=q_0=0$.
Moreover, the curves coincide approximately with the ones in Fig.~\ref{fig:cardy_T125_J1} (a) which were obtained
by integrating Newton equations for finite $N$.

The fluctuation-dissipation relation~\cite{Kubo} is satisfied with the temperature of the initial condition, that is the same as the one of the final state.
This fact can be proven in general for Newtonian evolution of initial configurations drawn from Gibbs-Boltzmann equilibrium~\cite{ArBiCu}.
Indeed in panel (b) we display the parametric plot $\chi(t_1,t_2)=\int_{t_2}^{t_1} dt' \, R(t_1,t')$
{\it vs} $C(t_1,t_2)$ for two waiting times $t_2$ and, with a dashed line, the equilibrium result $-1/T' \; C$ finding
perfect agreement within our numerical accuracy.
As a further confirmation of the validity of the fluctuation-dissipation theorem, we show the
response function, $R(t_1,t_2)$, and $-(1/T^{\prime}) \,  \partial_{t_1}C(t_1,t_2)$,
both plotted against $t_1-t_2$, for two different values of $t_2$ in panel (c).
The two quantities coincide almost perfectly. In the same panel
we checked that the response function coincides with the one derived by taking the inverse of the theoretical Fourier transform of the
stationary asymptotic response, given by Eq.~(\ref{eq:response_fourier_transform-text}).
The theoretical curve is indicated as $R_{\mathrm{st}}$.

The Lagrange multiplier and the potential and kinetic energies remain
constant throughout the evolution of the system and equal to their predicted values,
apart from small deviations due to the numerical errors introduced by the
numerical integration scheme, see panels (d) and (f), and App.~\ref{app:discrete-time}.
The plot showing $z(t)$ proves that the relative error in this quantity is at most of order $10^{-7}$.
All these results are compatible with Gibbs-Boltzmann equilibrium in the paramagnetic phase.
We do not show the time evolution of the off-diagonal correlation with the initial configuration, $C_2(t,0)$, since
it is identically zero at all times.



In panel (f) we show the Fourier transforms of the correlation and response functions,
$\hat{C}(\omega,t_2)$ and $\hat{R}(\omega,t_2)$ respectively (the transform is performed on the variable $\tau=t_1-t_2$ with $t_2$ fixed),
for two different values of $t_2$.
Note that we are showing only the real part of $\hat{C}(\omega,t_2)$, since we implicitly assume that
$C(t_2+\tau,t_2)=C(t_2-\tau,t_2)$.
The black solid lines represent the theoretical prediction for the real and imaginary parts of the Fourier transform of the response function
in the stationary regime, $\hat{R}_{\mathrm{st}}(\omega)$, given by Eq.~(\ref{eq:response_fourier_transform-text}), the inverse Fourier transform of which
is plotted in (c). In panel (g), the ratio $-\mathrm{Im}{\hat{R}(\omega)}/(\omega \hat{C}(\omega))$ together with the
prediction $1/T^{\prime}$ from FDT indicated by a horizontal dashed line are shown. Note the deviation from the
flat result at the right edge of the frequency spectrum. This is due to the fact that the ratio approaches zero over zero and the
numerical error incurred for those large frequencies is much amplified. At the left end of the spectrum, the more interesting low frequency regime,
the oscillations are only present for the $t=0$ curve.

In Fig.~\ref{fig:cardy_T125_J1} we show results obtained by solving the dynamics of each mode
in a finite $N$ system
with the method explained in Sec.~\ref{subsec:parametric-osc}.
In panel (a) we see that the correlations  with the initial condition quickly relax to $0$,
as expected in the PM phase. They do  with a weak size-dependence in the long time-delay tails. We only show the correlation
with the initial configuration since we have checked that the time-delayed one is stationary. Also included in this
panel is the same correlation function computed using the Schwinger-Dyson equation valid in the $N\to\infty$ limit. We
see perfect agreement with the finite $N$ results at short times and small deviations at longer times.
In Fig.~\ref{fig:cardy_T125_J1}~(b) we observe that the global kinetic, potential and total energy densities are constant, as expected.
The Lagrange multiplier is studied in panel (c) where we plot it subtracting
$z_{\rm eq}^{(N)}$, calculated as the solution to Eq.~(\ref{eq:finite_n_sph_constraint}) (a non-linear equation) that in the $N\rightarrow\infty$
limit yields $z_{\rm eq}^{(N)}\rightarrow T'+ J^2/T'$.
The very weak (oscillatory) deviation from $z_{\rm eq}^{(N)}$ decreases with the size of the time-step used
in the numerical solution of the dynamic equations
(in the $N\to\infty$ case we have a  similar effect, see App.~\ref{app:discrete-time}).

The mode-by-mode analysis of the finite $N$ dynamics
is performed in Fig.~\ref{fig:cardy_T125_J1}~(d) and~(e).
Two panels display the time-delay dependence of the correlation function of the
first and last mode.
In Fig.~\ref{fig:cardy_T125_J1}~(f) we display the mode temperatures $T_{\mu}(t)$ at the initial time and after
a long time evolution. The mode temperatures coincide with the expected equilibrium value, except for the largest modes, where there is a very small deviation.
These variations  represent small numerical errors due to the finite time-step discretisation used numerically, and are hard to improve algorithmically
unless by using a still smaller integration step. We have also checked (not shown) that the mode correlations $C_{\mu}(t_1,t_2)$ and the mode response function
$R_{\mu}(t_1,t_2)$ satisfy the fluctuation-dissipation relation with a temperature given by $T^{\prime}$ for all modes.



\subsubsection{Dynamics in the condensed phase}

We now turn to the constant energy dynamics in the condensed, low temperature equilibrium phase.

\begin{figure}[h!]
\vspace{0.5cm}
\begin{center}
\includegraphics[scale=0.4]{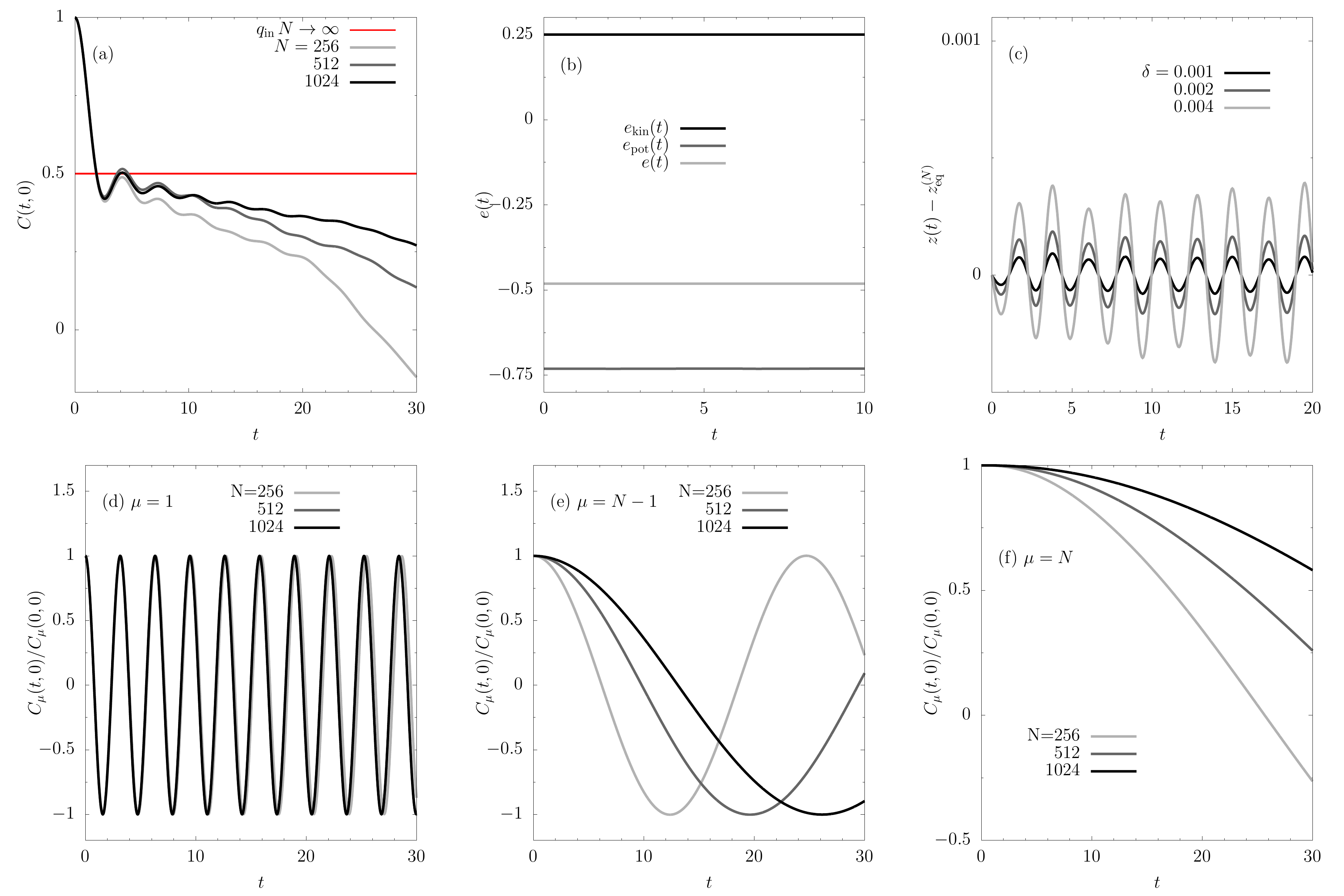}
\end{center}
\caption{\small {\bf Constant energy dynamics in the condensed phase.}
Results from the integration of the Newton equations for the individual modes at $T'=0.5$ for a single disorder realisation.
(a) Dynamics of the global correlation function with the initial condition for different system sizes.
(b) Evolution of the energetic contributions and the total energy.
(c) Dynamics of the Lagrange multiplier for different system sizes referred to the
equilibrium $z_{\rm eq}^{(N)}$ (that is slightly larger than $\lambda_{\rm max}$, see the text).
(d)-(f) Mode correlation function $C_{\mu}(t_1,t_2)$, normalised by their values at equal times, for $\mu=1, \, N-1, \, N$ and different system sizes
indicated in the key. The mode correlations are stationary, so we only show results for $t_2=0$.
}
\label{fig:p2_T05_J1}
\end{figure}

\begin{figure}[h!]
\vspace{0.5cm}
\begin{center}
  \includegraphics[scale=0.7]{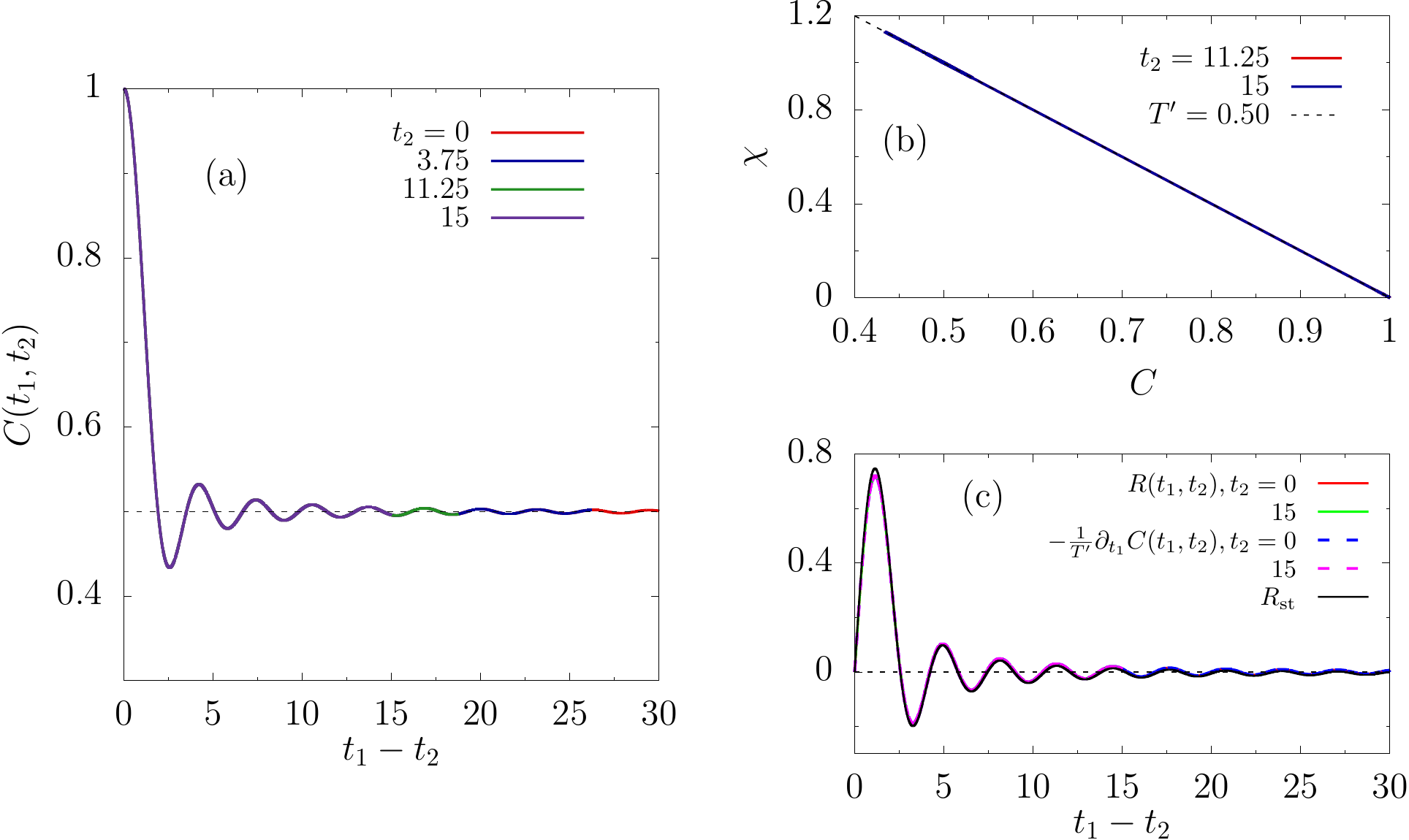}%

  \includegraphics[scale=0.7]{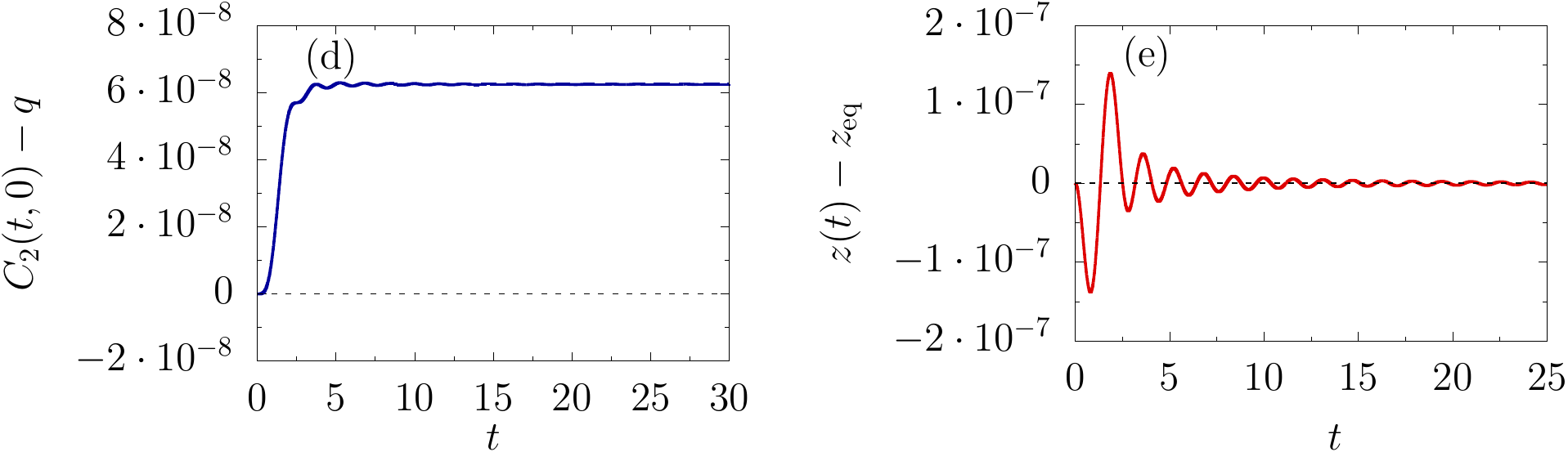}%

  \includegraphics[scale=0.7]{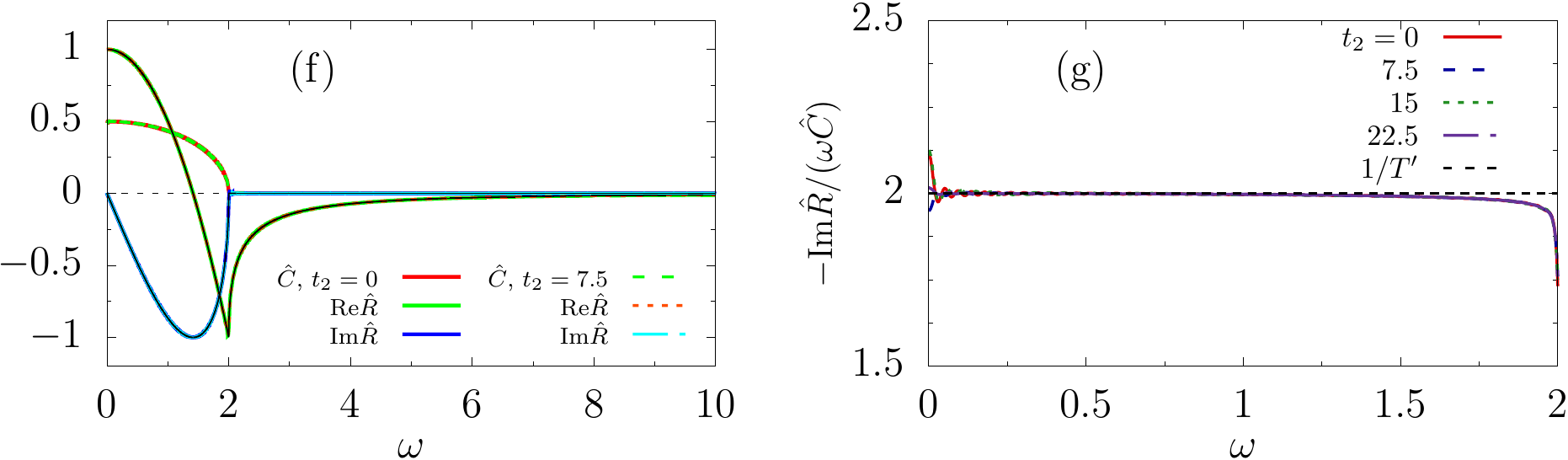}%
\end{center}
\caption{\small {\bf Constant energy dynamics of the $N\to\infty$ system in the condensed phase.} $\Delta e=0$, ensured by $J=J_0$.
$T'= 0.5 < T^0_c = 1$. (a) Stationary dynamics of the two-time correlation
function. The asymptotic limit is  $q =0.5$, shown as a dotted horizontal line.
(b) Linear-response vs. correlation parametric plot.
The dashed line shows the FDT with the initial temperature.
(c) The linear response function, $R(t_1,t_2)$, and the quantity $-(1/T^{\prime})\partial_{t_1}C(t_1,t_2)$,
both plotted against $t_1-t_2$, for two values of $t_2$.
The curve indicated by $R_{\mathrm{st}}$ is the (numerical) inverse Fourier transform of the theoretical
prediction given by Eq.~(\ref{eq:response_fourier_transform-text}) with parameters $m=1$, $J=1$ and $z_{f}=z_{\mathrm{eq}}=2$.
(d) Difference between $C_2(t,0)$ and $q$.
(e) The difference between the time-dependent Lagrange multiplier, $z(t)$, and the expected asymptotic value
$z_{\mathrm{eq}}=2J$.
(f) The Fourier transforms of the correlation and linear response functions, for two different values of $t_2$ indicated in the key.
The black solid lines represent the theoretical prediction for the real and imaginary parts of the Fourier transform of the response function
in the stationary regime, $\hat{R}_{\mathrm{st}}(\omega)$, given by Eq.~(\ref{eq:response_fourier_transform-text}), the inverse transform
of which is plotted in (c). In panel (g), the ratio $-\mathrm{Im}{\hat{R}(\omega)}/(\omega \hat{C}(\omega))$ together with the
prediction $1/T^{\prime}$ from FDT plotted with a dashed horizontal line.
}
\label{fig:T0p50_eq}
\end{figure}

In Fig.~\ref{fig:p2_T05_J1} we show the mode dynamics for initial conditions in equilibrium at
$T'=0.5$.  From Fig.~\ref{fig:p2_T05_J1}~(b) we notice that the total energy is conserved and that the kinetic and potential contributions are also constant,
consistent with thermal equilibrium in the isolated system. In Fig.~\ref{fig:p2_T05_J1}~(c) we show  the Lagrange multiplier, which should be constant
in equilibrium. In the numerical solution, the Lagrange multiplier exhibits oscillations around the initial value.
Their amplitude decreases consistently with the integration step $\delta$, implying that for $\delta\rightarrow0$ we
recover the expected constant behaviour. The two-time global
correlation $C(t_1,t_2)$ is stationary for all system sizes (not shown), so we focus on the particular case
with $t_2=0$. We can see from Fig.~\ref{fig:p2_T05_J1} (a) that, at variance with the
paramagnetic case, the dynamics of the correlation function $C(t_1,0)$ has a strong dependence on the system size.
After a fast decay from the initial value, the correlation shows a plateau,
the lifetime of which increases with system size, approaching asymptotically the value predicted by the $N\to\infty$
treatment that is shown with a (red) horizontal line. The source of this size dependence is the behaviour of
the largest mode $\mu=N$. In Fig.~\ref{fig:p2_T05_J1}~(f) we show the time dependence of the largest mode correlation function
$C_N(t,0)$ for different system sizes. We observe that its oscillation frequency decreases as we increase the system size.
A similar finite size effect is seen in the dynamics of the next-to-largest mode in panel~(g).
Since the largest modes dominate the long-time dynamics, this effect
causes the size dependence of the plateau lifetime.
For $N\rightarrow \infty$ the oscillation frequency of the $N$-th mode goes to zero, allowing for the presence of an infinite plateau,
see Fig.~\ref{fig:T0p50_eq}. The modes lying in the middle and other end of the spectrum have almost no size dependence, as shown in~(d).


\begin{figure}[h!]
\vspace{0.5cm}
\begin{center}
\includegraphics[scale=0.43]{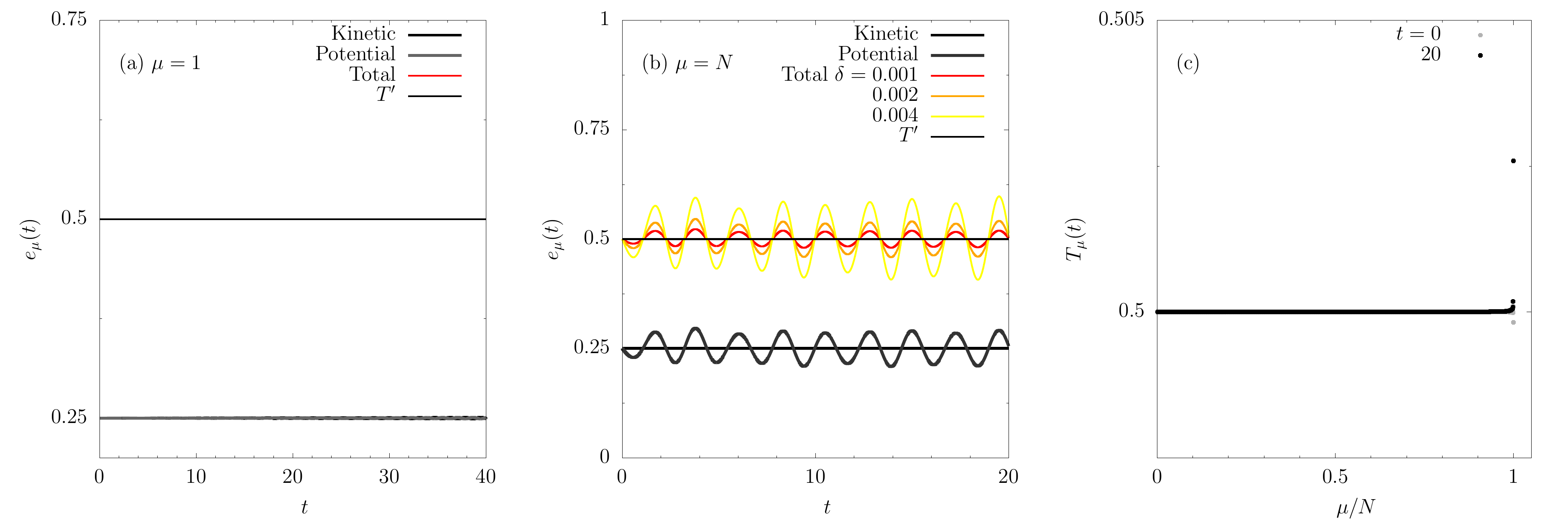}
\end{center}
\caption{\small
{\bf Constant energy dynamics in the condensed phase.}
(a), (b) Mode energies for $T'=0.5$ in equilibrium for different mode indices in a system with $N=1024$.
(c) Mode temperatures at different times.  The total energy (red) is identical to $T'$ (grey) and they both
equal $0.5$. The potential and kinetic energies are equal to $T'/2=0.25$.
}
\label{fig:modetemp_T05_J1}
\end{figure}


Figure~\ref{fig:modetemp_T05_J1} investigates the mode kinetic and potential energies and the
mode temperatures that can be extracted from them.
In Fig.~\ref{fig:modetemp_T05_J1} (c) we show the mode temperatures $T_{\mu}(t)$ at two measuring times, as a function
of the mode index, what we will call temperature spectrum later.
We observe deviations from the expected behaviour $T^{\rm tot}_{\mu}=T'\;\forall\mu$ only
close to $\mu=N$. To gain insight into the mode temperature deviations, in
panels~(a) and (b) we show the time evolution of the kinetic, potential and total energies for the first and last modes. For the lowest modes, the
kinetic and potential energies are constant,  equal, and their sum is identical to $T'$, consistently with equilibrium dynamics.
However, for the modes that are closer to $\mu=N$ the energies weakly oscillate with an amplitude that decreases with the integration step and
should vanish for $\delta \to 0$. This implies that in such limit, all mode temperatures should be equal to the equilibrium temperature,
even for modes close to the right edge of the spectrum.

We now turn to the analysis of the dynamics in the $N\to\infty$ limit.
In Fig.~\ref{fig:T0p50_eq} we show the results obtained through the numerical integration of the Schwinger-Dyson  equations for the
two-time correlation and response functions, and the same choice of parameters, that is $T^{\prime}=0.5$ and $J=1$.
Stationarity is satisfied as well as the FDT with the initial temperature, see (b) and (c).
The main difference with the case in Fig.~\ref{fig:T1p25_eq} is that
the correlation functions, both with the initial condition and with the configuration at a waiting-time $t_2$,
relax to a non-vanishing value (a).
Within numerical accuracy we observe $q_0 = q \simeq 1-T^{\prime}/J = 0.5$ and this value as well as the potential and
kinetic energies (not shown) are consistent with equilibrium at $T'$.
Also in this case, we checked that the response $R(t_2+\tau,t_2)$,
for $t_2 \ge 0$, coincides with the (numerical) inverse Fourier transform of the
theoretical prediction given by Eq.~(\ref{eq:response_fourier_transform-text}) for the Fourier transform
of the stationary asymptotic $R$, see panel (c).


In panel (d) we show the time-evolution of the off-diagonal correlation, $C_2(t,0)$. In the case of equilibrium dynamics, $C_2(t,0)$ should be constant
and equal to $q_{\mathrm{in}}$. As one can see, the value of $C_2(t,0)$ obtained by numerical integration is not exactly a constant function, but
it approaches a constant in the long time limit which differs from $q=q_{\mathrm{in}}=0.5$ by a very small amount. This deviation
is only due to the approximations introduced by the numerical integration scheme.
The same can be said about the behaviour of the numerical $z(t)$, see panel (e). Its value oscillates around the
expected equilibrium value $z_{\mathrm{eq}}=2 J = 2$ at $T^{\prime}=0.5$,  with oscillations amplitude
of order $10^{-7}$ for short times  and decreasing with time.

We next show the Fourier transforms of the correlation and response functions,
for two different values of $t_2$ in Fig.~\ref{fig:T1p25_eq} (f).
Again, note that we are showing only the real part of $\hat{C}(\omega,t_2)$, since we implicitly assume that
$C(t_2+\tau,t_2)=C(t_2-\tau,t_2)$.
The black solid lines represent the theoretical prediction for the real and imaginary parts of the Fourier transform of the response function
in the stationary regime, $\hat{R}_{\mathrm{st}}(\omega)$, given by Eq.~(\ref{eq:response_fourier_transform-text}), the inverse transform
of which
is plotted in (c). In panel (g) we display the ratio $-\mathrm{Im}{\hat{R}(\omega)}/(\omega \hat{C}(\omega))$ together with the
prediction $1/T^{\prime}$ from FDT indicated by a dashed horizontal line. In all presentations we find good agreement with the validity of FDT
with the proviso that in the  plot in (g) the high frequency regime is contaminated by the numerical error, and the
low frequency regime by the fact that we can perform the Fourier transform on a finite time window only, and this causes the
dependence on $t_2$ shown in the plot.


\subsection{Instantaneous quenches}
\label{sec:numerical-quenches}

We shall now vary the initial temperature $T'$ and the coupling $J$ of the Hamiltonian that drives the time evolution
to consider specific instantaneous quenching processes that inject or extract energy. The
aim is to  illustrate the analytical results of the previous Section and put them to the test. As we have already announced the structure of
the phase diagram, we will consider representative quenches in the four sectors, labeled I ($y>1$ and $y>x$), II ($y>1$ and $y<x$), III ($y<1$ and $y<x$) and IV
($y<1$ and $y>x$).

For each of the quenches performed we investigate if the system reaches a stationary state by checking whether:
\begin{itemize}
\item
The one-time quantities $z(t)$ and $e^{f}_{\mathrm{tot}}=\lim_{t \rightarrow +\infty} e_{\mathrm{tot}}(t)$
approach constants that we measure and compare to
the ones predicted in Sec.~\ref{sec:asymptotic}.
\item
The kinetic and potential energy average over time to constant
$\overline{e}_{\rm kin}$ and $\overline{e}_{\rm pot}$ that we also compare to the ones predicted
in Sec.~\ref{sec:asymptotic}.
 \item
The two-time correlation and linear response become stationary after a sufficiently long time, that is
$C(t_1,t_2) \sim C_{\mathrm{st}}(t_1-t_2)$ and $R(t_1,t_2) \sim R_{\mathrm{st}}(t_1-t_2)$
for $t_2$ large.
\item
The long-time limits $q_0=\lim_{t \rightarrow +\infty} C(t,0)$ and $q_2=\lim_{t \rightarrow +\infty} C_2(t,0)$
are equal or differ.
\end{itemize}
We also evaluate whether:
\begin{itemize}
 \item The susceptibility $\chi(t_1,t_2)=\int^{t_1}_{t_2}  \mathrm{d}t \; R(t_1,t)$
and $C(t_1,t_2)$ satisfy the FDT relation $\chi(t_1,t_2)= (1 - C(t_1,t_2) )/T_f$,
with a single temperature $T_f$.
\item The response function $R(t_1,t_2)$ coincides with  $-1/T_f \, \partial_{t_1} C(t_1,t_2)$,
for different values of $t_2$, and with the (numerical) inverse Fourier transform of the analytic form given in Eq.~(\ref{eq:response_fourier_transform-text}).
\end{itemize}
In all cases an asymptotic stationary regime is reached but it is not characterised by a single temperature. Consequently, we  study the
spectrum of mode temperatures as defined from the (time-averaged) kinetic and potential mode energies and we
compare it to the global fluctuation dissipation ratio in the frequency domain in each type of quench. This is motivated by the suggestion in~\cite{FoGaKoCu16,deNardis17}
(for quenches of isolated quantum integrable  systems)
that these should be related.

\subsubsection{Sector I: a paramagnetic initial condition}

\begin{figure}[h!]
\begin{center}
  \includegraphics[scale= 0.6]{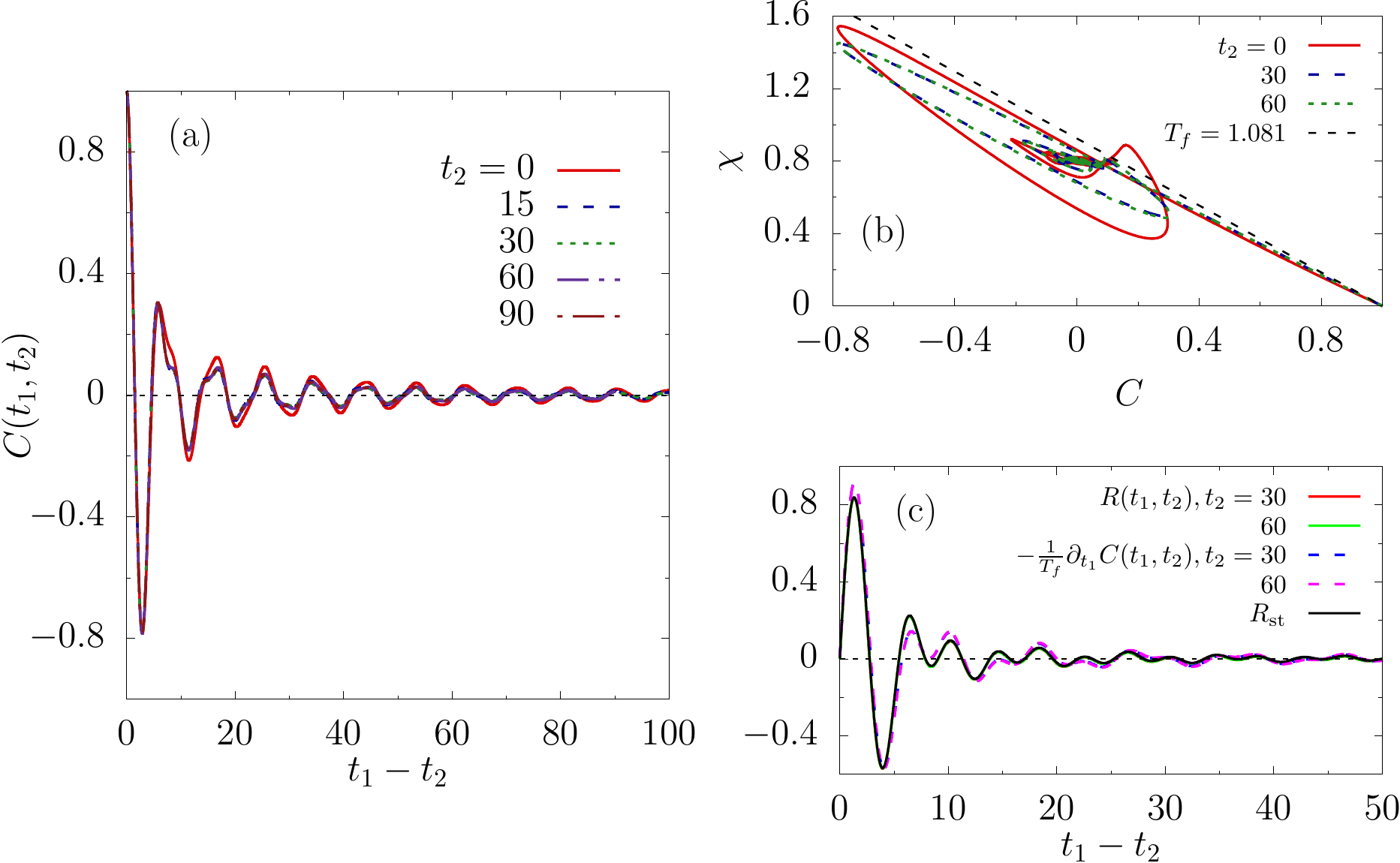}\quad%
\end{center}
  \vspace{0.1cm}
\begin{center}
  \includegraphics[scale=0.6]{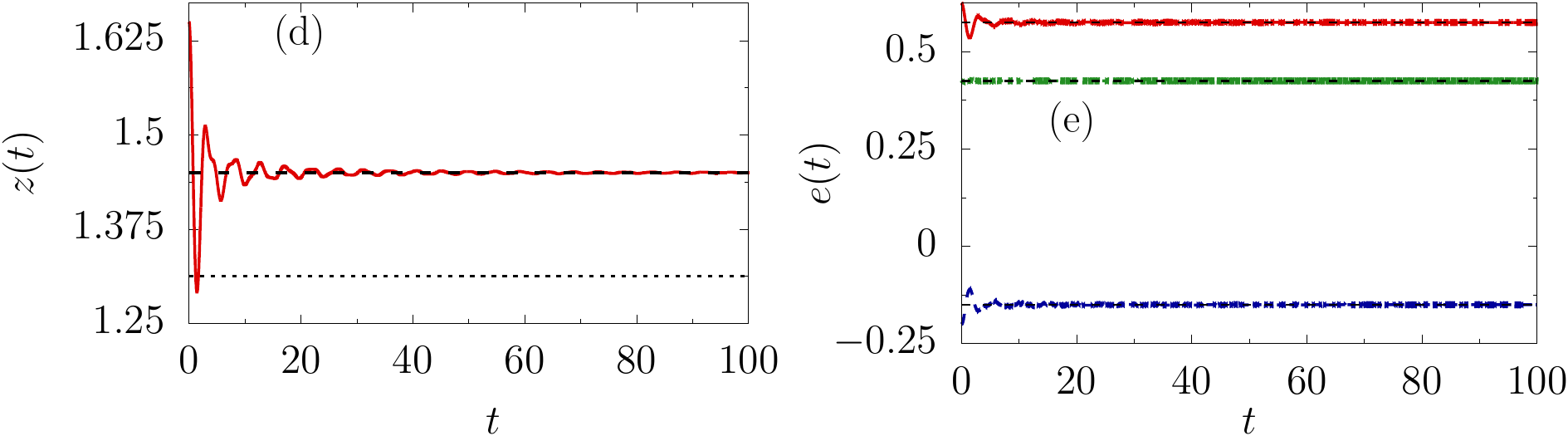}
\end{center}
  \vspace{0.1cm}
\begin{center}
  \includegraphics[scale=0.6]{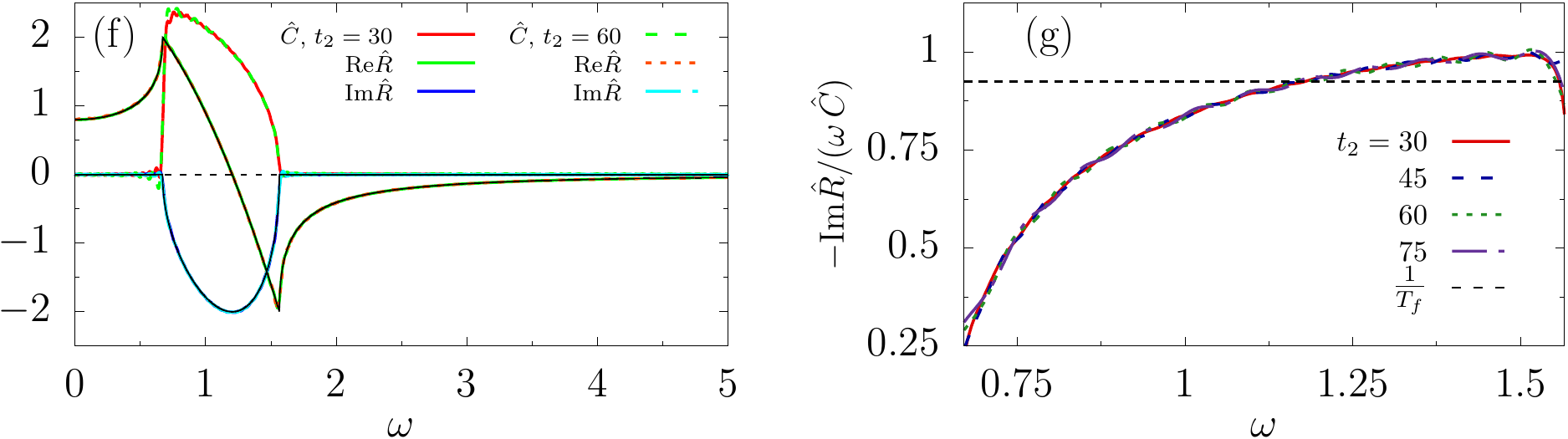}%
\end{center}
\caption{\small%
{\bf Sector I. Energy injection on a paramagnetic initial state.}
$T'=1.25 \, J_0$, $J=0.50 \, J_0$
and $\Delta e=0.20$.
(a) The time-delayed correlation function satisfies stationarity for $t_2 \gg 1$.
The horizontal dashed line is at $q=0$.
(b) The parametric plot of the linear response function, $\chi(t_{2}+\tau,t_{2})$, against the correlation, $C(t_{2}+\tau,t_2)$,
for two waiting times $t_2$.
The black dashed line shows the FDT relation with $T_f\simeq 1.081$ from Eq.~(\ref{eq:Tf_from_para_to_para}). The numerical data exclude this possibility
for the full range of $C$.
In (c) $R(t_1,t_2)$ and $-1/T_f \, \partial_{t_1} C(t_1,t_2)$  as a function of $t_1-t_2$.
In this scale we see no difference between the two functions; the presentation is misleading, since there is a difference, see (b).
(d) Time evolution of the Lagrange multiplier, $z(t)$, along with the constants $T_f+J^2/T_f \simeq 1.31$ (below)
and $T' + J^2/T' = 1.45 $ (above)
represented by dashed horizontal lines.
(e) From top to bottom: the kinetic energy, the total energy and the potential energy densities
 in very good agreement with their expected values.
(f) The Fourier transforms
of the correlation and response functions with respect to time delay.
The black solid lines represent  the real and imaginary parts of
$\hat{R}_{\mathrm{st}}(\omega)$, given by Eq.~(\ref{eq:response_fourier_transform-text}).
In panel (g), the ratio $-\mathrm{Im}\hat R(\omega)/(\omega \hat C(\omega))$ together
with $1/T_f \simeq 0.92$ that is off the data but not very far away from the constant
at the limit $\omega\to\omega_c^+$ (recall that the downward trend close $\omega_c^+$ is due to numerical inaccuracy).
We note that $1/T_{\rm kin} \simeq 0.87$ is still farther away from the high frequency limit.
Results for the same parameters and finite $N$ are shown in Fig.~\ref{fig:modetemp_T125_J05}.
}
\label{fig:T2p00_J0p50}
\end{figure}

\begin{figure}[h!]
\vspace{0.25cm}
\begin{minipage}{.6\textwidth}
\begin{center}
\includegraphics[scale=0.4]{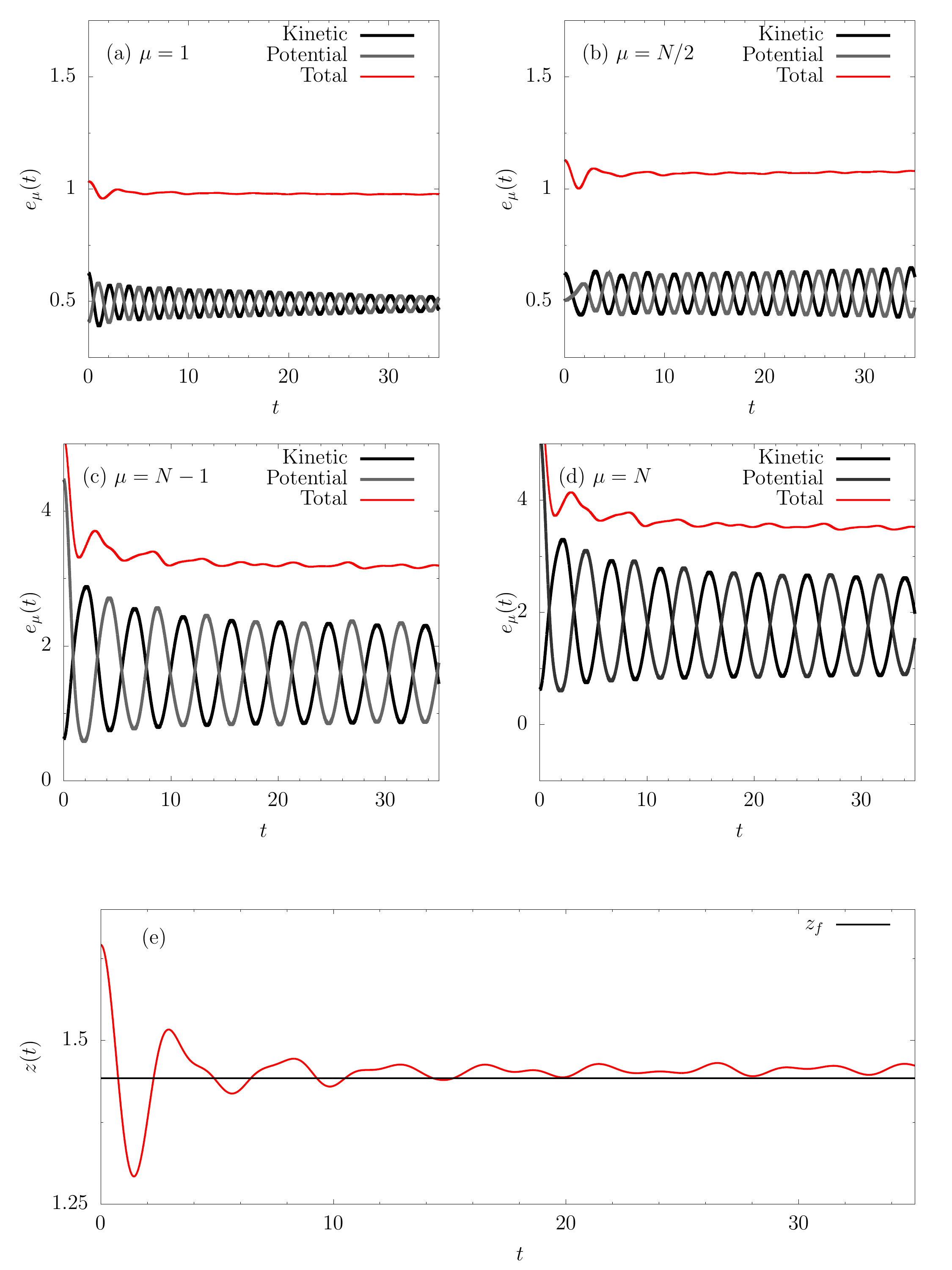}
\end{center}
\end{minipage}
\hspace{-0.5cm}
\begin{minipage}{.4\textwidth}
\includegraphics[scale=0.48]{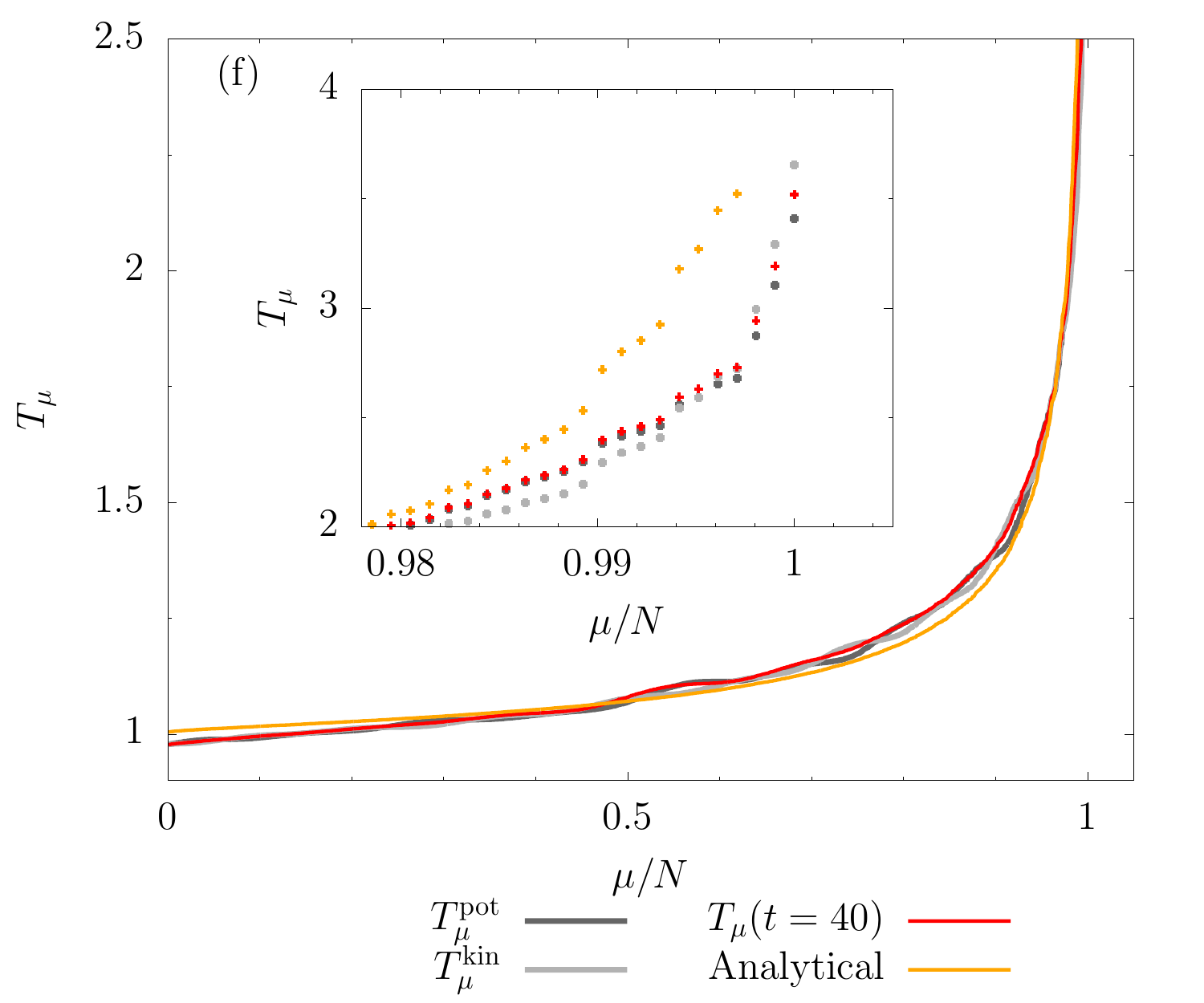}
\\
\includegraphics[scale=0.58]{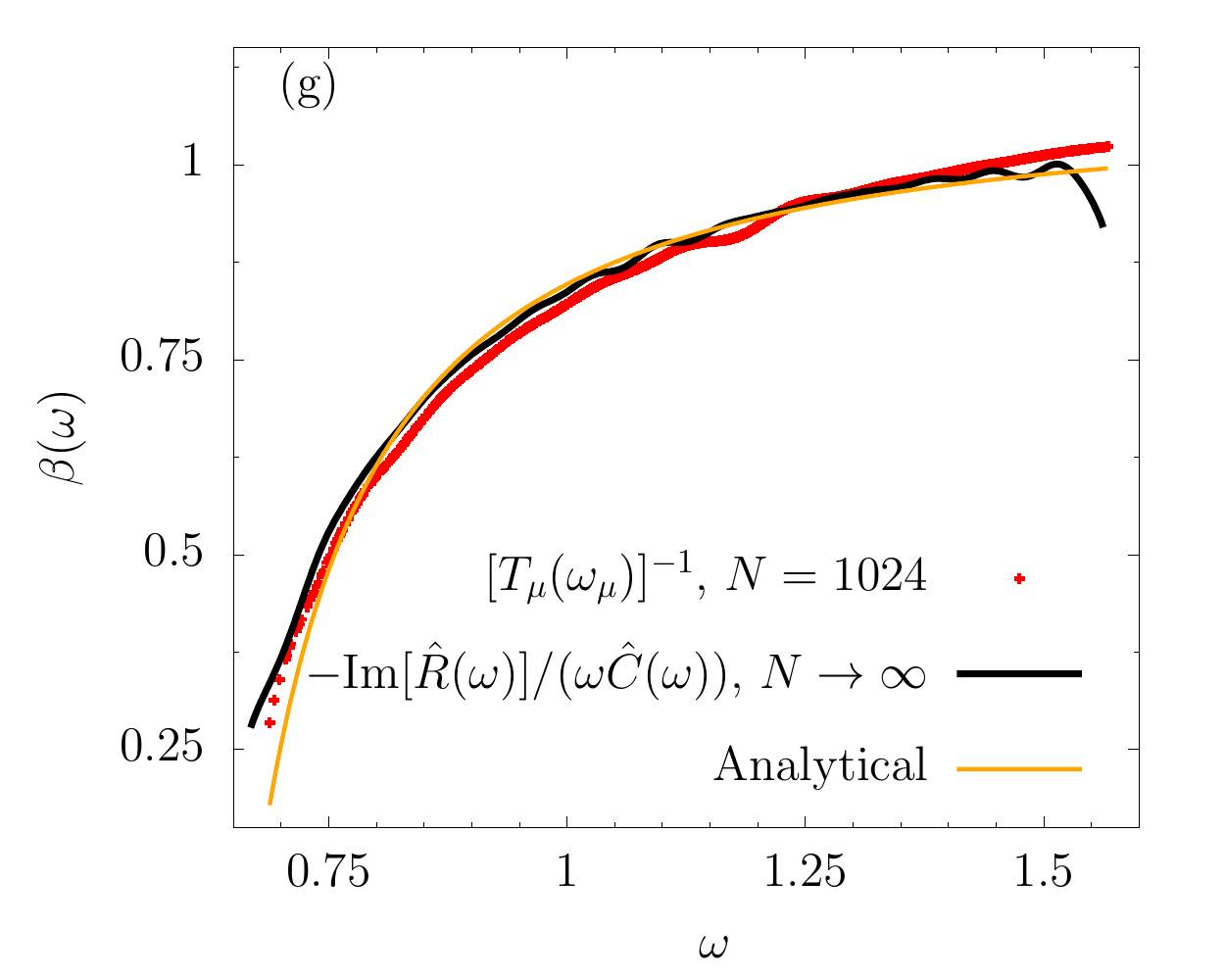}
\end{minipage}
\caption{\small
{\bf Sector I. Energy injection on a paramagnetic initial state.}
A system with $N=1024$ and parameters $T'=1.25 \, J_0$ and $J=0.5 \, J_0$,
leading to $\Delta e =0.20$.
(a)-(d) The time-dependent energy of four selected modes. (e) The Lagrange multiplier compared to the $N\to\infty$ prediction
$z_f=T'+\lambda^2_{\rm max} /(4T')$. There is a small deviation most likely due to numerical error.
(f) Mode temperatures extracted from the use of equipartition, $T_\mu^{\rm kin, pot} = 2 \overline{e}_\mu^{\rm kin,pot}$.
Inset: Detail of the behaviour of the largest modes.
(g) Comparison between the modes inverse temperatures and the inverse effective temperature from the fluctuation
dissipation relation of the $N\to\infty$ equations in the frequency domain (both measured in units of $J_0$).
In (f) and (g) the yellow curves are given by the
approximate prediction for $T_\mu$ in Eq.~(\ref{eq:mode-temp-prediction}). The frequency interval where
$\mbox{Im} \hat R(\omega)\neq 0$ is
$[\omega_-, \omega_+]$ with $\omega_-\simeq 0.67$ and $\omega_+\simeq 1.57$ in this case.
}
\label{fig:modetemp_T125_J05}
\end{figure}

\begin{figure}[h!]
\vspace{0.5cm}
\begin{center}
\includegraphics[scale=0.45]{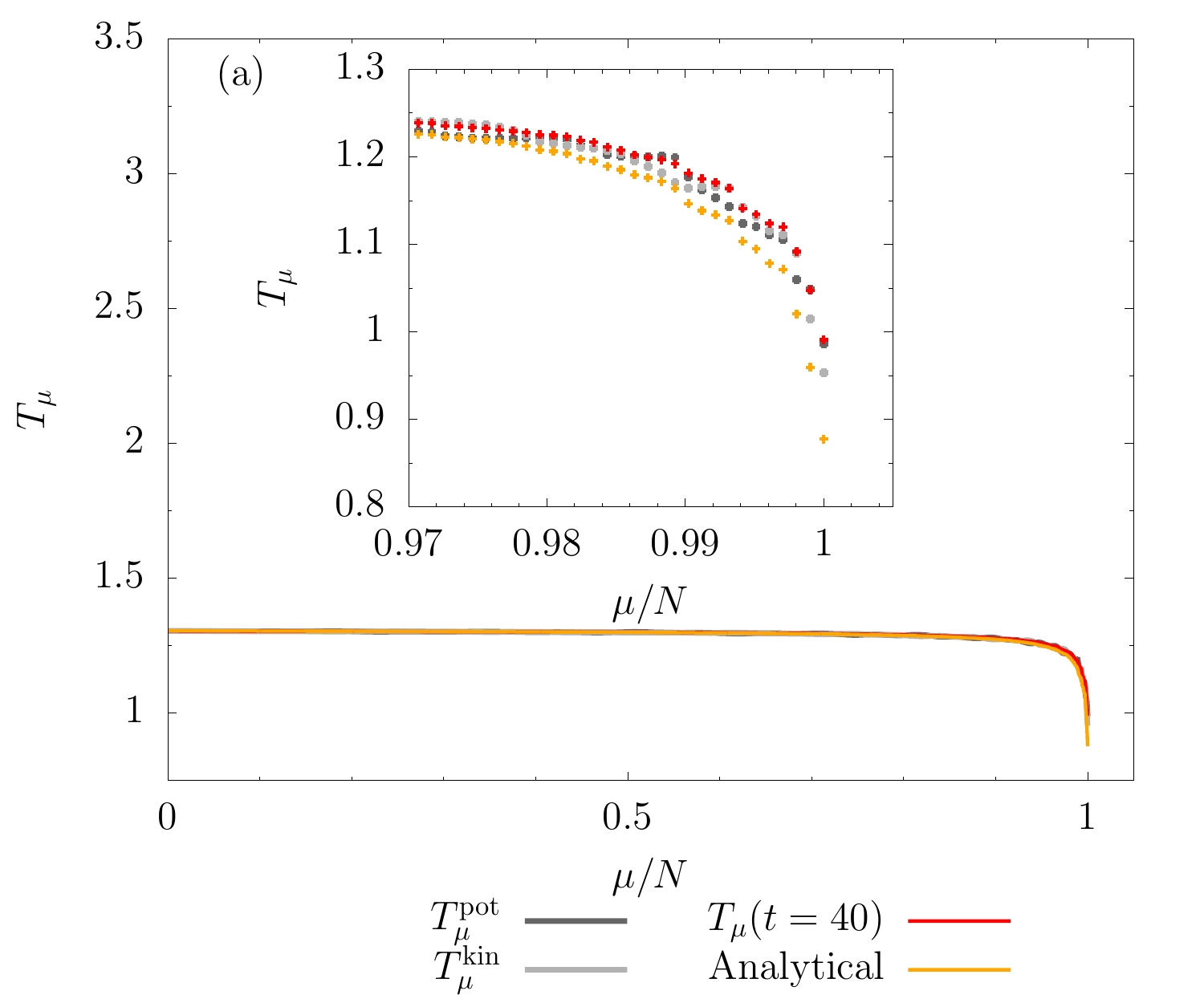}
\includegraphics[scale=0.55]{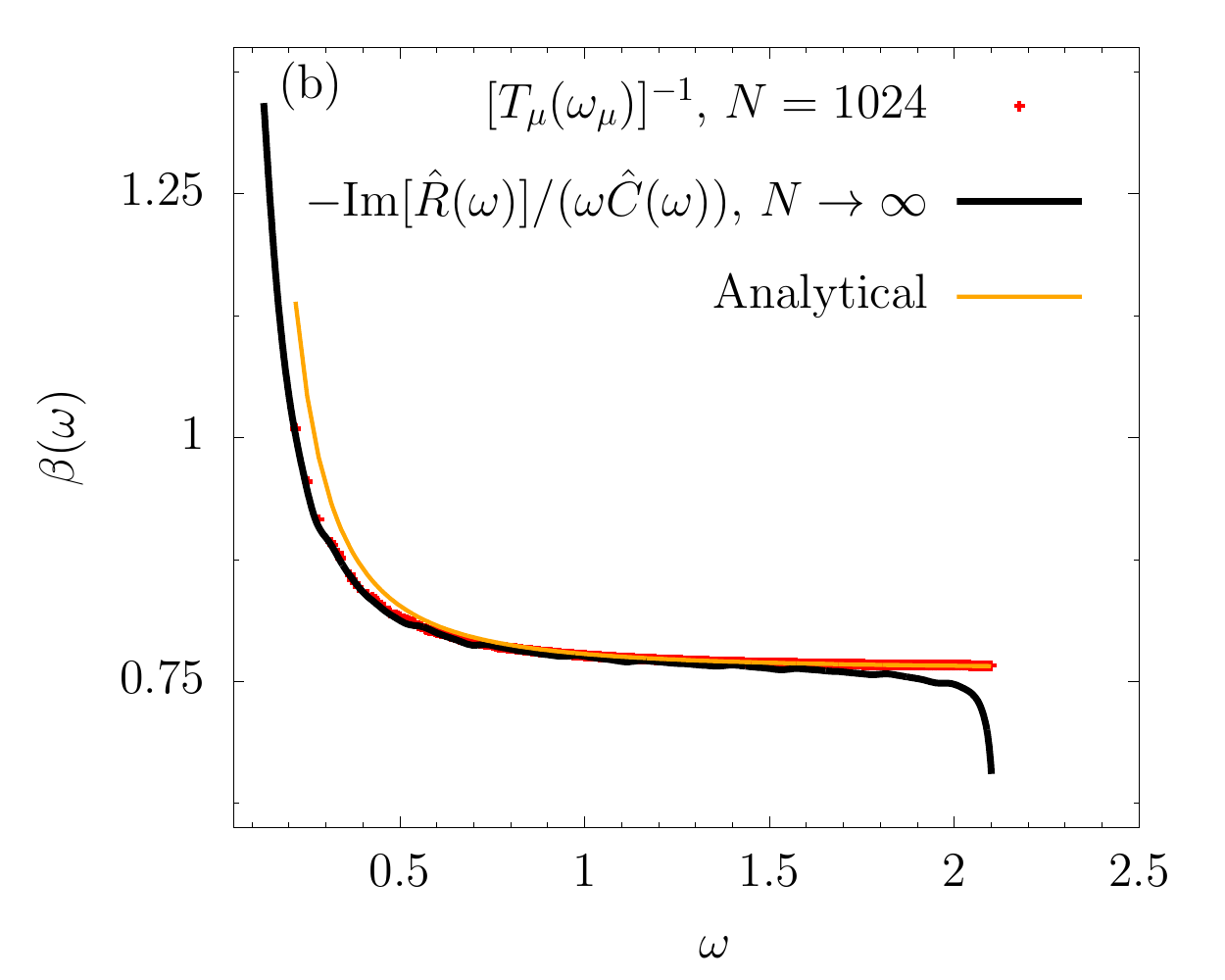}
\end{center}
\caption{\small
{\bf Sector I. Energy extraction from a paramagnetic initial state.} $T'=1.25 \, J_0$ and $J=1.1 \, J_0$ such that $y>x>1$.
(a) Non-equilibrium temperature profile with the temperatures of the largest modes being smaller than the rest
for energy extraction. Inset: Zoom over the behaviour of the largest modes.
(b) Comparison of the inverse mode temperatures with the frequency-dependent effective inverse temperature of the global fluctuation dissipation
relation. Also shown with a yellow solid line is the approximate analytic prediction in Eq.~(\ref{eq:mode-temp-prediction}).
The frequency interval where $\mbox{Im} \hat R(\omega)\neq 0$ is
$[\omega_-, \omega_+]$ with $\omega_- = 0$ and $\omega_+\simeq 2.10$ in this case.}
\label{fig:modetemp_T125_J14}
\end{figure}

In this Subsection we summarise the dynamics of a system prepared in equilibrium in the
paramagnetic state $y>1$ and quenched to a value of $J$ such that $y>x$. This is what we called Sector I in the
phase diagram. Two cases need to be further distinguished within this Sector. For $x<1$ energy is injected in the quench
and this problem is treated in Figs.~\ref{fig:T2p00_J0p50} and \ref{fig:modetemp_T125_J05}. For $y>x>1$, instead,
a small amount of energy is extracted from the system and we explain the difference that this implies in Fig.~\ref{fig:modetemp_T125_J14} and the discussion
around it.

The first observation is that we confirm that, in both cases $x>1$ and $x<1$, $\hat R(\omega=0)=1/T'$ and $z_f = T' + J^2/T'$.
The latter claim can be verified in Fig.~\ref{fig:T2p00_J0p50} (d) for $x>1$ and $N\to\infty$, for example.
Next we check that the dynamics approach a stationary asymptotic state in which the global one-time
quantities are constant, as seen for example in panels (d) and (e) in the same figure,
and the two-time correlation and linear response depend on the time delay only, see panels (a) and (c).
The last question concerns the fluctuation dissipation relation between linear response and correlation function.
We used the parametric plot (b) to demonstrate that the evolution does not approach a
state characterised by a single temperature but, instead, $\chi(C)$ is curved and not even single valued.
Having said this, the comparison of the time-dependent $R(t_1,t_2)$ and $-\partial_{t_1} C(t_1,t_2)$ in panel (c) could have fooled us into
the belief that the FDT holds with a single temperature. Indeed, the difference between the two
functions is not visible  in this scale.
The Fourier analysis in (f) demonstrates that the frequency dependence of the
real and imaginary parts of the linear response
are highly  non-trivial, though respecting the limits in the frequency interval where it does not vanish
derived in Sec.~\ref{subsubsec:thelinearresponse}. The fluctuation-dissipation ratio is shown in (g) and we will come back to its analysis
below, where we present the mode dependent results for finite $N$.


We next study the evolution under the same parameters in a single system with $N=1024$ modes.  In the first
four panels in Fig.~\ref{fig:modetemp_T125_J05} we display the time dependence of the kinetic and potential energies, $e_\mu^{\rm kin}(t)$
and $\epsilon_\mu^{\rm pot}(t)$. These functions oscillate as a function of time with amplitude that slowly decreases in the
cases $\mu=1, \mu=N-1, \mu =N$, while it slowly increases for
$\mu=2$, in this time window. The total energy of each mode shown with a solid red line displays a small downward drift towards, one may expect, a constant value.
The Lagrange multiplier shows a residual time variation around a value that is slightly above the
$N\to\infty$ prediction
but is very close to its finite $N$ modified value $T' + \lambda_{\rm max}^2/(4T')$ (shown with a solid horizontal
line). We then determine the mode temperatures from the equipartition of the kinetic and potential energies averaged
over a sufficiently long time window. All modes are in equilibrium with themselves  in the sense that
the potential, kinetic and total mode temperatures coincide except for small deviations present in the largest modes.
Panel~(f) shows the non-equilibrium temperature profile with the largest modes having higher temperature than the others.
$T_f$, the temperature obtained assuming a paramagnetic final state in equilibrium at  a single temperature, see Eq.~(\ref{eq:Tf_from_para_to_para}),
is clearly different from the mode temperatures and is not the average of them either (not shown), confirming that under these quenches
the system does not equilibrate to the paramagnetic
state. Finally, panel (g) displays the comparison between the mode inverse temperatures and the
frequency dependence of the fluctuation-dissipation inverse temperature. The agreement is very good
(except at the edge of the spectrum where both numerator and denominator vanish and it is very hard to control the limiting
behaviour even in the equilibrium case, see Fig.~\ref{fig:T1p25_eq}).
The (yellow) continuous line is the approximate theoretical prediction in Eq.~(\ref{eq:mode-temp-prediction}) that captures the numerical behaviour
rather accurately. We recall that it was derived assuming that the Lagrange multiplier takes its asymptotic constant value
immediately after the quench, at time $t=0^+$, and that  the ensuing dynamics is the one of independent harmonic
oscillators with mode-dependent frequencies.

As already mentioned, parameters in Sector I also permit quenches with energy extraction, realised by $x>1$. We repeated the analysis
of the $N\to\infty$ equations for parameters in this regime, choosing, for example, $T'=1.2 \, J_0$ and $J= 1.1 \, J_0$. We do not
present the data here as there are no fundamental variations with respect to what we have already discussed for the energy
injection case. Having said so, the mode temperatures do present an interesting difference that we discuss
with the support of Fig.~\ref{fig:modetemp_T125_J14}. Also in this case, the temperatures of the modes are approximately the same
for the low lying modes while a mode dependence appears close to the edge of the spectrum. However,
the temperatures  of the largest modes are, in this case, lower than the temperatures of the lower modes, see panel (a) and its inset.
We ascribe this feature to the
fact that the quench extracts energy from the system.
In panel (b) we confront the mode inverse temperatures to the ones extracted from the fluctuation dissipation ratio in the frequency
domain and, once again, the agreement between numerical curves for $N\to\infty$ and finite $N$ data is very good. Moreover,
the data are also in good agreement with the outcome of the assumption $z(0^+) =z_f$ that leads to Eq.~(\ref{eq:mode-temp-prediction}), shown
with a yellow solid line.


\vspace{0.5cm}

\subsubsection{Sector II: large energy extraction from a paramagnetic initial state}

For these parameter values the Lagrange multiplier approaches $z_f=2J$. We confirmed this claim with the study of
several cases in this Sector (see Fig.~\ref{fig:response}). Concerning the behaviour of the other global observables,
energies, correlation and linear response, and the mode-dependent ones, we differentiate three cases lying in Sector II:
$y > \sqrt{x}$, $y=\sqrt{x}$ and $y<\sqrt{x}$, all with $y>1$ and $y<x$.

\vspace{0.5cm}

\noindent
{\it $y>\sqrt{x}$}

\vspace{0.25cm}

 An example of the decay of the two-time global correlation function
can be seen in panel (j) in Fig.~\ref{fig:corr_cmp}. It is stationary and it rapidly approaches zero with
progressively decaying oscillations around this value. The linear response and correlation function are not
related by an FDT with a single temperature (not shown) and in this respect there are no important differences
regarding what we have just shown for energy extraction processes in Sector I. For these reasons we
chose not to show more data for these parameters.

\vspace{0.5cm}

\noindent
{\it $y=\sqrt{x}$}

\vspace{0.25cm}

A prediction from the asymptotic analysis of the Schwinger-Dyson equations, the fact that on the curve $y=\sqrt{x}$ and $y\geq 1$ FDT is satisfied
with $T_f=J$, is made explicit in Fig.~\ref{fig:q-fdt} where the parametric plot $T_f \chi(t_1-t_2)$ against $C(t_1-t_2)$ is
constructed for four pairs of $(x,y=\sqrt{x})$ given in the key. The dotted line is the FDT prediction with $T_f=J$. The agreement
between numerics and analytics is very good. Additional support on the fact that the dynamics after the quench occur as in equilibrium at $T_f$ is
given in panel (b) in the same figure where quenched and equilibrium correlations are indistinguishable. The latter are obtained by drawing the
initial conditions drawn from equilibrium at $T'=T_f=J$ and running the code with the same parameter $J$. Coincidence of a similar quality (not shown)
is found for the other three sets of parameters used in~(a).


\begin{figure}[h!]
\vspace{0.5cm}
\begin{center}
\includegraphics[scale=0.66]{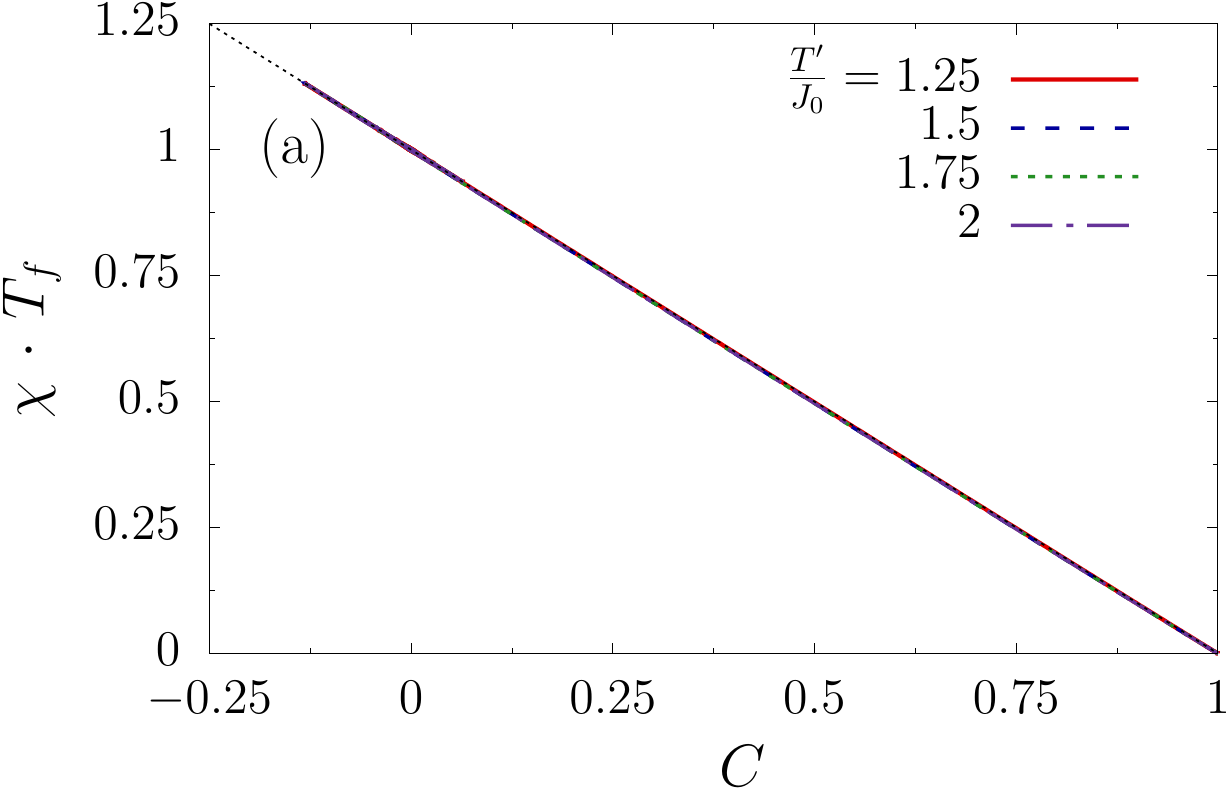}
\hspace{0.35cm}
\includegraphics[scale=0.59]{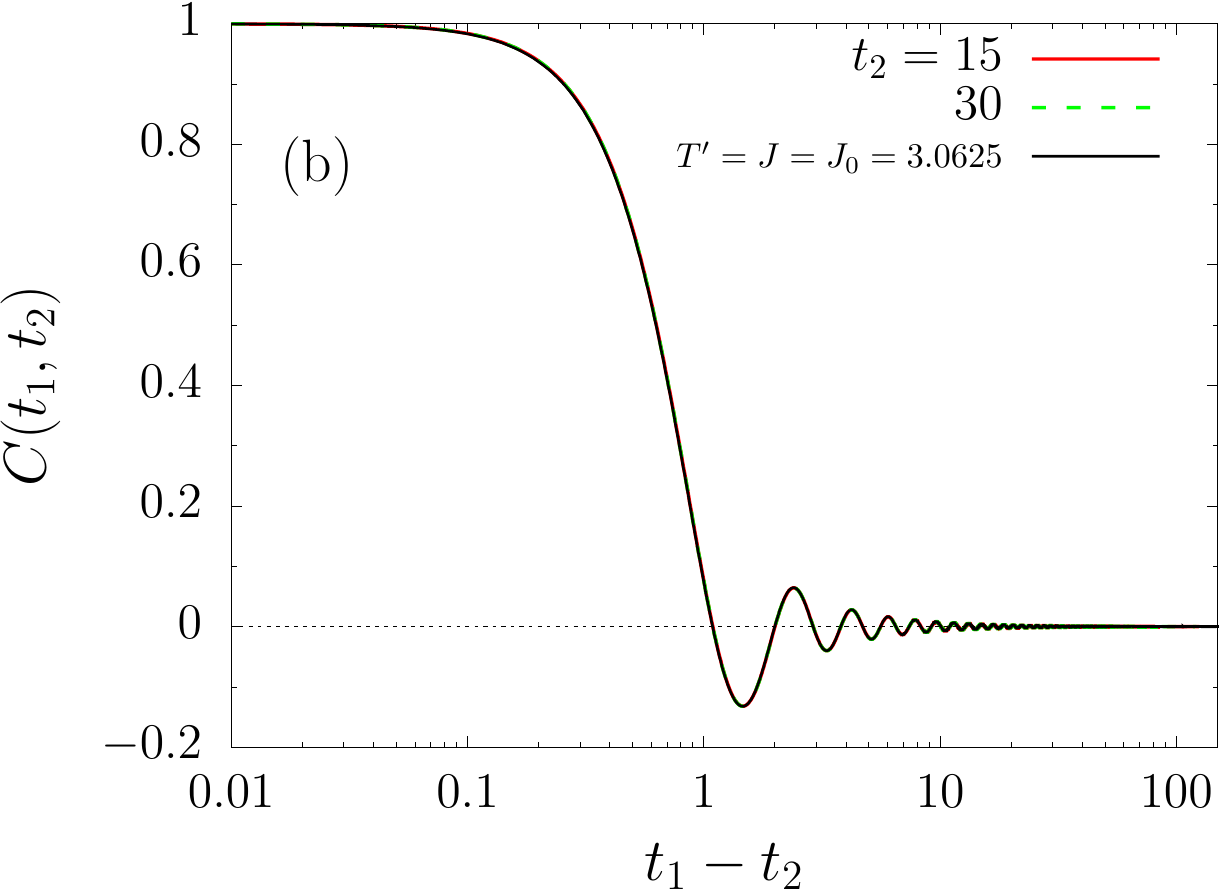}
\end{center}
\caption{\small
{\bf Sector II. Energy extraction from a paramagnetic initial state} for parameters lying on
the special curve $y=\sqrt{x}$ of the phase diagram. (a) Check of FDT at
 $T_f=J$ for four sets of parameters on this curve.
(b) Comparison between the time-delayed correlation after a quench with $T'=1.75$ and $J=3.065$, and
the equilibrium (no quench) correlation at $T_f=J=3.0625$. The agreement is perfect.
}
\label{fig:q-fdt}
\end{figure}


\begin{figure}[h!]
\begin{center}
  \includegraphics[scale= 0.65]{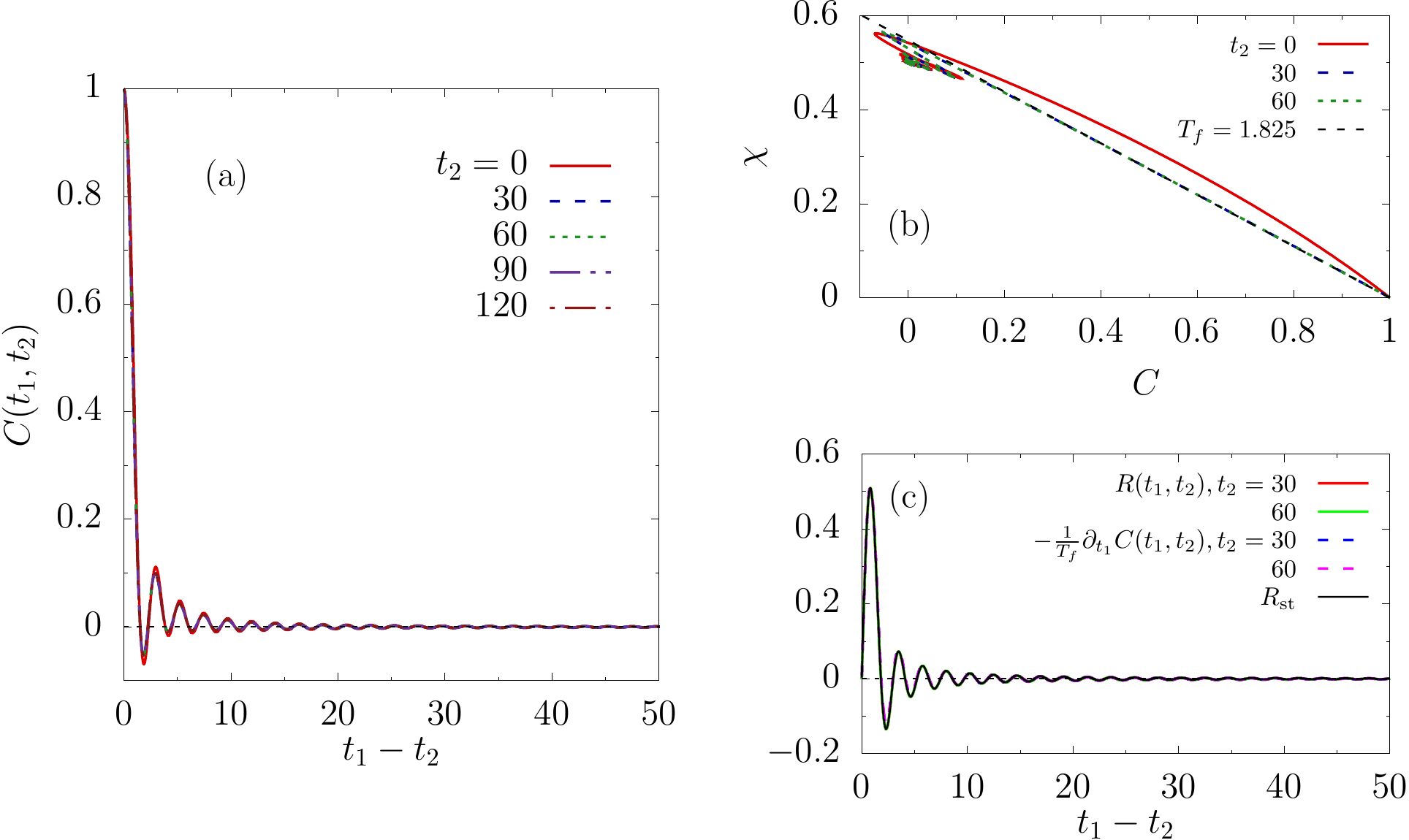}
  \end{center}
  \begin{center}
\includegraphics[scale=0.45]{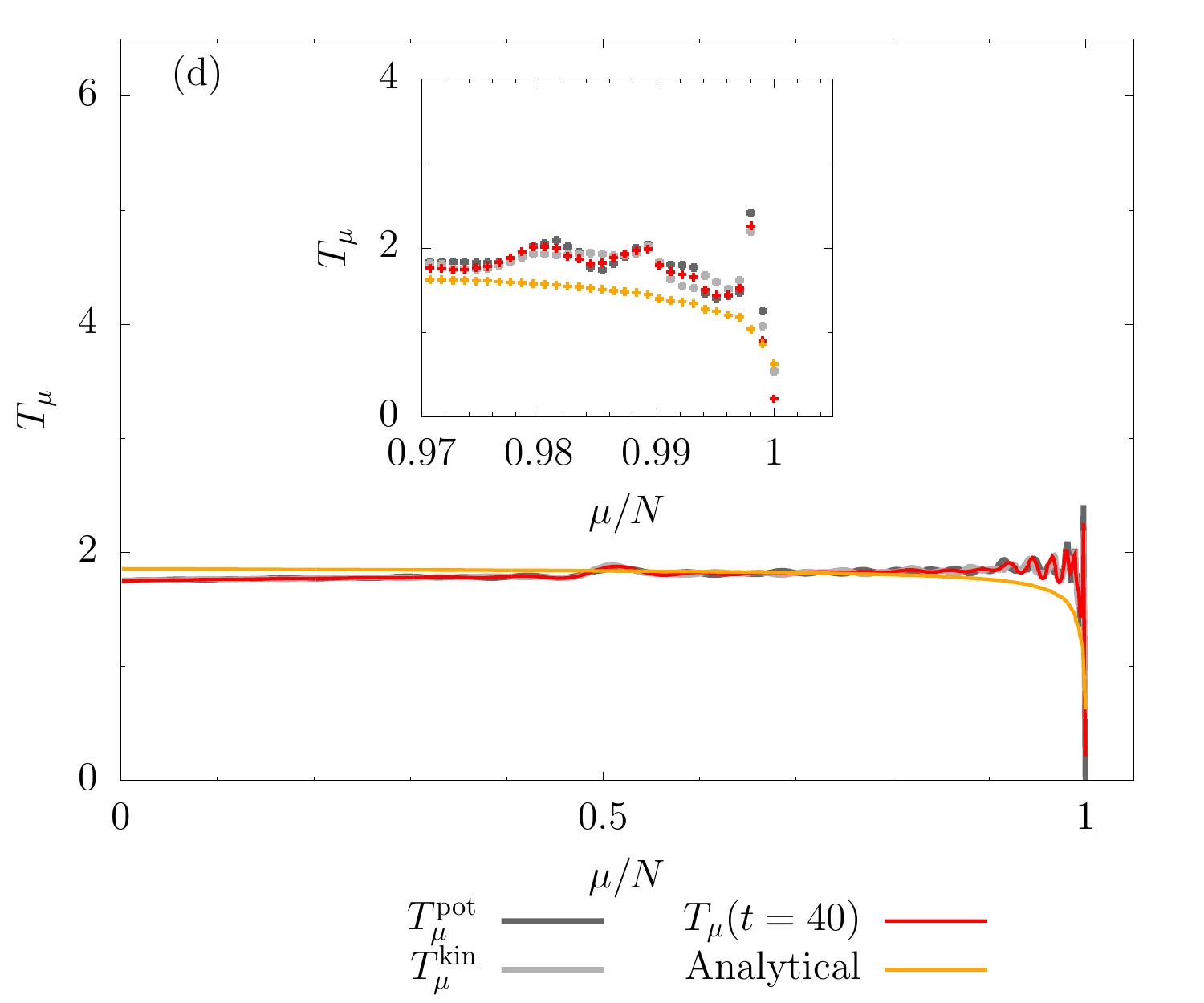}
\includegraphics[scale=0.45]{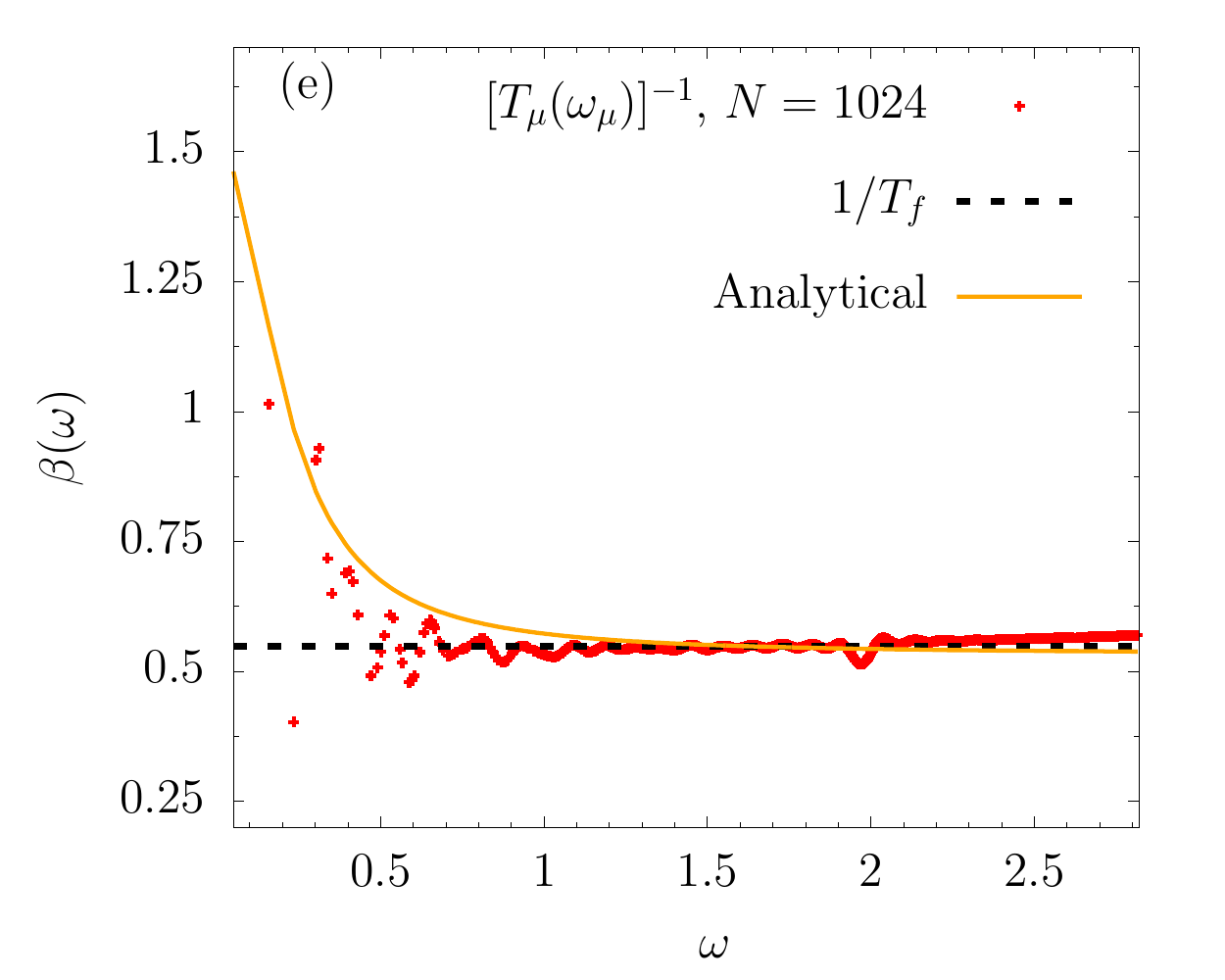}
\end{center}
\caption{\small {\bf Sector II. Energy extraction from a paramagnetic initial state ($y< \sqrt{x}$)}
$T'=1.25 > T^0_c$, and post-quench coupling $J=2$ leading to $\Delta e=-0.4$.
(a) The correlation function, $C(t_1,t_2)$ against $t_1-t_2$ for different  $t_2$.
$C(t_1,t_2)$ vanishes as $t_1-t_2 \rightarrow +\infty$.
(b) The parametric plot $\chi(\tau,t_{2})$ against $C(t_{2}+\tau,t_2)$,
for two different values of the waiting time $t_2$.
The black dashed line is the FDT at $T_f\simeq 1.825 \ J_0$, see Eq.~(\ref{eq:Tf_from_para_to_ageing}).
(c) $R$ and $-1/T_f \, \partial_{t_1} C$ as a function of $t_1-t_2$, with the same $T_f$ as in (b).
$R_{\mathrm{st}}$ is reproduced from Eq.~(\ref{eq:response_fourier_transform-text}) with $m=1$, $J=2$ and $z_{f}=4$.
Again, in this representation the difference with $-1/T_f \, \partial_{t_1} C$ is not visible, see panel (b) for a better understanding.
(d) Almost all modes in the bulk of the spectrum have the same temperature, and it is very close to $T_f$ shown with an
horizontal dashed line.
Inset: Detail of the largest modes.
(e) The inverse mode temperatures (red data) and the outcome of the harmonic oscillator approximation (yellow curve) in Eq.~(\ref{eq:mode-temp-prediction}).
}
\label{fig:T1p25_J2p00}
\end{figure}


\vspace{0.5cm}

\noindent
{\it $y<\sqrt{x}$}

\vspace{0.25cm}

We chose to show data for $T'=1.25 \, J_0$ and $J=2 \, J_0$, parameters that lie in the region $y<\sqrt{x}$ of the phase
diagram where the naive analysis of the $N\to\infty$ equations allows for ageing behaviour.
The $N\to\infty$ and finite $N$ data are shown in Fig.~\ref{fig:T1p25_J2p00}.
The Lagrange multiplier and kinetic and potential energies approach their expected asymptotic values with an oscillatory
behaviour and smoothly decaying amplitude (not shown).  The two-time correlation is stationary and shows no ageing.  The decorrelation from the initial
configuration and from a later configuration at time $t_2$ behave very similarly. The time-delayed $C(t_1,t_2)$ shows a rapid decay towards a value
close to $0.1$ around which it oscillates once and next decays towards zero with damped oscillations (a). The linear
response function shows a similar effect in the sense of having a fast variation at short time differences and a slower
one later (c). The value of $R$ obtained from the numerical integration agrees very well with $R_{\mathrm{st}}$ from Eq.~(\ref{eq:response_fourier_transform-text}).
The most interesting results concern the comparison between the linear response and the correlation
function in the parametric plot in panel (b). The very short time delay behaviour, for $C$ close to $1$ and $\chi$ close to
$0$ seems to be described by the slope dictated by $T_f$, the value of the temperature deduced from an
ageing like asymptotic scenario. However, the curves deviate from this straight line for smaller $C$ and larger $\chi$. When the
correlation reaches a value close to $0.1$ corresponding to its first oscillation, the parametric plot approaches a
flat form that ensures the limit $\chi_{\rm st}=1/J$. This second behaviour is reminiscent of what happens
in the ageing relaxation of the same model~\cite{CuDe95a}.


\begin{figure}[h!!]
\vspace{0.5cm}
\begin{center}
  \includegraphics[scale= 0.7]{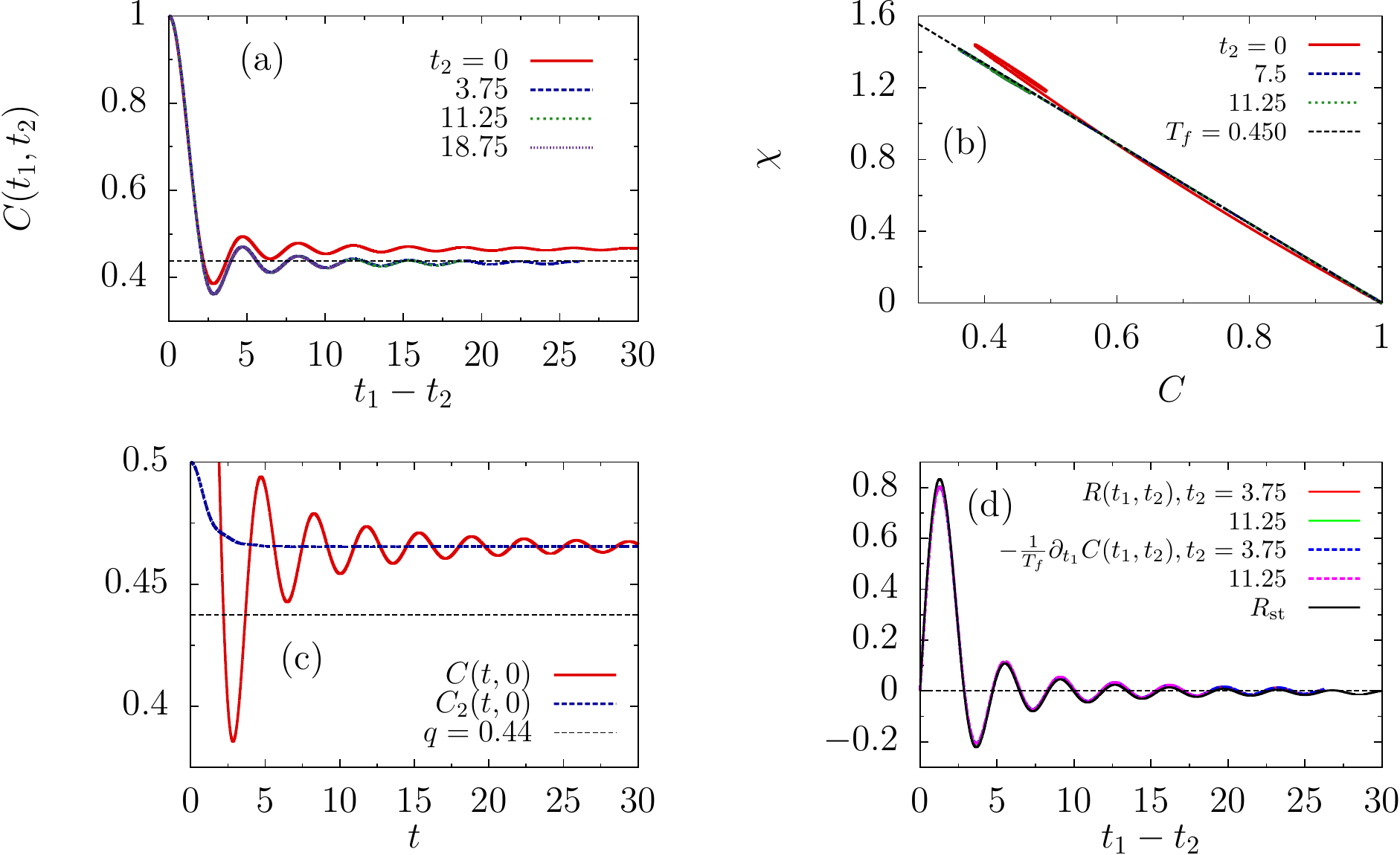}\quad%
  \end{center}
  \vspace{0.1cm}
  \begin{center}
   \includegraphics[scale=0.7]{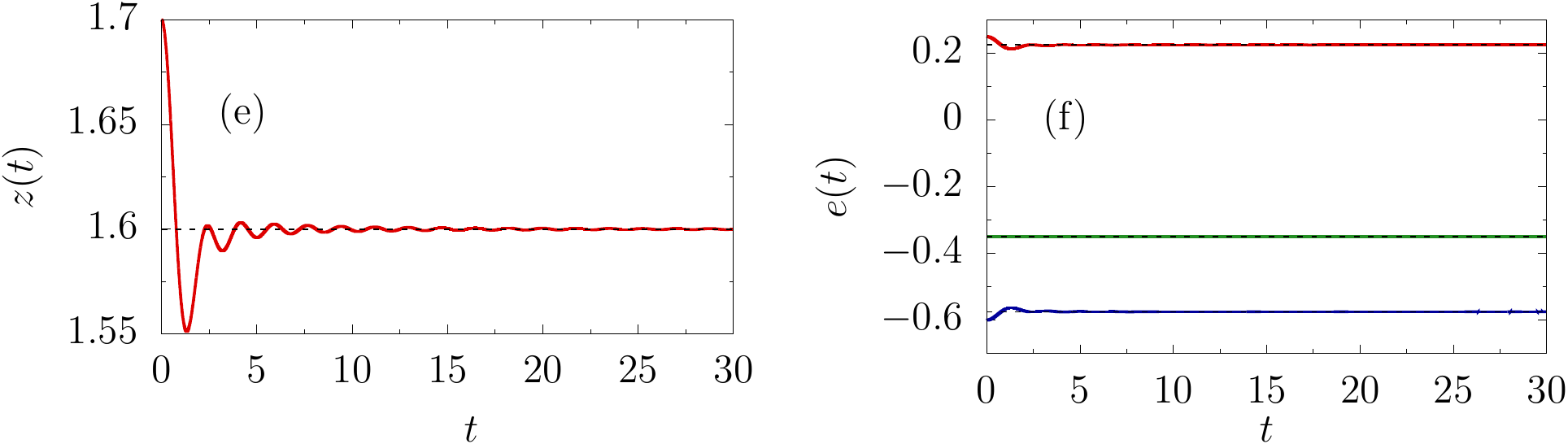}
\end{center}
\caption{\small%
{\bf Sector III. Energy injection, shallow quench from condensed to condensed.}
$T'=0.50 < T^0_c$ and $J=0.80$. The small energy injection,
$\Delta e=0.15$, is not sufficient to drive the system out of a condensed state.
(a) Dynamics of the correlation
function. The horizontal dashed line is at $q=1-T_f/J\simeq0.44$, with $T_f$ from Eq.~(\ref{eq:Tf_from_cond_to_cond}).
(b) $\chi(t_1,t_{2})$ against $C(t_{1},t_2)$
for fixed $t_2$ and using $t_1-t_2$ as a parameter.
The black dashed line shows the FDT with $T_f=0.45$.
In (d) $R$, $R_{\mathrm{st}}$ from  Eq.~(\ref{eq:response_fourier_transform-text}), and  $-1/T_f \, \partial_{t_1} C$  as a function of $t_1-t_2$.
(e) Time evolution of the Lagrange multiplier, $z(t)$, and $z_f=2J=1.6$ with a dashed line.
(f) From top to bottom: the kinetic energy (in good agreement with  $e^f_{\rm kin}=T_f/2$),
the total energy (constant in time) and the potential energy (in good agreement
with $e^f_{\rm pot}=-J^2/(2 T_f) (1-q^2)$).
}
\label{fig:T0p50_J0p80}
\end{figure}


The asymptotic analysis of the $N\to\infty$ equations allow for an ageing solution with a diverging effective temperature for
correlation values below the plateau at $q$, in this region of the parameter space.
(In the $p=3$ model, for similar sets of parameters an ageing solution though with a finite $T_{\rm eff}$ is realised~\cite{CuLoNe17}.) The numerical solution of
the complete set of Schwinger-Dyson equations demonstrates a behaviour with some features that are similar to the approximate asymptotic solution, but no signature of ageing.
Indeed, the parametric plot could be interpreted as formed by two pieces, one in which the form is non-trivial
close to $C=1$ and another one that is, on average, flat, separated by the correlation value at which the first oscillation occurs. A flat piece in the
parametric plot means that the integrated response, and all other functions such as the potential energy,
do not get contributions from these time scales and it corresponds to an infinite~\cite{CuDe95a} effective temperature~\cite{CuKuPe97,Cu11}. In practice, the parametric
plot is not locally flat for small values of $C$ but, as we can see from the mode analysis in  Fig.~\ref{fig:T1p25_J2p00}~(e), the temperatures of the corresponding modes
do take very large values.


\subsubsection{Sector III: initial and final condensed states}

In Fig.~\ref{fig:T0p50_J0p80} we start discussing the behaviour of the $N\to\infty$ model  in Sector III.
We use $T'=0.5$ and $J=0.8$, a quench that injects a {\it small} amount of energy from the system.
Panel (a) proves that the two time correlation approaches a non-vanishing value asymptotically. The
horizontal dashed line is at $q=1-T_f/J\simeq 0.44$, the theoretical value derived from the assumption of
equilibration {\it \`a la} Gibbs-Boltzmann.
 Further information about the decay of the correlation functions is given in (c) where we
show $C(t,0)$ and the off-diagonal correlation with the initial configuration $C_2(t,0)$ against time.
$C_2$ starts at $q_{\mathrm{in}}=0.5$ and decreases monotonically. $C(t,0)$ quickly  decays from $1$ with superimposed oscillations.
Both curves should reach $q_2=q_0\neq q$ asymptotically and the data are compatible with this claim.

Panel (a) in Fig.~\ref{fig:q-fdt} shows the value of the asymptotic potential energy as a function of $J/J_0$ for the same three values of $T'/J_0$.
The agreement between $e_{\rm pot}^f$ and $T'/4 + T'J(4J_0)-4J$, the parameter dependence found from the energy conservation
and the asymptotic value of $z$  is extremely good.

For quenches with $y = T^{\prime}/J_0 < 1$ and $x>x_c(y)=y$, the asymptotic values of $C(t_1,t_2)$ and $C(t_1,0)=C_2(t_1,0)$, $q$ and $q_0$, respectively, do not vanish.
However, it is not always easy to extract their functional dependence on $x$ and $y$,
especially for  parameters that get away from the shallow quenches $x\simeq 1$. Figure~\ref{fig:q-fdt}~(b)
shows  $q$ and $q_0$ against $J/J_0$ for three values of $T'/J_0$, together with the
{\it naive} single temperature prediction $q=1-T_f/J$ with $2T_f=2T_{\rm kin}= T'(1+J/J_0)$ drawn with black solid lines. The agreement is good close to $x=1$
though deviations are clear close to the critical line $x_c(y)$ and for large energy extraction deep inside this parameter sector.
Note that the prediction of equilibration {\it \`a la} Gibbs-Boltzmann is such that $q\neq 0$ at $x_c(y)$ and this fact is not clear at all from the data
(we have extracted $q$ from a very short plateau in the correlation, that continues to decrease
possibly to zero, in these cases).
Also shown in this plot is the asymptotic value of $q_0$. One clearly finds that $q_0 >q$ for $x<1$ (injection) and $q_0<q$ for $x>1$ (extraction).
We will come back to the behaviour of the correlation function below.


\begin{figure}[h!]
\vspace{0.5cm}
\begin{center}
\includegraphics[scale=0.57]{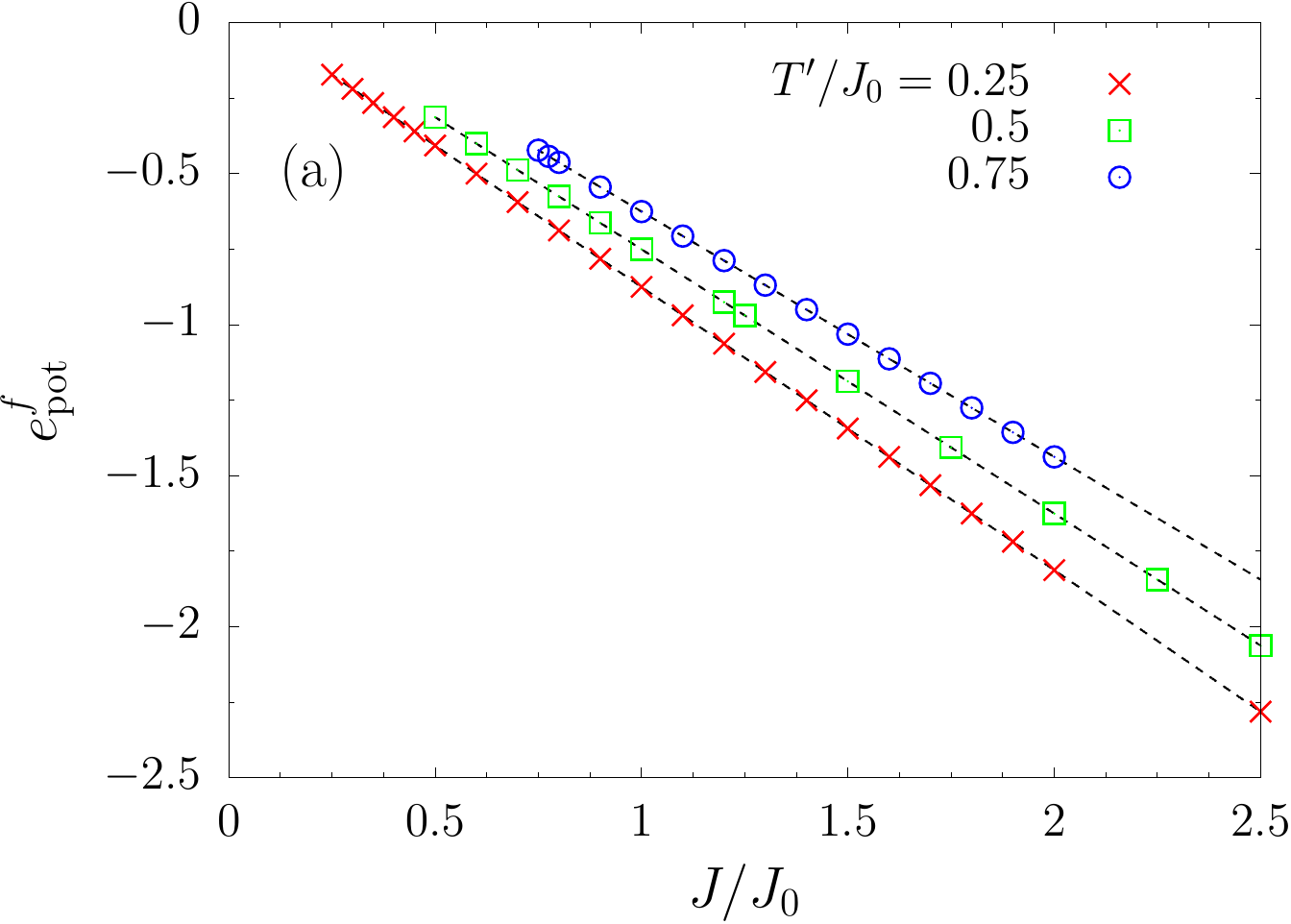}
\hspace{0.5cm}
\includegraphics[scale=0.61]{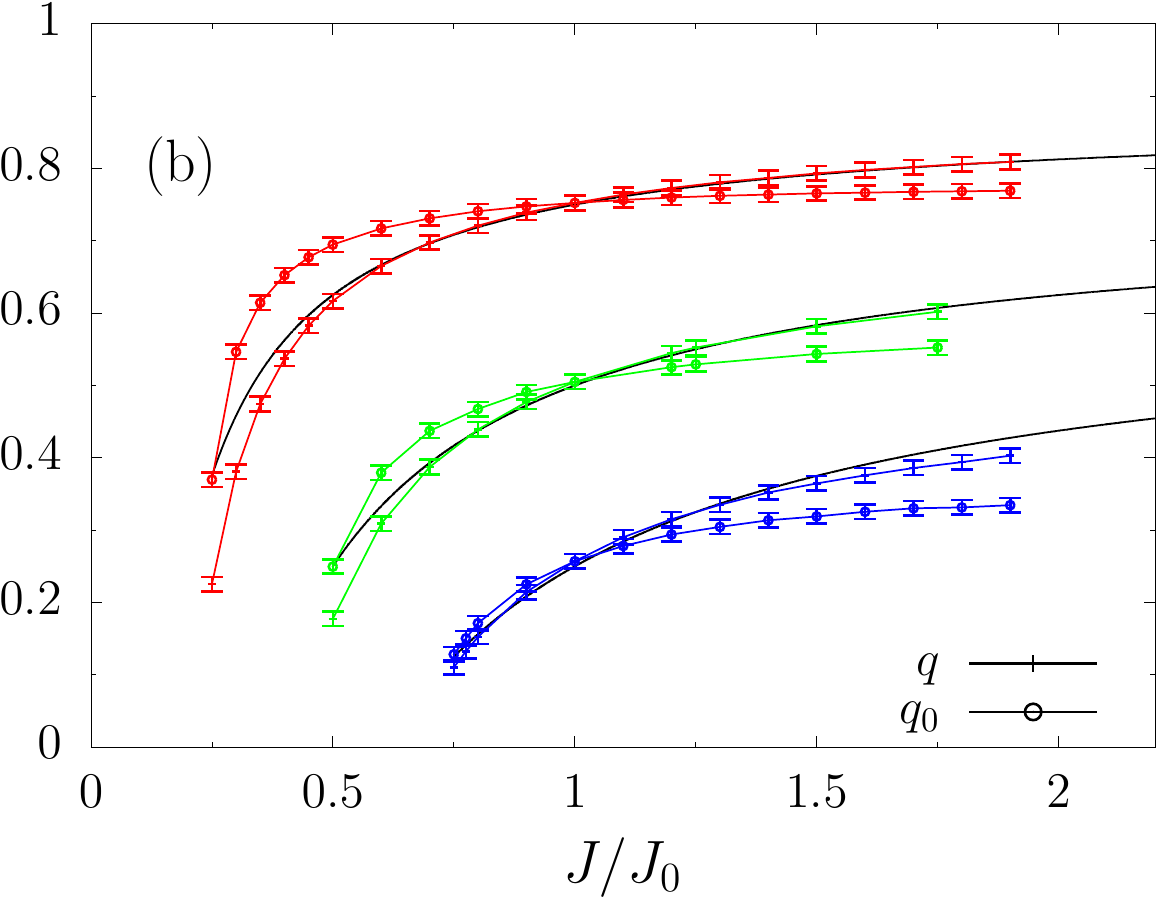}
\end{center}
\caption{\small {\bf Sector III, shallow and deep quenches from condensed to condensed.}
The three datasets correspond to $T^{\prime}/J_0=0.25$ (red), $0.5$ (green), $0.75$ (blue).
(a) The final potential energy as a function of $J/J_0$. The dotted lines are
$T'/4 + T'J(4J_0)-4J$.
(b) The estimated asymptotic value of the correlation function,
$q= \lim_{{(t_1 - t_2) \rightarrow +\infty}, t_2 \gg 1} C(t_1,t_2)$ (simple points), and
$q_0= \lim_{t \rightarrow +\infty} C(t,0)$ (circles),
against $J/J_0$ for the three values of $T^{\prime}/J_0 < 1$
(increasing from top to bottom). Data are equipped with error bars.
The solid lines are the equilibrium predictions for $q$.
Notice  the deviations close to $x_c=y$ and for $x$ far from $1$.
The long time limits of $C(t_1,t_2)$ and $C(t_1,0)$ are ordered according to
$q>q_0$ for $J<J_0$ and $q<q_0$ for $J>J_0$.
}
\label{fig:q-fdt}
\end{figure}


Panel~(b) in Fig.~\ref{fig:T0p50_J0p80} shows the parametric $\chi(C)$ curve where one sees that the $t=0$ (red) data are
very different from the ones for long $t_2$. The data for long $t_2$ suggest that FDT has established at temperature $T_f$. Another test of FDT,
now  in the time
domain, is shown in (d) with the comparison between the linear response and the time derivative of the correlation.
The other panels show the asymptotic values of the
Lagrange multiplier (e) and kinetic and potential energies (f). These yield further support to the asymptotic value of
the $q$ parameter and $T_f$ estimated analytically under the Gibbs-Boltzmann assumption,
since they demonstrate perfect agreement with the asymptotic contributions to the total energy. Nevertheless,
we  will revisit this claim below when studying deeper quenches in the same sector.


\begin{figure}[h!]
\vspace{0.5cm}
\begin{minipage}{.6\textwidth}
\begin{center}
\includegraphics[scale=0.4]{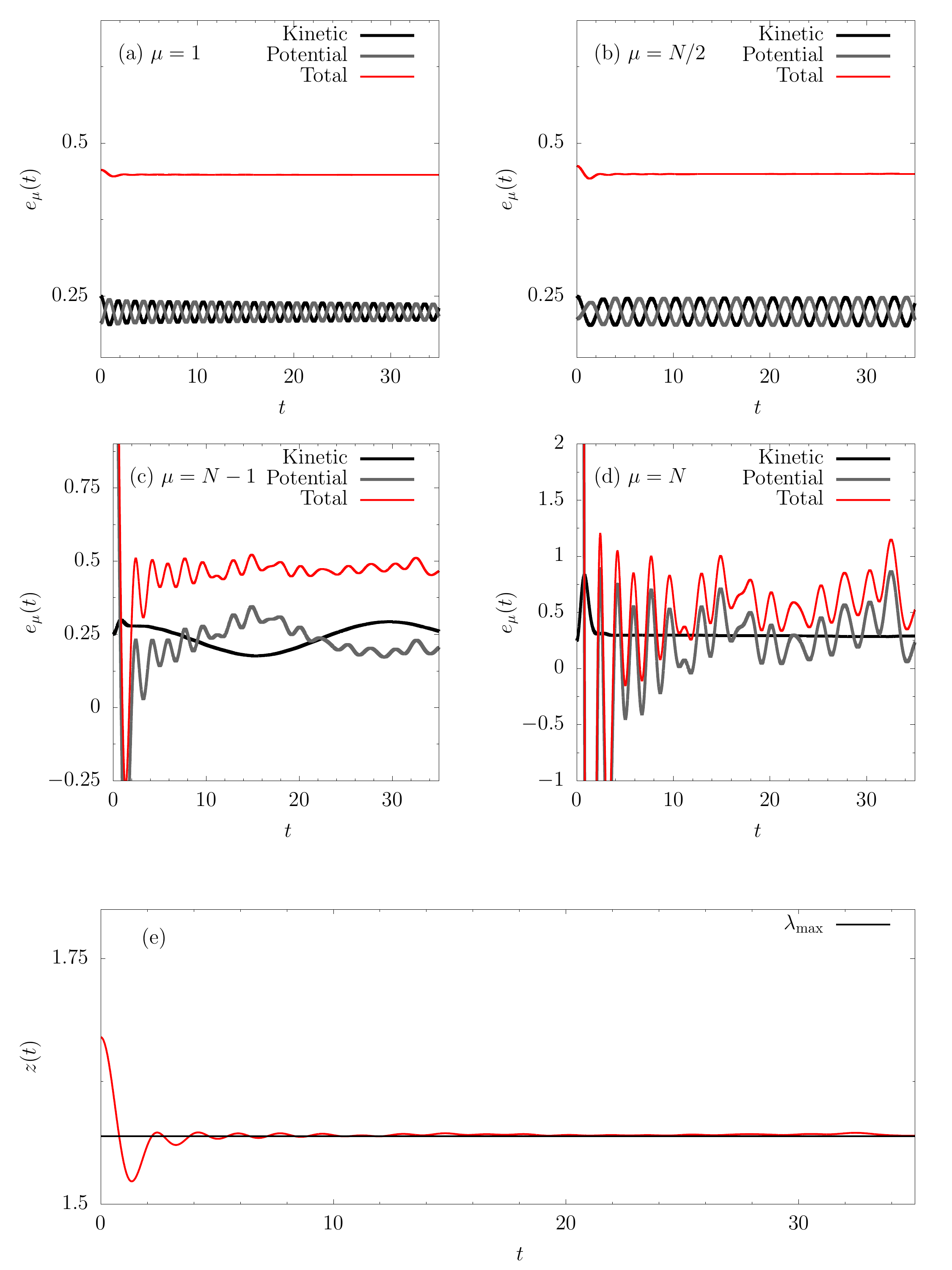}
\end{center}
\end{minipage}
\begin{minipage}{.6\textwidth}
\includegraphics[scale=0.43]{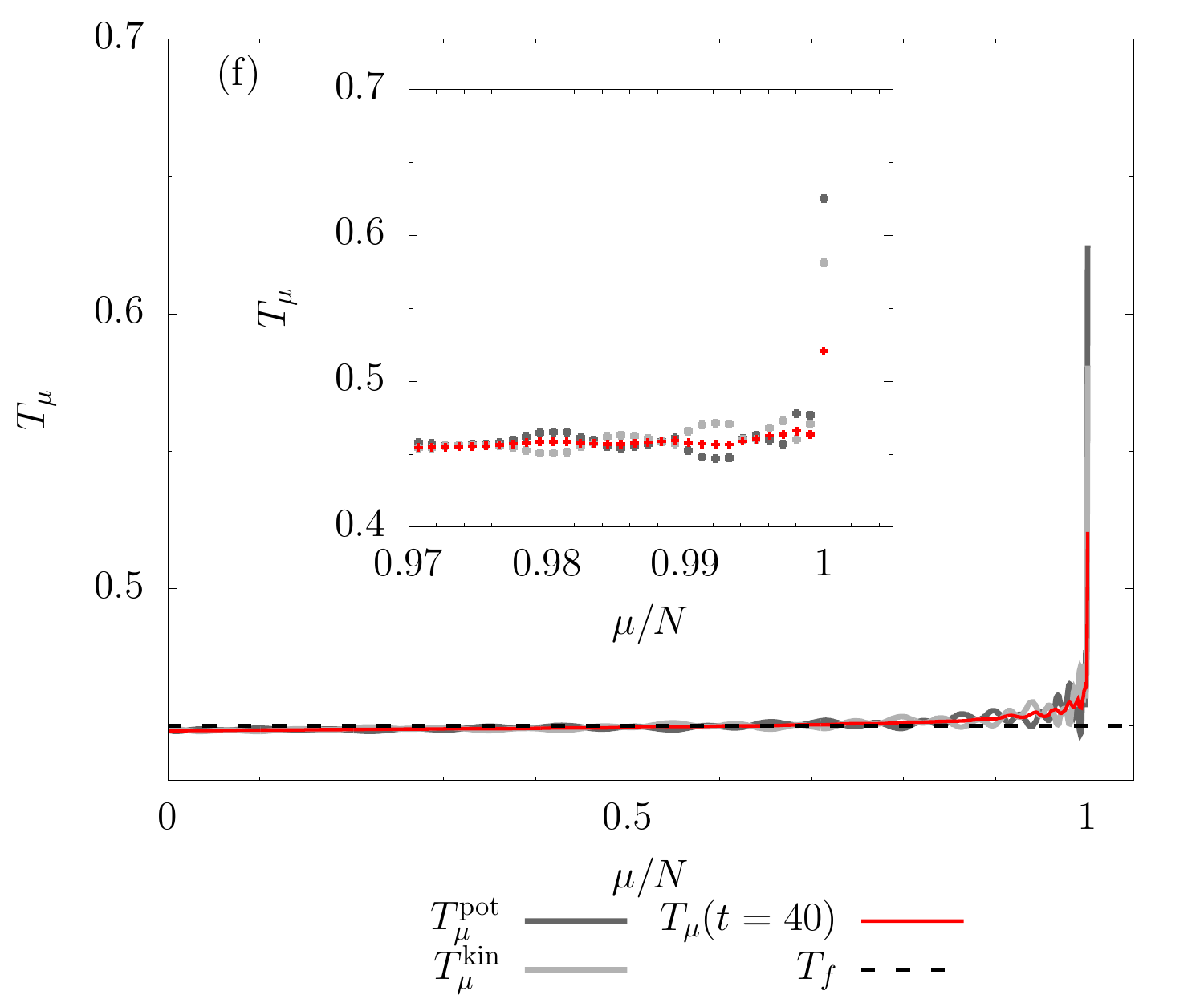}
\\
\includegraphics[scale=0.52]{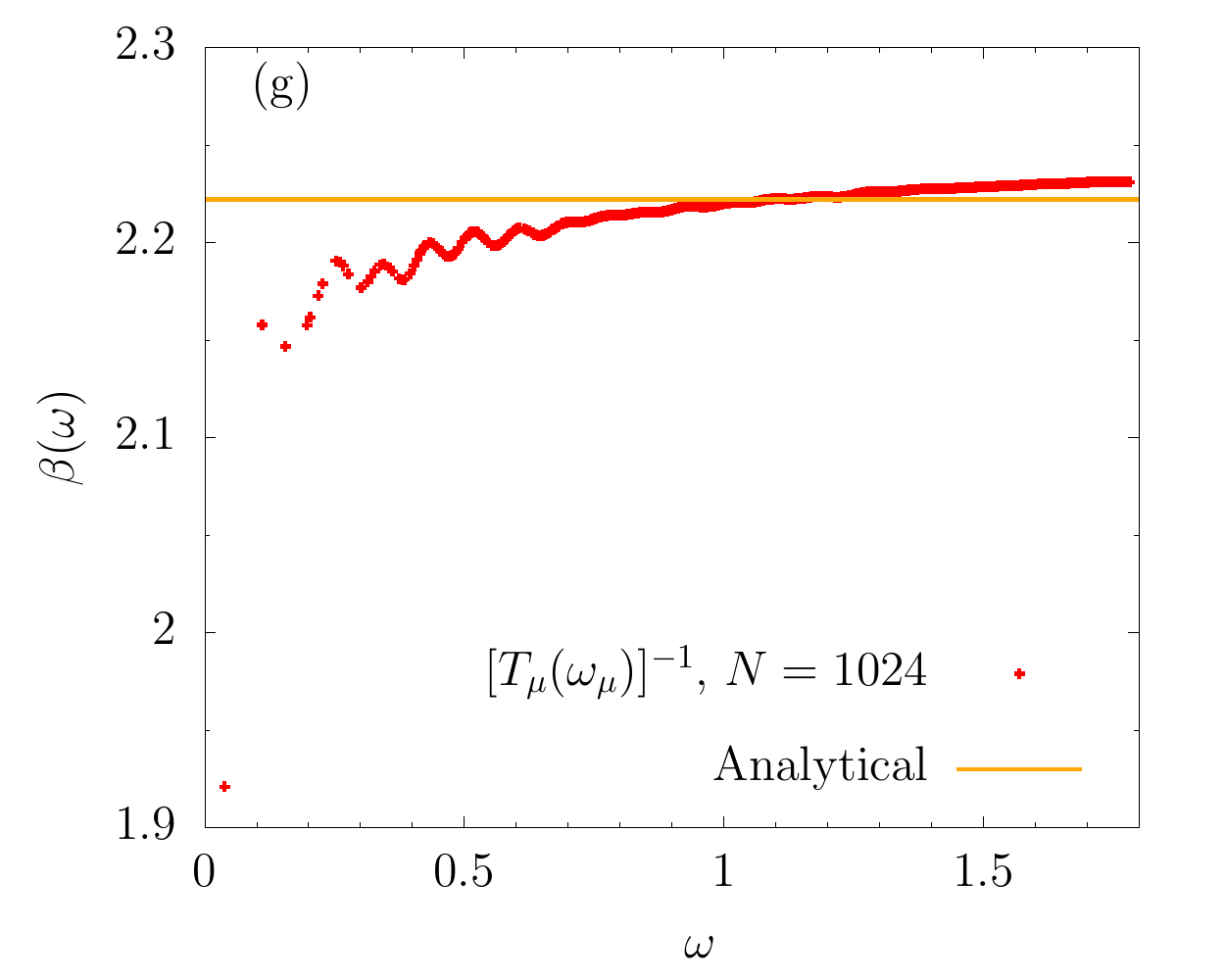}
\end{minipage}
\caption{\small {\bf Sector III, condensed initial conditions and small energy change.} $T'=0.5$ and $J=0.8$.
(a)-(d) Mode energies in a system with $N=1024$.
The largest modes are neatly out of equilibrium.
(e) The Lagrange multiplier.
(f) Mode temperatures, with a zoom over the  largest modes in the inset.
Close to equilibrium like profile apart from the deviations close to the edge of the spectrum.
(g) Comparison of the inverse mode temperatures with the ones of independent harmonic oscillators.
}
\label{fig:sector3-fig-finiteN}
\end{figure}


The companion curves for finite $N$ are in Fig.~\ref{fig:sector3-fig-finiteN}. First of all, panels (a)-(d)
display the time dependence of the $\mu=1, \, N/2, \ N-1, \ N$ mode energies in a system with $N=1024$.
While the modes $\mu=1, \ N/2$ show the usual oscillatory behaviour of a harmonic oscillator, the largest modes
$\mu=N-1$ and $\mu=N$ are clearly  out of equilibrium. The Lagrange multiplier is approximately constant and equal to the
largest eigenvalue, within numerical accuracy.
The spectrum of mode temperatures is plotted in (f) with a zoom over the  largest modes in its inset.
The profile is an equilibrium one, with $T_\mu$ being independent of $\mu$,
apart from the deviations close to the edge of the spectrum.
Finally, (g) shows a comparison between the inverse mode temperatures
and the $N\to\infty$ frequency dependent effective temperature extracted from the FDR.
$N\to\infty$ and finite $N$ results coincide (except at high frequencies where the result  is biased
by the numerical limitations in the computation of the Fourier transform). Higher modes
(low frequency) have temperatures slightly below the temperature $T_f$
while lower modes (high frequencies) have temperatures slightly above $T_f$, that is
shown with an horizontal yellow line. This fact is another warning concerning the claim of complete equilibration to a
Gibbs-Boltzmann measure.

\begin{figure}
\begin{center}
\includegraphics[scale=0.4]{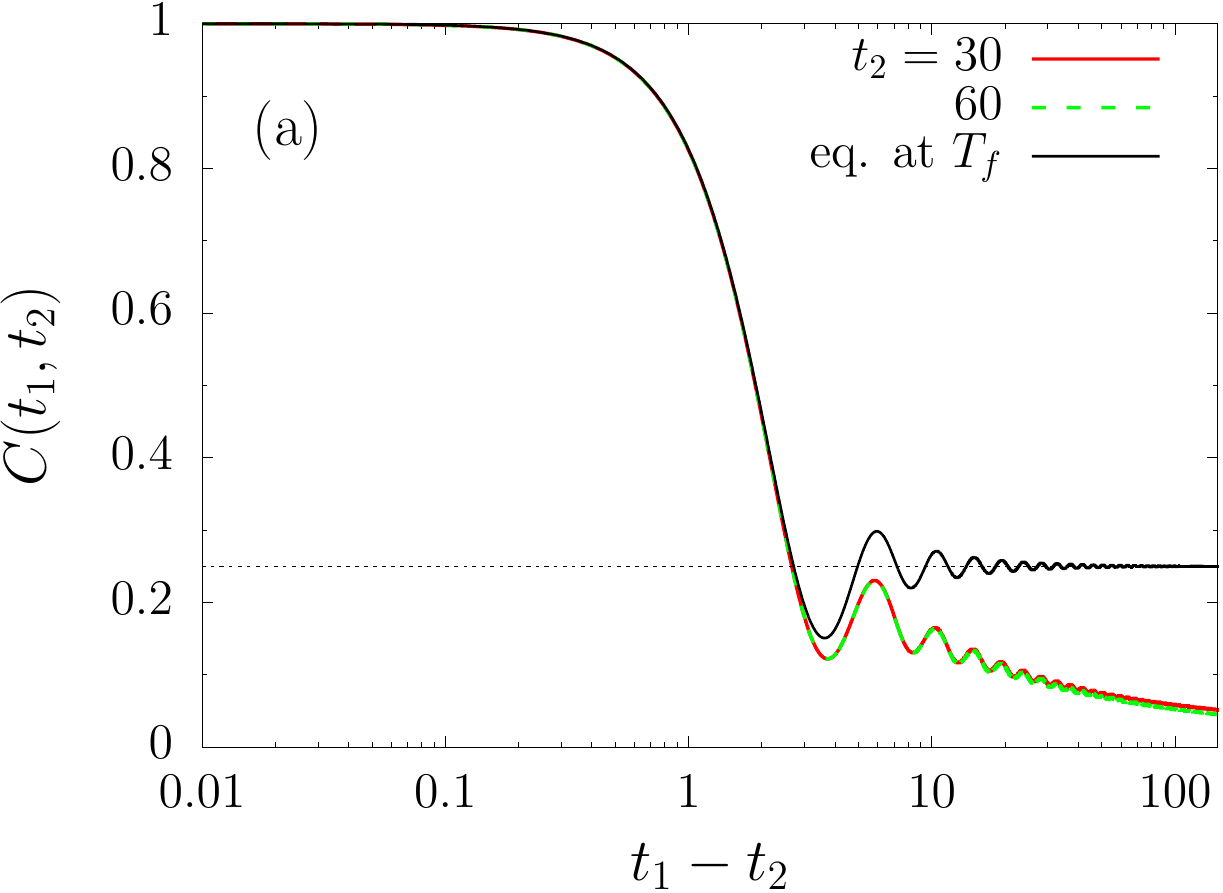}\quad%
\includegraphics[scale=0.4]{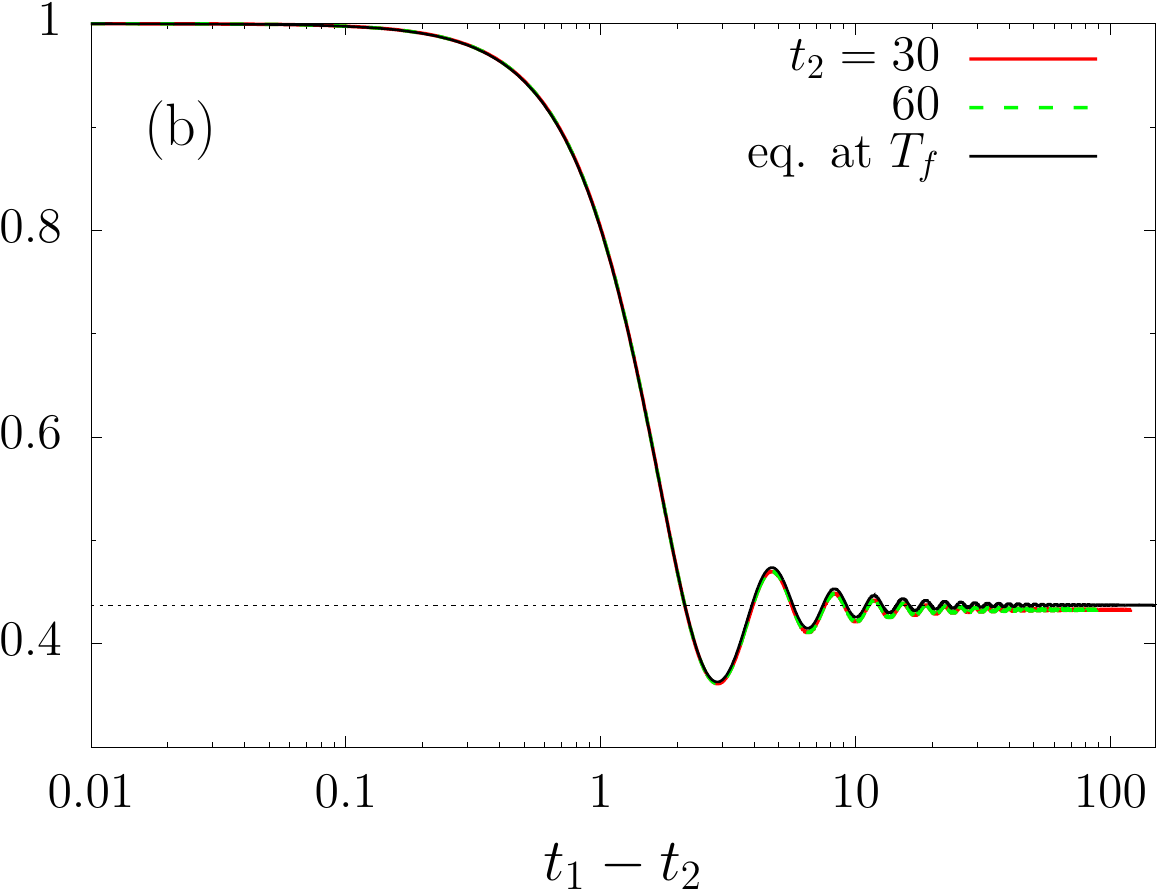}\quad%
\includegraphics[scale=0.4]{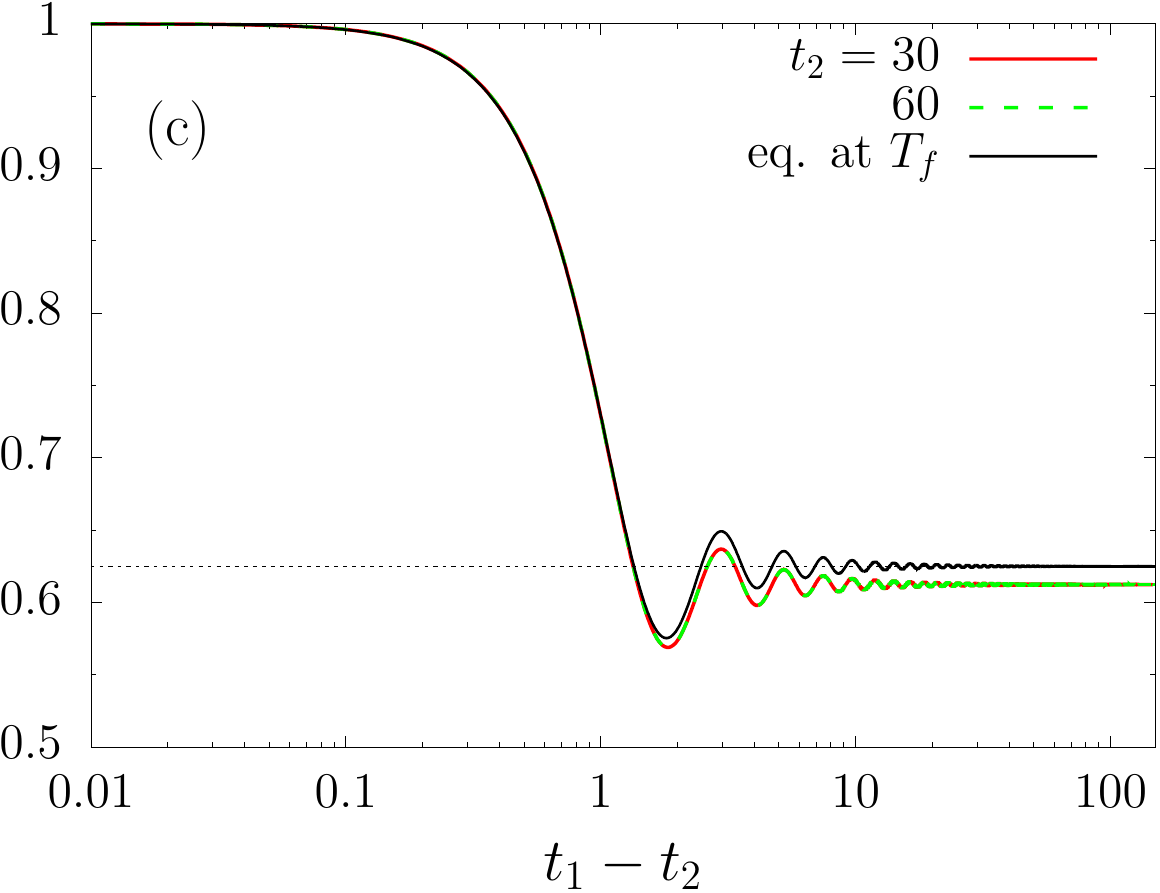}
\end{center}
\caption{\small {\bf Lack of Gibbs-Boltzmann equilibration in Sector III. }
The time-delay correlation function after a quench from $J_0$ to $J$ in Sector III (red and green curves)
and in equilibrium at parameters $J$ and $T_f$, the single temperature that would correspond to
Gibbs-Boltzmann equilibrium (black curves). The three panels are for $J=0.5$ (quench to the critical line), $J=0.8$ (quench close
to $x=1$) and $J=2$ (deep quench). The dotted horizontal line is $q=1-T_f/J$.}
\label{fig:noGB}
\end{figure}

Indeed, although the data for the shallow quench that we have just described might have suggested equilibration
to a Gibbs-Boltzmann probability distribution, the detailed comparison of the full time-delay dependence
of the correlation function after a quench and in equilibrium (no quench) at parameters $J$ and $T_f$ that
are the ones that the equilibrium measure would have, prove that such a steady state is not
reached by the dynamics.  This statement is proven in Fig.~\ref{fig:noGB} where we display the self-correlations
stemming from the two procedures for three choices of quenches: to the critical line $x=y$ (a case that will be
further studied in the next Subsubsection), the shallow quench, and a deep quench.


We note that the comparison of the linear response functions after a quench
and in equilibrium yield identical results. It is the correlation function the one that
deviates from Gibbs-Boltzmann equilibrium.


\subsubsection{Transition between Sectors III \& IV}
\label{subsubsec:from_cond_to_crit}

\begin{figure}[h!]
\vspace{0.5cm}
\begin{center}
  \includegraphics[scale= 0.7]{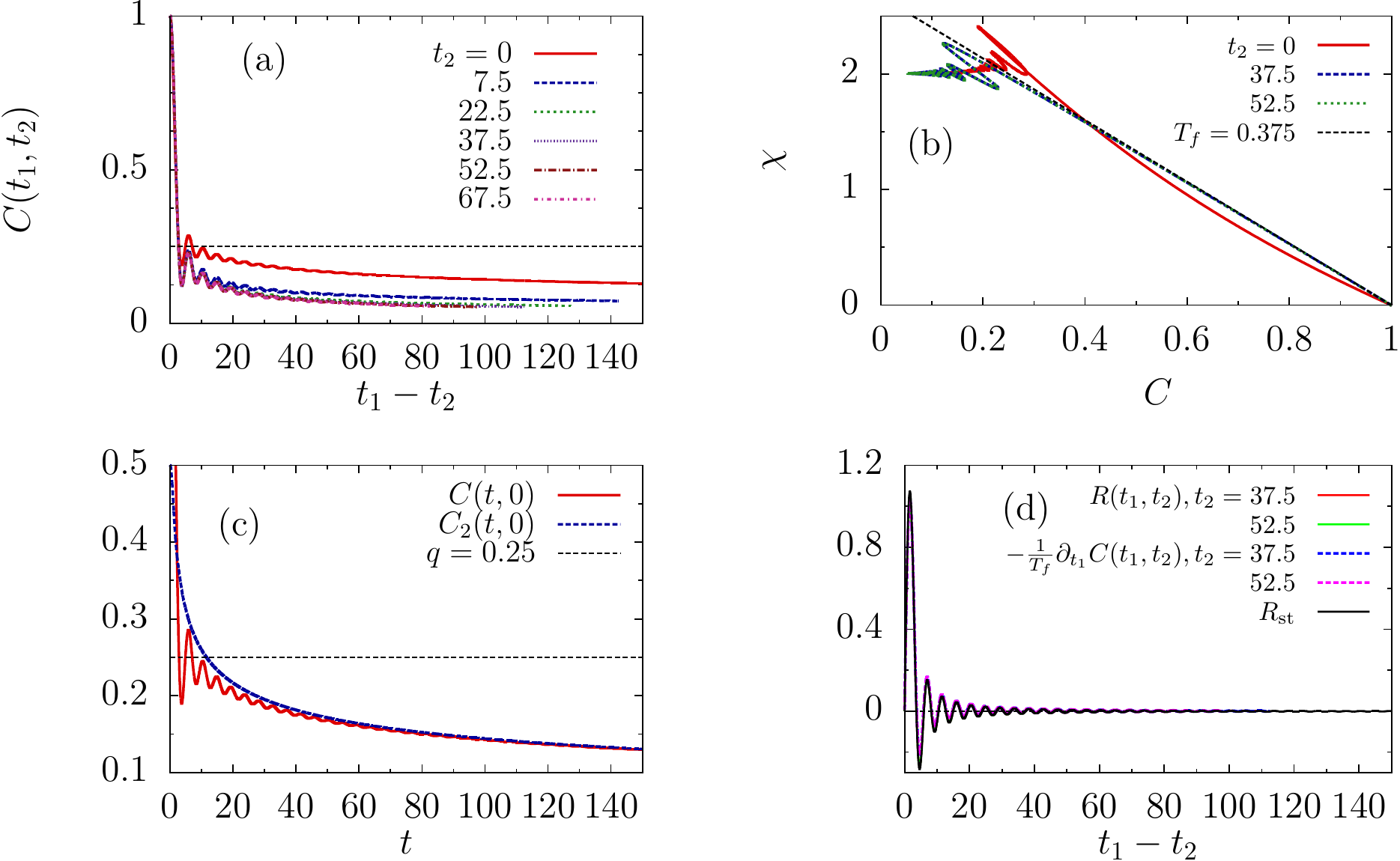}\quad%
  \end{center}
  \vspace{0.1cm}
  \begin{center}
   \includegraphics[scale=0.7]{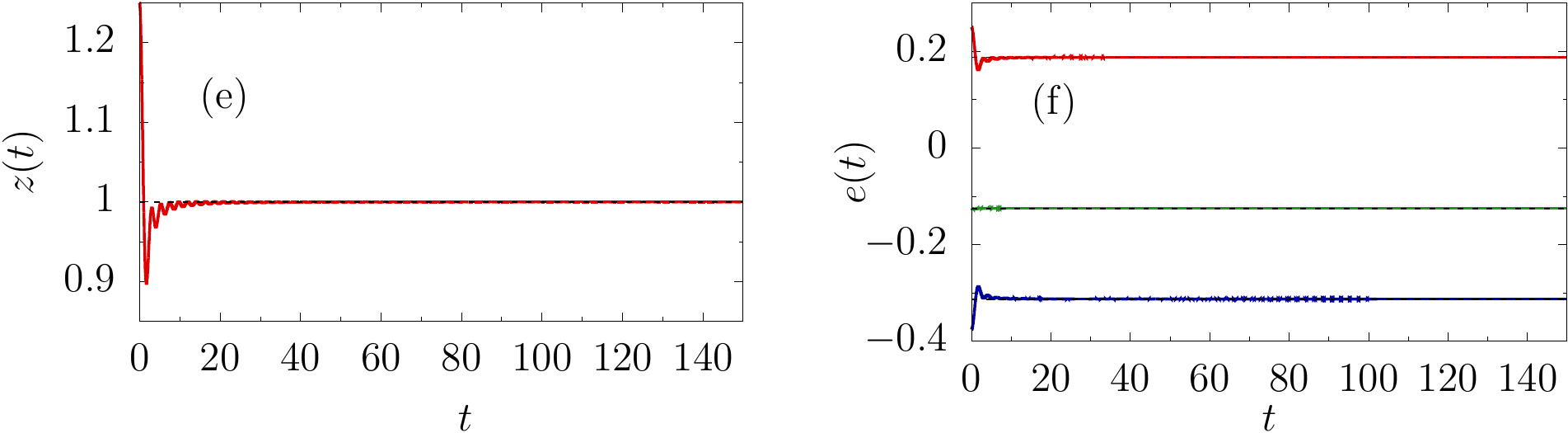}
\end{center}
\caption{\small%
{\bf Energy injection from condensed to the critical line $y=x$.}
$T'=0.50 < T^0_c$,  $J=0.50$ and $\Delta e=0.375$.
(a) The correlation
function and $q=1-T_f/J=0.25$ (horizontal dotted line).
(b) $\chi(t_1,t_{2})$ against $C(t_{1},t_2)$,
for fixed $t_2$ and using $t_1-t_2$ as a parameter.
The black dashed line is the FDT with $T_f=0.375$.
(d) The curve indicated by $R_{\mathrm{st}}$ is the (numerical) inverse Fourier transform of the theoretical
prediction given by Eq.~(\ref{eq:response_fourier_transform-text}).
(e) Time evolution of the Lagrange multiplier, $z(t)$,
and $2J=1$ with a dashed horizontal line.
(f) From top to bottom: the kinetic energy (in good agreement with  $e^f_{\rm kin}=T_f/2 \simeq 0.187$),
the total energy (constant in time with value $e_f\simeq-0.125$) and the potential energy (in apparent agreement
with $e^f_{\rm pot}=-\frac{J^2}{2 T_f} (1-q^2) \simeq -0.312$, though see the text for a revision of this claim).
}
\label{fig:T0p50_J0p50}
\end{figure}

Figure~\ref{fig:noGB} (a) has already shown non-equilibrium dynamics on the critical line $y=x$ for $y<1$.
We give here further evidence for this fact.
In Fig.~\ref{fig:T0p50_J0p50} we show results for $T'=0.5 < T^0_c$ and $J=0.5$, that is to say, point c on the phase diagram in
Fig.~\ref{fig:phase_diagram_new}.
This quench injects a relatively small amount of energy into the system, $\Delta e\simeq 0.375$
and takes the parameters to be on the critical line $y=x$.

The self correlation is shown in panel (a) of Fig.~\ref{fig:T0p50_J0p50}.
Stationarity is clear for short time delays $t_1-t_2$, for $t_2>0$ and there is
some remanent waiting time dependence at these short $t_2$. This double behaviour is
reminiscent of what is seen in the relaxational dynamics where a sharp separation
of time-scales exists. The data suggest that
$q_0=\lim_{t\rightarrow \infty} C(t,0)$ is different from
$q=\lim_{t_1-t_2\rightarrow \infty} \lim_{t_2\to\infty}C(t_1,t_2)$, although it seems hard to determine these
values from the numerics with good precision. In the event of a two step relaxation of $C(t_1,t_2)$ with an approach to a
plateau at $q$ and a further decay from it to zero, the plateau should be
at $q=1-T_f/J=0.25$, shown with a dashed horizontal line. The data still lie below
this value and we infer that they might not converge to a plateau but simply decay to zero.

Further information about the decay of the correlation functions is given in (c) where we
show $C(t,0)$ and the off-diagonal correlation with the initial configuration $C_2(t,0)$ against time.
$C_2$ starts at $q_{\mathrm{in}}=0.5$ and decreases monotonically. $C(t,0)$ quickly  decays from $1$ with superimposed oscillations.
Both curves seem to join and slowly and monotonically decay to zero.

The parametric plot of the linear response function, $\chi(t_1,t_{2})$, against the correlation, $C(t_{1},t_2)$,
for fixed $t_2$ and using $t_1-t_2$ as a parameter, for three different values of the waiting time $t_2$,
is shown in panel (b). The $\chi(C)$ curve for $t=0$ does not have any special form. The curves for
late $t_2$ are close to the straight line $1/T_f$ for time-delays that correspond to the first oscillations
of the correlation and linear response, they then oscillate, and for longer time-delays the parametric
construction becomes flat, with $\chi$ approaching, for $C\to 0$, the expected static susceptibility
$1/J=2$. Again, this second flat regime is reminiscent of the one found for the relaxation stochastic dynamics
at and below the critical temperature~\cite{Calabrese-Gambassi,Corberi-etal07}.

The fluctuation-dissipation theorem relating $\chi$ to $C$ in a stationary regime
with target temperature $T_f=0.375$ obtained
from Eq.~(\ref{eq:Tf_from_cond_to_cond}), is indicated as a straight dashed line.
The presentation in panel (d) confirms the agreement between
linear response and time variation of the time-delayed correlation function
dictated by the FDT at temperature $T_f$ for $C\geq 0.2$, say. However, it clearly breaks for smaller values of $C$.
As for the other quenches, we checked that the response $R(t_2+\tau,t_2)$
coincides with the one derived by anti-transforming the theoretical Fourier transform of the
stationary asymptotic response, given by Eq.~(\ref{eq:response_fourier_transform-text}), $R_{\mathrm{st}}$.

In (e) we provide a close look at the time evolution of  $z(t)$.
As one can see, $z(t)$ approaches the predicted value $z_f=2J$.
From the time evolution of the energy (f), we observe that the total energy is constant in time,
as it should be, while the kinetic and potential energies quickly approach their asymptotic values, which
coincide within numerical accuracy with the ones predicted in Sec.~\ref{app:energies}. These values could be mistaken for
$e^{f}_{\rm kin}=T_f/2$ and $e^{f}_{\rm pot}=-J^2(1-q^2)/(2T_f)$,
with $q=1-T_f/J=0.25$, the Gibbs-Boltzmann equilibrium predictions, though, the system is not in equilibrium.



\begin{figure}[h!]
\vspace{0.5cm}
\begin{center}
  \includegraphics[scale= 0.7]{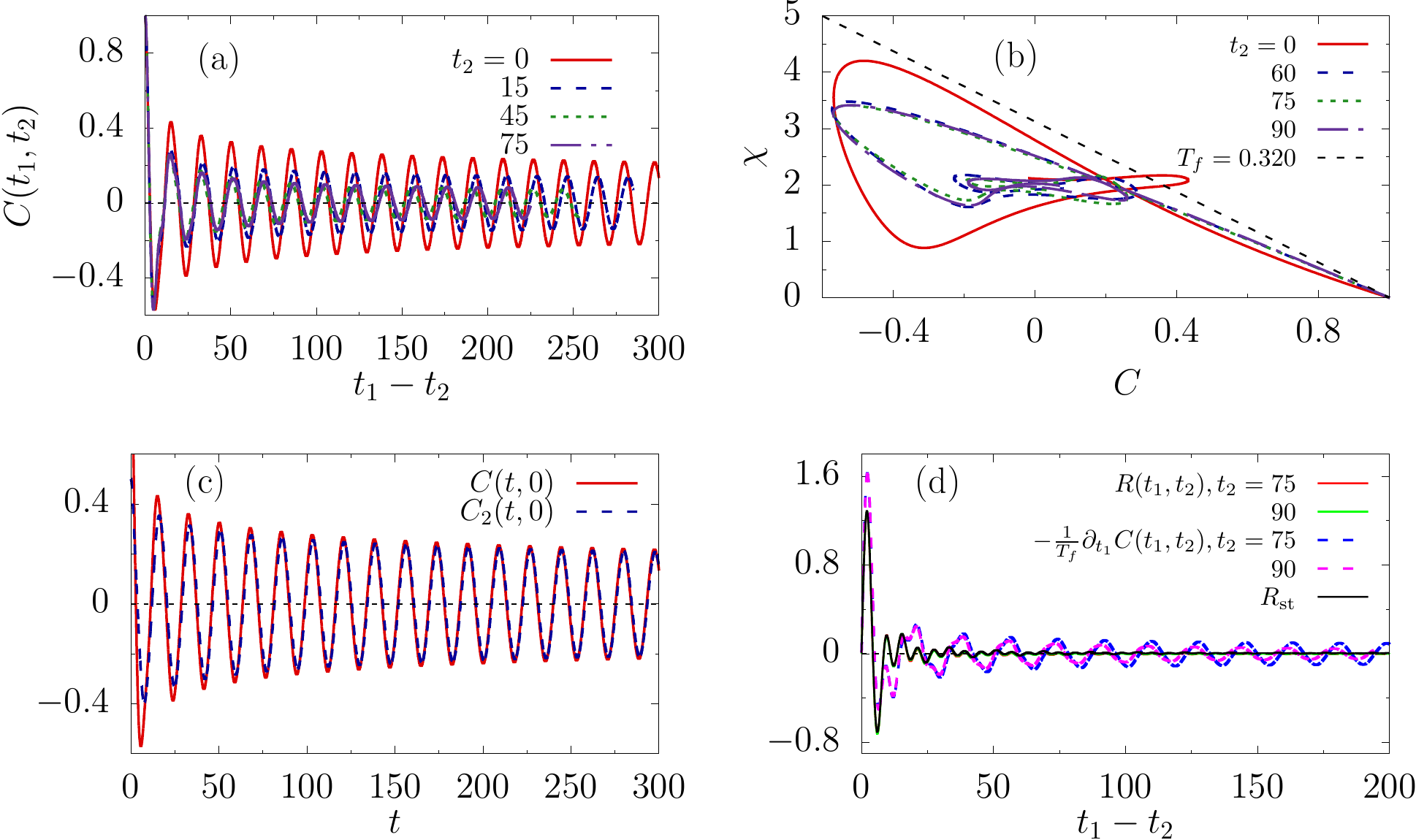}
   \includegraphics[scale=0.7]{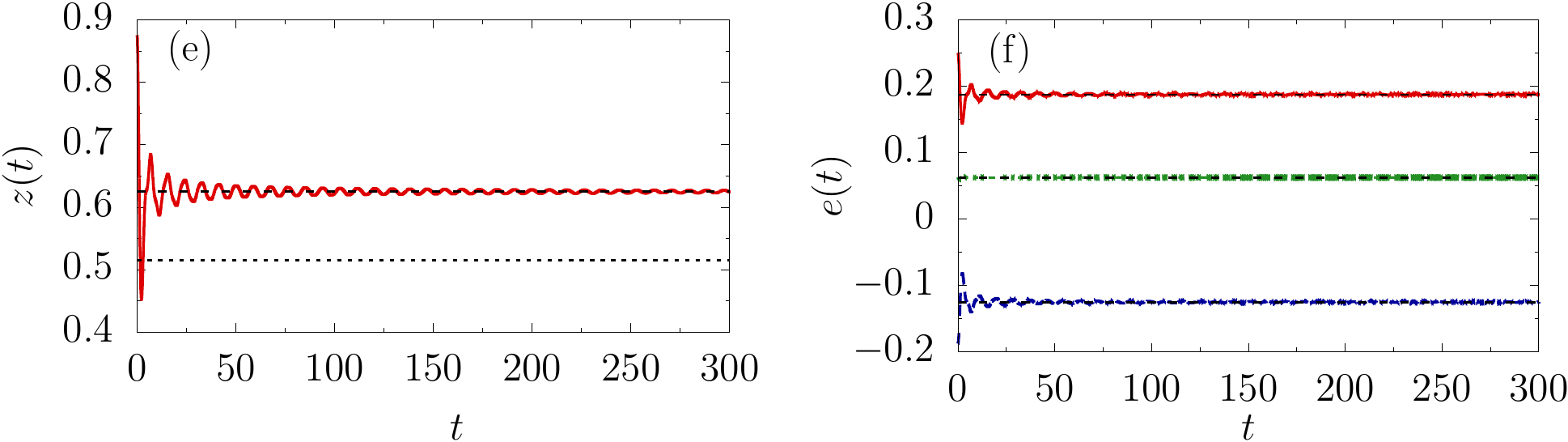}
    \includegraphics[scale=0.7]{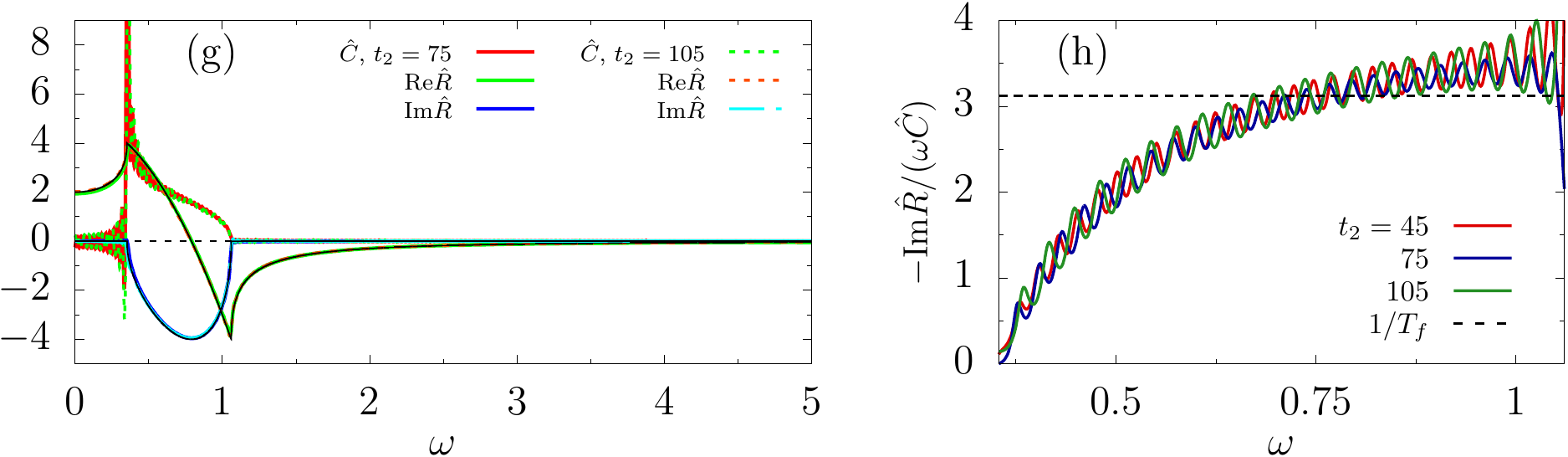}
\end{center}
\caption{\small {\bf Sector IV. Large energy injection on a condensed state.}
$T'=0.50 < T^0_c$, $J=0.25$ and
$\Delta e=0.5625$. (a) Dynamics of the correlation
function. The horizontal line is at $q=0$.
(b) $\chi(\tau+t_2,t_{2})$ against $C(\tau+t_2,t_2)$
for four waiting times $t_2$ specified in the key. The black dotted line is the
FDT at $T_f\simeq0.320$.
(c) Comparison between $C(t,0)$ and $C_2(t,0)$. $C_2$ starts at $q_{\mathrm{in}}=0.5$ and oscillates around $0$, even though the amplitude of oscillations is slowly
decreasing with time.
(d) $R$, $R_{\rm st}$ and $-1/T_f \, \partial_{t_1} C$ as functions of $t_1-t_2$.
(e) The Lagrange multiplier $z(t)$, $T_f+J^2/T_f$ (dotted line) and  $z_f=T'+J^2/T'$ (dashed line).
(f) From top to bottom: kinetic,
total (constant), and potential energy densities.
(g) Fourier transforms of the correlation and response functions for two $t_2$ indicated in the key.
The black solid lines are the theoretical predictions for the real and imaginary part of the Fourier transform of the linear
response function, given by Eq.~(\ref{eq:response_fourier_transform-text}), with
parameters $m=1$, $J=0.25$ and $z_{f}=0.625$.
(h) The ratio $-\mathrm{Im}{\hat{R}(\omega)}/(\omega\hat{C}(\omega))$ where the various curves correspond to different waiting times
$t_2$. The dashed line is at $1/T_f \simeq 3.125$ and $1/T_{\rm kin}=2.67$ would be even below it.
}
\label{fig:T0p50_J0p25}
\end{figure}


 \subsubsection{Sector IV: large energy injection on a condensed state}
\label{subsubsec:from_cond_to_pm}

In  Fig.~\ref{fig:T0p50_J0p25} we show results for $T'=0.5 < T^0_c$ and $J=0.25$.
This quench injects a large amount of energy into the system, $\Delta e=0.5625$,
which is sufficient to take it out of the initial condensed state. Had the system
reached an equilibrium paramagnetic state asymptotically after the quench its temperature
would be $T_f \simeq 0.320 $, so that $T_f/J \simeq 1.281$, from
Eq.~(\ref{eq:Tf_from_cond_to_para}). This, however, is not consistent within the
$N\to\infty$ analysis and, accordingly, it is not realised dynamically.

\begin{figure}[h!]
\begin{center}
  \includegraphics[scale= 0.7]{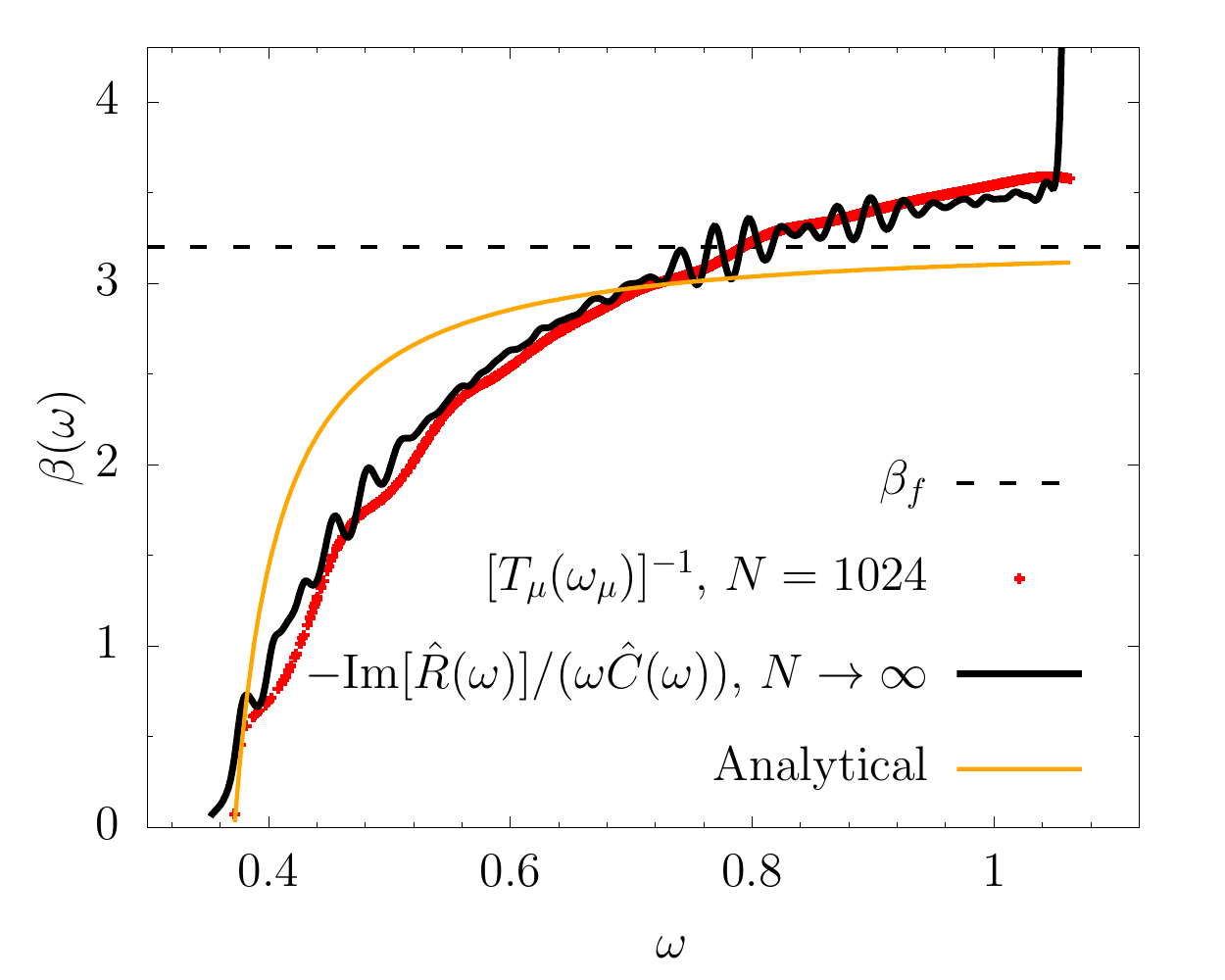}
\end{center}
\caption{\small {\bf Sector IV. Large energy injection on a condensed state.}
Same parameters as in Fig.~\ref{fig:T0p50_J0p25}. Comparison between the
mode inverse temperature for a finite size system (red), the inverse temperature for
the fluctuation dissipation ratio in the infinite size limit (black curve) and the
analytical expression for the mode inverse temperatures for independent
harmonic oscillators (yellow curve). All measured in units of $J_0$.
The frequency interval in which Im$\hat R(\omega)$
is non-zero is $(\omega_-, \omega_+) = (1/(2\sqrt{2}), 3/(2\sqrt{2})) \simeq (0.35, 1.06)$.
}
\label{fig:T0p50_J0p25-finiteN}
\end{figure}

The self correlations shown in (a)
present large oscillations with weakly decaying amplitude around the expected asymptotic value
$\lim_{t_1-t_2\rightarrow \infty} \lim_{t_2\to\infty} C(t_1,t_2) = 0$
for all values of $t_2$. $C(t_1,t_2)$ satisfies stationarity for sufficiently late $t_2$,
as one can see from the plot where the curves for $t_2$ coincide, within numerical accuracy.
The parametric plot of the susceptibility $\chi$ against the correlation $C$,
shown in (b), is rather complex and very far from linear.
The FDT relating $\chi$ to $C$ in a stationary regime
with putative target temperature $T_f \simeq 0.320$ is indicated as a straight dashed line.
This behaviour is confirmed by the time evolution of the stationary response function, $R(t_1,t_2)$, see panel (d).
$R(t_1,t_2)$ does not coincide with $-\frac{1}{T_f} \partial_{t_1}C(t_1,t_2)$, with $T_f$ the target temperature.
In panel (c) we also show the time evolution of the off-diagonal correlation with the initial configuration, $C_2(t,0)$.
In contrast to the previous cases, $C_2(t,0)$ is oscillating in time around the value $q_2=0$.
Panel (e) demonstrates that $z(t)$ is far from $z_f=T_f + J^2/T_f$ relative to an equilibrium paramagnetic state
but oscillates around $T'+J^2/T'$, consistently with our claims so far.
From panel (f)  we observe that the kinetic and the potential energies  relax to
the predicted asymptotic values specified in Sec.~\ref{app:energies}.
The Fourier transforms of the correlation and response functions for two different values of $t_2$, as indicated in the key,
are shown in panel (g).
The black solid lines represent the theoretical prediction for the real and imaginary part of the Fourier transform of the response function
in the stationary regime, $\hat{R}_{\mathrm{st}}(\omega)$, given by Eq.~(\ref{eq:response_fourier_transform-text}), with
parameters $m=1$, $J=0.25$ and $z_{f}=0.625$. They are in excellent agreement with the numerical results.
In panel (h) we show the ratio $-\mathrm{Im}{\hat{R}(\omega)}/(\omega\hat{C}(\omega))$. The dashed line is
$-1/T_f$ and is clearly off the data. The different curves were computed for various waiting times $t_2$
given in the key. In Fig.~\ref{fig:T0p50_J0p25-finiteN} we compare the FDR (averaged over different waiting
times to get rid of the undesired oscillations) to the mode temperatures of the finite $N$ system. The
correspondence between the two ways of extracting the temperatures is good. The yellow
line is the approximate prediction in Eq.~(\ref{eq:mode-temp-prediction}) stemming from the independent harmonic
oscillator approaximation that for this kind of quench is quite far from the
numerical results though it captures the qualitative features.

\vspace{0.5cm}

\section{Integrals of motion}
\label{sec:integrals-of-motion}

In  Sec.~\ref{sec:Neumann-sec} we recalled the relation between the $p=2$ spherical disordered model
and the classical integrable model introduced by Neumann. In this Section we will present some results concerning
the behaviour of the integrals of motion. A key issue we address here is how these influence the
statistical properties in the steady state.

\subsection{The integrals of motion landscape}

In Sec.~\ref{sect:p2-statics} we studied the potential energy landscape of the $p=2$ disordered
model and we found that it has $N$ extrema corresponding to $\vec s=\pm \sqrt{N} \vec v_\mu$ with $\vec v_\mu$
the $N$ eigenvectors of the interaction matrix $J_{ij}$. These directions turn out to be the
extrema of the integrals of motion landscape as well~\cite{Babelon-private}. This claim is
proven easily. Indeed,
\begin{eqnarray}
\frac{\partial I_\mu}{\partial p_\eta} = 0
\; ,
\end{eqnarray}
where  we labeled the constants of motion defined in Eq.~(\ref{eq:ImuUhlenbeck}) with $\mu$, the eigenvalue index,
implies the following two conditions
\begin{eqnarray}
\begin{array}{rcll}
0 &=& \displaystyle{s_\mu^2 p_\eta - s_\mu p_\mu s_\eta}  & \qquad\qquad \mbox{for} \quad \mu \neq \eta
\; ,
\vspace{0.1cm}
\nonumber\\
0 &=& \displaystyle{ \frac{1}{mN} \sum_{\nu (\neq \eta)} \frac{s_\nu^2 p_\eta -  p_\nu s_\nu s_\eta}{\lambda_\nu -\lambda_\eta} } & \qquad\qquad \mbox{for} \quad \mu = \eta
\; .
\end{array}
\end{eqnarray}
It is clear that the first relation makes the second one hold identically. Moreover, replaced in the definition of the
$I_\mu$s one finds
\begin{equation}
- \frac{1}{2} \lambda_\mu I_\mu = -\frac{1}{2}\lambda_\mu s_\mu^2
\end{equation}
that has to be extremised under the global spherical constraint on the $s_\mu$. This is just the analysis of the potential energy landscape that
we performed in Sec.~\ref{sect:p2-statics}, leading to ${\vec s}^*=\pm \sqrt{N} \vec v_\mu$ and $z^*=\lambda_\mu$ for all the
saddles in the landscape.

\subsection{Averaged values}

In our setting we use random initial conditions and we average over them. We will therefore focus on the averaged
integrals of motion,
\begin{equation}
\langle I_\mu \rangle = \langle s_\mu^2 \rangle +\frac{1}{mN} \sum_{\nu(\neq \mu)}  \frac{\langle s_\mu^2 p_\nu^2 \rangle + \langle s_\nu^2 p_\mu^2 \rangle
- 2 \langle s_\mu p_\nu s_\nu p_\mu \rangle}{\lambda_\nu-\lambda_\mu}
\; .
\end{equation}
Right after the instantaneous quench the initial values $\langle s_\mu^2(0^+) \rangle$ and $\langle p_\mu^2(0^+) \rangle$ are the ones right before the
quench, $\langle s_\mu^2(0^-) \rangle$ and $\langle p_\mu^2(0^-) \rangle$.
Under the quench the eigenvalues transform as $\lambda_\mu = J/J_0 \; \lambda_\mu^{(0)}$. We can therefore
compute the $\langle I_\mu(0^+)\rangle$ using these values.
Owing to the fact that the initial
conditions are drawn from an equilibrium probability
density, we have $\langle s^2_\mu(0^+)  p^2_\nu(0^+)  \rangle = \langle s^2_\mu(0^+)  \rangle \langle p^2_\nu(0^+) \rangle$
and
$\langle p_\mu(0^+)  p_\nu(0^+) \rangle=\langle s_\mu(0^+)  s_\nu(0^+) \rangle = \langle s_\mu(0^+)  p_\nu(0^+) \rangle = 0$
for all $\mu \neq \nu$, and $\langle s_\mu(0^+)  p_\mu(0^+) \rangle =0$ for all $\mu$. The constants are then
\begin{equation}
\langle I_\mu(0^+) \rangle = \langle s_\mu^2(0^+) \rangle +\frac{1}{mN} \sum_{\nu(\neq \mu)}  \frac{\langle s_\mu^2(0^+) \rangle \langle p_\nu^2(0^+) \rangle + \langle s_\nu^2(0^+) \rangle
\langle p_\mu^2(0^+) \rangle}{\lambda_\nu-\lambda_\mu}
\; .
\end{equation}

\vspace{0.25cm}

\noindent
{\it $y<1$: condensed initial states.}

\vspace{0.25cm}

In the cases in which $y<1$ the initial state is condensed and the integrals of motion of the modes $\mu\neq N$ and $\mu=N$ scale
very differently with $N$. For the modes in the bulk
\begin{eqnarray}
\langle I_{\mu\neq N}(0^+) \rangle &=&
\frac{T'}{z(0^-)-\lambda^{(0)}_\mu}
+
\frac{T'}{N}  \frac{\langle s_N^2(0^+)  \rangle}{\lambda_N-\lambda_\mu}
+
\frac{{T'}^2}{N} \sum_{\nu(\neq \mu, N)} \frac{1}{z(0^-)-\lambda_\nu^{(0)}} \frac{1}{ \lambda_\nu-\lambda_\mu}
\nonumber\\
&&
+ \frac{{T'}^2}{N} \sum_{\nu(\neq \mu)}  \frac{1}{z(0^-)-\lambda_\mu^{(0)}} \frac{1}{\lambda_\nu-\lambda_\mu}
\; .
\label{eq:Imu0plus}
\end{eqnarray}
This expression involves two sums that will appear again and again in the rest of this Section,
and depend only on the pre-quench parameters $J_0$ and $T'$,
\begin{eqnarray}
\begin{array}{ccl}
&&
\displaystyle{S^{(1)}_\mu \equiv \frac{1}{N} \sum_{\nu (\neq \mu)}  \frac{1}{\lambda_\nu^{(0)} - \lambda_\mu^{(0)}}}
=
\frac{1}{J_0}  \frac{1}{N} \sum_{\nu (\neq \mu)}  \frac{1}{\lambda_\nu^{(n)} - \lambda_\mu^{(n)}}
\equiv \frac{1}{J_0} S_\mu^{(1n)}
\; ,
\nonumber\\
&&
\displaystyle{S^{(2)}_\mu \equiv \frac{1}{N} \sum_{\nu (\neq \mu, N)}
\frac{1}{z(0^-)-\lambda_\nu^{(0)}} \frac{1}{\lambda_\nu^{(0)} - \lambda_\mu^{(0)}}}
\; .
\end{array}
\end{eqnarray}
For $\mu\neq N$ the second sum reads
\begin{equation}\label{eq:smu1_smu2}
S_{\mu \neq N}^{(2)} = \frac{1}{z(0^-)-\lambda_\mu^{(0)}} \left(\frac{1}{J_0} + S_\mu^{(1)}\right) =
 \frac{1}{z(0^-)-\lambda_\mu^{(0)}} \frac{1}{J_0} \left(1 + S_\mu^{(1n)}\right)
\; .
\end{equation}
The superscript $n$ means that the eigenvalues have been rescaled by $J_0$ in such a way that
they vary in the interval $(-2,2)$ and $S_\mu^{(1n)}$ is just a number.
Using these definitions we find
\begin{eqnarray}
\langle I_{\mu\neq N}(0^+) \rangle &=&
\frac{T'}{z(0^-)-\lambda^{(0)}_\mu}
+
\frac{J_0}{J} \left[
 \frac{T' q_{\rm in}}{\lambda^{(0)}_N-\lambda^{(0)}_\mu}
+
{T'}^2  \ S_\mu^{(2)}
+
\frac{{T'}^2 }{z(0^-)-\lambda_\mu^{(0)}} \ S_\mu^{(1)}
\right]
\; .
\label{eq:Imu0plus}
\end{eqnarray}
This expression has a rather simple dependence on $J$ that only appears as a prefactor
in front of the sum of three terms within the square brackets. (This fact justifies the similarity of
the bulk numerical values of $\langle I_\mu\rangle$ in, for example, panels
(c) and (d) in Fig.~\ref{fig:neu_diff}.) In the large $N$ limit $z(0^-) \to \lambda^{(0)}_N$ and
\begin{eqnarray}
\langle I_{\mu\neq N}(0^+) \rangle &=&
\frac{T'}{\lambda_N^{(0)}-\lambda^{(0)}_\mu}
\left[
1+ \frac{J_0}{J} +  \frac{2T'}{J} \ S_\mu^{(1n)}
\right]
\; .
\label{eq:Imuyplus-compact}
\end{eqnarray}


\begin{figure}[h!]
\vspace{0.5cm}
\begin{center}
\includegraphics[scale=0.5]{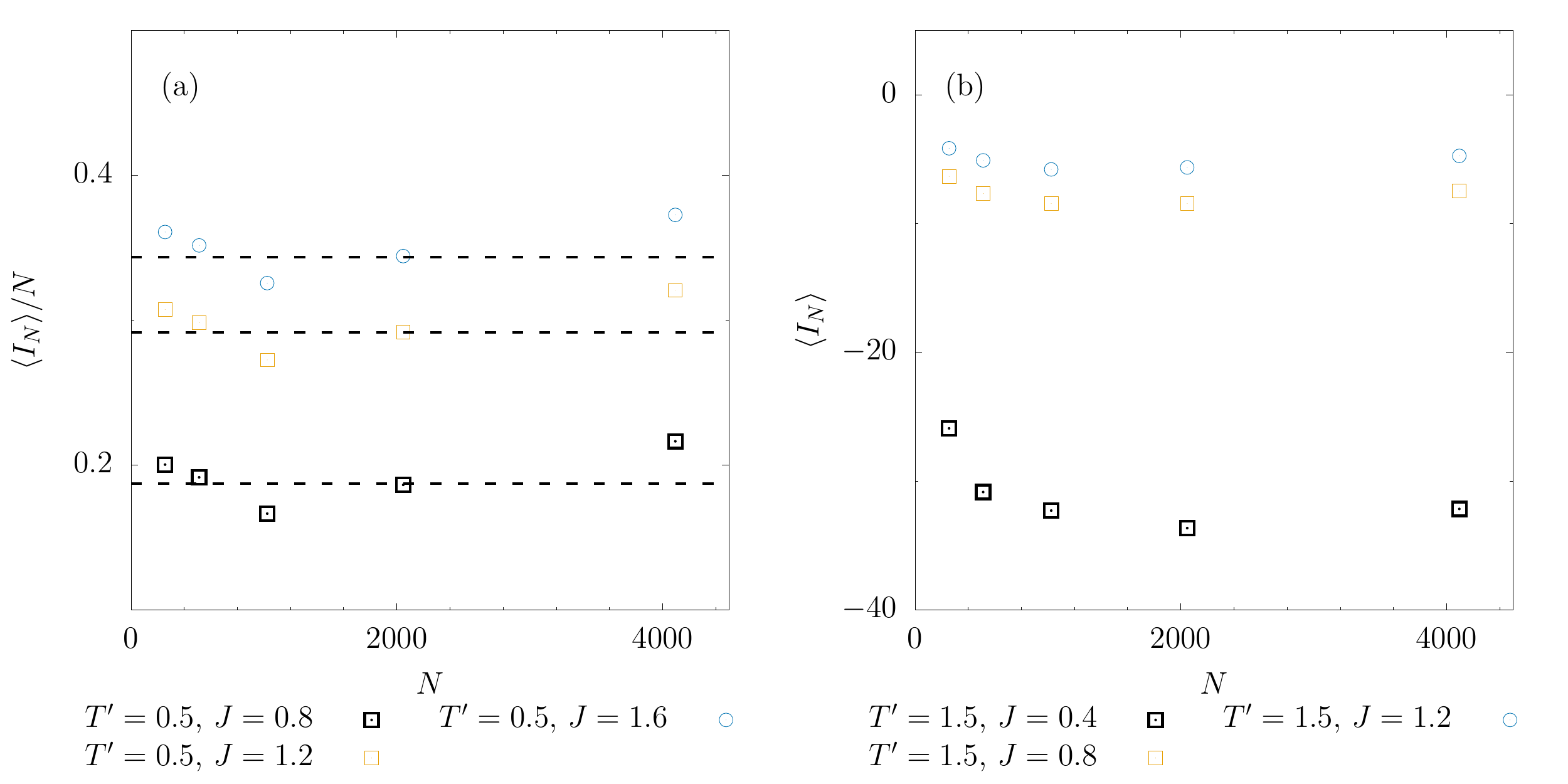}
\end{center}
\caption{\small
{\bf Uhlenbeck's integral of motion} $\langle I_{N}\rangle$ for the largest mode, $\mu=N$, as a function of system size for (a) three quenches with
$T^{\prime}/J_0<1$, and (b) three quenches with $T^{\prime}/J_0>1$. We observe a clear linear scaling with system size in the case of a condensed initial state, while
the result is order 1 for a PM initial state. In panel (a) the dashed lines are the slopes predicted by Eq.~(\ref{eq:in_cond}).
}
\label{fig:neu_in}
\end{figure}


For the largest mode $\mu=N$, if $y<1$, we obtain
\begin{eqnarray}
\langle I_N(0^+) \rangle &=&
q_{\rm in} \left( 1+ \frac{T' J_0}{J}  S_N^{(1)}\right) \, N
+
\frac{{T'}^2 J_0}{J} \ S_N^{(2)}
\; .
\label{eq:IN0plus}
\end{eqnarray}
In the limit $N\to\infty$ we can use $S_N^{(1)} \to -1/J_0$ and the known form of $q_{\rm in}$
to find
\begin{equation}\label{eq:in_cond}
\langle I_N(0^+) \rangle \mapsto \left(1-\frac{T'}{J_0}\right)\left(1-\frac{T'}{J}\right) \; N  + \frac{{T'}^2 J_0}{J} \ S_N^{(2)}(J_0)
\; .
\end{equation}
The parameter dependence of the slope in the right-hand-side seen as a function of $N$
is verified numerically in Fig.~\ref{fig:neu_in} (a).
The good match between this form and the numerics indicates that  $S_N^{(2)}$ must be negligible with respect to the first term that is  $O(N)$;
indeed, we have checked numerically that $S_N^{(2)}$ is $O(1)$.

\vspace{0.25cm}

\noindent
{\it $y<1$: paramagnetic initial states.}

\vspace{0.25cm}

If, instead, $y>1$, there is no condensed mode and the integrals of motion  are
\begin{eqnarray}\label{eq:imu_para}
\langle I_\mu(0^+)\rangle
&=&
\frac{T'}{z(0^-) - \lambda^{(0)}_\mu}
\left( 1+ \frac{T' J_0}{J} S_\mu^{(1)}\right)
+ \frac{{T'}^2 J_0}{J} S_\mu^{(2)}
\nonumber\\
&=&
\frac{J_0}{z(0^-) - \lambda^{(0)}_\mu} \left[ \frac{T'}{J_0} +\frac{T'^2}{JJ_0}
+
\frac{2{T'}^2}{JJ_0} \
S_\mu^{(1n)}\right]
\end{eqnarray}
and all of order $1$. The particular case $\mu=N$ is displayed in the panel (b) of Fig.~\ref{fig:neu_in}, showing no
dependence on $N$ for $N\stackrel{>}{\sim} 2000$, as expected.

\vspace{0.25cm}

\noindent
{\it Summary.}

\vspace{0.25cm}

Summarising, Fig.~\ref{fig:neu_in} displays the integral of motion associated to the mode at the edge of the spectrum for $y<1$ (a) and $y>1$ (b). All data points
were obtained using a single realisation of the random matrix. For the
condensed initial conditions we clearly see the linear scaling with $N$. For the non-condensed ones
the variations show a weak $N$ dependence for
small $N$ plus a possible variation due to the fluctuations in the realisation of the eigenvalues. The result is distinctly finite in this case.

As for the time-evolution of these averaged quantities, we have verified (not shown) that each of them are
conserved $\langle I_\mu(0^+)\rangle = \langle I_\mu(t)\rangle$ for all $\mu$, and that they satisfy the
two constraints $\sum_\mu I_\mu(0^+) =  \sum_\mu I_\mu(t) = N$ and  $\sum_\mu \lambda_\mu I_\mu(0^+) =  \sum_\mu \lambda_\mu I_\mu(t) = -2e_{\rm tot} N$,
with $e_{\rm tot}$ the total energy density.

\subsubsection{Gibbs-Boltzmann equilibrium?}

The analysis of the constants of motion should shed light on the ``distance'' from  complete
equilibration to a Gibbs-Boltzmann probability density, especially  in Sector III where shallow quenches
in the $N\to\infty$ Schwinger-Dyson formalism suggested proximity from this description.
We here compare the actual values of the $\langle I_\mu\rangle$ to the
ones a system in Gibbs-Boltzmann equilibrium at a single temperature $T_f$ would have.

\vspace{0.25cm}

\noindent
{\it The modes in the bulk}

\vspace{0.25cm}

In a final Gibbs-Boltzmann equilibrium state the integrals of motion should read
\begin{eqnarray}\label{eq:imuf}
\overline{\langle I_{\mu\neq N} \rangle}_{T_f}
&=&
\frac{T_f}{z_f-\lambda_\mu}
+
 \frac{T^2_fJ_0}{J} \ \frac{1}{z_f-\lambda_\mu} \ S_\mu^{(1)}
+
T_f \  \frac{q_f}{\lambda_N-\lambda_\mu}
+
\frac{{T_f}^2 J^2_0}{J^2} \; S^{(2)}_\mu
\; ,
\label{eq:ImuTf}
\end{eqnarray}
where we allowed for the condensation of the largest mode. We wish to compare this expression to
the one in Eq.~(\ref{eq:Imuyplus-compact}).
After some lengthy calculations, using the $N\to\infty$ values
\begin{equation}
q_{\rm in} = 1 - \frac{T'}{J_0}
\; ,
\qquad
z_f = 2J = \frac{J}{J_0} z(0^-) \; ,
\qquad
T_f = \frac{T'}{2} \left( 1 + \frac{J}{J_0}\right)
\;,
\qquad
q_f =1-\frac{T_f}{J}
\; ,
\end{equation}
 we find that the difference $\Delta I_{\mu\neq N} = \langle I_{\mu\neq N}(0^+) \rangle
- \overline{\langle I_{\mu\neq N}\rangle}_{T_f}$ is
\begin{eqnarray}
\Delta I_{\mu\neq N} &=&
- \frac{{T'}^2 }{2} \left( \frac{J_0}{J}-1\right)^2
 \ \frac{1}{\lambda_N^{(0)}-\lambda^{(0)}_\mu} \  S_\mu^{(1)}
\end{eqnarray}
a finite, non-zero,  value for all $\mu\neq N$.

\begin{figure}[h!]
\begin{center}
\includegraphics[scale=0.55]{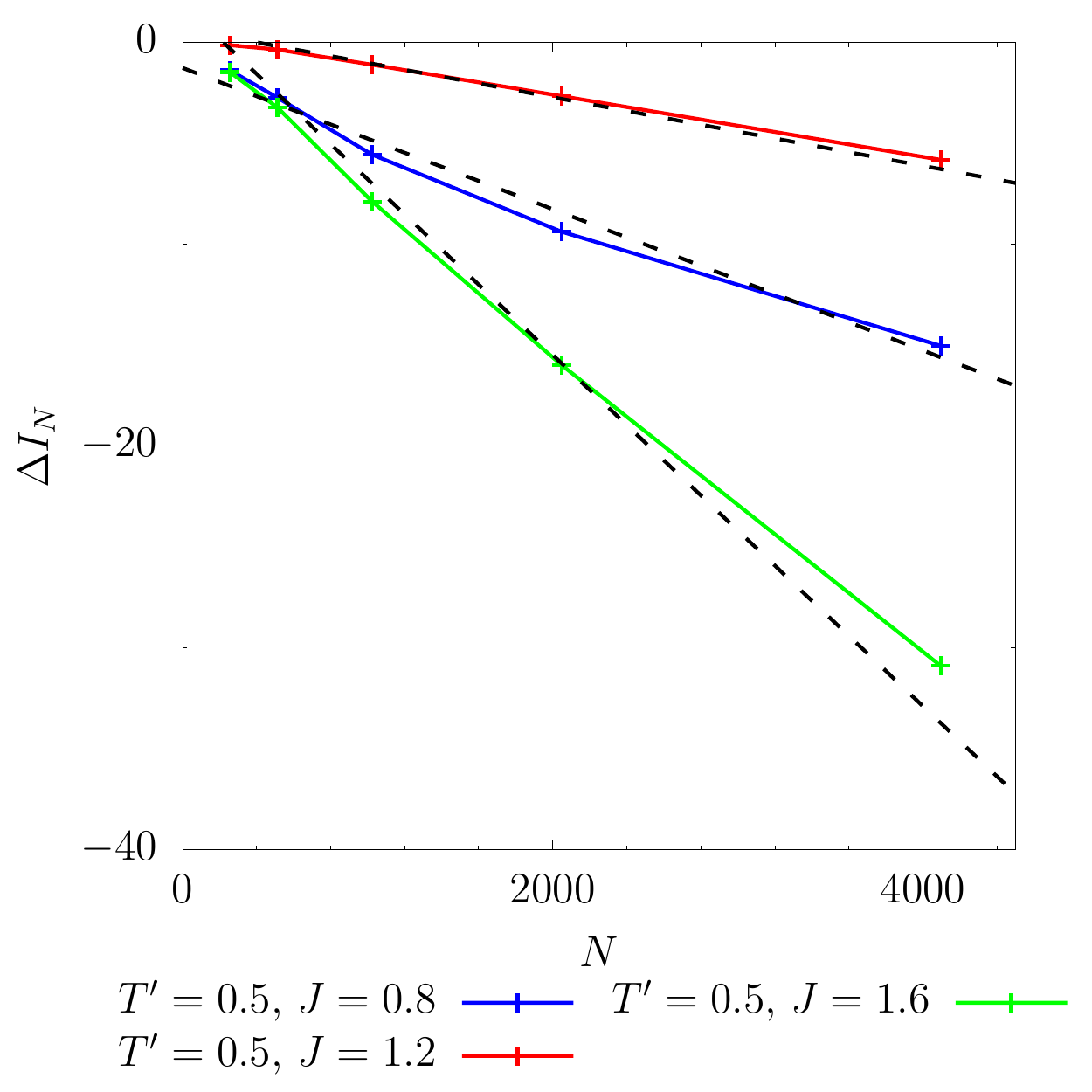}
\end{center}
\caption{\small {\bf Scaling of $\Delta I_{N}$ in sector III}.
The data points at five values of $N$ are shown with points that are joined by coloured straight lines.
The dashed black lines are the predictions of Eq.~(\ref{eq:slope_in}).
}
\label{fig:in_scaling}
\end{figure}

\vspace{0.25cm}

\noindent
{\it The $N$th mode}

\vspace{0.25cm}

The $N$th mode has a different scaling with system size. We have already computed $\langle I_N(0^+)\rangle$
in Eq.~(\ref{eq:in_cond})
and we want to compare it to what it should read in equilibrium at $T_f$:
\begin{eqnarray}
\overline{\langle I_N\rangle}_f &=& q_f  \left( 1+ \frac{T_f J_0}{J} \ S_N^{(1)} \right) N
+\frac{T^2_f J^2_0}{J^2} \ S_N^{(2)}
\; .
\end{eqnarray}
The difference between the two, $\Delta I_{N} \equiv \langle I_{N}(0^+)\rangle - \overline{\langle I_{N}\rangle}_{T_f}$,
is
\begin{eqnarray}
\Delta I_{N}
&=&
\left[ q_{\rm in} - q_f + (q_{\rm in} T' - q_f T_f) \frac{J_0}{J} \ S_N^{(1)} \right] N + \left( {T'}^2 - T_f^2 \frac{J_0}{J} \right) \frac{J_0}{J} \ S_N^{(2)}
\; .
\end{eqnarray}
The first term diverges linearly with $N$. We have already argued that the second one is $O(1)$, see the
discussion after Eq.~(\ref{eq:in_cond}). Therefore, we
focus on the slope of the difference, seen as a function of $N$. After replacing $q_f$, $q_{\rm in}$ and $T_f$ by their expressions
in terms of $J$, $J_0$ and $T'$ in the large $N$ limit,  and $S_N^{(1)} \to -1/J_0$,
\begin{equation}
\label{eq:slope_in}
\frac{\Delta I_{N}}{N} \to - \frac{{T'}^2}{4J_0^2} \left( \frac{J_0}{J} - 1\right)^2
\; .
\end{equation}
This form is validated by the numerical data in Fig.~\ref{fig:in_scaling} for three
pairs of $T'$ and $J$, all in Sector III.

Quite clearly, the differences between the actual $\langle I_\mu\rangle$ and the ones in equilibrium at
a temperature $T_f$ are proportional to $J_0/J-1$, a factor that vanishes for $J_0=J$. Otherwise,
the differences are finite for $\mu\neq N$ and proportional to $N$ for the last mode, making the
mode-by-mode difference non-vanishing for $J\neq J_0$. This fact  confirms, then, the lack of
equilibration to a Gibbs-Boltzmann measure with a single $T_f$.

\vspace{0.25cm}

\noindent
{\it The special line $y=\sqrt{x}$}
\vspace{0.25cm}

On the curve $y=\sqrt{x}$ the asymptotic analysis for $N\rightarrow\infty$ predicts thermalization at temperature $T_f=J$.
The numerical analysis of the Schwinger-Dyson equations confirms this prediction as the correlation and linear response
are linked by FDT and the time-dependence of the correlation function after an instantaneous quench is identical (within numerical accuracy) to the
one found in equilibrium at this temperature.
We shall briefly analyse the behaviour of the constants of motion in this case.

On the special line $y=\sqrt{x}$, ${T'}^2 = J J_0$
and Eq.~(\ref{eq:imu_para}) yields the following expression for the averaged integrals of motion
\begin{equation}
\langle I_{\mu}(0^+)\rangle=\frac{1}{z(0^-)/J_0-\overline{\lambda}_{\mu}}\left(  \sqrt{\frac{J}{J_0}}+1+2S^{(1n)}_{\mu}  \right)
\; ,
\end{equation}
where $\overline{\lambda}_{\mu}=\lambda^{(0)}_{\mu}/J_0$ are the normalised eigenvalues that vary in the interval $(-2,2)$.
From Eq.~(\ref{eq:imuf}) with $q_f=0$, $z_f=2J$ valid in the $N\rightarrow\infty$ limit, and using
Eq.~(\ref{eq:smu1_smu2}) the constants of motion in an equilibrium state at temperature $T_f$ are
\begin{equation}
\overline{\langle I_{\mu}\rangle}_{T_f}=\frac{1}{2-\overline{\lambda}_{\mu}}\left(  2+2S^{(1n)}_{\mu}  \right)
\; .
\end{equation}

We observe that the two sets of values are very similar if $J$ is close to $J_0$. Numerically, we find that, in the special case $T^{\prime}=1.25 \ J_0$ and
$J=1.5625 \ J_0$, the $\Delta I_{\mu}$ is of order of $10^{-2}$. We conclude that even in this case, in which the asymptotic analysis predicts that the global,
mode-averaged quantities, behave as in equilibrium, a prediction that seems to be confirmed by the numerics,
the constants of motion are not exactly the same in the initial state and in the
thermal state the system would reach in case of thermalisation.

In order to properly interpret these results, it is important to keep in mind that, strictly speaking, the conserved energy dynamics of an isolated (finite size) system
should keep memory of the initial conditions, even if the system is non-integrable. In our problem, we see this information
encoded in the $I_\mu$s. More so, not even in the $N\to\infty$ limit this memory is
erased as the $\Delta I_{\mu}$s remain finite.

\subsubsection{Independent harmonic oscillators}

We have seen in Sec.~\ref{subsec:independent-harmonic-oscillators} that the $N\to\infty$ system decouples into independent
harmonic oscillators in the asymptotic long time limit (taken after $N\to\infty$) since $z(t) \to z_f$.
A natural
idea is to check whether we can identify the integrals of motion $\langle I_\mu(0^+)\rangle$ with the ones that
an ensemble of harmonic oscillators with spring constants $m \omega_\mu^2 = z_f-\lambda_\mu$ in equilibrium at the temperatures
$T_\mu$ would have.

On the one hand, we note that $\langle s_\mu^2(t)\rangle$ and $\langle p_\mu^2(t)\rangle$ are not constant
for harmonic oscillators but their time averages are, so we evaluate $\overline{\langle I_\mu \rangle} $ finding, in this framework,
\begin{eqnarray}
\overline{\langle I_\mu \rangle}_{\rm osc} &=&
\frac{T_\mu}{z_f-\lambda_\mu}
+
\frac{T_\mu}{N} \sum_{\nu(\neq \mu)}  \left[ \frac{T_\nu}{z_f-\lambda_\mu} + \frac{T_\nu}{z_f-\lambda_\nu}  \right] \frac{1}{\lambda_\nu-\lambda_\mu}
\; .
\end{eqnarray}
In the $N\to\infty$ limit the value of $z_f$ is expected to be  $T'+J^2/T'$ for $x<y$ and $2J$ for $y<x$.
Imposing the identity
between the $\overline{\langle I_\mu \rangle}_{\rm osc}$  and the $\langle I_\mu(0^+)\rangle$ given in Eqs.~(\ref{eq:Imu0plus}) and (\ref{eq:IN0plus}) should yield
a better estimate of  the temperatures $T_\mu$ than the one explained in
Sec.~\ref{subsec:independent-harmonic-oscillators}. We leave this
analysis aside for the moment.

On the other hand, once the oscillators have decoupled and the Lagrange multiplier stabilised, the mode total energies, or the time-averaged
kinetic and potential energies {\it are} also constants of motion. In the next Subsection we investigate to what extent the $\langle I_\mu\rangle$
are proportional to $e_\mu^{\rm tot}= 2\overline{e_\mu^{\rm kin}}$.

\subsubsection{The integrals of motion and the mode temperatures}

\begin{figure}[t!]
\vspace{0.5cm}
\begin{center}
\includegraphics[scale=0.5]{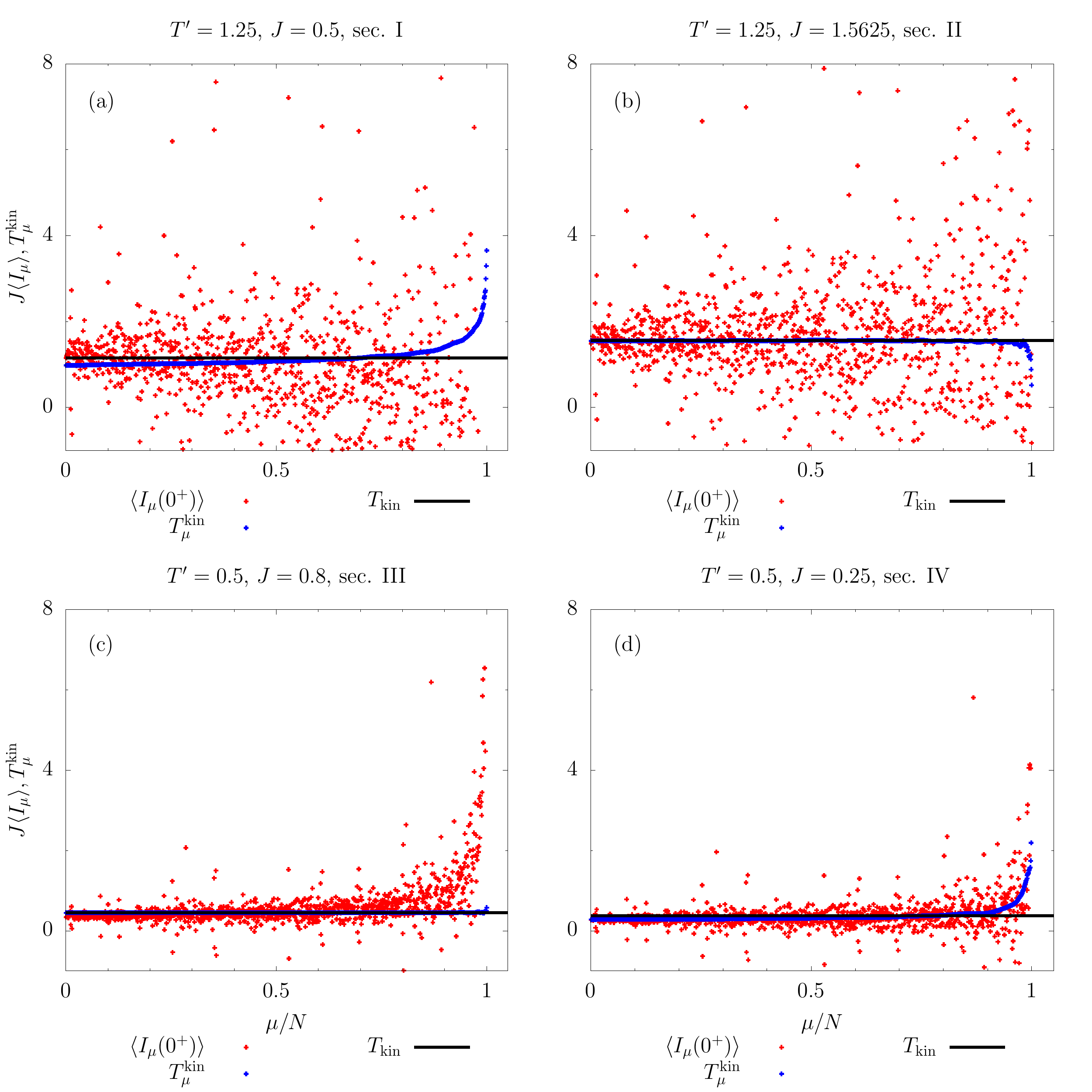}
\end{center}
\caption{\small
Comparison between the averaged Uhlenbeck's integrals of motion $J\langle I_{\mu}\rangle$ (red data points) and the
mode temperatures $T^{\mathrm{kin}}_{\mu}=2\overline{e_\mu^{\rm kin}}$ (blue datapoints) of each mode
in the four sectors of the phase diagram
(we have checked that $2\overline{\epsilon_\mu^{\rm pot}}$ yield equivalent results).
We used the same vertical scale in all plots, leaving aside the values of $\langle I_\mu \rangle$ close to the edge in~(c)
for which,  in particular, $\langle I_N\rangle \simeq 180$. There is no such divergence at the edge of the spectrum in the other panels.
The black horizontal lines represent the global kinetic temperatures $T_{\rm kin}$ in the three sectors. They are located at
$T_{\rm kin}
\simeq 1.15$ (a),
$T_{\rm kin}
\simeq 1.56$ (b), $T_{\rm kin}
\simeq0.45$ (c) and $T_{\rm kin}
\simeq 0.37$~(d), according to  Eq.~(\ref{eq:kin-temps}). We have also checked that $N^{-1} \sum_\mu T_\mu^{\rm kin} = T_{\rm kin}$.
}
\label{fig:neu_diff}
\end{figure}

Figure~\ref{fig:neu_diff} shows the spectrum of integrals of motion $J\langle I_{\mu}\rangle$ (red data points)
together with the mode kinetic temperatures  $T^{\mathrm{kin}}_{\mu}=2\overline{e_\mu^{\rm kin}}$ (that, we have checked,
are in agreement within numerical accuracy with the potential ones $T^{\mathrm{pot}}_{\mu}$ away from the edge of
the spectrum), for parameters in the four sectors of the dynamic phase
diagram. We show the data using the same vertical scale in all cases. Although the data for $J\langle I_\mu \rangle$ are
noisier than the ones for $T_\mu^{\rm kin}$,  the two are in fairly good agreement in the bulk of the
spectrum, far from the edge, in all panels. In the same figures we include the values of the global kinetic
temperature, $T_{\rm kin}$ defined in Eq.~(\ref{eq:kin-temps}) and we see that the mode temperatures are very close to
it again away from the edge of the spectrum and satisfying the constraint $N^{-1} \sum_\mu T_\mu^{\rm kin}=T_{\rm kin}$.

The parameters in Sector II, panel (b), are on the curve $y=\sqrt{x}$ on which the data for the global correlation and
linear response show equilibrium at a single
temperature $T_f=J$. The blue data points for the mode kinetic energies are precisely on this
value. The integrals of motion scatter, however, quite a lot around this value. Importantly enough,
the difference $\Delta I_{\mu}$ is very small in this case.

In sector III, panel~(c), the data for $J \langle I_\mu\rangle$ tend to
be relatively flat in the bulk of the spectrum and very close to $T_{\rm kin}=T_f=T'(1+J/J_0)/2$.
This result can be derived from Eq.~(\ref{eq:Imu0plus}) taking advantage of a simple rearrangement of the two
sums, $S_\mu^{(2)} = (S_\mu^{(1)}-S_N^{(1)})/(z(0^-)-\lambda_\mu^{(0)})$ valid for $\mu\neq N$.
In particular, taking $\lambda_\mu^{(0)}$ right at the middle of the spectrum, $S_\mu^{(1)}=0$ by symmetry and
$J\langle I_0\rangle = T_f$ for $N\to\infty$.
The incipient divergence close to the right edge of the spectrum, with the deviation from this
constant value, is also clear.
By using a maximal value of 8 in the vertical axes we have explicitly left aside the points for $\mu\simeq N$ that take much
larger values. For instance, $J \langle I_N \rangle\simeq 180$.   The approximate
mode temperatures in Eq.~(\ref{eq:mode-temp-prediction}) are all identical to $T_f=T_{\rm kin}$ in this Sector, consistently
with the numerical data away from the edge. The fact that the system is not in proper Gibbs-Boltzmann equilibrium
is due to the fact that the higher lying integrals of motion do not comply with this temperature.


The data for $J \langle I_\mu\rangle$ in Sector III look very similar to the ones found in Sector IV, see panel (d).
Indeed, the data almost
coincide far from the edge, since they
are both determined by Eq.~(\ref{eq:Imu0plus}) that has a weak dependence on $J$ (the only  parameter that takes a different
value in panels (c) and (d)). Close to the edge, the $J \langle I_\mu\rangle$ differ since in III (c) there is scaling with $N$ while in IV (d) there is
not. The mode kinetic energies are mode independent and identical to $T'(1+J/J_0)/2$ contrary to what the
approximation in the last line of Eq.~(\ref{eq:mode-temp-prediction}) tells. This means that for these quenches the assumption in
Eq.~(\ref{eq:finite_N_app}) fails.

We reckon that the kinetic temperature $T_{\rm kin} =2 \overline{e}_{\rm kin}$ should be equal to the sum
of the mode kinetic temperatures  $T_{\rm kin} = 1/N \, \sum_{\mu=1}^N T_\mu^{\rm kin}$.
We have  checked numerically this property.

We leave a more detailed analysis of the comparison between the two and their use to build a GGE for a future publication.

\subsection{Fluctuations of the integrals of motion in the equal energy hypersurface}

The fact that the integrable system reaches a state that is very close to thermal equilibrium in sector III, the sector with the lowest total energy in the phase diagram, allows us
to infer some properties of the phase space structure of the model. Only a system for which it is possible to visit all configurations with the same energy in the course of its
dynamical evolution is capable of reaching thermal equilibrium. An integrable model cannot achieve this goal since it
is bound to wander in a region of the phase space compatible with the values that the integrals of motion (IOM)
take on the initial configuration. The dynamics is constrained inside the phase space region composed by
configurations which have the same values of the $I_{\mu}s$ $\forall\mu$. Such regions can be labeled with
the values of the $I_{\mu}$s (which also define the energy of the group of configurations since $-\sum \lambda_{\mu} I_{\mu}=2H$),
and we shall call them iso-IOMs-regions in the following.


\begin{figure}[h!]
\vspace{0.5cm}
\begin{center}
\includegraphics[scale=0.5]{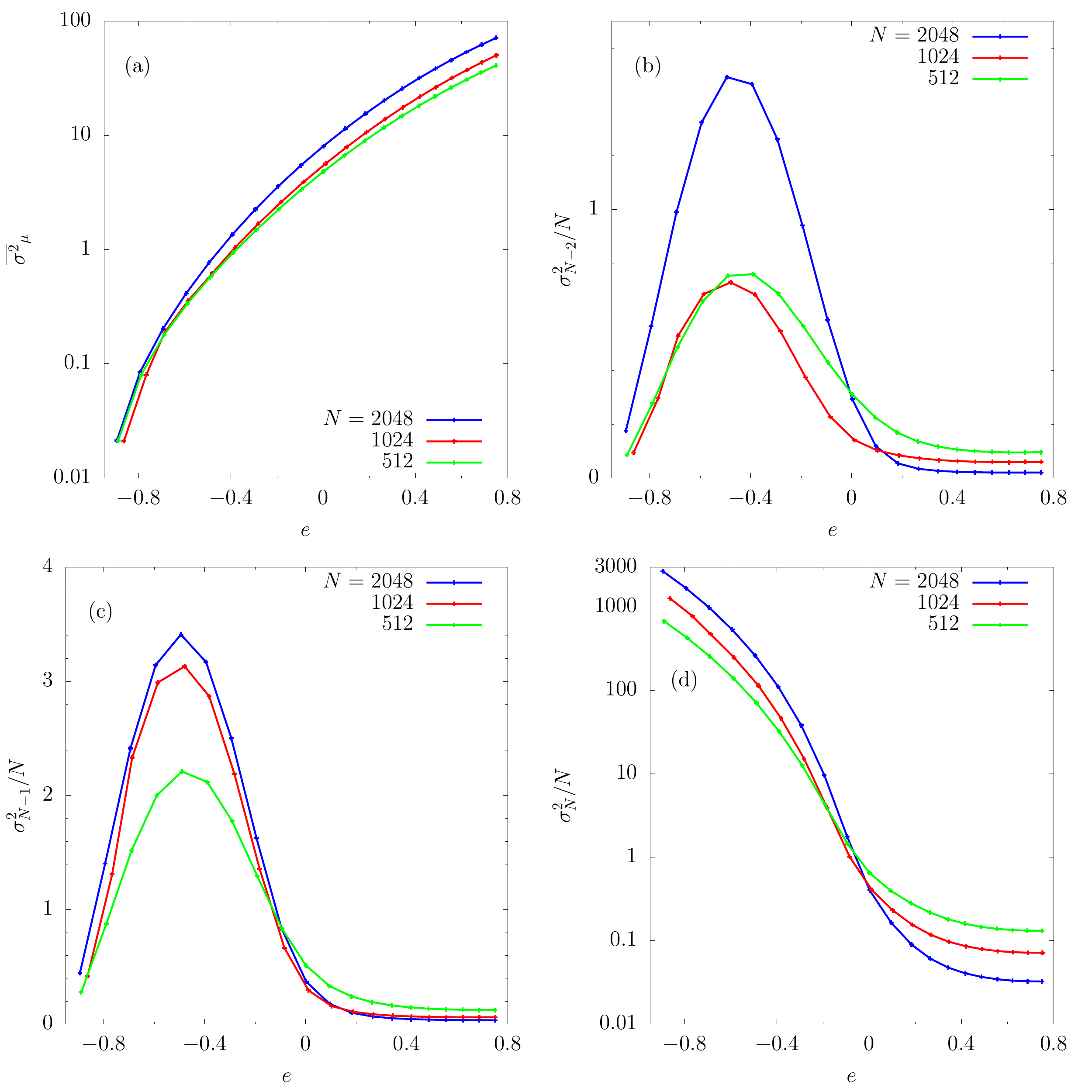}
\end{center}
\caption{\small {\bf Variance $\sigma^2_{\mu}(e)$ for different system sizes.}
(a) Behaviour of the average over the first three fourths of the spectrum.
(b), (c) $\sigma^2_{\mu}/N$ for modes near the edge of the spectrum. (d) $\sigma^2_{N}/N$.
Note the logarithmic scale in the vertical axis in panels (a) and (d).
}
\label{fig:neumann_var_scaling}
\end{figure}

A close-to-thermalised dynamics in an integrable system
should be indicative of a substantial overlap
between the constant energy manifold and the equal IOMs
region in phase space.  This claim is, however, highly non-trivial since
the equal energy manifold is $2N-1$ dimensional while the iso-IOMs-region has only $2N-N=N$ dimensions.
In the large $N$ limit, the equal energy manifold is huge with respect to the equal IOMs one
(for $N=2$ the constant energy region is a volume and the equal IOMs-configuration is a surface).

Our hypothesis for the dynamics of the Neumann model with parameters close to $x=1$ in phase III,
is that the constant energy manifolds have a substantial overlap with any iso-IOMs-region that include a configuration with the given energy.
In order to test this guess, we studied the following quantity:
\begin{equation}
\sigma^2_{\mu}(e)=\int \prod^{N}_{i=1}ds_{i}\,dp_{i}\,\delta(H[s_{i},p_{i}]/N-e)(I_{\mu}[s_{i},p_{i}]-\langle I_{\mu}[s_{i},p_{i}]\rangle)^2
\; ,
\end{equation}
where the average is a microcanonical one given by
\begin{equation}
\langle I_{\mu}[s_{i},p_{i}]\rangle=\int \prod^{N}_{i=1}ds_{i}\,dp_{i}\,\delta(H[s_{i},p_{i}]/N-e)I_{\mu}[s_{i},p_{i}]
\; .
\end{equation}
The quantity $\sigma^2_{\mu}(e)$ measures how large are the fluctuations in the value of a given IOM $I_{\mu}$ in the set of configurations with the same energy $e$.
According to the discussion at the beginning of this Section, if, for a given energy, we observe a small value in $\sigma_{\mu}(e)$, this  indicates a tendency of the integrable
system to thermalise.  In order to perform the averages over equal energy configurations, we replace the microcanonical average by a canonical one, introducing a Lagrange
multiplier $\beta$ and a measure $\exp(-\beta H)/Z$, fixing the average energy density of the ensemble. For large $N$, the fluctuations of the energy average are small and
we get a good approximation to the microcanonical mean. The advantage of the canonical measure is that the $\langle I_{\mu}\rangle$s are expressed in terms of canonical
averages of the same kind as those which we were using in the previous Sections to describe the initial state of the
dynamics. Moreover, once $z$ is fixed, the Hamiltonian is quadratic, and this  allows one
to express the higher order average $\langle I^2_{\mu}\rangle$, which includes products of $4$, $6$ and $8$ phase space variables, in terms of quadratic averages
$\langle s^2_{i}\rangle$ and $\langle p^2_{i}\rangle$. A straightforward numerical calculation of $\sigma^2_{\mu}(e)$ is then possible.
We show numerical results in Fig.~\ref{fig:neumann_var_scaling}. In  panel (a) we plot $\overline{\sigma^2_{\mu}}$, the average of
$\sigma^2_{\mu}(e)$ in the first three fourths of the spectrum
\begin{equation}
\overline{\sigma^2_{\mu}}=\sum^{3N/4}_{\mu=1}\sigma^2_{\mu}(e)
\; .
\end{equation}
We can clearly observe that it is very small for low energies and that it increases by several
orders of magnitude as we increase the energy density of the system.
Such is the behaviour of $\sigma^2_{\mu}(e)$ for the great majority of the modes. In panels~(b) and~(c) we show the behaviour of
$\sigma^2_{\mu}(e)$ for $\mu$ close to $N$, the edge of the spectrum. We observe that $\sigma^2_{\mu}(e)$ exhibits a maximum at low energies. Finally, in panel~(d),
$\sigma^2_{\mu=N}(e)$ is plotted. At very low energies, $\sigma^2_{N}(e)$ is very large and scales with $N$.
As we increase energy, $\sigma^2_{\mu}(e)$ decreases abruptly until it reaches a value that is inversely proportional to $N$.


Summarising, we observe that, far from the edge of the spectrum, the fluctuations in the IOMs are very small for sufficiently low energies. This means that the low energy configurations of the model have very similar values of the $I_{\mu}$s, at least for $\mu$ far from the edge of the spectrum. For modes close or at the edge of the spectrum fluctuations can be very large. These results suggest that the iso-IOMs-regions which lie at low energies have very similar values for the $I_{\mu}$s for $\mu$ far from $N$, and only differ in the values of the $I_{\mu}$s close to the edge of the spectrum. In fact, in sector III, the sector with the lowest energies in the phase diagram, we have verified in previous Section that the initial state and the final thermal state at temperature $T_f$ which partially describes the long-time dynamics have very similar values of the IOMs far from the edge of the spectrum, with discrepancies only near the edge of the spectrum.

\section{Conclusions}
\label{sec:conclusions}

This paper continues our study of the conserved energy dynamics following sudden quenches in classical disordered isolated models
ruled by Newton dynamics.

\begin{figure}[h!]
\vspace{0.5cm}
\begin{center}
\includegraphics[scale=0.6]{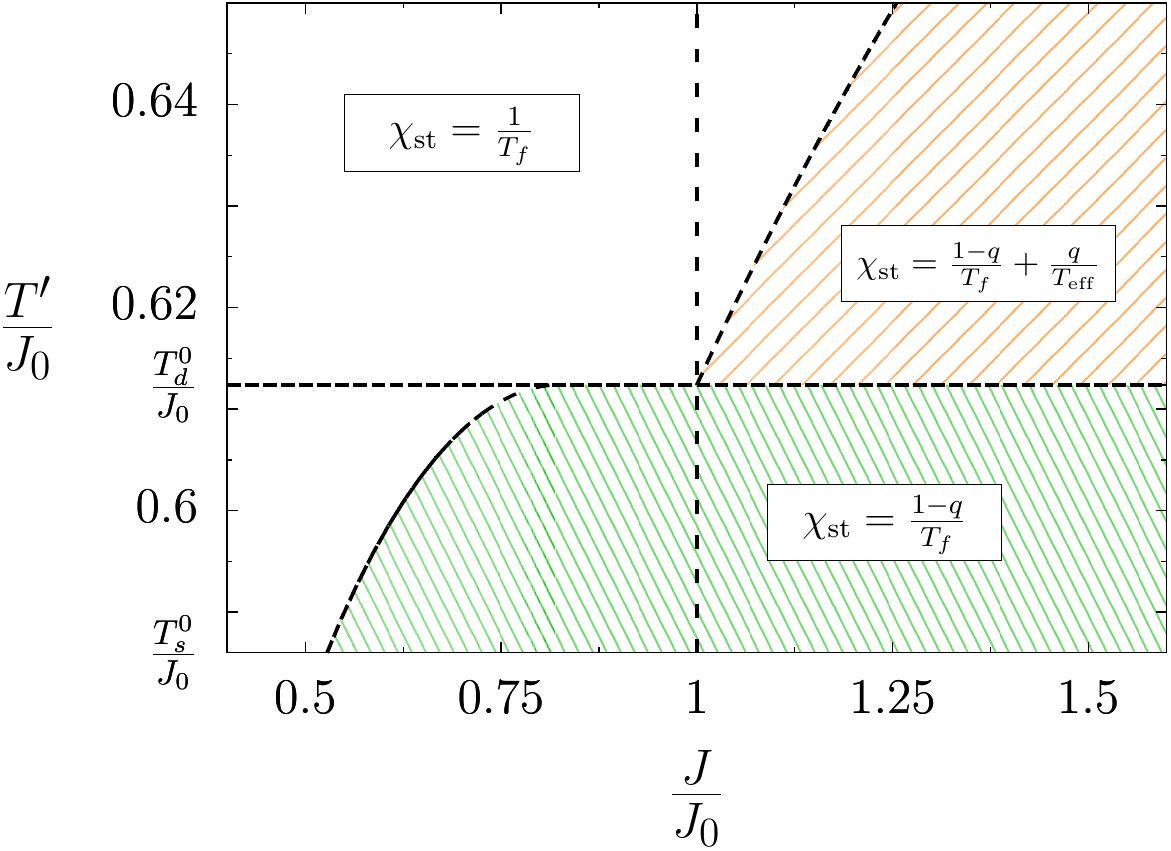}
\hspace{0.75cm}
\includegraphics[scale=0.62]{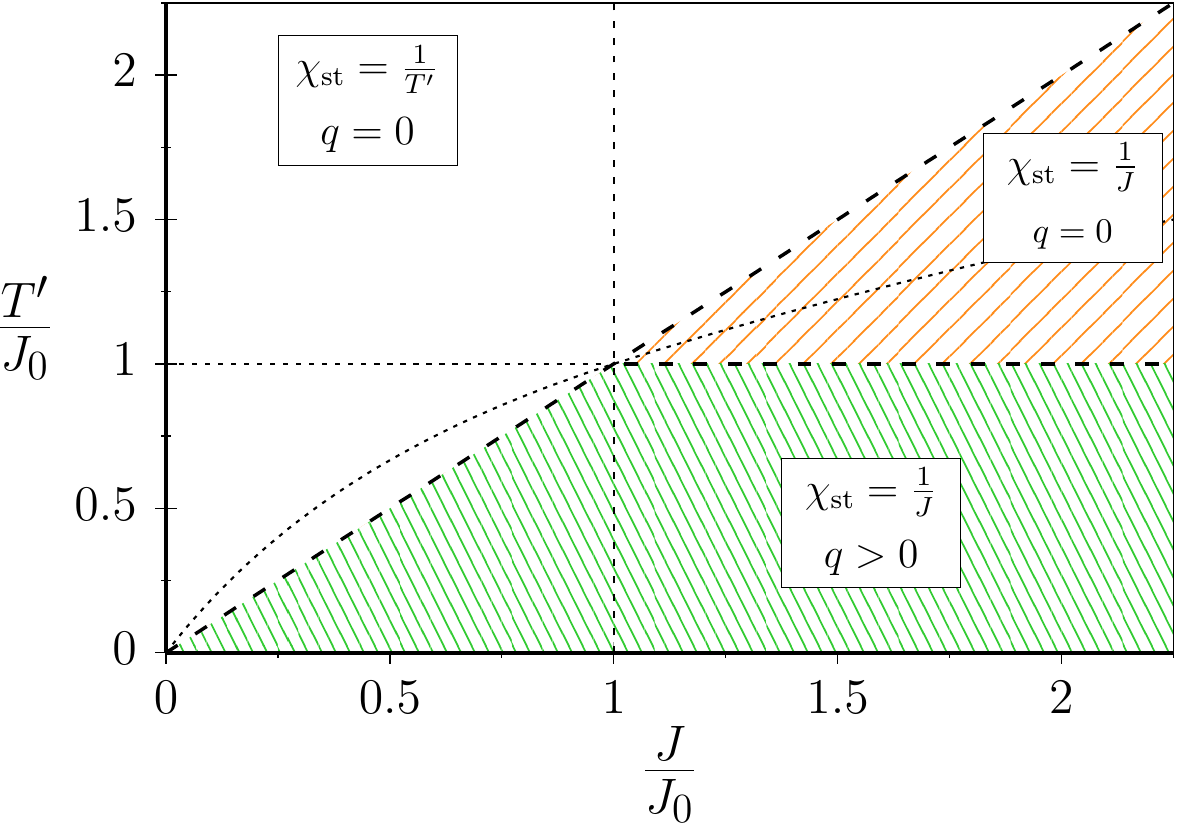}
\end{center}
\caption{\small {\bf Dynamic phase diagrams.}
(a) $p=3$. (b) $p=2$.
}
\label{fig:dyn-phase-diagram-p2-p3}
\end{figure}

In the $p=3$ strongly interacting case~\cite{CuLoNe17} all quenches reach an asymptotic regime in which a single (or double)
temperature dynamical regime establishes.
The systems either equilibrate to a paramagnetic state with a proper temperature, they remain confined in a metastable state
with restricted Gibbs-Boltzmann equilibrium at a single temperature, or age indefinitely after a quench to the threshold with the
dynamics being characterised by one temperature at short time delays and another one at long time delays, similarly to what happens
in the relaxational case~\cite{CuKu93,CuKu95,Ba97,CaCa05,LesHouches}. The two temperatures $T_f$ and $T_{\rm eff}$ depend on the
pre and post quench parameters in ways that were determined in~\cite{CuLoNe17}. For the sake of comparison, the dynamic phase diagram of the
$p=3$ isolated model is reproduced
in Fig.~\ref{fig:dyn-phase-diagram-p2-p3} (a).


In the $p=2$ model the conserved energy dynamics are quite different both
from the relaxational ones~\cite{ShSi81,CidePa88,CuDe95a,CuDe95b,FyPeSc15}
and the isolated $p=3$ interacting case~\cite{CuLoNe17}. This is due to its integrability, made
explicit by its relation to the Neumann integrable model.
The main results in this paper are the following.

\begin{itemize}

\item
We identified the dynamic phase diagram according to the asymptotic behaviour of the
static susceptibility, the Lagrange multiplier imposing the spherical constraint (an action density), and the
long time-delay limit of the two-time correlation function. The phase diagram is shown in
Fig.~\ref{fig:dyn-phase-diagram-p2-p3} (b), and it can be compared to the one of the
$p=3$ isolated case shown on its left side.

\item
In the analysis of the Schwinger-Dyson equations we distinguished four sectors in the
phase diagram depending on the initial state (being condensed or not) and the
final value of the static susceptibility. We reduced  these four sectors to
three phases, indicated with different colours in the Figure, in which the dynamic behaviour
is different. Basically, they are distinguished by two ``order parameters'', $\chi_{\rm st}$ and
$q$, the static susceptibility and the asymptotic value of the two-time correlation.

\item
In none of the phases the system equilibrates to
a Gibbs-Boltzmann measure. Accordingly, there is no single temperature characterising
the values taken by different observables in the long time limits, not even after
being averaged over long time intervals.

\item
There is one case, quenches from the condensed equilibrium state in which energy is
extracted or injected in small amounts, in which the dynamics of the
{\it global} observables, those averaged over all modes, are {\it close} to the ones at a single temperature $T_f$.
However, a closer look into the mode
dynamics and its observables exhibits the fact that the modes are not in Gibbs-Boltzmann equilibrium.
Moreover, for deep quenches in the same sector III one clearly sees that standard thermal equilibration is
not reached.

\item
Another special case is provided by quenches with $T'>J_0$ and ${T'}^2= JJ_0$. On this
special curve the global observables satisfy thermal equilibrium properties at $T_f=J$.

\end{itemize}

Much has been learnt from the evolution of the system with finite number of degrees of freedom using a
formalism that allows one to show that in the infinite size and long time limit (taken after
the former one) the modes decouple and become independent harmonic oscillators.
Once this regime is reached, the mode energies can be associated to mode temperatures, {\it via} standard equipartition,
and a temperature spectrum obtained. A naive approximation to determine their  dependence on
the control parameters was explained and leads to a variation
from sector to sector of the phase diagram. We compared these forms with the numerical measurements and the agreement is
quite good in all cases.

Having obtained the total energies of the modes, and from them the mode temperatures,  we put to the test the recently proposed relation between them and
the frequency dependent effective temperature stemming from the fluctuation-dissipation relation of the
spin observable~\cite{FoGaKoCu16,deNardis17}. We found excellent agreement between the two in all
phases of the phase diagram.

The $p=2$ spherical system turns out to be equivalent to the Neumann classical mechanics integrable model.
We stress the fact that in the field of classical integrable systems, the model of Neumann was
usually defined and studied having only a few degrees of freedom. Here, as we are interested in searching for
a statistical description of the post-quench dynamics, we dealt with the limit of large,
and even diverging, number of degrees of freedom.

The $N-1$ integrals motion of the Neumann model have been identified by K. Uhlenbeck~\cite{Uhlenbeck,BaTa92}.
After a trivial extension that allows us to deal with the large $N$ limit, we studied their scaling properties with
system size. In cases in which the initial state is condensed, the integrals of motion associated to the
edge of the spectrum also scale with $N$. The distance between their values and the ones they would have
taken in equilibrium at a single temperature $T_f$ gave us a rough measure of distance from Gibbs-Boltzmann
equilibrium. Importantly enough, in the particular case in which  the global correlation and linear response
behave as in thermal equilibrium at $T_f=J$, that is to say, parameters on the curve ${T'}^2=J J_0$ in sector or phase II, the
integrals of motion are not identical to the ones expected in equilibrium. This proves that not even in this case
the system is able to fully equilibrate.

The $N-1$ integrals of motion
could be used to build a putative Generalised Gibbs Ensemble,
or they may be a guideline to choose the ones with good scaling  properties.
We will investigate this problem in a sequel to this publication.


\vspace{1cm}

\noindent
{\bf Acknowledgements}

\vspace{0.25cm}

We are deeply indebted to O. Babelon for letting us know that we were dealing with Neumann's model, and for illuminating discussions.
LFC thanks M. Serbyn for useful discussions.
We acknowledge financial support from ECOS-Sud A14E01 and PICS 506691 (CNRS-CONICET Argentina).
 LFC is a member of Institut Universitaire de France.

\clearpage

\appendix
\renewcommand{\theequation}{\thesection.\arabic{equation}}

\section{Asymptotic analysis in the $N\to\infty$ limit}
\label{app:asymptotic}
\setcounter{equation}{0}

In this Appendix we give additional details on the study of the
full set of equations (\ref{eq:dyn-eqs-R})-(\ref{eq:dyn-eqs-z}) that couple the correlation $C$ and linear response $R$
functions derived in the $N\to\infty$ limit.
Based on a number of hypotheses that we carefully list below, we analyse the behaviour of the model in the
long times limit.

\subsection{Stationary dynamics}

Consider the system in equilibrium at $T'$ with parameters $J_0,\, m_0$ and let it
evolve in isolation with parameters $J, \, m$. We will assume that the dynamics approach a steady state in which
one-time quantities approach a constant. This assumption does not apply to certain quenches of the isolated system. Still,
we investigate the consequences of  the stationary assumption.

\subsubsection{The asymptotic values}

Let us assume that the limiting value of the Lagrange multiplier is a constant
\begin{equation}
\lim_{t_1\to\infty} z(t_1) = z_f
\; .
\end{equation}

Recalling the definitions given in the main part of the paper, the limits of the correlation functions are
\begin{equation}
q = \lim_{t_1-t_2\to\infty} \lim_{t_2\to \infty} C(t_1,t_2)
\; ,
\qquad
q_0 =  \lim_{t_1\to \infty} C(t_1,0)
\; ,
\qquad
q_2 = \lim_{t_1\to\infty}C_2(t_1,0)
\; .
\end{equation}
The linear response was analysed in the main body of the paper and, independently of the
quench parameters it is given by
\begin{equation}
\hat R(\omega)
=
\frac{1}{2 J^2} \left[ (-m\omega^2 + z_f) \pm \sqrt{(-m\omega^2 + z_f)^2 - 4 J^2} \right].
\label{eq:response_fourier_transform}
\end{equation}
in the frequency domain, where the Fourier transform has been computed with respect to the
time difference $t_1-t_2$.
This result only assumes a long-time limit in which $z_f$ is time-independent.

\subsubsection{The parameters $q_0$ and $q_2$}

We could estimate the asymptotic value of $z(t_1)$ taking the long $t_1$ limit of Eq.~(\ref{eq:dyn-eqs-z});
however, without the use of FDT we cannot compute the integral involved.
\comments{
\begin{equation}
$
z_f = T_f + \frac{J^2}{T_f} \; (1-q^2) + \frac{JJ_0}{T'}
(q_0^2-q_2^2)
\; .
$
\label{eq:zf}
\end{equation}
Equation~(\ref{eq:zf}) is the equivalent of Eq.~(77) in~\cite{CuLoNe17}
We can now see what this equation imply for the values of $q_0$ and $q_2$ if we insert the two possible solutions
for the parameter $q$ derived from the analysis of the linear response function. In both cases we find
}

We are tempted to propose
\begin{equation}
q_0 = q_2
\; .
\end{equation}
The interpretation of this result in terms of the evolution of real replicas is that the asymptotic
value of the self-correlation between times $t_1$ and $0$, $q_0$, is the same as the asymptotic value of the correlation
between two replicas evaluated at the same times $t_1$ and $0$ with $t_1$ diverging.

\subsubsection{The two-time correlation function}
\label{app:asymptotic-corr}

Allowing for the two-time correlation function not to decay to zero but to a finite value $q$, we separate this contribution explicitly and we write
$C(t_1,t_2) = q + C_{\rm st}(t_1-t_2)$ with $\lim_{t_1-t_2\to\infty} C_{\rm st}(t_1-t_2)=0$. We also use $\lim_{t_1\to\infty} C_2(t_1,0) =q_2$
leaving the possibility of there being a different asymptotic value for this quantity open.

The dynamic equation now becomes an equation on $\tilde C$ that  reads
\begin{eqnarray}
\hspace{-0.5cm}
\left( m \partial^2_{t_1-t_2 } + z_f \right) (q + C_{\rm st}(t_1-t_2))
\!\! & \!\! = \!\! & \!\!
\frac{JJ_0}{T'} (q_0^2 - q_2^2)
+ 2 J^2 q \; \lim_{t_1\to\infty} \int_0^{t_1} dt' \, R(t_1-t')
\nonumber\\
& &
+
J^2 \int_0^{t_1} \!\! ds \; R(t_1-t_2+s)
C_{\rm st}(-s)
+
J^2 \int_0^{t_2} \!\! ds \; C_{\rm st}(t_1-t_2+s) R(s)
\; .
\label{Cst}
\end{eqnarray}
Using the causal properties of the linear response, one can extend the lower limit of the last integral to $-\infty$ and safely
 take the upper limit to infinity since we are interested in the long time limit $t_2\to\infty$ while keeping $t_1-t_2$ fixed.
 The Fourier transform of this term with respect to $t_1-t_2$ yields $R(-\omega) C_{\rm st}(\omega)$ that using
 $R(-\omega) = R^*(\omega)$ simply becomes $R^*(\omega) C_{\rm st}(\omega)$. Proceeding in a similar way with the
 first integral one finds that it equals $\hat R(\omega) C_{\rm st}(\omega)$. Thus,
 \begin{eqnarray}
\left(- m \omega^2 +z_f - J^2 \hat R(\omega) \right) C_{\rm st}(\omega)
&=&
\left(-z_f q +
\frac{JJ_0}{T'} (q_0^2 - q_2^2)
+  2 J^2 \, q \lim_{t_1\to\infty} \int_0^{t_1} \!\! dt' \, R(t_1-t')
\right)
\delta(\omega)
\nonumber\\
&&
+ J^2 \,  C_{\rm st}(\omega) R^*(\omega)
\; .
\end{eqnarray}
The factor between parenthesis in the first term on the right-hand-side can be written as
\begin{equation}
\displaystyle{
-z_f q +
\frac{JJ_0}{T'} (q_0^2 - q_2^2)
+
2 J^2 \,  q  \lim_{t_1\to\infty} \int_0^{t_1} dt' \, R(t_1-t') =0
\; .
}
\label{cond2}
\end{equation}
If we now assume $q_0=q_2$, this equation imposes
\begin{eqnarray}
&& \mbox{if} \;\; q\neq 0 \qquad \mbox{then} \;\;  z_f = 2J \;\; \mbox{and} \;\; \lim_{t_1\to\infty} \int_0^{t_1} dt' \, R(t_1-t') =  1/J
\; ,
\\
&& \mbox{otherwise} \;\; q=0
\; .
\end{eqnarray}
The remaining equation, using Eq.~(\ref{Rstat}) to replace the parenthesis on the left-hand-side by $1/\hat R(\omega)$ is recast as
\begin{equation}
J^2  C_{\rm st}(\omega) |\hat R(\omega)|^2 = C_{\rm st}(\omega) \;.
\end{equation}
At each frequency this equation has two possible solutions
\begin{eqnarray}
C_{\rm st}(\omega) \neq 0 \;\;\mbox{and}\;\; |\hat R(\omega)|^2=1/J^2
\qquad\mbox{or}\qquad
C_{\rm st}(\omega) = 0
\; .
\end{eqnarray}

In the frequency domain in which the linear response (\ref{eq:response_fourier_transform}) is complex, that is
Im$\hat R(\omega)\neq 0$, one
can easily check that it satisfies $|\hat R(\omega)|^2=1/J^2$. This holds for any set of parameters $x,y$.

In the cases in which the linear response is real, it is not always true that its square equals $1/J^2$. For instance,
at zero frequency in cases in which $x>y$, $\hat R(\omega=0)=1/J$, verifying that its square is $1/J^2$. However,
for $x<y$, $\hat R(\omega=0)=1/T'$ and its square is different from $1/J^2$. In these cases, the Fourier transform
of the decaying part of the correlation, $\hat C(\omega)$ vanishes.  We have tested this statement numerically
and several examples can be seen in the main part of the paper.

\subsubsection{The correlation with the initial condition}
\label{app:asymptotic-corr0}

The asymptotic limit of the equation for $C(t_1,0)$ implies
\begin{equation}
z_f q_0 = J^2 q_0 \lim_{t_1\to\infty} \int_0^{t_1} dt' \, R(t_1-t') + \frac{JJ_0}{T'} (q_0 -q_{\mathrm{in}} q_2)
\label{eq:q0}
\end{equation}
where
\begin{eqnarray}
q_{\mathrm{in}}
&=&
\left\{
\begin{array}{ll}
1 - \frac{T'}{J_0}
& \qquad T<T^0_s
\\
0
& \qquad T>T^0_s
\end{array}
\right.
\end{eqnarray}
and $T^0_s$ the critical temperature of the initial potential energy
\comments{
\begin{equation}
z_f q_0 = J^2q_0 \lim_{t_1\to\infty} \int_0^{t_1} dt' \, R(t_1,t') + \frac{JJ_0}{T'} (q_0 -q_{\mathrm{in}} q_2)
\label{eq:q0}
\end{equation}
}
Equation~(\ref{eq:q0}) is the equivalent of Eq.~(82) in~\cite{CuLoNe17}.

This equation admits the solution $q_0=q_2=0$ or
\begin{equation}
z_f = J^2 \lim_{t_1\to\infty} \int_0^{t_1} dt' \, R(t_1-t') + \frac{JJ_0}{T'} (1-q_{\rm in})
\label{eq:zfnueva0}
\end{equation}
if we assume $q_0=q_2$.
Recalling that $q_{\rm in} =1 -T'/J_0$, the remaining equation becomes
\begin{equation}
z_f = J^2 \lim_{t_1\to\infty} \int_0^{t_1} dt' \, R(t_1-t') + J
\label{eq:zfnueva}
\end{equation}
and is consistent for
$z_f=2J$ and $\lim_{t_1\to\infty} \int_0^{t_1} dt' \, R(t_1-t') =1/J$.

\subsection{The off-diagonal correlations in the no quench problem}
\label{app:asymptotic-offdiag}

In the no quench case we can check whether $C_2(t_1,0)=q_2=q_{\rm in}$ for all $t_1$ is a solution of the
corresponding equation. Plugging this form in the evolution equation for $C_2$
we find that, at $t_1=0$, either $q_{\rm in} =0$ (the paramagnetic case) or
$z_f(0)=2J^2/T' (1-q_{\rm in})$ and after replacing $q_{\rm in}$ by its expression as a function of $T'/J$
the correct equilibrium $z_f=2J$ is recovered.
However, for  $t_1=\delta$ and later times, $C_2(t_1,0)$ cannot remain constant for initial conditions with
$q_{\rm in}\neq 0$, due to the non-trivial contribution of the term $J^2 \int_0^{t_1} dt' \; R(t_1,t') q_{\rm in} =  J^2/m \, q_{\rm in} \; \delta$.

If we assume that $C_2(t_1,0)$ approaches a constant $q_2$ possibly different from $q_{\rm in}$, the equation for $C_2(t_1,0)$ in the long $t_1$ limit reads
\begin{equation}
z_f q_2 =
J^2q_2 \lim_{t_1\to\infty} \int_0^{t_1} dt' \, R(t_1,t')
+\frac{J J_0}{T'} [q_0 q_{\mathrm{in}} + (1-2q_{\mathrm{in}}) q_2]
\;
\end{equation}
We have already shown $q_0=q_2$. Therefore, this equation has solution $q_2=0$ or it becomes
Eq.~(\ref{eq:zfnueva}).

In the no quench case we can use FDT and obtain
\begin{equation}
z_f q_2 =
\frac{J^2}{T'} q_2 (1-q)
+\frac{J J_0}{T'} [q_0 q_{\mathrm{in}} + (1-2q_{\mathrm{in}}) q_2]
\end{equation}
This is a new equation with no parallel in~\cite{CuLoNe17} since in this reference we only studied the
dynamics starting from paramagnetic-like initial states.

\subsection{Energy conservation}
\label{app:energy-conservation}

In this Section we impose that the kinetic and potential energies of the
asymptotic final state correspond to the ones of an equilibrium paramagnetic
of condensed state at temperature $T_f$. This means that
$e_{\rm pot} = -J^2/T_f$ in the former case and $e_{\rm pot} = -J^2/T_f (1-q^2)$ with
$q=1-T_f/J$ in the latter.
We then derive, from the conservation of the total energy, the temperature $T_f$ as a function of the control parameters
$T', \, J_0, \, J$ or, in other words, $x$ and $y$.
Here, we will see the internal limits of validity of these assumptions. Elsewhere we
will determine where it is  realised numerically.

\subsubsection{Final temperature}
\label{app:final-temp-anal}

For condensed (first line in the left) and paramagnetic (second line in the left) initial conditions
going to condensed (first line in the right) and paramagnetic-like (second line in the right) states
the energy conservation law reads
\begin{equation}
\frac{m}{m_0} \frac{T'}{2}  - \frac{J J_0}{2T'}
\left[
\begin{array}{c}
 1  - \left( 1-\frac{T'}{J_0}\right)^2
 \\
 1
 \end{array}
 \right]
=
\frac{T_f}{2}
 - \frac{J^2}{2T_f}
 \left[
\begin{array}{c}
 1  - \left( 1-\frac{T_f}{J}\right)^2
 \\
 1
 \end{array}
 \right]
- \frac{J J_0}{2T'} (q_0^2 - q_2^2)
\; .
\end{equation}
We have already determined  $q_0=q_2$ so the last term vanishes identically. From this condition we find
the following expressions for $T_f$ depending on the kind of quench performed:
\begin{itemize}
\item
From condensed to condensed
\begin{equation}
T_f = \frac{J}{2} \left( \frac{m J_0}{m_0 J} + 1 \right) \frac{T'}{J_0} \qquad\qquad \mbox{(III)}
\; .
\label{eq:Tf_from_cond_to_cond}
\end{equation}
\item
From condensed to paramagnetic
\begin{equation}
T_f = \frac{J}{2} \frac{J_0}{T'}
\left\{
\frac{mJ_0}{m_0 J} \left( \frac{T'}{J_0} \right)^2
- (1-q_{\rm in}^2)
+ \sqrt{
\left[ \frac{mJ_0}{m_0 J} \left( \frac{T'}{J_0} \right)^2 - (1-q_{\rm in}^2)  \right]^2 + \left( \frac{2T'}{J_0}\right)^2
}
\right\}
\;\; \qquad \mbox{(IV)}
\; .
\label{eq:Tf_from_cond_to_para}
\end{equation}
\item
From paramagnetic to paramagnetic
\begin{equation}
T_f = \frac{J}{2} \frac{J_0}{T'}
\left\{
\frac{mJ_0}{m_0 J} \left( \frac{T'}{J_0} \right)^2
- 1
+ \sqrt{
\left[ \frac{mJ_0}{m_0 J} \left( \frac{T'}{J_0} \right)^2 - 1 \right]^2 + \left( \frac{2T'}{J_0}\right)^2
}
\right\}
\qquad \mbox{(I) and (IIa)}
\; .
\label{eq:Tf_from_para_to_para}
\end{equation}
\item
From paramagnetic to ageing
\begin{equation}
T_f = J \left[ 1 + \frac{m J_0}{m_0 J} \frac{T'}{2J_0}  - \frac{J_0}{2T'} \right]
\qquad \mbox{(IIb)}
\; .
\\
\label{eq:Tf_from_para_to_ageing}
\end{equation}
\end{itemize}
We have chosen to single out in these expression the parameters $T'/J_0$ and $(m J_0)/(m_0 J)$ that
we have used in~\cite{CuLoNe17} to characterise the dynamic phase diagram of the $p\geq 3$ model. In the two cases (condensed to
condensed and paramagnetic to paramagnetic) in which the no quench limit can be taken we
naturally recover $T_f=T'$. In the ageing case $T_f$ should be the temperature of the
stationary regime. As we will show from the numerical solution to the full dynamic equations, and the mode by mode analysis,
not all these cases are realised.

Below we prove analytically that a state with a single $T_f$ characterising the fluctuation-dissipation relation
can be realised for particular choices of the parameters only.
 Moreover, these conditions are not restrictive enough, as the numerical solution of the full equations show
 that FDT is not realised for quenches allowed by them.

\subsubsection{Limits of validity}
\label{app:critical-lines}

Let us consider the cases $y\geq 1$ and $y\leq 1$ separately.

\vspace{0.25cm}

\noindent
{\it The case $y \geq 1$: Quench from a paramagnetic equilibrium state}

\vspace{0.25cm}

 The double condition $0\leq T_f \leq J$ translates into
\begin{equation}
-1 \leq \frac{1}{2} \frac{mJ_0}{m_0J}  \frac{T'}{J_0} - \frac{1}{2} \frac{J_0}{T'} \leq 0
\end{equation}
that in terms of $x$ and $y$ reads
\begin{equation}
-1 \leq  \frac{y}{2x}- \frac{1}{2y} \leq 0
\end{equation}
and yields
\begin{equation}
0 \leq  y^2-x + 2 xy
\qquad\mbox{and}\qquad
y^2\leq x
\label{eq:upper-bound0}
\; .
\end{equation}
The first condition in Eq.~(\ref{eq:upper-bound0}) is satisfied for
\begin{equation}
0 \leq (y-y_+) (y-y_-)
\end{equation}
with $y_+$ and $y_-$ the two roots of the quadratic equation, that is
\begin{equation}
y_{\pm} = -x \pm \sqrt{x^2 + x}
\; ,
\end{equation}
a positive and a negative value.
One should then have
\begin{equation}
y \geq y_+	
\qquad\qquad
\mbox{or}
\qquad\qquad
y \leq y_-
\; .
\end{equation}
As $y\geq 1$, $y\leq y_-$ is not possible. Instead, $y> y_+$ is trivially satisfied. The second condition in Eq.~(\ref{eq:upper-bound0}) is the only restrictive
one and reads
\begin{equation}
y \leq \sqrt{x}\equiv f(x)
\; .
\end{equation}
This upper bound is the dashed line representing $f(x)$ in the phase diagram for $y>1$.
Note that this curve has no finite limit.

\vspace{0.25cm}

\noindent
{\it The case $y \leq 1$: Quench from a condensed equilibrium state}

\vspace{0.25cm}

Using $q_{\rm in} =1 -T'/J_0$, one has
\begin{equation}
\frac{T_f}{J} = \frac{1}{2} \frac{J_0}{J} \frac{T'}{J_0} - \frac{1}{2} \frac{T'}{J_0}
\end{equation}
and the condition to use reads
\begin{equation}
0 \leq \frac{y}{2} \left( \frac{1}{x}  + 1 \right)   \leq 1
\end{equation}
The lower bound is trivially satisfied while the upper bound implies
\begin{equation}
y \leq \frac{ 2x}{1+x} \equiv g(x) \qquad\qquad \mbox{for} \;\; y\leq 1
\; .
\end{equation}
The dashed line for $y<1$ in the phase diagram represents $g(x)$.

In the condensed case the energy conservation implies
\begin{equation}
T_f = J \; \frac{y(1+x)}{2x} \; , \qquad q=1-\frac{y}{2} - \frac{y}{2x} \qquad\mbox{and}\qquad \overline e_{\rm pot} = -J \left(1-\frac{x}{4y}- \frac{y}{4} \right)
\; .
\label{eq:Tf-condensed}
\end{equation}
We note that $q$ vanishes on the curve $y=2x/(1+x)$, that it equals one for $y=0$ and that, for fixed $x>1$ it increases
from $(x-1)/(2x)$ at $y=1$ to $1$ at $y=0$. Moreover, $q$ is different from zero on the lines $y=x$ and  $y=1$.

\subsubsection{Consistency with FDT at $T_f$}
\label{app:special-cases}

We now reason as follows. We know, from the numerical analysis of the Schwinger-Dyson set of equations, that
$\chi_{\rm st} = \lim_{t\to\infty} \int_0^t dt' R(t-t') = 1/T'$ for $x<y$ and $\chi_{\rm st} = \int_0^t R(t) = 1/J$ for $x>y$. If we
evaluate the integral of the linear response using FDT at a single temperature we obtain
\begin{eqnarray}
\chi_{\rm st} = \lim_{t\to\infty} \int_0^t dt' R(t-t') = \lim_{t\to\infty} \int_0^t dt' \frac{1}{T_f} \frac{\partial}{\partial t'} C_{\rm st}(t-t') =
\left\{
\begin{array}{ll}
\displaystyle{\frac{1}{T_f}} & \qquad y>1
\vspace{0.1cm}
\\
\displaystyle{\frac{1}{T_f} (1-q) }= \frac{1}{J} & \qquad y<1
\end{array}
\right.
\end{eqnarray}
Using these results we can check whether
there is a set of $x,y$ for which $T_f$ derived in the previous Section from the conservation of the energy is consistent with
these conditions. We list below the conclusions drawn in the
four parameter Sectors of the phase diagram.

\vspace{0.25cm}

\noindent{\it $y>1$ and $x<y$} (Sector I)

\vspace{0.25cm}

$T_f$ can equal $T'$ only for $x=1$ that implies no-quench.

\vspace{0.25cm}

\noindent{\it
$y>1$ and $x>y$} (Sector II)

\vspace{0.25cm}

Only on  the curve $y=\sqrt{x}$ and $y>1$ FDT holds at $T_f$.
The validity of FDT at $T_f$ for parameters lying on the curve $y = \sqrt{x}$ is verified
 numerically in Fig.~\ref{fig:q-fdt}. For all other values of $x,y$ in this Sector FDT cannot be satisfied.

\vspace{0.25cm}

\noindent{\it $y<1$ and $x>y$} (Sector III)

\vspace{0.25cm}

In this Sector we do not find any contradiction. This reasoning suggests that the
dynamics may satisfy FDT in this Sector. The no-quench case $x=1$ for $y<1$ is obviously
included.

\vspace{0.25cm}

\noindent{\it $y<1$ and $x<y$} (Sector IV)

\vspace{0.25cm}

There is no solution and FDT at a single temperature is excluded.


\section{The harmonic oscillator}
\label{app:harmonic-oscillator}
\setcounter{equation}{0}

Take, as an example, a system constituted of a single point-like particle with mass $m$,
under a harmonic potential $V(x)$. The dynamics of this problem is given by the
familiar equation
\begin{equation}
m \ddot x + m \omega^2 x = 0
\label{eq:harmonic-oscillator}
\; ,
\end{equation}
with initial conditions $x(0) = x_0$ and $p(0)= p_0$.

\subsection{Equilibrium initial conditions}

Let us take initial conditions in canonical equilibrium within a harmonic potential $V_0(x)$.
The probability distribution of the initial conditions is
\begin{equation}
P_0(x_0,p_0) = Z_0^{-1} \; e^{-\beta' H_0} =
Z_0^{-1} \; e^{-\beta' [\frac{p_0^2}{2m}  + V_0(x_0)]}
\end{equation}
with $\beta'=1/(k_BT')$ the inverse temperature, using the same notation adopted in the
main part of the paper for the initial temperature $T'$.
The averaged kinetic energy of the ensemble of initial states sampled with this
probability distribution function is
\begin{equation}
\frac{1}{2m} \langle p_0^2 \rangle = \frac{k_BT'}{2}
\; ,
\end{equation}
the equipartition of the kinetic energy. We recall that angular brackets indicate average over the
initial conditions sampled as above. The averaged total energy is
\begin{equation}
\langle H_0\rangle = - \frac{\partial}{\partial \beta'} \ln Z_0(\beta')
\; .
\end{equation}

\subsection{Potential energy quench}

Make now a quench in the potential that corresponds to $V_0 \mapsto V$ and do it so
quickly that the phase space variables do not change and remain $p_0, x_0$. By performing this
abrupt change one injects or extracts a finite amount of energy,
\begin{equation}
\Delta E = H(x_0,p_0) - H_0(x_0,p_0) = V(x_0)-V_0(x_0)
\; .
\end{equation}
The energy surface on which the dynamics will take place is the one
of the post-quench energy $E(0^+)=p_0^2/(2m) +V(x_0)$.

We now focalise on $V$ being a harmonic potential.
The Newton evolution of each initial configuration is
\begin{eqnarray}
\begin{array}{lll}
&&
\displaystyle{
x(t) = x_0 \cos \omega t + \frac{p_0}{m\omega} \sin\omega t
}
\; ,
\label{eq:xt}
\\
&&
\displaystyle{
p(t) = -m \omega x_0\sin\omega t + p_0 \cos\omega t
}
\; .
\label{eq:pt}
\end{array}
\end{eqnarray}

Let us call $y=x(t)$ and $z=p(t)$ the position and momentum at a time $t$. The probability density of $y,z$ at time $t$ is
\begin{eqnarray*}
P(y,z, t)
\! \! & \!\! =  \!\! & \!\!
\int dx_0 \int dp_0 \; P_0(x_0,p_0) \; \delta(y-x_0 \cos\omega t - \frac{p_0}{m\omega} \sin\omega t) \nonumber\\
&& \qquad\qquad\qquad \times \; \delta(z+m \omega x_0\sin\omega t - p_0\cos\omega t)
\; .
\end{eqnarray*}
We use the second $\delta$ function to integrate over $p_0$,
\begin{eqnarray*}
P(y,z, t)
\! \! & \!\! =  \!\! & \!\!
\int dx_0 \; P_0\left(x_0,\frac{z}{\cos\omega t} + m \omega x_0 \tan\omega t \right) \frac{1}{\cos\omega t} \;
\nonumber\\
&& \qquad\qquad \times
\; \delta(y-x_0 \cos\omega t - \frac{z+m x_0\omega\sin\omega t}{m\omega\cos\omega t} \sin\omega t)
\; .
\end{eqnarray*}
The remaining $\delta$ function implies
\begin{equation}
y- \frac{z}{m\omega} \tan\omega t  -x_0 \left( \cos\omega t + \tan\omega t  \sin\omega t \right) =
y- \frac{z}{m\omega} \tan\omega t  -x_0 \frac{1}{\cos\omega t}
 = 0
\end{equation}
and we use it to integrate over $x_0$. Indeed, replacing $x_0=y\cos\omega t- \frac{z}{m\omega} \sin\omega t$ and
taking care of the Jacobian one finds
\begin{eqnarray}
P(y,z, t)
\! \! & \!\! =  \!\! & \!\!
P_0\left(y\cos\omega t- \frac{z}{m\omega} \sin\omega t, z \cos\omega t + m \omega y \sin\omega t
\right)
\; .
\label{eq:relation-btw-probas}
\end{eqnarray}

In the case in which no quench is performed the initial potential
has to be, then, harmonic $V_0(u)=m\omega^2 u^2/2$ and the equation above implies
\begin{eqnarray*}
\ln Z_0 + \ln P(y,z, t)
\! \! & \!\! =  \!\! & \!\!
-\frac{\beta_i}{2m} \left(z \cos\omega t + m \omega y \sin\omega t \right)^2
-\frac{\beta_i}{2} m\omega^2 \left(y\cos\omega t- \frac{z}{m\omega} \sin\omega t\right)^2
\\
\! \! & \!\! =  \!\! & \!\!
-\frac{\beta_i}{2m} z^2 -\frac{\beta_i}{2} m \omega^2 y^2
\; .
\end{eqnarray*}
The equilibrium distribution is conserved by the dynamics, as it should.

Imagine now that the quench corresponds to a change in the spring parameter of the quadratic potential
$\omega_0 \mapsto \omega$. The equation for $P(y,z)$ implies
\begin{eqnarray*}
\ln Z_0 + \ln P(y,z, t)
\! \! & \!\! =  \!\! & \!\!
-\frac{\beta_i}{2m} \left(z \cos\omega t + m \omega y \sin\omega t \right)^2
-\frac{\beta_i}{2} m\omega_0^2 \left(y\cos\omega t- \frac{z}{m\omega} \sin\omega t\right)^2
\; .
\end{eqnarray*}
Expanding the squares
\comments{
\begin{eqnarray*}
\ln Z_0 + \ln P(y,z, t)
\! \! & \!\! =  \!\! & \!\!
-\frac{\beta_i}{2m} \left(z^2 \cos^2\omega t + m^2 \omega^2 y^2 \sin^2\omega t + 2 z m \omega y \cos\omega t  \sin\omega t\right)
\nonumber\\
\! \! & \!\!   \!\! & \!\!
-\frac{\beta_i}{2} m\omega_0^2 \left(y^2\cos^2\omega t + \frac{z^2}{m^2\omega^2} \sin^2\omega t -  2 y \frac{z}{m\omega} \cos\omega t  \sin\omega t \right)
\; .
\end{eqnarray*}
}
and collecting terms
\begin{eqnarray*}
\ln Z_0 + \ln P(y,z, t)
\! \! & \!\! =  \!\! & \!\!
-\frac{\beta_i}{2} \left(
( \cos^2\omega t + \frac{\omega_0^2}{\omega^2} \sin^2\omega t ) \frac{z^2}{m}
+ m \omega^2
(\sin^2\omega t + \frac{\omega_0^2}{\omega^2}  \cos^2\omega t) y^2
\right.
\nonumber\\
\! \! & \!\!   \!\! & \!\!
\left.
\qquad + 2 z y m \omega  (1 + \frac{\omega_0^2}{\omega^2} ) \cos\omega t  \sin\omega t
\right)
\; .
\end{eqnarray*}
Although the measure $P$ is still Gaussian it does not have the same covariance as the initial $P_0$.

The averages and the variances of the position and momentum can be computed directly from the
solutions to the equations of motion. The averages vanish and for the variances one finds
\begin{eqnarray}
\begin{array}{lll}
&&
\displaystyle{
\sigma_x^2(t) = \langle x^2(t) \rangle = \langle x_0^2 \rangle \cos^2 \omega t + \langle p_0^2\rangle \ \frac{1}{m^2 \omega^2} \sin^2\omega t
}
\; ,
\\
&&
\displaystyle{
\sigma_p^2(t) = \langle p^2(t) \rangle = \langle x_0^2 \rangle \ m^2 \omega^2 \sin^2 \omega t + \langle p_0^2\rangle  \cos^2\omega t
}
\; .
\end{array}
\end{eqnarray}
Replacing now the averages of the initial values $\langle p_0^2\rangle/m = m\omega_0^2 \langle x_0^2 \rangle = T'$
\begin{eqnarray}
\begin{array}{lll}
m \omega^2 \sigma_x^2(t) \! & \! = \! & \!
\displaystyle{
m \omega^2 \langle x^2(t) \rangle = T' \left(  \frac{\omega^2}{\omega_0^2} \cos^2 \omega t +  \sin^2\omega t \right)
}
\; ,
\\
\frac{1}{m} \sigma_p^2(t) \! & \! = \! & \;\;\
\displaystyle{
\frac{1}{m}  \langle p^2(t) \rangle = T' \left( \cos^2\omega t +  \frac{\omega^2}{\omega_0^2} \sin^2 \omega t   \right)
}
\; .
\end{array}
\end{eqnarray}
One readily verifies that, as expected, the averaged total energy is conserved
\begin{eqnarray}
\langle E(t) \rangle = m \omega^2 \sigma_x^2(t) + \frac{1}{m} \sigma_p^2(t) = T' \left( 1 +  \frac{\omega^2}{\omega_0^2} \right)
\qquad \mbox{for} \;\;\; t>0
\end{eqnarray}
since each trajectory does conserve its initial energy. The averaged total energy is, however, different from the one right before the
quench, $T' = \langle e_{\rm tot}(t=0^-)\rangle \neq \langle e_{\rm tot}(t=0^+)\rangle = T' \left( 1 +  \omega^2/\omega_0^2 \right)$.

Time-independent values of the variances are found from the average over a long time window:
\begin{eqnarray}
m\omega^2 \overline{\sigma_x^2(t)} &=& \frac{T'}{2} \left( \frac{\omega^2}{\omega_0^2} + 1 \right)
\; , \qquad
\frac{1}{m}  \overline{\sigma_p^2(t)} = \frac{T'}{2} \left( \frac{\omega^2}{\omega_0^2} + 1 \right)
\; , \qquad
\overline{x(t) p(t)} = 0
\end{eqnarray}
and from  these one can identify the final temperature
\begin{equation}
T_f=\frac{T'}{2} \left( \frac{\omega^2}{\omega_0^2} + 1 \right)
\; .
\end{equation}

\subsection{Fluctuation dissipation relations}

The linear response of the harmonic oscillator defined in (\ref{eq:harmonic-oscillator}) to an infinitesimal perturbation
modifying its energy as $H \mapsto H- h x$, and therefore adding a linear term in $h$ to Eq.~(\ref{eq:harmonic-oscillator}) instantaneously at time $t_2$,
is
\begin{equation}
R(t_1,t_2) = \left. \frac{\delta x(t_1)}{\delta h(t_2)} \right|_{h=0} = \frac{\sin\omega (t_1-t_2)}{m \omega} \, \theta(t_1-t_2)
\end{equation}
while the product of the unperturbed position at the times $t_1$ and
$t_2$, after a long-time average over the reference time $t_2$, is
\begin{equation}
\overline C(t_1,t_2) = \overline{x(t_1) x(t_2)} = \frac{1}{2} \left( x_0^2 + \frac{p_0^2}{m^2\omega^2}\right) \cos\omega(t_1-t_2)
\; .
\end{equation}
The Fourier transform with respect to the time delay, $t_1-t_2 \to \nu$,  of these two expressions yield
\begin{eqnarray}
{\mbox Im} \tilde R(\nu) &=& \frac{\pi}{2m\omega} \ \delta(\nu-\omega)
\; ,
\\
\nu \tilde C(\nu) &=& \frac{\pi \nu}{m \omega^2} \left( m \omega^2 x_0^2 + \frac{p_0^2}{m}\right) \ \delta(\nu-\omega)
\; .
\end{eqnarray}
 Focusing on the frequency $\nu$ where both are non-zero due to the Dirac deltas, the ratio between the two
 yields the ratio between the prefactors
 \begin{equation}
\beta_{\rm FDR}(\nu) = \frac{2{\mbox Im} \tilde R(\nu) }{\nu \tilde C(\nu)} = \frac{1}{m \omega^2 x_0^2 + \frac{p_0^2}{m}} = \frac{1}{e_{\rm tot}(0^+)}
\qquad\qquad \mbox{for} \; \nu=\omega \; ,
 \end{equation}
 where $e_{\rm tot}$ is the total energy of the harmonic oscillator.

\subsection{The Generalized Gibbs Ensemble}

The harmonic oscillator dynamics conserves its total energy $e_{\rm tot}$ and the GGE density function is
\begin{equation}
p_{\rm GGE}(x,p) = Z^{-1} \; e^{-\beta_{\rm GGE} H(x,p)}
\end{equation}
with $Z= \int dx dp \; e^{-\beta_{\rm GGE} H(x,p)}$ the normalisation constant or partition function. The inverse temperature
of the GGE ensemble is determined by the condition
\begin{equation}
\langle H \rangle_{\rm GGE} = \int dx dp \; p_{\rm GGE}(x,p) H(x,p) = e_{\rm tot}(0^+)
\end{equation}
and with a simple calculation one finds
\begin{equation}
\beta_{\rm GGE} = \frac{1}{e_{\rm tot}(0^+)}
\; .
\end{equation}
Therefore,
\begin{equation}
\beta_{\rm FDR}(\nu=\omega) = \beta_{\rm GGE} \; .
\end{equation}
The generalisation of this result to an ensemble of $N$ harmonic oscillators is straightforward:
\begin{equation}
\beta_{\rm FDR}(\nu=\omega_\mu) = {\beta_{\rm GGE}}_\mu \qquad\qquad \mbox{for all} \; \mu \; .
\end{equation}


\section{Discrete time version}
\label{app:discrete-time}
\setcounter{equation}{0}

Let us now consider the discrete time version of the $C_2(t_1,0)$ and $z(t_1)$ equations and look for the
discretisation needed to recover the results $C_2(t_1,0)=q_{\rm in}$ and $z(t_1)=2J$ at low temperatures.

First of all, we use the Taylor expansions $R(\delta,0)=\delta/m$ and
$C(\delta,0) = 1 - \delta^2 T^{\prime}/m$, for $\delta \to 0$.

The discretized version of Eq.~(\ref{eq:dyn-eqs-C1}) evaluated at $t_1=n \delta$ is given by
\begin{eqnarray}
&&  \frac{m}{\delta^2} \left[ C_2( (n+1) \delta,0) - 2 C_2( n \delta,0) + C_2( (n-1) \delta,0)\right] =
 \nonumber\\
&&
\qquad
  -z(n\delta) C_2(n\delta,0) +
  \frac{J J_0}{T^{\prime}} \left[ q_{\mathrm{in}} C(n\delta,0) +(1-2q_{\mathrm{in}}) C_2(n\delta,0) \right] +
  J^2 \, \delta  \sum^{n}_{k=0} w^{(n)}_{k} R(n\delta,k\delta) C_2(k\delta,0)
  \qquad
\label{eq:C2_discrete}
\end{eqnarray}
where we approximated the integral by the discrete sum
\begin{equation}
 \int^{n\delta}_0 d t^{\prime} R(n\delta,t^{\prime}) C_2(t^{\prime},0) = \delta  \sum^{n}_{k=0} w^{(n)}_{k} R(n\delta,k\delta) C_2(k\delta,0)
\label{eq:integral_discretization}
\end{equation}
with $w^{(n)}_{k}$ coefficients the value pf which depend on the particular approximation used.
The most common method of approximation is the one given by the Newton-Cotes formulas, which, for example, gives
for $n=1$, $w^{(1)}_0=\frac{1}{2}$ and $w^{(1)}_1=\frac{1}{2}$ (Trapezoidal rule),
for $n=2$, $w^{(2)}_0=\frac{1}{3}$, $w^{(2)}_1=\frac{4}{3}$ and $w^{(2)}_2=\frac{1}{3}$ (Simpson's rule),
for $n=3$, $w^{(3)}_0=\frac{9}{8}$, $w^{(3)}_1=\frac{3}{8}$, $w^{(3)}_2=\frac{3}{8}$ and $w^{(3)}_3=\frac{9}{8}$
(Simpson's $\frac{3}{8}$ rule), and so on.

Consider now the equilibrium dynamics, that is, set $J=J_0$ and assume stationarity conditions for $C$ and $R$
\begin{eqnarray}
  C(t_1,t_2) = C_{\mathrm{st}}(t_1-t_2) \; ,
  \nonumber \\
  R(t_1,t_2) = R_{\mathrm{st}}(t_1-t_2)
  \; .
\label{eq:stationarity}
\end{eqnarray}
Moreover, suppose that $z(k\delta)=z=2\frac{J^2}{T^{\prime}} \, (1-q_{\mathrm{in}})$ and $C_2(k\delta,0)=q_{\mathrm{in}}$ for $k=0,1,...,n$.

We want $C_2(t,0)$ to be a constant, so we have to enforce $C_2( (n+1) \delta,0)=q_{\mathrm{in}}$.
Using these assumptions, Eq.~(\ref{eq:C2_discrete}) can be rewritten as
\begin{eqnarray}
 C_2( (n+1) \delta,0) \,& = &\, q_{\mathrm{in}} + q_{\mathrm{in}} \frac{\delta^2}{m} \ \left[
  -z\,+\, \frac{2 J^2}{T^{\prime}}(1- q_{\mathrm{in}}) - \frac{J^2}{T^{\prime}}(1- C_{\mathrm{st}}(n\delta)) \,+\,
  J^2 \, \delta \, \sum^{n}_{k=0} w^{(n)}_{k} R_{\mathrm{st}}( (n-k) \delta) \right]
    \nonumber  \\
  & = &\, q_{\mathrm{in}} + q_{\mathrm{in}} \frac{\delta^2 J^2}{m} \ \left[
  - \frac{1}{T^{\prime}}(1- C_{\mathrm{st}}(n\delta)) \,+\,
  \, \delta \, \sum^{n}_{k=0} w^{(n)}_{k} R_{\mathrm{st}}( (n-k) \delta) \right]
  \nonumber \\
 & = &\, q_{\mathrm{in}} + q_{\mathrm{in}} \frac{\delta^2 J^2}{m} \ \left[
  - \frac{1}{T^{\prime}}(1- C_{\mathrm{st}}(n\delta)) \,+\,
  \, \delta \, \sum^{n}_{k=0} u^{(n)}_{k} R_{\mathrm{st}}( k \delta) \right]
  \; ,
\label{eq:C2_discrete_eq}
\end{eqnarray}
where we used the notation $u^{(n)}_{k}=w^{(n)}_{n-k}$ (in the case of the coefficients of the Newton-Cotes formulas, one
has $w^{(n)}_{k}=w^{(n)}_{n-k}$, that is, the coefficients appearing in symmetric positions in the sum are equal).
Notice that, if $q_{\mathrm{in}}=0$, $C_2(t,0)$ is trivially constant, being always identical to zero independently
of the choice of the coefficients $w^{(n)}_{k} $.
To enforce $C_2( (n+1) \delta,0) \, = \, q_{\mathrm{in}} \ne 0$ we need to satisfy the following equation
\begin{equation}
  \, \delta \, \sum^{n}_{k=0} u^{(n)}_{k} R_{\mathrm{st}}( k \delta) = \frac{1}{T^{\prime}}(1- C_{\mathrm{st}}(n\delta))
\label{eq:cond_C2_const}
\end{equation}
that ensures that the terms between square brackets cancel identically.
Take the case $n=1$, for example. Eq.~(\ref{eq:cond_C2_const}) is equivalent to the condition
\begin{equation}
  C(\delta,0) = C_{\mathrm{st}}(\delta) = 1 - \delta \, T^{\prime} u^{(1)}_1 R(\delta,0)
  \; .
\label{eq:C_delta}
\end{equation}
The Taylor expansions $C_{\rm st}(\delta)=1-\delta^2 T'/m$ and $R_{\rm st}(\delta) \delta/m$
satisfy this equation
when we use $u^{(1)}_1=1$.

Take now the generic $n$ case and suppose that the discrete version of the FDT holds, namely
\begin{equation}
  R_{\mathrm{st}}(n\delta)
  = -\frac{1}{T^{\prime}} \frac{1}{\delta}\left[ C_{\mathrm{st}}(n\delta)-C_{\mathrm{st}}((n-1)\delta) \right]
  \; .
\label{eq:FDT_discrete}
\end{equation}
Equation~(\ref{eq:cond_C2_const}) reduces to
\begin{eqnarray}
   \, -\frac{1}{T^{\prime}} \, \sum^{n}_{k=1} u^{(n)}_{k}
   \left[ C_{\mathrm{st}}( k \delta) - C_{\mathrm{st}}( (k-1) \delta) \right] &=&
   \frac{1}{T^{\prime}}(1- C_{\mathrm{st}}(n\delta))
   \end{eqnarray}
   that can be rewritten as
   \begin{eqnarray}
    (u^{(n)}_n-1) C_{\mathrm{st}}( n \delta) + 1 - u^{(n)}_1 +
    \sum^{n-1}_{k=1} \left[ u^{(n)}_k - u^{(n)}_{k+1} \right] C_{\mathrm{st}}( k \delta) &=& 0
\label{eq:cond_C2_const-final}
\end{eqnarray}
 where we used $R_{\mathrm{st}}(0)=0$ and $C_{\mathrm{st}}(0)=1$.

As one can see, wether or not the left-hand-side indeed vanishes, implying  $C_2((n+1)\delta,0)=C_2(n\delta,0) = \ldots = C_2(0,0)$ via Eq.~(\ref{eq:C2_discrete_eq}),
depends on the particular choice of the coefficients $w^{(n)}_k$ used to approximate the integrals.
A simple way to satisfy Eq.~(\ref{eq:cond_C2_const-final}) is to use $w^{(n)}_k = 1$, for any $n,k$.
We adopt, therefore, this rule to express the sums that represent the integrals.

Let us now investigate what does this discretisation rule implies for the discrete time evolution of the Lagrange multiplier $z$.
The discrete version of Eq.~(\ref{eq:z-Nicolas}) evaluated at $t_1=n\delta$, and rewritten in a way that makes $z_{\rm eq}$ appear is
\begin{eqnarray}
 z(n\delta) & = & 2 e_{f} + \frac{2J^2}{T^{\prime}} (1 - q^2_{\mathrm{in}}) +
  \frac{2J^2}{T^{\prime}} \left[ C_{\mathrm{st}}(n\delta)^2 -1\right] \, +\,
  4J^2 \, \delta \, \sum^{n}_{k=0} w^{(n)}_k R_{\mathrm{st}}((n-k)\delta) C_{\mathrm{st}}((n-k)\delta) \nonumber \\
  & = & z_{\mathrm{eq}} +\frac{2J^2}{T^{\prime}} \left[ C_{\mathrm{st}}(n\delta)^2 -1\right] \, +\,
  4J^2 \, \delta \, \sum^{n}_{k=0} u^{(n)}_k R_{\mathrm{st}}(k\delta) C_{\mathrm{st}}(k\delta)
\label{eq:z_discrete}
\end{eqnarray}
where $z_{\mathrm{eq}}=2 e_{f} + \frac{2J^2}{T^{\prime}} (1 - q^2_{\mathrm{in}})=T^{\prime}+\frac{J^2}{T^{\prime}} (1 - q^2_{\mathrm{in}})$ is the
equilibrium value of the Lagrange multiplier (for any $q_{\mathrm{in}}$), $u^{(n)}_k=w^{(n)}_{n-k}$,
and we have assumed stationarity for $C$ and $R$.

To enforce $z(n\delta)=z_{\mathrm{eq}}$, we need to satisfy the following equation
\begin{equation}
 \frac{1}{2T^{\prime}} \left[ C_{\mathrm{st}}(n\delta)^2 -1\right] \, +\,
 \delta \, \sum^{n}_{k=0} u^{(n)}_k R_{\mathrm{st}}(k\delta) C_{\mathrm{st}}(k\delta) =0
 \; .
 \label{eq:cond_z_const}
\end{equation}

If one supposes that the discrete version of the FDT, Eq.~(\ref{eq:FDT_discrete}),
holds then Eq.~(\ref{eq:cond_z_const}) reduces to
\begin{eqnarray}
 \frac{1}{2T^{\prime}} \left[ C_{\mathrm{st}}(n\delta)^2 -1\right] \,
 -\frac{1}{T^{\prime}} \, \sum^{n}_{k=1} u^{(n)}_k
 \left[ C_{\mathrm{st}}(k\delta)- C_{\mathrm{st}}((k-1)\delta) \right] C_{\mathrm{st}}(k\delta) =0
 \end{eqnarray}
 that can be rewritten as
 \begin{eqnarray}
  (1 - 2 u^{(n)}_n ) C_{\mathrm{st}}(n\delta)^2 -1 +2 u^{(n)}_1 C_{\mathrm{st}}(\delta)
 - 2  \, \sum^{n-1}_{k=1}
  \left[ u^{(n)}_k C_{\mathrm{st}}(k\delta) - u^{(n)}_{k+1} C_{\mathrm{st}}((k+1)\delta) \right]
  C_{\mathrm{st}}(k\delta) =0
  \; .
 \label{eq:cond_z_const-2}
\end{eqnarray}

If we choose $u^{(n)}_k=1$ for any $n,k$, we obtain
\begin{eqnarray}
  C_{\mathrm{st}}(n\delta)^2 +1 -2 C_{\mathrm{st}}(\delta)
 + \, \sum^{n-1}_{k=1} \left \{ C_{\mathrm{st}}(k\delta)^2 - C_{\mathrm{st}}((k+1)\delta)^2 +
  \left[ C_{\mathrm{st}}((k+1)\delta) - C_{\mathrm{st}}(k\delta) \right]^2 \right \} = 0
  \end{eqnarray}
  that simplifies to
  \begin{eqnarray}
  \sum^{n-1}_{k=0} \left[ C_{\mathrm{st}}((k+1)\delta) - C_{\mathrm{st}}(k\delta) \right]^2 =0
  \; .
 \label{eq:cond_z_const-final}
\end{eqnarray}

\newpage

\bibliographystyle{phaip}

\end{document}